%% file: manuscript.tex
\documentclass[]{pasj01}

\Received{2020/03/26}
\Accepted{2020/07/15}
 
\usepackage{url}
\usepackage{color}
\usepackage{multirow}
\usepackage{multicol}
 
\begin{document} 

\title{ 
A 16 deg$^2$ survey of emission-line galaxies at z$<$1.6 from HSC-SSP PDR2 and CHORUS
}

\author{Masao \textsc{Hayashi}\altaffilmark{1}}%
\email{masao.hayashi@nao.ac.jp}
\author{Rhythm \textsc{Shimakawa}\altaffilmark{1,2}}
\author{Masayuki \textsc{Tanaka}\altaffilmark{1,3}}
\author{Masato \textsc{Onodera}\altaffilmark{2}}
\author{Yusei \textsc{Koyama}\altaffilmark{2,3}}
\author{Akio K. \textsc{Inoue}\altaffilmark{4,5}}
\author{Yutaka \textsc{Komiyama}\altaffilmark{1,3}}
\author{Chien-Hsiu \textsc{Lee}\altaffilmark{6}}
\author{Yen-Ting \textsc{Lin}\altaffilmark{7}}
\author{Kiyoto \textsc{Yabe}\altaffilmark{8}}
\altaffiltext{1}{National Astronomical Observatory of Japan, 2-21-1 Osawa, Mitaka, Tokyo 181-8588, Japan}
\altaffiltext{2}{Subaru Telescope, National Astronomical Observatory of Japan, National Institutes of Natural Sciences, 650 North A'ohoku Place, Hilo, HI 96720, USA}
\altaffiltext{3}{Department of Astronomy, School of Science, Graduate University for Advanced Studies (SOKENDAI), 2-21-1, Osawa, Mitaka, Tokyo 181-8588, Japan}
\altaffiltext{4}{Department of Physics, School of Advanced Science and Engineering, Waseda University, 3-4-1, Okubo, Shinjuku, Tokyo 169-8555, Japan}
\altaffiltext{5}{Waseda Research Institute for Science and Engineering, 3-4-1, Okubo, Shinjuku, Tokyo 169-8555, Japan}
\altaffiltext{6}{NSF’s National Optical-Infrared Astronomy Research Laboratory, USA}
\altaffiltext{7}{Academia Sinica Institute of Astronomy and Astrophysics, PO Box 23-141, Taipei 10617, Taiwan}
\altaffiltext{8}{Kavli Institute for the Physics and Mathematics of the Universe (Kavli IPMU, WPI), The University of Tokyo, 5-1-5 Kashiwanoha, Kashiwa, Chiba 277-8583, Japan}
%


\maketitle

\begin{abstract}
  We have conducted a comprehensive survey of emission-line galaxies
  at $z\lesssim1.6$ based on narrowband (NB) imaging data taken with
  Hyper Suprime-Cam (HSC) on the Subaru telescope. In this paper, we
  update the catalogs of H$\alpha$, [OIII], and [OII] emission-line
  galaxies using the data from the second Public Data Release (PDR2)
  of Subaru Strategic Program (SSP) of HSC and Cosmic HydrOgen
  Reionization Unveiled with Subaru (CHORUS) survey along with the
  spectroscopic redshifts for 2,019 emission-line galaxies selected
  with the PDR1 data. 
  The wider effective coverage of NB816 and NB921, 16.3 deg$^2$ and
  16.9 deg$^2$ respectively, are available in the Deep and UltraDeep
  layers of HSC-SSP from the PDR2. The CHORUS survey provides us with
  data with additional three NBs (NB527, NB718, and NB973) in the COSMOS
  field in the UltraDeep layer (1.37 deg$^2$). The five NB datasets
  allow us to investigate the star-forming galaxies presenting
  emission-lines at 14 specific redshifts ranging from $z\sim1.6$ down
  to $z\sim0.05$.  
  We revisit the distribution of large-scale structures and luminosity
  functions (LFs) for the emission-line galaxies with the large
  samples of 75,377 emission-line galaxies selected. 
  The redshift revolution of LFs shows that the star formation rate
  densities (SFRDs) decreases monotonically from $z\sim1.6$, which
  is consistent with the cosmic SFRD ever known. Our samples of
  emission-line galaxies covering a sufficiently large survey volume
  are useful to investigate the evolution of star-forming galaxies
  since the cosmic noon in a wide range of environments including
  galaxy clusters, filaments, and voids.  
\end{abstract}


\section{Introduction}
\label{sec:intro}
Our global understanding of evolution of star-forming galaxies has
reached a consensus; the cosmic star formation rate (SFR) density
peaks at $z=$ 1--3 and then gradually declines towards the local
Universe (e.g., \cite{HopkinsBeacom2006,MadauDickinson2014}).   
It is also well-known that star-forming galaxies maintain a tight
correlation between SFR and stellar mass from $z\sim6$ all the way to
the local Universe, while SFR at a given stellar mass becomes lower 
at decreasing redshifts (e.g., \cite{Daddi2007,Elbaz2007,Noeske2007,Speagle2014,Whitaker2014,Tomczak2016}). 
However, physics governing the evolution of the activity in individual
star-forming galaxies is not yet fully understood. Given that a
fraction of quiescent galaxies is dependent on both stellar mass and 
environment (e.g., \cite{Peng2010}) and there is a small fraction of
galaxies in a starburst phase (e.g., \cite{Rodighiero2011}),
star-formation history of individual star-forming galaxies is
complicated. In order to reveal the contribution of each process
stimulating and/or quenching star-formation activity to the evolution
of the individual galaxies, first of all, it is essential to conduct a
comprehensive survey of star-forming galaxies at the redshifts
$z\lesssim3$ that covers a wide range in terms of star formation
activity, stellar mass and environment.    

Narrowband (NB) imaging is effective in surveying star-forming
galaxies with nebular emission lines at a specific redshift in a
homogeneous field-of-view (FoV) without any bias for target
selection. The distribution of emission-line galaxies selected is
insensitive to the projection effect thanks to small redshift ranges
surveyed by NB imaging.    
Therefore, a large sample of emission-line galaxies is useful to
investigate the properties of star-forming galaxies as well as the
dependence of galaxy properties on environment in each specific
redshift surveyed 
(e.g., \cite{
Ly2007,Ly2011,Ly2012,Lee2012,Dale2010,Drake2013,Sobral2009,Sobral2011,Sobral2012,Sobral2013,Sobral2014,Sobral2015,Sobral2016,Sobral2016b,Khostovan2015,Khostovan2016,Khostovan2018,Khostovan2020,Matthee2017,Stroe2014,Stroe2015a,Stroe2015,Stroe2017,Hayashi2015,Hayashi2018a,Koyama2018,Coughlin2018,Harish2020,Bongiovanni2020,Ramon-Perez2019,Ramon-Perez2019b,Nadolny2020}).     
However, a wide-field imaging is required to survey a volume large
enough so that galaxy properties investigated are independent of
cosmic variance. 
The $\sim10$ deg$^2$ NB imaging surveys by \citet{Sobral2015}
and \citet{Stroe2015} show that a large survey volume of 
$>$3.5--5.0$\times10^{5}$ Mpc$^3$ can overcome cosmic variance and
then the luminosity functions of emission-line galaxies are derived
with an error of $<10$\%.  
From a theoretical point of view,
\citet{Ogura2020} investigate the influence of the field
variance in H$\alpha$ emission-line galaxies at $z=0.4$ using a
semi-analytic model for the galaxy formation, the New Numerical Galaxy
Catalog ($\nu^2$GC), and find that a survey area of more than 15 
deg$^2$ is required to restrict the uncertainties in the luminosity
functions of H$\alpha$ emitters at $z=0.4$ to less than
$\sim10\%$. 

\begin{figure}  
  \begin{center}
    \includegraphics[width=0.5\textwidth]{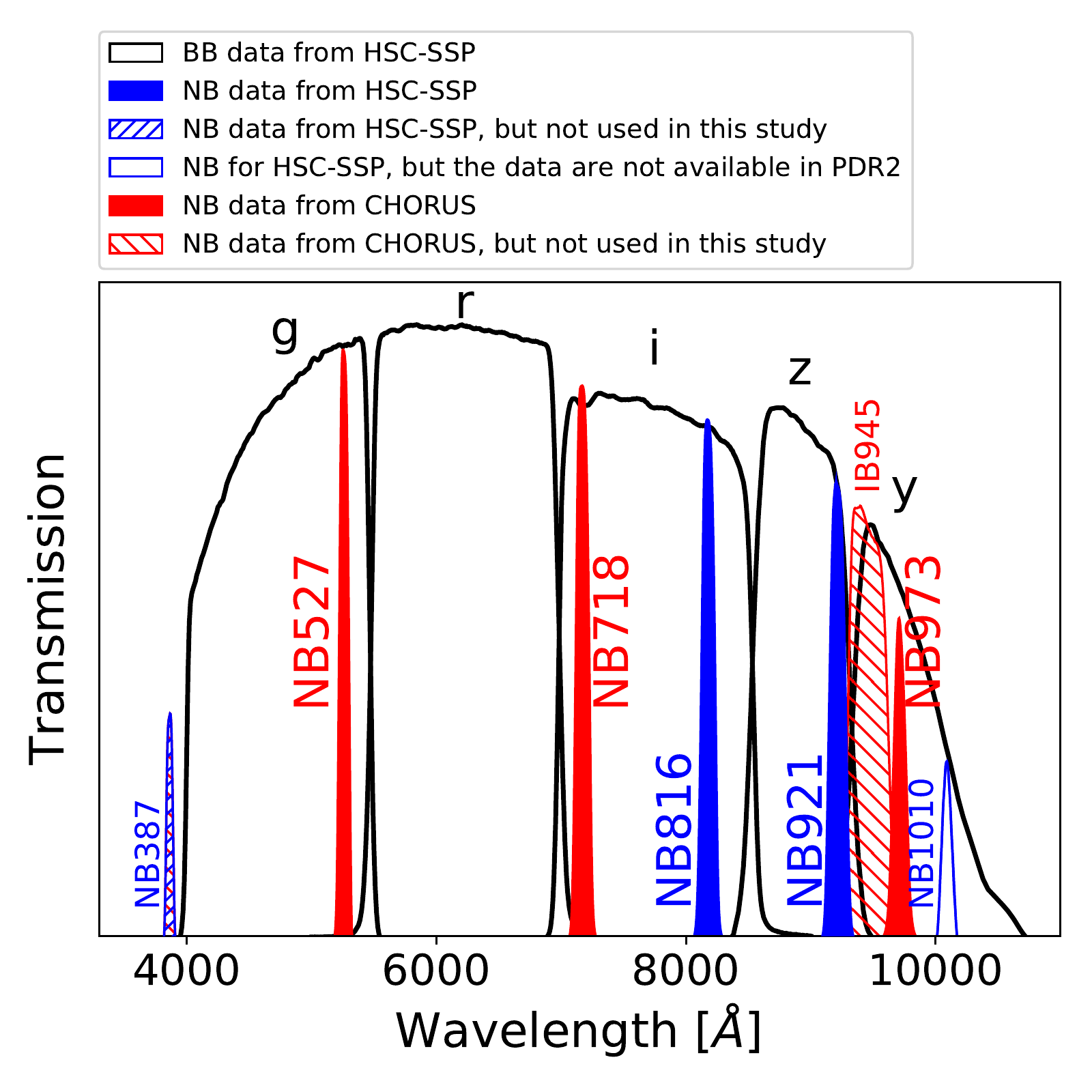}
  \end{center}
  \caption{The set of broadband (BB) and narrowband (NB) filters used
    in the HSC-SSP and CHORUS surveys. The response function of the
    filters and the quantum efficiency of CCD are taken into account
    for the transmission. The transmission curves in blue show the NB
    filters used in the HSC-SSP, while the red ones are used in the
    CHORUS survey. The data from the filled NB filters are used in
    this study. The data from the open NB filter is not available in
    the PDR2. The data from the hatched filters are not used in this
    study (see the text for the details), although the data are available.   
  }\label{fig:filters}   
\end{figure}

The Subaru Strategic Program (SSP) with Hyper Suprime-Cam (HSC)
provides us with the data useful for the study of galaxy
evolution. The HSC-SSP is an ongoing survey since 2014 \citep{HSCSSP}, 
and the deep imaging is conducted in five broadband (BB) filters:
$grizy$, plus four NB filters: NB387, NB816, NB921, and NB1010
(figure~\ref{fig:filters}) in the Deep (D) and UltraDeep (UD) layers
over 28 deg$^2$ in total. The HSC is an instrument on the 8.2-m Subaru
Telescope with the capability to cover 1.77 deg$^2$ in a single pointing
\citep{HSC,HSCdewar,HSCfilters,HSConsiteqa}. In addition to the wide
FoV and good image quality of the data, an assortment of many NB filters is
one of the uniqueness of HSC. Since the dataset of the HSC-SSP is
huge, constructing catalogs of emission-line galaxies in a consistent
manner is crucial for systematic studies of galaxy evolution.   

The first public data release (PDR1) of HSC-SSP data was published on
2017 February 28 \citep{HSCSSPDR1}.  
\citet{Hayashi2018a} use the data from two NB filters, NB816 and
NB921, available in the PDR1 of HSC-SSP to search for 
galaxies with nebular emission of H$\alpha$,
[OIII]($\lambda\lambda=4960,5008$\AA), or
[OII]($\lambda\lambda=3727,3730$\AA) at $z\lesssim1.5$.      
Since the NB816 (NB921) data cover 5.7 (16.2) deg$^2$ area,
the HSC-SSP survey provides us with one of the largest samples of
emission-line galaxies at $z<1.5$ ever constructed.   
The catalogs\footnote{
  The catalogs are available in
  \url{https://hsc-release.mtk.nao.ac.jp/}.  
  The name of the catalogs is {\tt pdr1\_udeep.nbemitter} for
  emission-line galaxies in the fields of UltraDeep layer and 
  {\tt pdr1\_deep.nbemitter} for those of Deep layer. 
}
include 8054 H$\alpha$ emitters at $z \approx$ 0.25 and 0.40, 8656
[OIII] emitters at $z \approx$ 0.63 and 0.84, and 16877 [OII] emitters
at $z \approx$ 1.19 and 1.47. 
The spatial distribution of the emitters shows large-scale structures
over $\gtrsim$ 50 Mpc region that consist of star-forming galaxies in
various environments covering from cores of galaxy clusters to
voids \citep{Hayashi2018a,Koyama2018}. Thanks to the large survey
volume of more than $5\times10^5$ Mpc$^3$, the luminosity functions of
the emitters overcome the field-to-field variance and are determined
with small uncertainty.  

However, since the survey is not completed yet as of the PDR1, there
remains room for improvement in search for emission-line galaxies. 
Only data from two NBs are available in the PDR1, and the field
coverage of NB816 data (5.7 deg$^2$) is limited to one-third of NB921
data (16.2 deg$^2$).  
Furthermore, we realize that the PDR1 
catalogs of emission-line galaxies are biased towards galaxies with
large equivalent width (EW) of nebular emission line, compared with
the other deep surveys \citep{Ly2007,Sobral2013}; an observed EW of
48$\rm \AA$ for NB816 emitters and 56$\rm \AA$ for NB921 emitters,
respectively.   
\citet{Ly2007} selected NB816 and NB921 emitters with the observed
EWs of $>$33 and $>$15 \AA, respectively. 
\citet{Sobral2013} selected NB921 emitters with the rest-frame EWs
greater than 25 \AA, namely the H$\alpha$ emitters with smaller EWs
were selected compared with the PDR1 catalog.  

The second public data release (PDR2) of HSC-SSP data was published on
2019 May 31 \citep{HSCSSPDR2}. The NB data from the PDR2 are deeper
and wider than the PDR1 data. The data with three other NB filters
(NB527, NB718, and NB973) are also available in the UD-COSMOS field
(1.7 deg$^2$) from the Subaru open-use intensive program named Cosmic
HydrOgen Reionization Unveiled with Subaru (CHORUS, \cite{CHORUS}). We
have conducted follow-up spectroscopic campaigns and confirmed about
2000 emission-line galaxies selected from the PDR1 data. Based on the
newly available imaging data as well as the spectroscopic data, we aim
to update the catalogs of emission-line galaxies at $z\lesssim1.6$ in
this study. 
The deeper and wider data allow us to select more representative
star-forming galaxies in a wide range of environments covering rare
ones in high or low density regions. With the catalogs, we can
investigate the redshift evolution of star-formation activity of
star-forming galaxies and large-scale structures. 

The outline of this paper is as follows.
In \S~\ref{sec:data}, the NB data from HSC-SSP PDR2 and CHORUS
surveys as well as the spectroscopic data from the follow-up
observations of the PDR1 galaxies are described.
Emission-line galaxies at $z\lesssim$ 1.6 are selected in
\S~\ref{sec:ELGs}, and we mention therein how we improve the selection
of emission-line galaxies. 
In \S~\ref{sec:LFs}, luminosity functions of the emission-line
galaxies are investigated. 
We discuss the redshift evolution of the luminosity functions and
the cosmic star formation rate densities by integrating the luminosity
functions in \S~\ref{sec:discussions}. 
Finally, conclusions are given in \S~\ref{sec:conclusions}.
Throughout this paper, 
magnitudes are presented in the AB system \citep{OkeGunn1983}.
The cosmological parameters of $H_0=70$ km s$^{-1}$ Mpc$^{-1}$,
$\Omega_m=0.3$ and $\Omega_\Lambda=0.7$, along with
\citet{Chabrier2003} initial mass function (IMF), are adopted.  

\input{table1.tex}

\begin{table}
  \tbl{Representative regions in each field used for calculation of
    detection completeness and 1$\sigma$ sky noise.}{%
    \begin{tabular}{cccc}   
      \hline
      D/UD & field & tract & patch \\ 
      \hline
      UltraDeep & SXDS     & 8523  & 105,106,107,205,206,207,305,306,307 \\
      UltraDeep & COSMOS   & 9813  & 303,304,305,403,404,405,503,504,505 \\
      Deep      & XMM-LSS  & 8524  & 203,204,205,303,304,305,403,404,405 \\
      Deep      & E-COSMOS & 9814  & 606,607,608,706,707,708,806,807,808 \\
      Deep      & ELAIS-N1 & 17271 & 405,406,407,505,506,507,605,606,607 \\
      Deep      & DEEP2-3  & 9464  & 402,403,404,502,503,504,602,603,604 \\
      \hline
  \end{tabular}}\label{tbl:9patches}
\end{table}

\section{Data}
\label{sec:data}

\subsection{HSC imaging data}

We use the data from five NB filters, NB527, NB718, NB816, NB921, and
NB973 (see also figure~\ref{fig:filters}), taken through the HSC-SSP
and CHORUS programs. Table~\ref{tbl:NBdataArea} summarizes the NB
data we use in this study. The BB data are all from the HSC-SSP PDR2.   

\subsubsection{SSP public data release 2}
\label{sec:data.ssp}

\begin{figure*}
 \begin{center}
   \includegraphics[width=0.45\textwidth, bb=0 0 461 346]{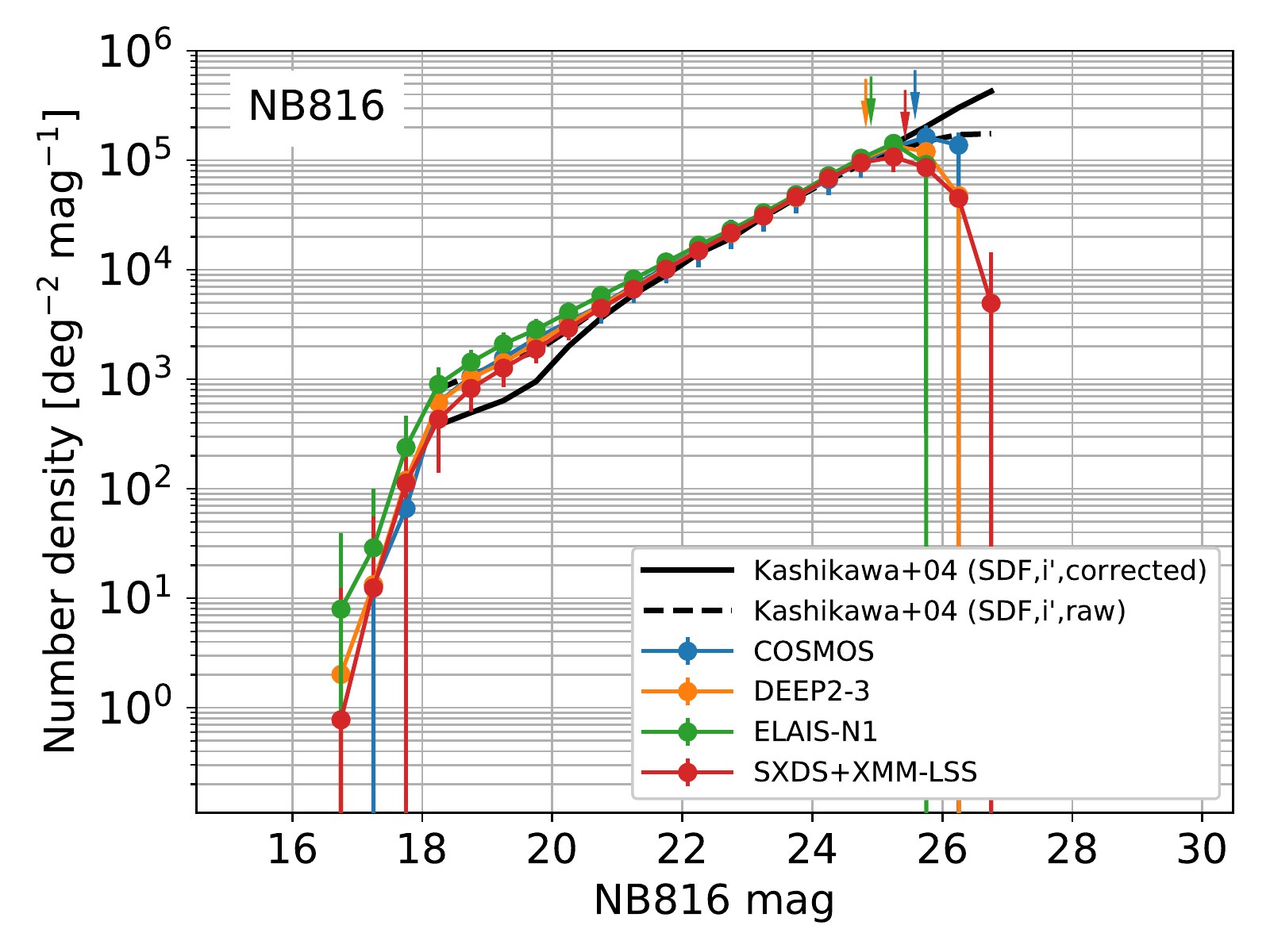} 
   \includegraphics[width=0.45\textwidth, bb=0 0 461 346]{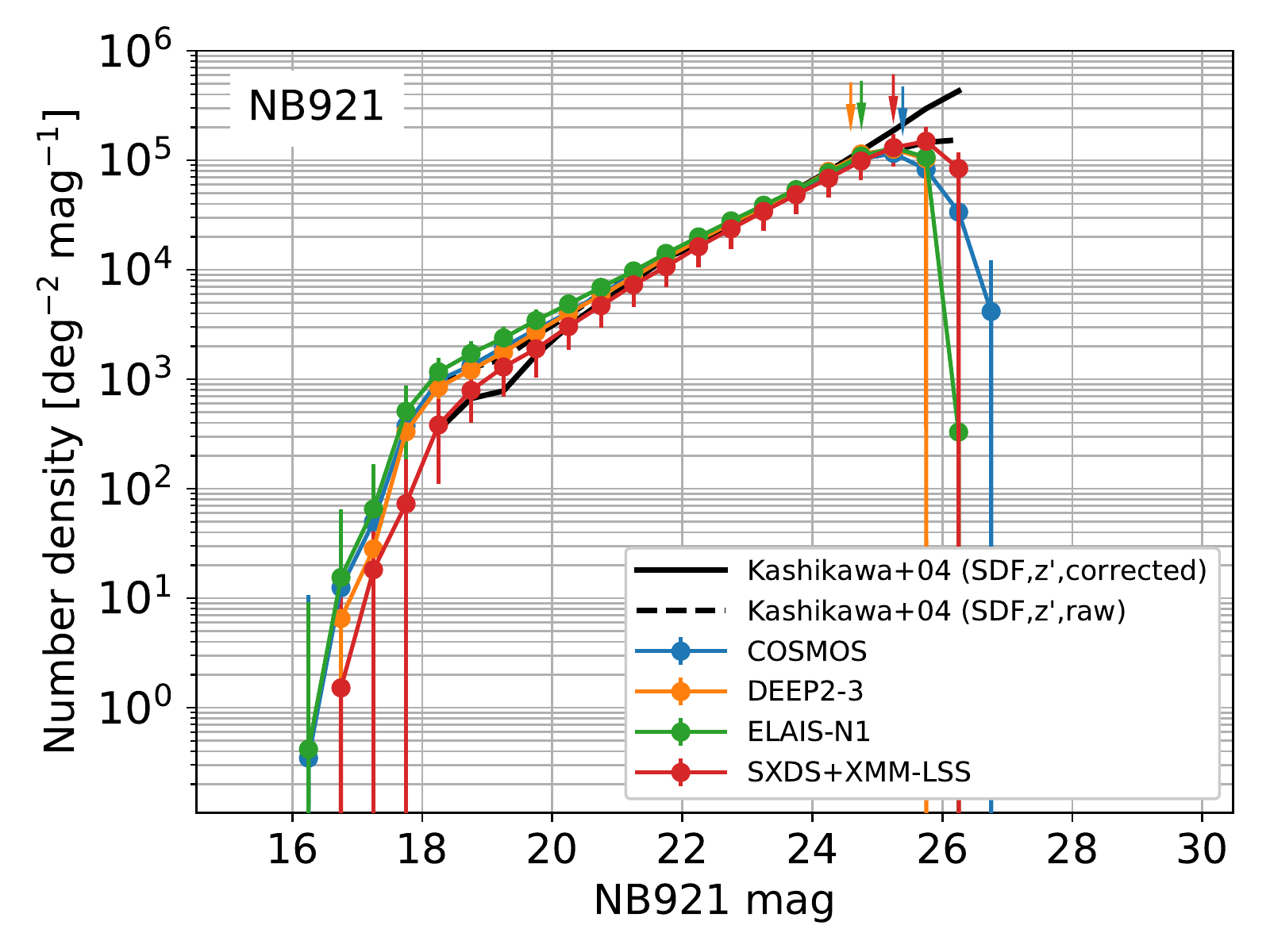} 
 \end{center}
 \caption{The number counts of sources detected in NB816 or NB921,
   namely stars and galaxies are all included. The horizontal axis is
   {\tt convolvedflux\_2\_20\_mag}, which is magnitude by 2 arcsec
   aperture photometry on the images with PSF matched to 1.1 arcsec.  
   The number counts are corrected for detection completeness and only
   the sources with more than 50\% of detection completeness are used
   for the plots.
   The symbols in different colors show the number counts in each four
   field. The error bars show the standard deviation of the number
   counts in individual patch regions. The black lines are from the
   literature \citep{Kashikawa2004}, where the dashed line is the
   number count of all sources. The arrows show the 5$\sigma$ limiting
   magnitudes.}\label{fig:NumCountSSP}   
\end{figure*}

The HSC-SSP PDR2 publishes the data observed with 174 nights between
March 2014 and January 2018 \citep{HSCSSPDR2}. The PDR2 is equivalent
to the internal S18A release for the HSC-SSP collaboration. Among the
four NB filters planned in the HSC-SSP, NB387 data are newly available
in this release, while NB1010 data are not yet, but will be available
in the next release, i.e., PDR3. The coverage of NB816 and NB921 data
that are already available in the PDR1 increases to be 26 deg$^2$ covered
by the 13 pointings and 28 deg$^2$ covered by the 14 pointings,
respectively \citep{HSCSSPDR2}. The survey area in the Deep layer,
where the NB data are available, consists of four separate fields:
XMM-LSS, E-COSMOS, ELAIS-N1, and DEEP2-3. Among them, XMM-LSS and E-COSMOS
fields each encompasses the UltraDeep layer covered by a single pointing
of HSC, which are named SXDS and COSMOS, respectively. Note that
the UD-COSMOS in the UD layer is similar to the well-known COSMOS field \citep{Scoville2007},
and the E-COSMOS in the D layer is a extended field around the original
COSMOS. Unless specifically mentioned, we use ``COSMOS" as a field
covering the extended area around the original COSMOS field in this
paper, because the data in both UD-COSMOS and E-COSMOS are jointly
processed and cataloged. In this study, we use the NB816 and NB921 data from the
HSC-SSP, because only [OII]($\lambda\lambda=3727,3730$\AA) from
galaxies at $z\sim0.038$ can enter the NB387 filter among strong
nebular emission lines of H$\alpha$, [OIII], and [OII] we focus on now
and the redshift is too low to get enough survey volume. Also, it is
difficult to estimate the stellar continuum level underlying emission
lines at the wavelength of NB387 with only the HSC BB filters
(figure~\ref{fig:filters}). The $u$-band data are essential to select
emission-line galaxies detected in NB387 \citep{Nakajima2012,Konno2016,Sobral2017,Stroe2017b}.  

The HSC-SSP PDR2 data are processed with the pipeline, {\tt hscPipe}
version 6 \citep{hscPipe,HSCSSPDR2}, which includes the data
reduction, source detection, and photometry. Readers should refer to
\citet{hscPipe} for the details of the pipeline and to
\citet{HSCSSPDR2} for improvements, new features, and remaining issues
specific to this version of {\tt hscPipe}. The most notable points
closely related to this study are the improvement of sky subtraction
and joint processing of the data in the D and UD layers. Note that in
the PDR1, the data in the COSMOS and SXDS+XMM-LSS fields were
processed separately for the D and UD layers, although there are
overlapped regions between the two layers. As with the data in the
PDR1, the survey area is divided into predefined gridded regions known
as {\tt tract}, each covering $\sim1.7\times1.7$ deg$^2$. A single
{\tt tract} is divided into 9 × 9 sub-regions, each one called a 
{\tt patch} and covering $\sim12\times12$ arcmin$^2$. 

\begin{figure*}
 \begin{center}
   \includegraphics[width=0.45\textwidth, bb=0 0 461 346]{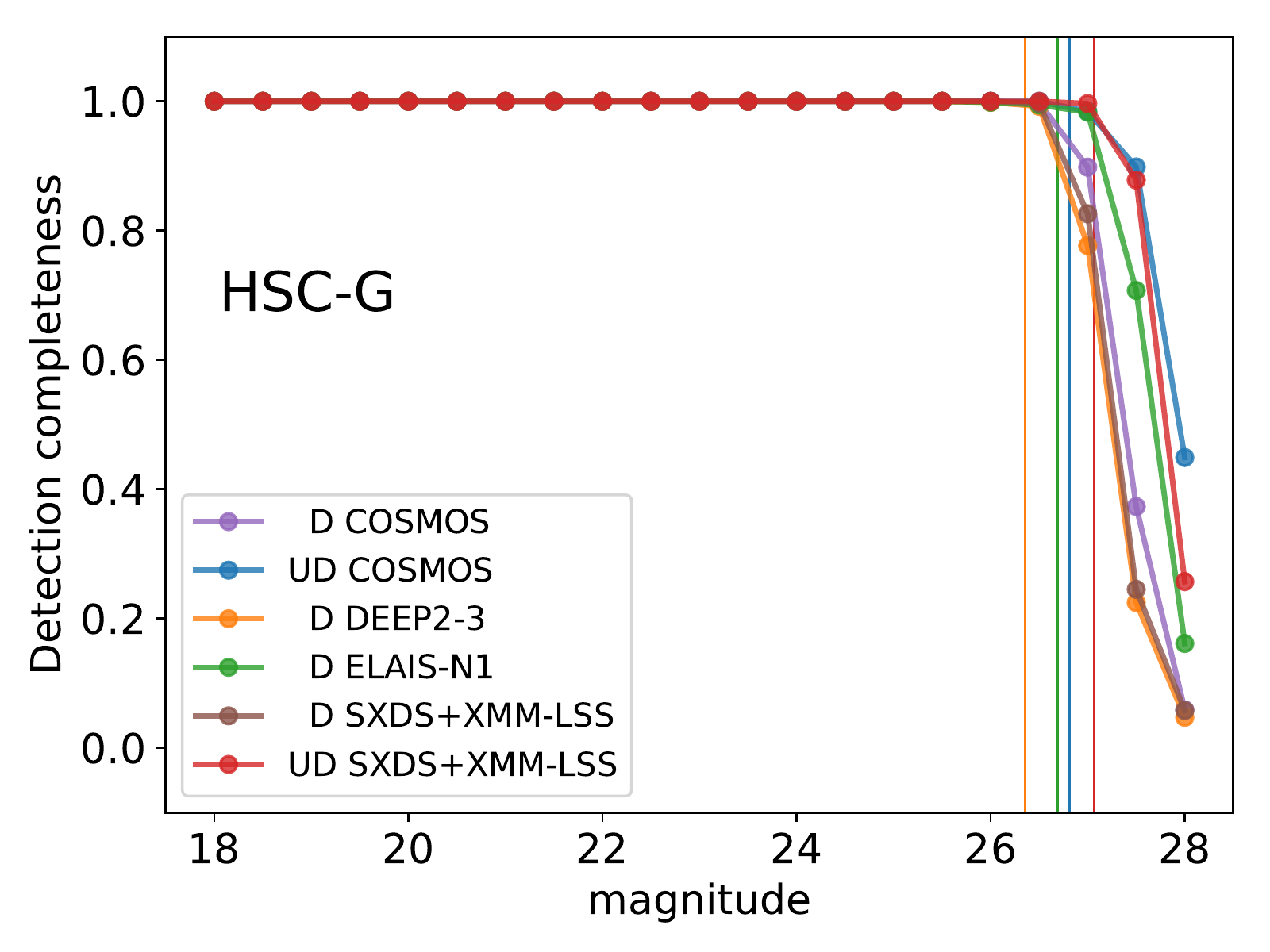} 
   \includegraphics[width=0.45\textwidth, bb=0 0 461 346]{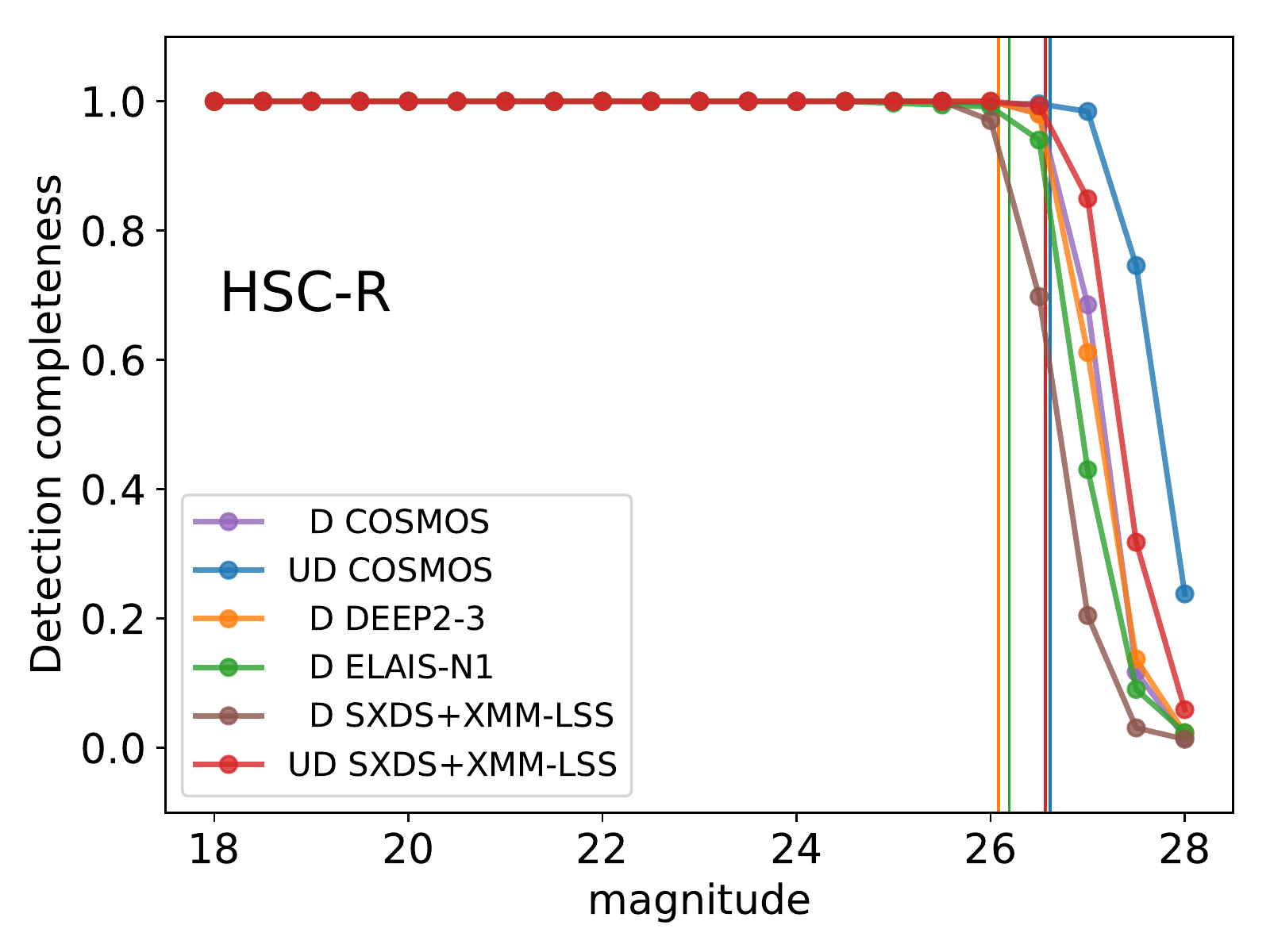} 
   \includegraphics[width=0.45\textwidth, bb=0 0 461 346]{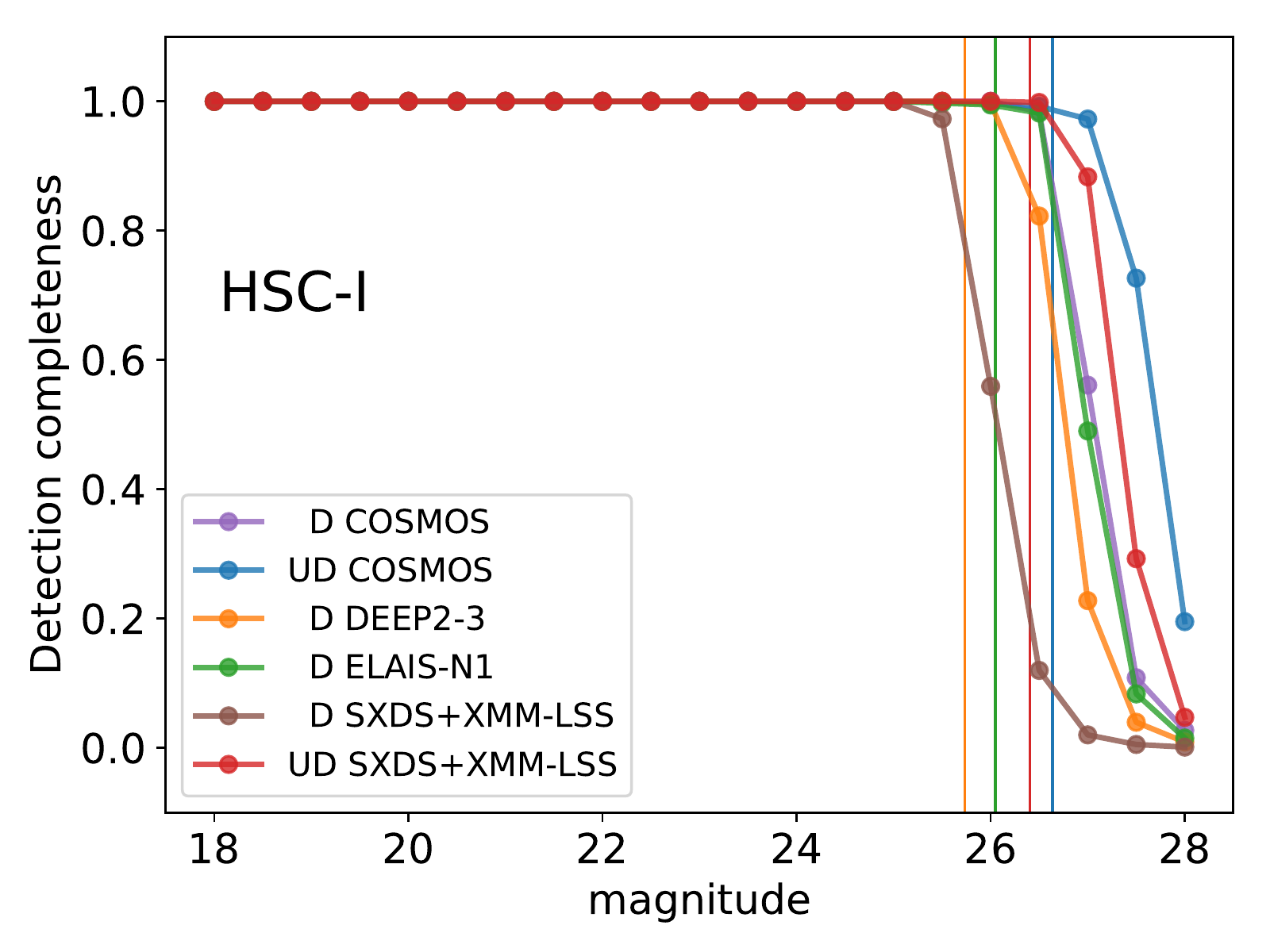} 
   \includegraphics[width=0.45\textwidth, bb=0 0 461 346]{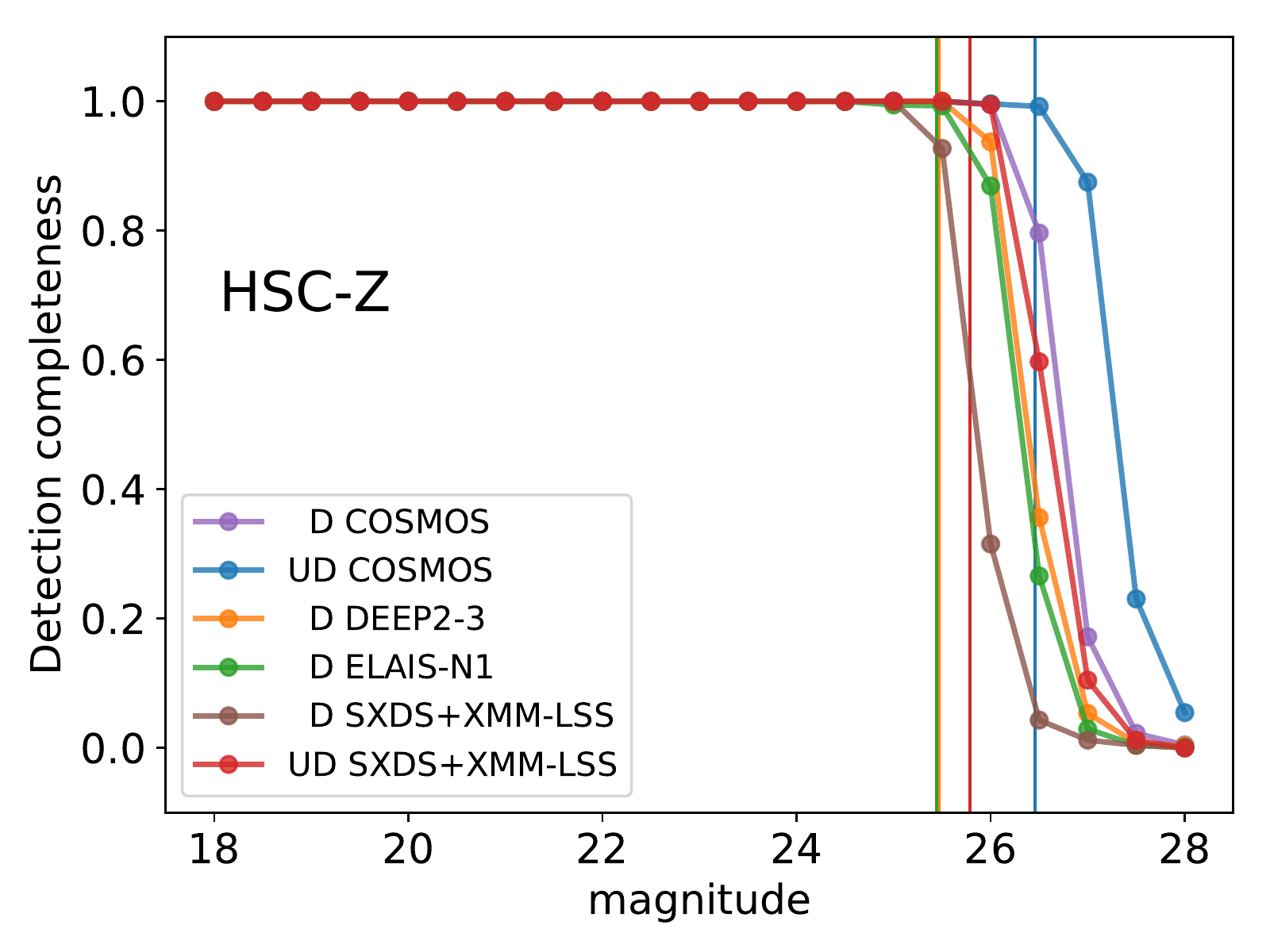} 
   \includegraphics[width=0.45\textwidth, bb=0 0 461 346]{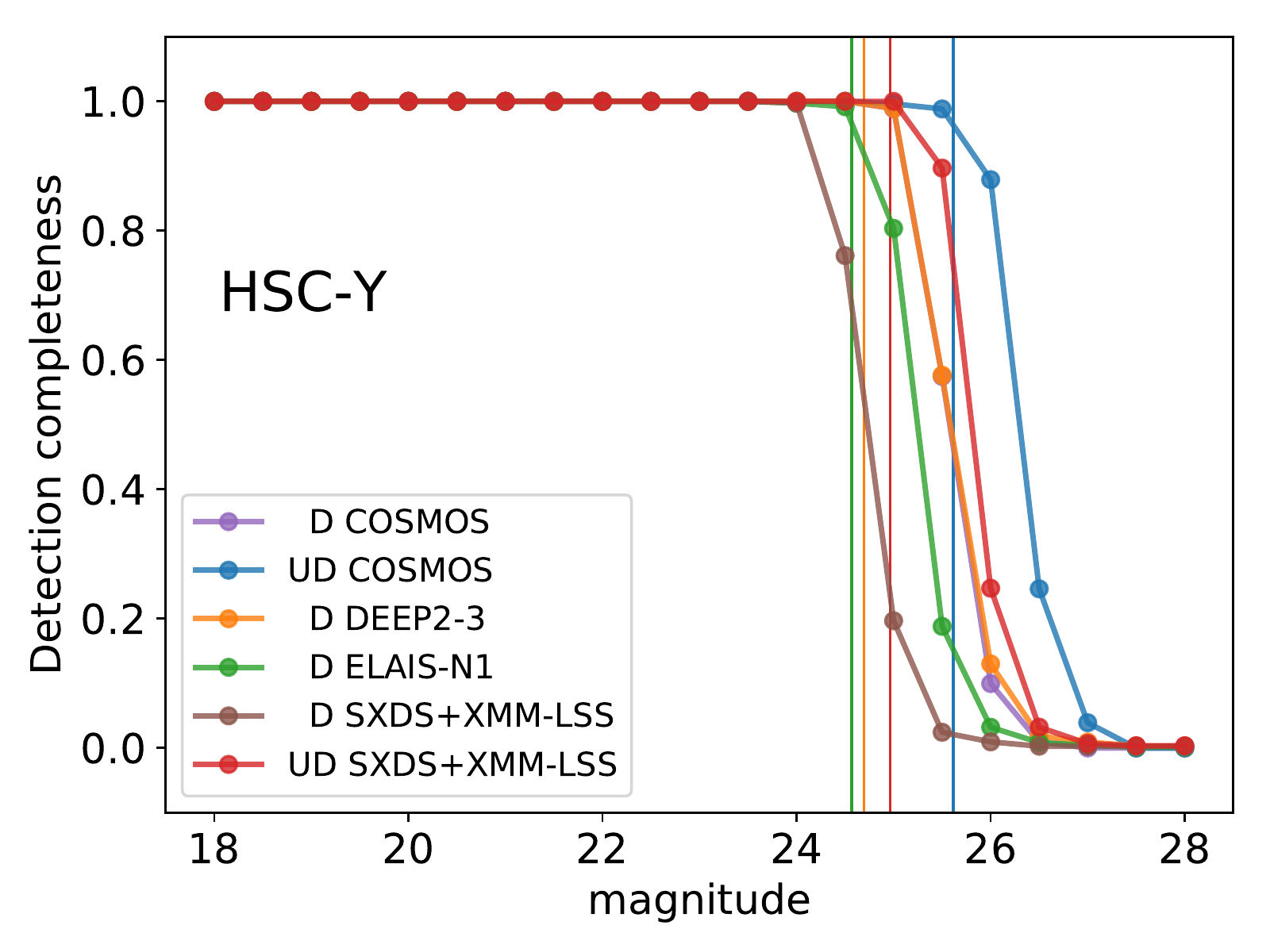} 
   \includegraphics[width=0.45\textwidth, bb=0 0 461 346]{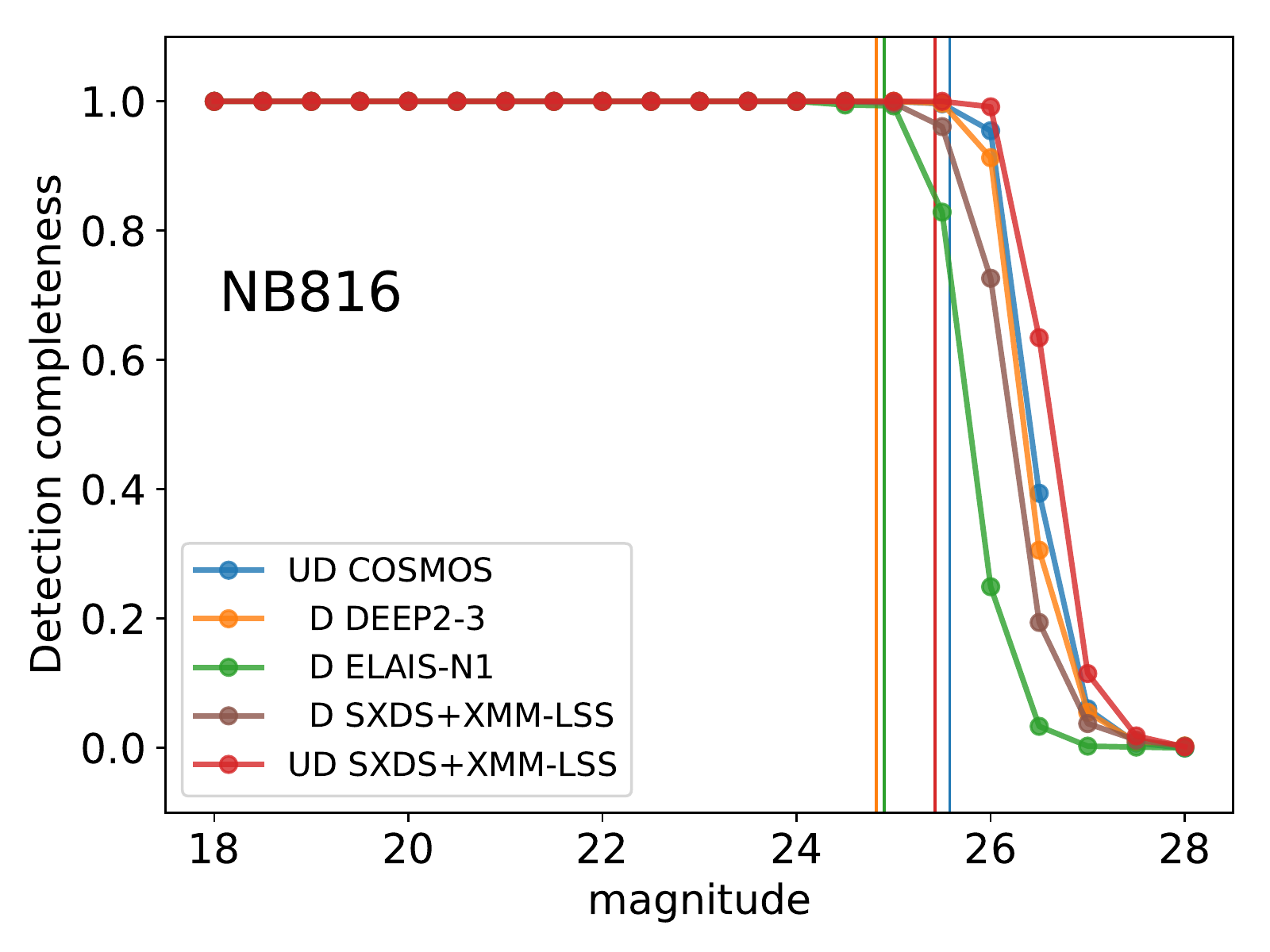} 
   \includegraphics[width=0.45\textwidth, bb=0 0 461 346]{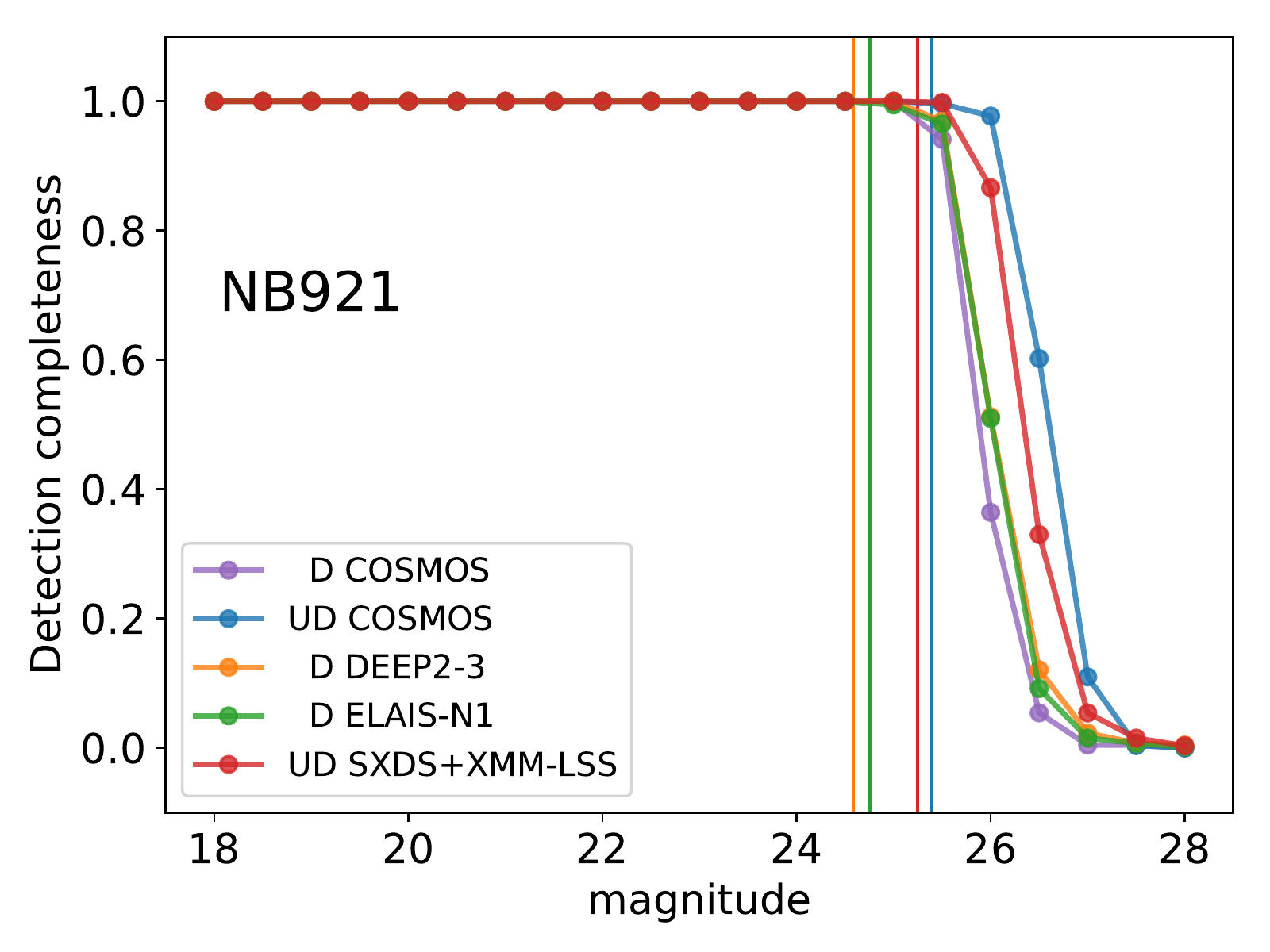} 
 \end{center}
 \caption{Detection completeness for five BBs, NB816 and NB921
   filters. The vertical lines show the 5$\sigma$ limiting magnitudes
   which are measured from the random photometry with 2 arcsec
   aperture in sky region (see text for the details). For the COSMOS
   and SXDS+XMM-LSS fields, the 5$\sigma$ limiting magnitudes are
   measured in the UltraDeep region. The magnitudes where a detection
   completeness starts to drop are consistent with the 5$\sigma$
   limiting magnitudes.}\label{fig:DetectionCompletenessSSP} 
\end{figure*} 

\begin{figure*}
 \begin{center}
   \includegraphics[width=0.45\textwidth, bb=0 0 461 346]{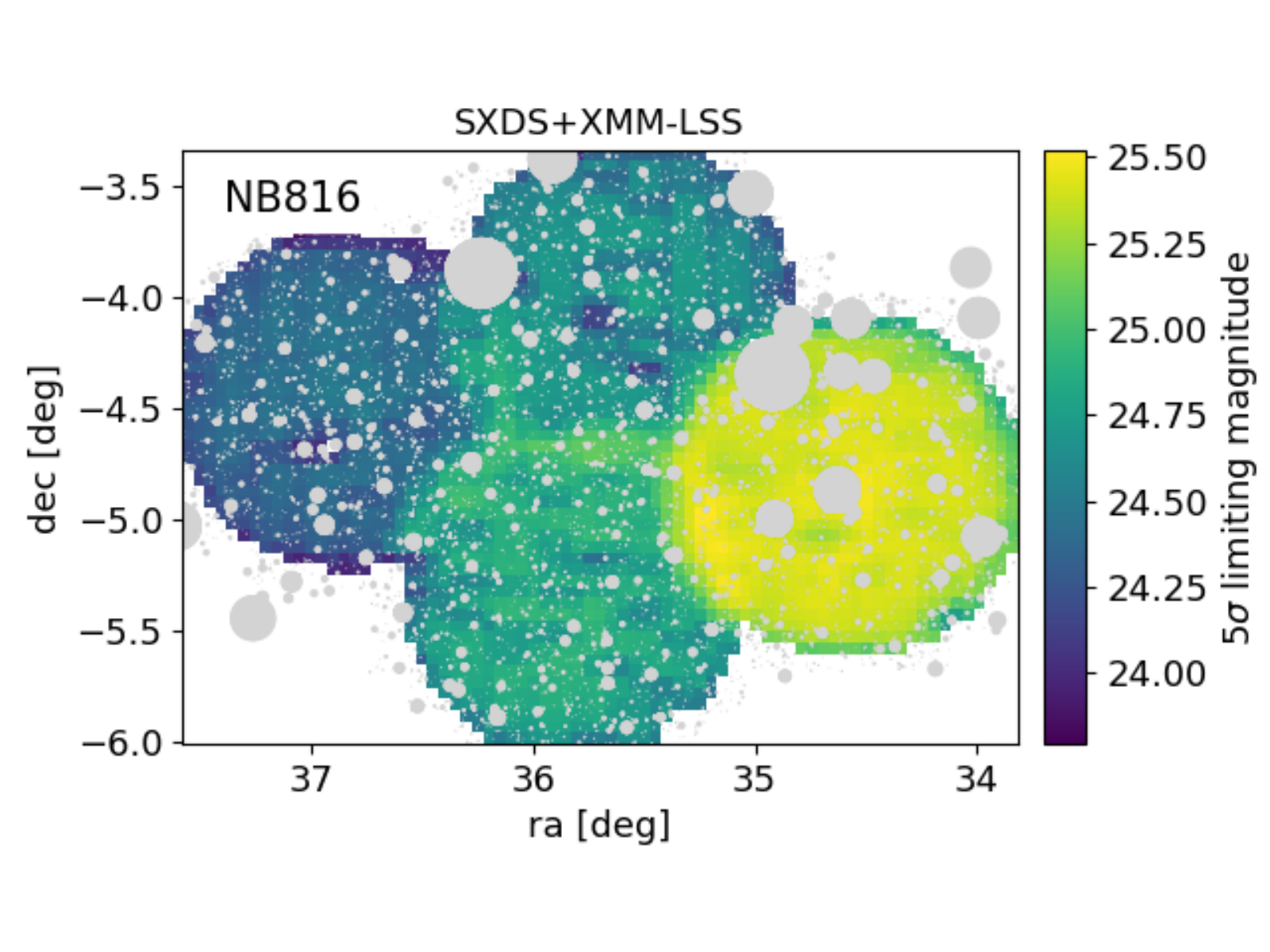} 
   \includegraphics[width=0.45\textwidth, bb=0 0 461 346]{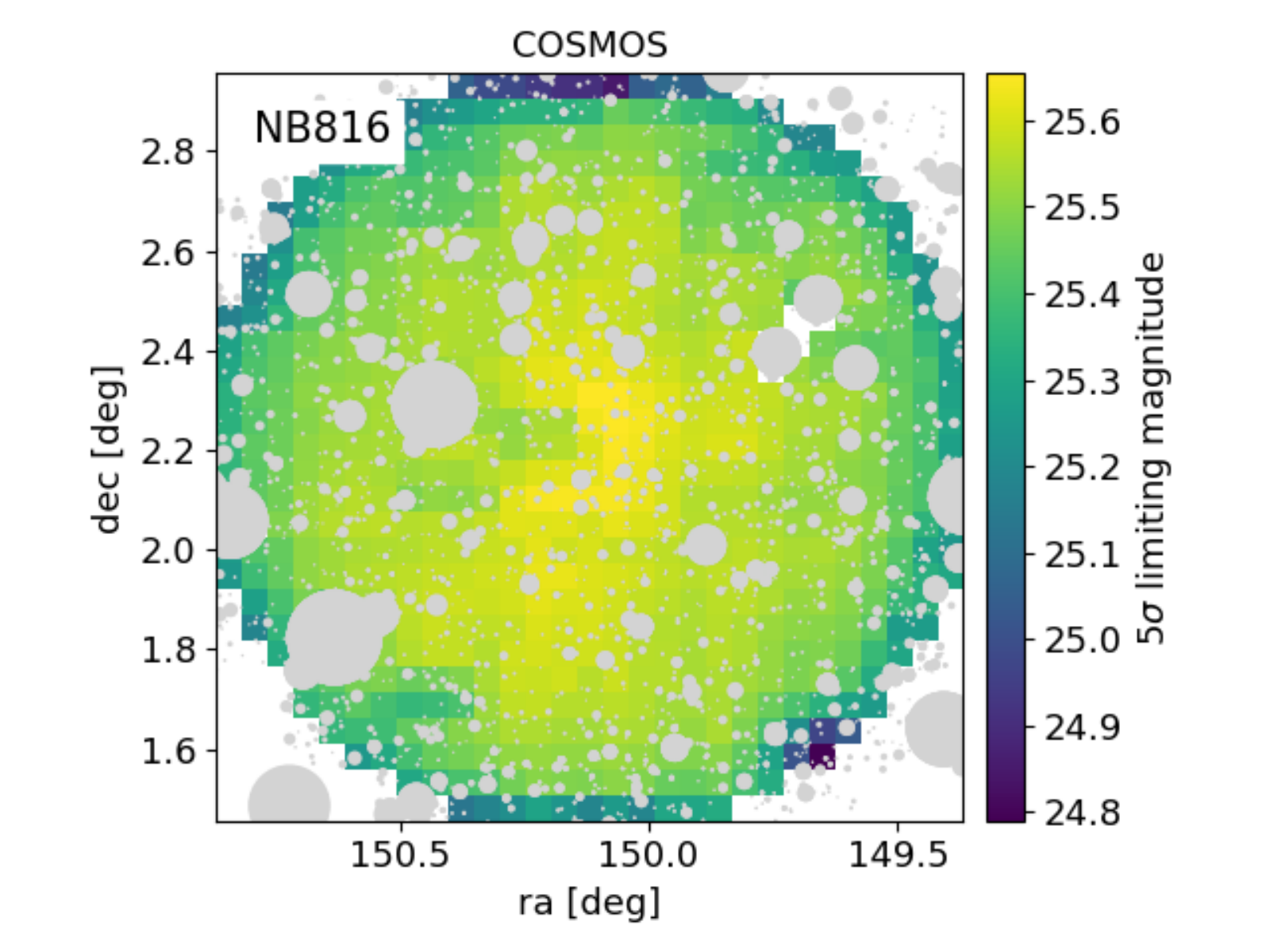} 
   \includegraphics[width=0.45\textwidth, bb=0 0 461 346]{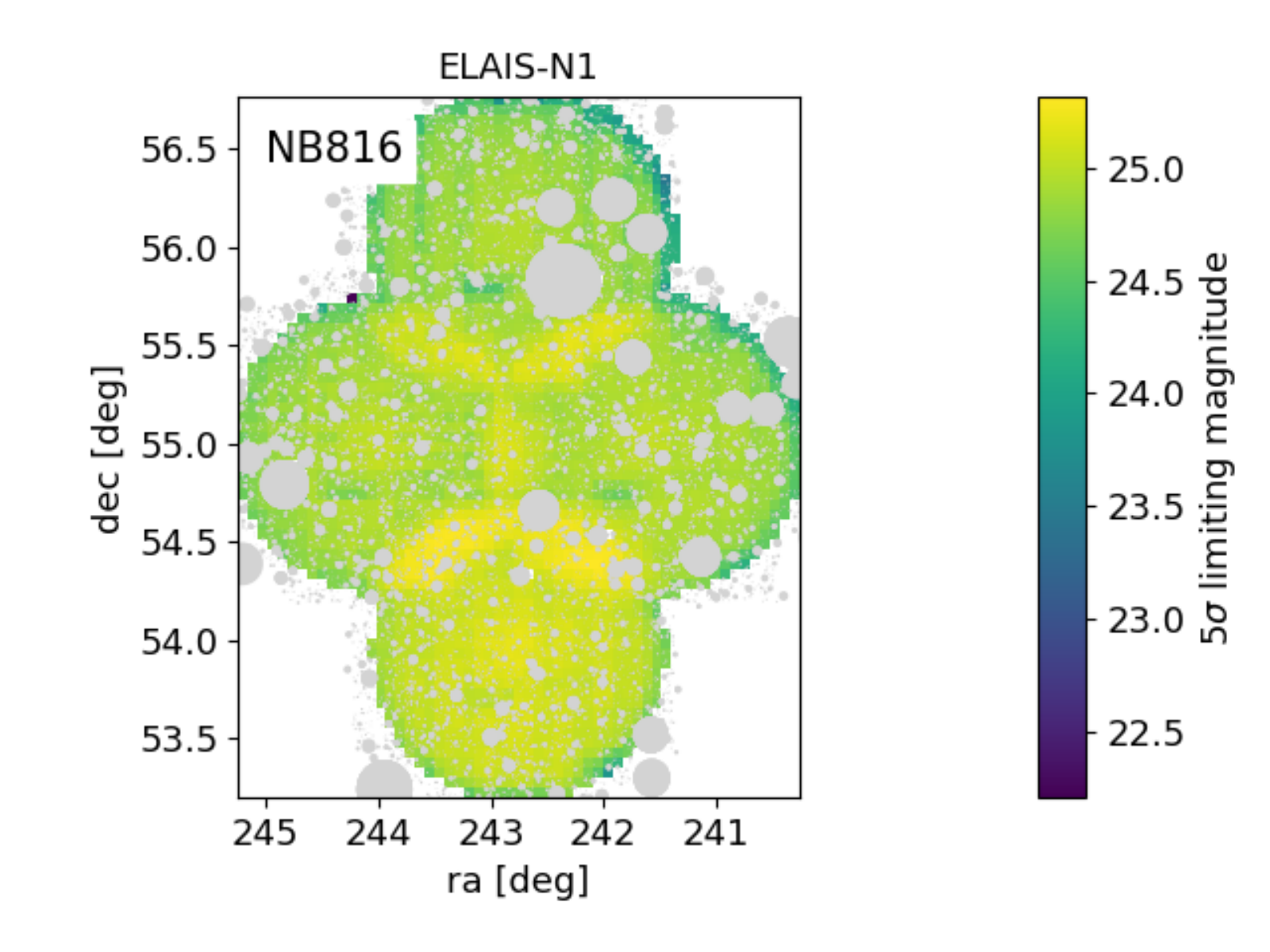} 
   \includegraphics[width=0.45\textwidth, bb=0 0 461 346]{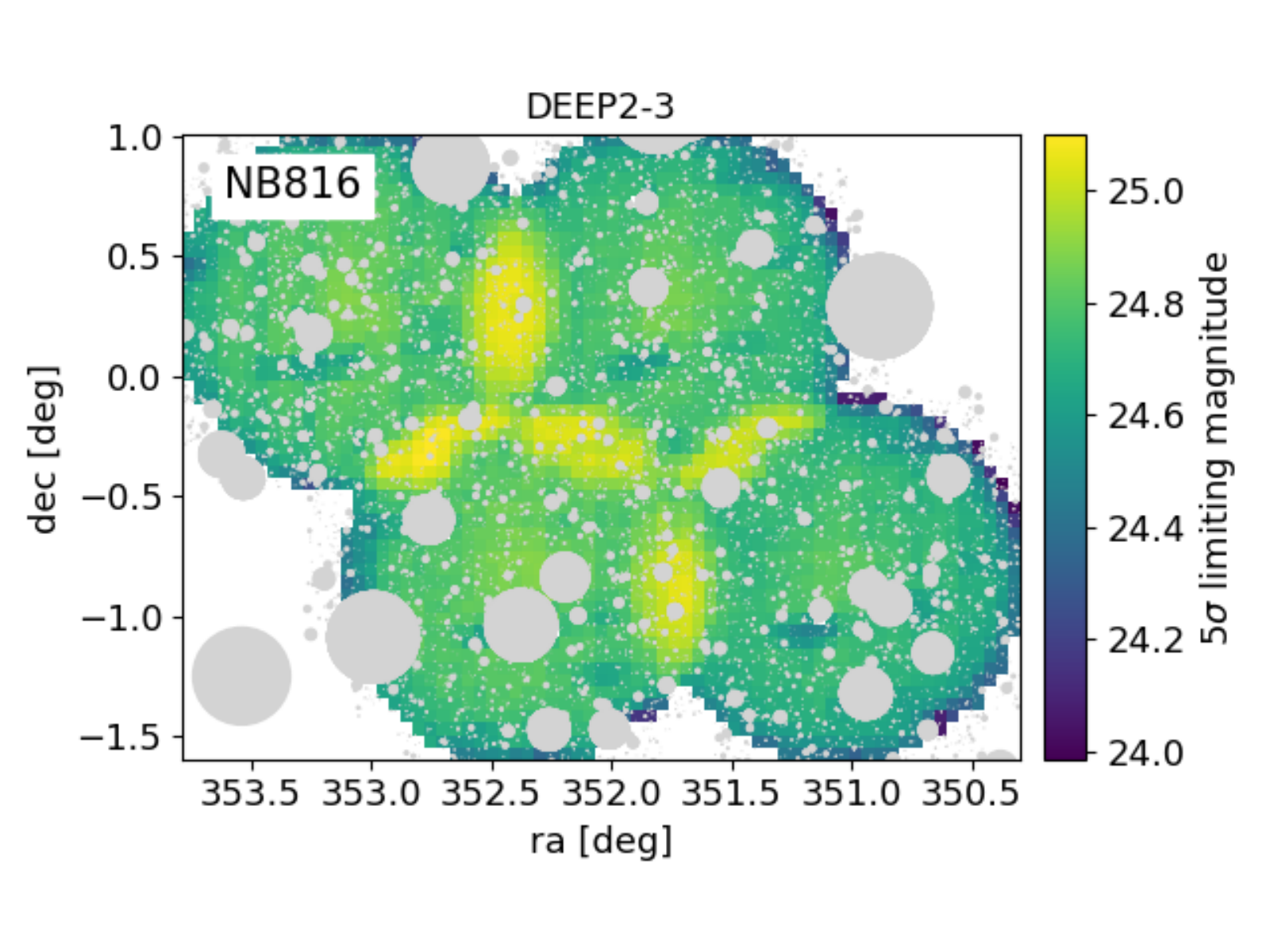} 
   \includegraphics[width=0.45\textwidth, bb=0 0 461 346]{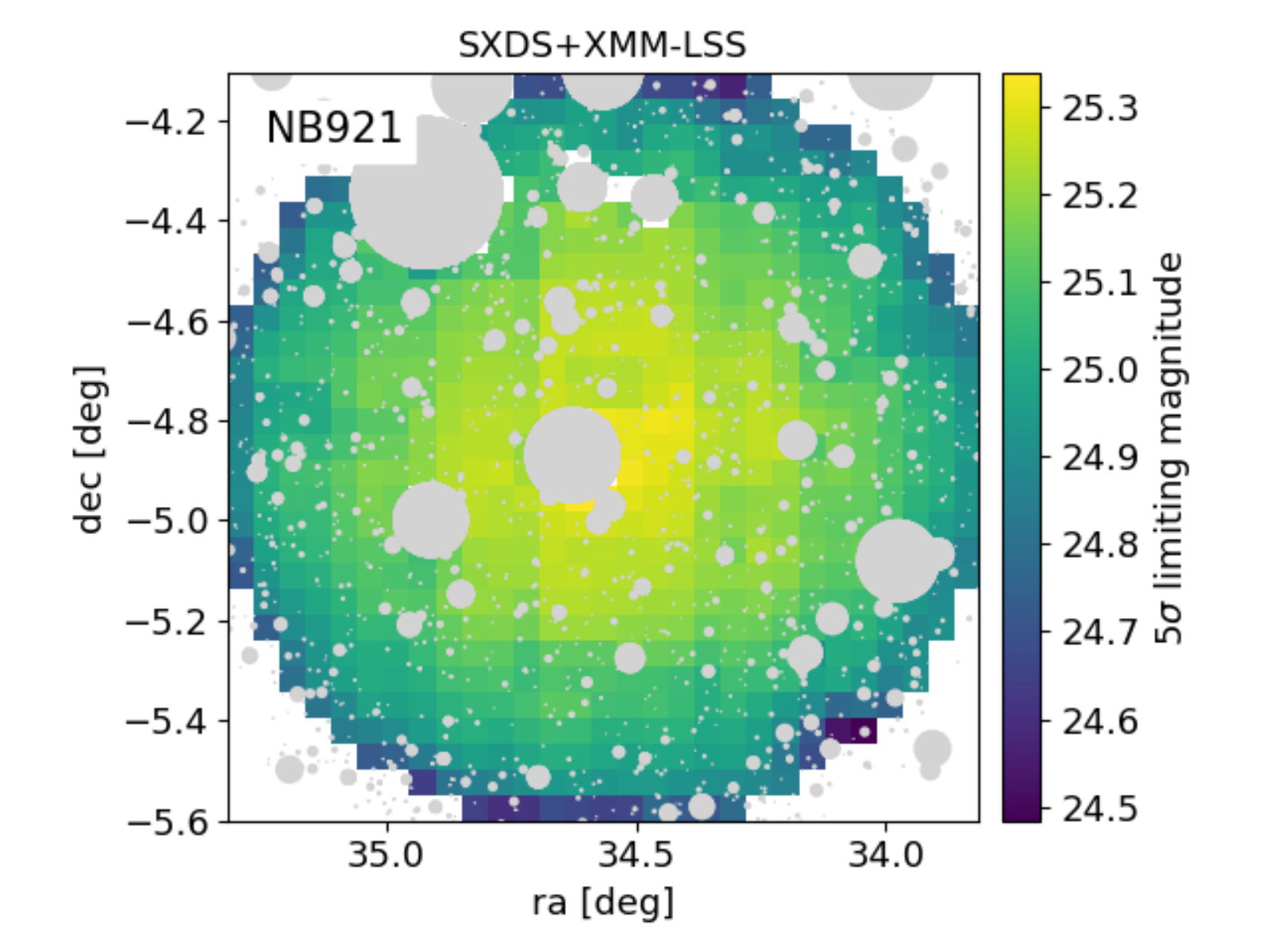} 
   \includegraphics[width=0.45\textwidth, bb=0 0 461 346]{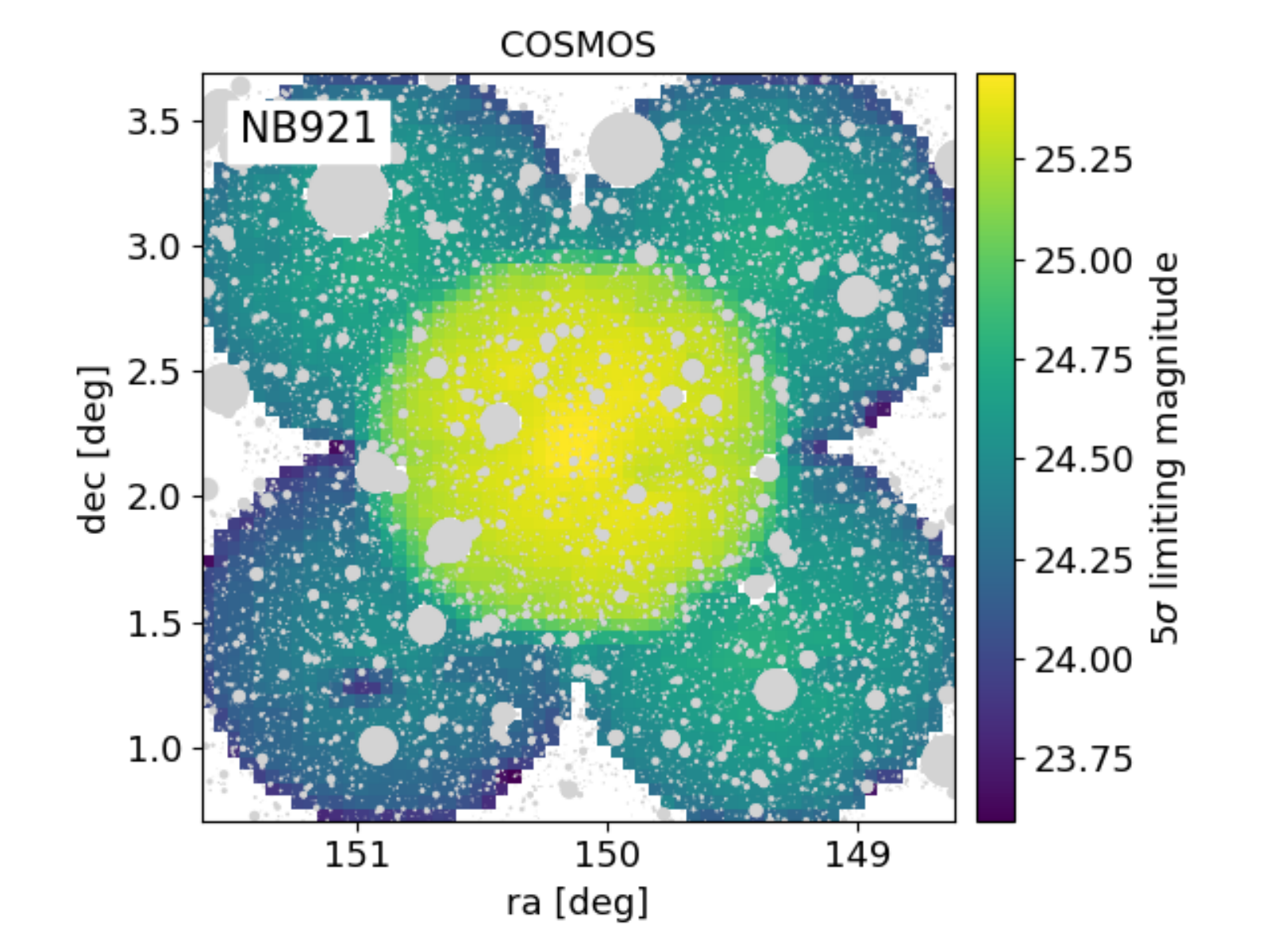} 
   \includegraphics[width=0.45\textwidth, bb=0 0 461 346]{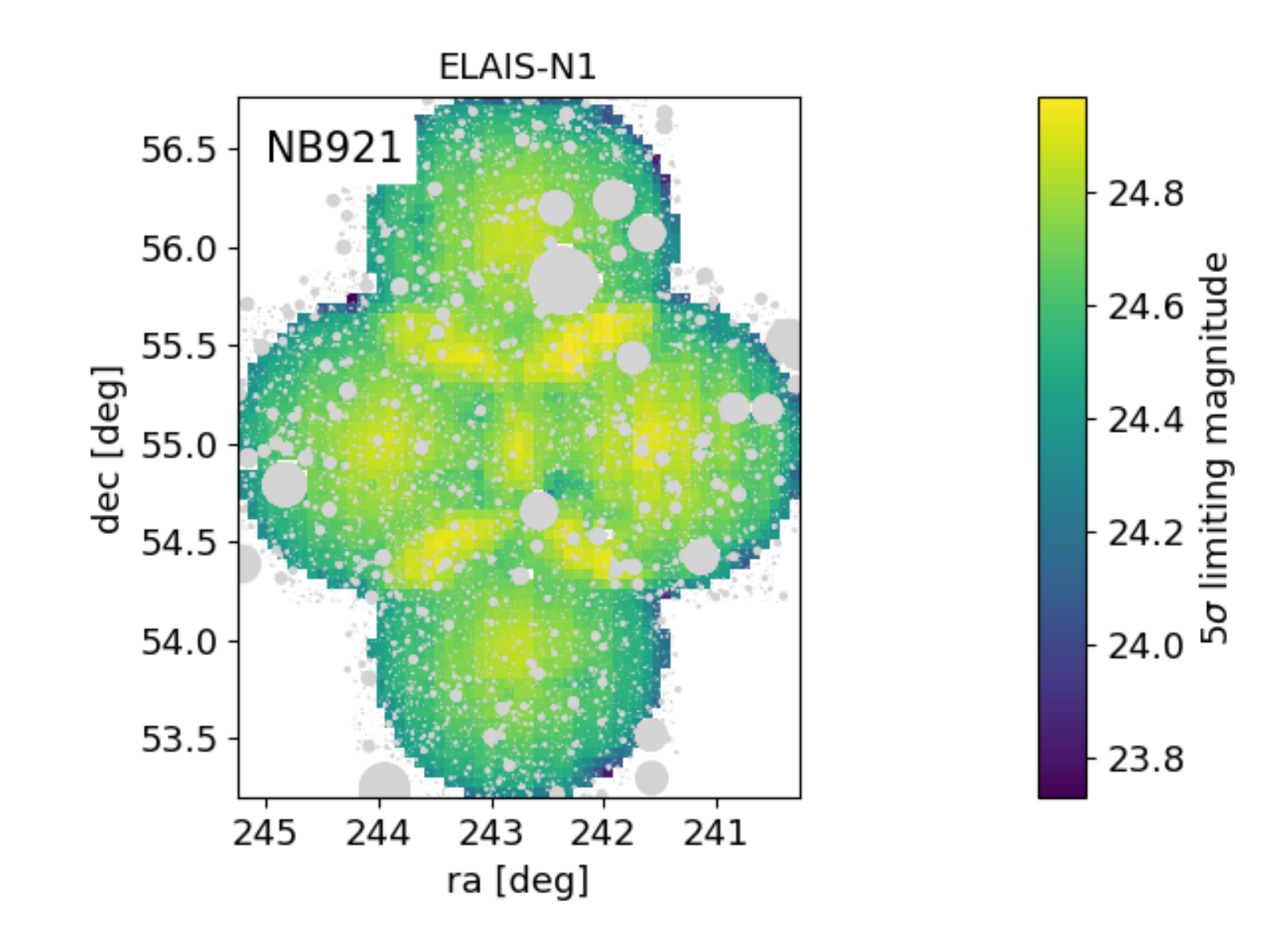} 
   \includegraphics[width=0.45\textwidth, bb=0 0 461 346]{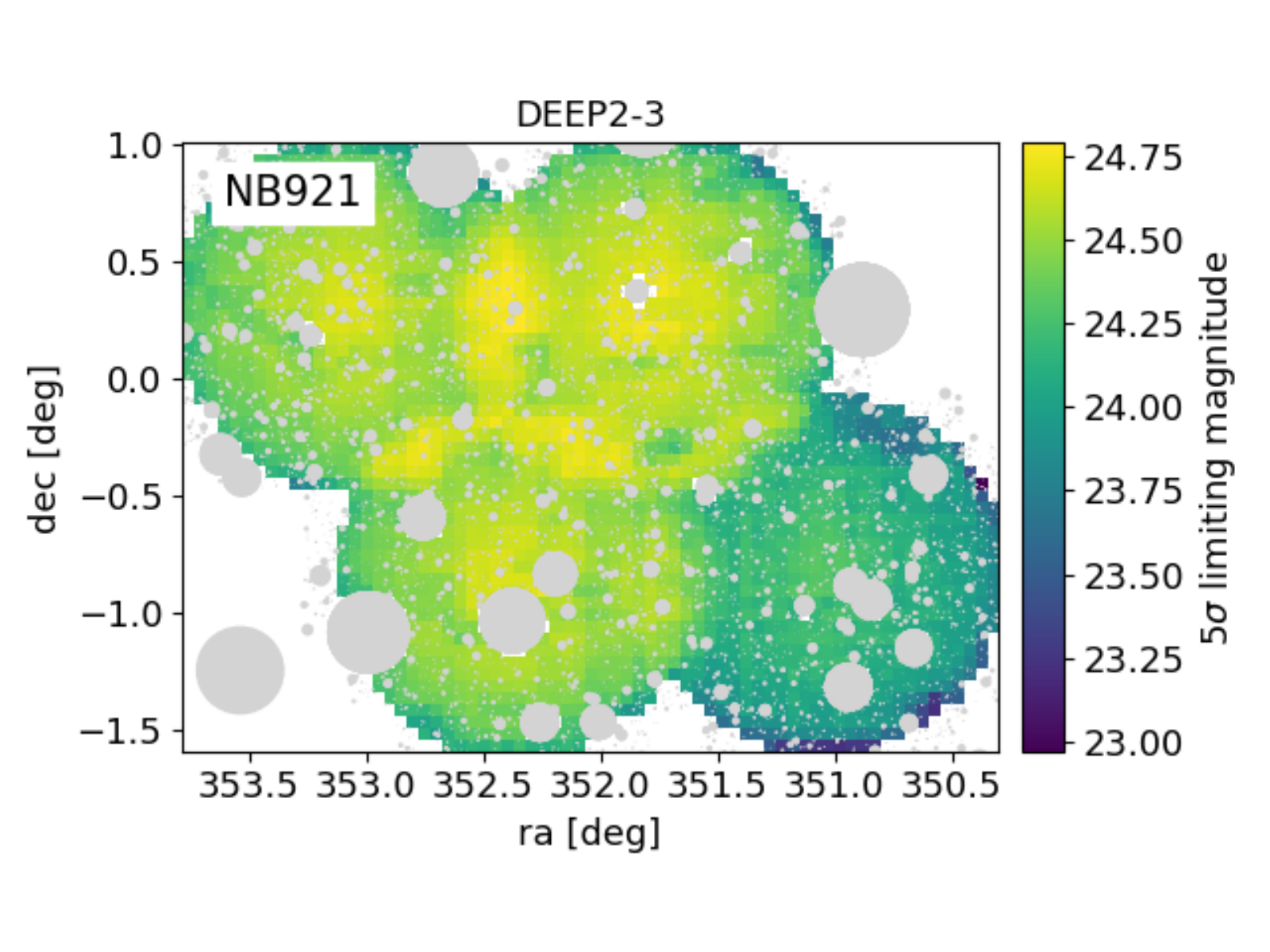} 
 \end{center}
 \caption{The map of 5$\sigma$ limiting magnitudes in NB816 and
   NB921. 
   The gray regions are masked for bright objects. The flux density
   errors in 2 arcsec aperture photometry on the images with PSF
   matched to 1.1 arcsec are used for the estimate of limiting magnitude.
}\label{fig:LimMagSSP}
\end{figure*}

The source catalogs in individual D/UD fields are extracted from the
Catalog Archive Server (CAS) of the HSC-SSP. To select NB921-detected
sources, we apply the following criteria in {\tt SQL} query:  

\noindent
{\tt 
\footnotesize 
forced.isprimary=True,\\ 
meas.(g,r,i,z,y,n921)\_inputcount\_flag\_noinputs=False,\\ 
meas.(g,r,i,z,y,n921)\_inputcount\_value>2,\\ 
meas.merge\_peak\_n921=True,\\ 
meas2.n921\_sdssshape\_flag\_badcentroid=False,\\ 
meas.(z,n921)\_pixelflags\_bad=False,\\ 
meas.(z,n921)\_pixelflags\_edge=False,\\ 
meas.(z,n921)\_pixelflags\_saturatedcenter=False,\\ 
meas.n921\_cmodel\_flag=False,\\ 
meas2.n921\_psfflux\_flag=False,\\ 
meas2.n921\_psfflux\_flag\_badcentroid=False,\\
meas2.n921\_psfflux\_flux!=`NaN',\\
meas2.n921\_psfflux\_flux/meas2.n921\_psfflux\_fluxsigma>=5.0,\\
masks.(g,r,i,z,y)\_mask\_s18a\_bright\_objectcenter=False.\\
}%
\noindent
For NB816-detected sources, we use {\tt i}-band and {\tt n816} for a combination
of BB and NB. 

The flag of bright object mask applied in the extraction of sources
from the CAS is from the revised masks which are released as part of
the incremental release 1 (see section 6.6.2 of \cite{HSCSSPDR2} as
well as \cite{HSCBrightStarMask}). However, we find that the size of
masks for bright stars is not large enough to remove false detections
around the prominent stellar haloes. 
As a workaround, we enlarge a size of masked region around the bright
stars by a factor of 1.5, if the original size of masked region is
more than 3.0 arcmin in radius.   
Otherwise, we use the original masked regions around the other
stars. The appropriate masked regions are essential not only to remove
the spurious objects but also to calculate the proper
area and volume for investigating the number counts and luminosity
functions of the galaxies selected. We use the catalog of random
points (one of the value-added products of HSC-SSP PDR2, see the
section 5.4 of \cite{HSCSSPDR2}) to calculate the effective survey
area in each field, where the random points within the mask regions
modified are removed. The results are summarized in table~\ref{tbl:NBdataArea}.   

To validate the source detection in the PDR2, we investigate the
number counts of sources detected in NB816 or NB921
(figure~\ref{fig:NumCountSSP}). The figure shows that the number
densities of NB-detected sources including stars and galaxies are
consistent with the previous studies \citep{Kashikawa2004}. We use 
{\tt convolvedflux\_2\_20\_mag}, which is the 2-arcsec aperture
photometry on the images with point spread function (PSF) matched to 1.1
arcsec. Note that the number counts are corrected for detection
completeness and only the sources with more than 50\% of detection
completeness are used for the plots. The detection completeness are
estimated by running the  {\tt hscPipe} on the images where synthetic
objects with a PSF profile are embedded in the position  
of {\tt sky objects}\footnote{\label{foot:sky}
  These are for the measurements on the blank sky position, which can
  be extracted from the CAS by applying the flag of 
  {\tt merge\_peak\_sky=True}. See the section 6.6.8 of \cite{HSCSSPDR2}. 
} and then calculating the recovery rate of the embedded synthetic
objects with magnitudes ranging from 18.0 to 28.0 mag. We investigate
the detection rate for the synthetic objects with magnitude bins of
width 0.5 mag. Because doing the analysis in all regions is too
time-consuming and unrealistic, we perform the measurements in nine
representative {\tt patch} regions in each field; the list of 
{\tt patch} regions is shown in table~\ref{tbl:9patches}. The
figure~\ref{fig:DetectionCompletenessSSP} shows the detection
completeness as a function of magnitude.  

The limiting magnitudes in each band are estimated from the
photometric errors in the catalogs. However, we find that the absolute
values of the photometric errors output by {\tt hscPipe} are
underestimated, while the relative values among the objects seem to be
valid. Therefore, we rescale the photometric errors to match with the 
1$\sigma$ sky noise. The 1$\sigma$ sky noise is estimated by fitting
a Gaussian profile to the distribution of the flux densities at the
random sky positions. The measurements at the sky position are stored
as {\tt sky objects} in the CAS. The 5$\sigma$ limiting magnitudes 
measured by this method are consistent with the magnitude where 
the number count is bent or the detection completeness starts to 
steeply decrease towards zero (see figures~\ref{fig:NumCountSSP} and \ref{fig:DetectionCompletenessSSP}). 
This shows that rescaling the photometric errors is valid. 
Figure~\ref{fig:LimMagSSP} shows the map of 5$\sigma$ limiting
magnitudes of the individual NB data in each field.

\subsubsection{CHORUS}

CHORUS is a Subaru open-use intensive program (S16B-001I, PI:
A.~K.~Inoue) supplemented with a normal program (S18B-004) to use
14.6 nights in total between 2017 January and 2018 December to conduct
HSC imaging with four NB and one intermediate-band filters (NB387,
NB527, NB718, IB945, and NB973) in the HSC-SSP UD-COSMOS field. The
main science of the CHORUS project is to understand reionization of
the Universe at $z>3$. Therefore, deep NB imaging data are available from this
survey. As by-products, the deep data are useful to search for
emission-line galaxies at relatively lower redshifts of $z\lesssim1.6$. 
Readers should refer to \citet{CHORUS} for the details of the CHORUS
survey such as survey design, observing runs, data quality, and data
release. Among the five filters available from the CHORUS survey, we
use the data from three filters of NB527, NB718, and NB973 in this
study. This is because the same things for the HSC-SSP NB387 apply to
the CHORUS NB387 and response function of IB945 is a factor of $\sim$3 
wider than the other NB filters (324$\AA$).   

The data reduction, source detection, and photometry are conducted
with {\tt hscPipe} version 6 in the same manner as the HSC-SSP. 
Since the data and catalogs from the CHORUS survey are stored in the
database of the HSC-SSP internal release, the extract of NB-detected
sources from the CAS, the mask of the survey area, measurements of
detection completeness, and rescaling the photometric errors are also
carried out in the same manner as the HSC-SSP. Although \citet{CHORUS}
also perform the similar validation, we notice that the methods
\citet{CHORUS} apply are slightly different from ours. To keep
consistency between the NB data from HSC-SSP and CHORUS surveys within
this study, it would be worth showing the results in this
paper. The number counts, the detection completeness, and the map of
$5\sigma$ limiting magnitudes are shown in appendix 
(figures~\ref{fig:NumCountCHORUS}--~\ref{fig:LimMagCHORUS}).  

\subsection{Spectroscopic data}
\label{sec:specz}

\subsubsection{Literature}
\label{sec:specz_literature}
The HSC-SSP CAS contains a catalog of public spectroscopic redshifts
collected from literature\footnote{%
  zCOSMOS DR3 \citep{Lilly2009},
  UDSz \citep{Bradshaw2013,McLure2013},
  3D-HST \citep{Skelton2014,Momcheva2016},
  FMOS-COSMOS \citep{Silverman2015,Kashino2019},
  VVDS \citep{LeFevre2013},
  VIPERS PDR1 \citep{Garilli2014},
  SDSS DR12 \citep{Alam2015},
  the SDSS IV quasi-stellar object catalog \citep{Paris2018}, 
  GAMA PDR2 \citep{Liske2015}, 
  WiggleZ DR1 \citep{Drinkwater2010},
  DEEP2 DR4 \citep{Davis2003,Newman2013},
  DEEP3 \citep{Cooper2011,Cooper2012}, 
  and PRIMUS DR1 \citep{Coil2011,Cool2013}.
}, which is one of the value-added products of HSC-SSP PDR2. The
catalog of public spectroscopic redshifts is updated since the
PDR1. Among the redshifts, we use only the secure ones with a flag of
{\tt specz\_flag\_homogeneous = True} in this study (see the section
5.4 of \cite{HSCSSPDR2}), where 73 (98)\% of the galaxies with the
public spectroscopic redshifts have $i$-band magnitudes brighter than
22.5 (24.0). In addition, we use the spectroscopic redshifts from the
Keck/DEIMOS survey in the COSMOS field, which are given by \citet{Hasinger2018}.  

\subsubsection{Subaru/FOCAS}
\label{sec:specz_Subaru}
We have confirmed 86 emission-line galaxies by spectroscopy with
Subaru/FOCAS \citep{Kashikawa2002}, which were observed in the three
programs: 44 NB921 emission-line galaxies (19 in the COSMOS field and
25 in the ELAIS-N1 field) and 28 photo-z/color selected galaxies
(these are filler targets in COSMOS) are confirmed by the observing
run on 14--15 March 2017 (S17A-083, PI: M. Hayashi), 9 galaxies (4
NB816 emitters and 5 NB921 emitters) in the DEEP2-3 field are
confirmed by the observing run in September 2016 (S16B-029, PI:
T. Shibuya, \cite{Shibuya2018}), and 5 NB921 emitters (four in COSMOS
and one in ELAIS-N1) are confirmed by the run in March -- May 2017
(S16B-071I, PI: Y. Matsuoka, \cite{Matsuoka2018}).  
For our own run (S17A-083), we use 300R grism + SO58 order
cut filter. The sky condition in the run on 14 March 2017 was
good. Although there were thin clouds over the sky on 15 March 2017,
we were able to obtain the spectra. The seeing ranges from 0.60 to
0.79 arcsec. 

We reduce the data in the standard manner using {\tt FOCASRED} which
is the  {\tt IRAF} scripts package for the Subaru/FOCAS data
reduction. The reduction procedures are basically the same procedures
as described in \citet{Hayashi2019}. The details of the other two
observations and their data reduction are described in these papers
\citep{Shibuya2018,Matsuoka2018}, respectively. 

The redshifts are determined by fitting a Gaussian profile to the
emission lines detected in the 1-D spectra. Among the 86 galaxies
confirmed, multiple emission lines are detected from 73 galaxies,
whose spectroscopic redshifts are robust. Although the other 13
galaxies have a single line detected in the individual spectra, we
identify the galaxies with assistance of the photometric redshifts and
colors.    

\subsubsection{AAT/AAOmega+2dF}
\label{sec:specz_AAT}
Most of the spectroscopic confirmation are performed by the
Multi-Object Spectroscopy (MOS) observations on 2018 November 10--14
with AAOmega+2dF \citep{Sharp2006,Lewis2002} on the Anglo-Australian
Telescope (AAT). The AAOmega+2dF has 400 fibers in 2 degree FoV and
the light from objects through the fibers is fed to the blue and red
arms with a dichroic mirror. We use the 580V grating for blue and 385R
grating for red. Since we set the dichroic wavelength to 670nm, the
spectra covering 4500-9800$\AA$ are simultaneously obtained. We target
NB816 or NB921-selected emission-line galaxies in the SXDS field as
well as H$\alpha$ emitters at $z\sim0.4$ in the DEEP2-3 field. The
targets are basically selected from  the PDR1 catalog, and
supplemented from the PDR1 emitter candidates with lower equivalent
width of emission line. We give higher priority to galaxies with
larger emission line flux. As a result, we observed more than 3000
galaxies with NB-detected line fluxes of $>$ 4.5$\times$10$^{-17}$
erg~s$^{-1}$~cm$^{-2}$ (362 H$\alpha$ emitters at
$z=0.25/0.40$, 907 [OIII] emitters at $z=0.63/0.84$, and 1,418 [OII]
emitters at $z=1.19/1.47$) in the SXDS field, 
which is a complete sample down to the line flux of $\gtrsim$ 1.0$\times$10$^{-16}$
erg~s$^{-1}$~cm$^{-2}$ in the surveyed volume at each redshift. Among the
five nights allocated, it was clear sky for three nights. We observed
8 fiber configurations in the SXDS field and 2 configurations in the
DEEP2-3 field. The integration time is 80 -- 120 min, depending on the
fiber configurations. 

The spectra are reduced in the standard manner by the pipeline {\tt 2dfdr}
version 6.46, while referring to the procedures in the Australian Dark
Energy Survey (OzDES, \cite{Yuan2015}). Only the information of
redshifts is required from the spectra to identify the NB-selected
emission-line galaxies in this study. More details of the spectra will
be described in a forthcoming paper where the spectroscopic
properties of the emitters are discussed.   

\begin{figure}
 \begin{center}
   \includegraphics[width=0.43\textwidth, bb=0 0 461 346, trim=50 0 50 20]{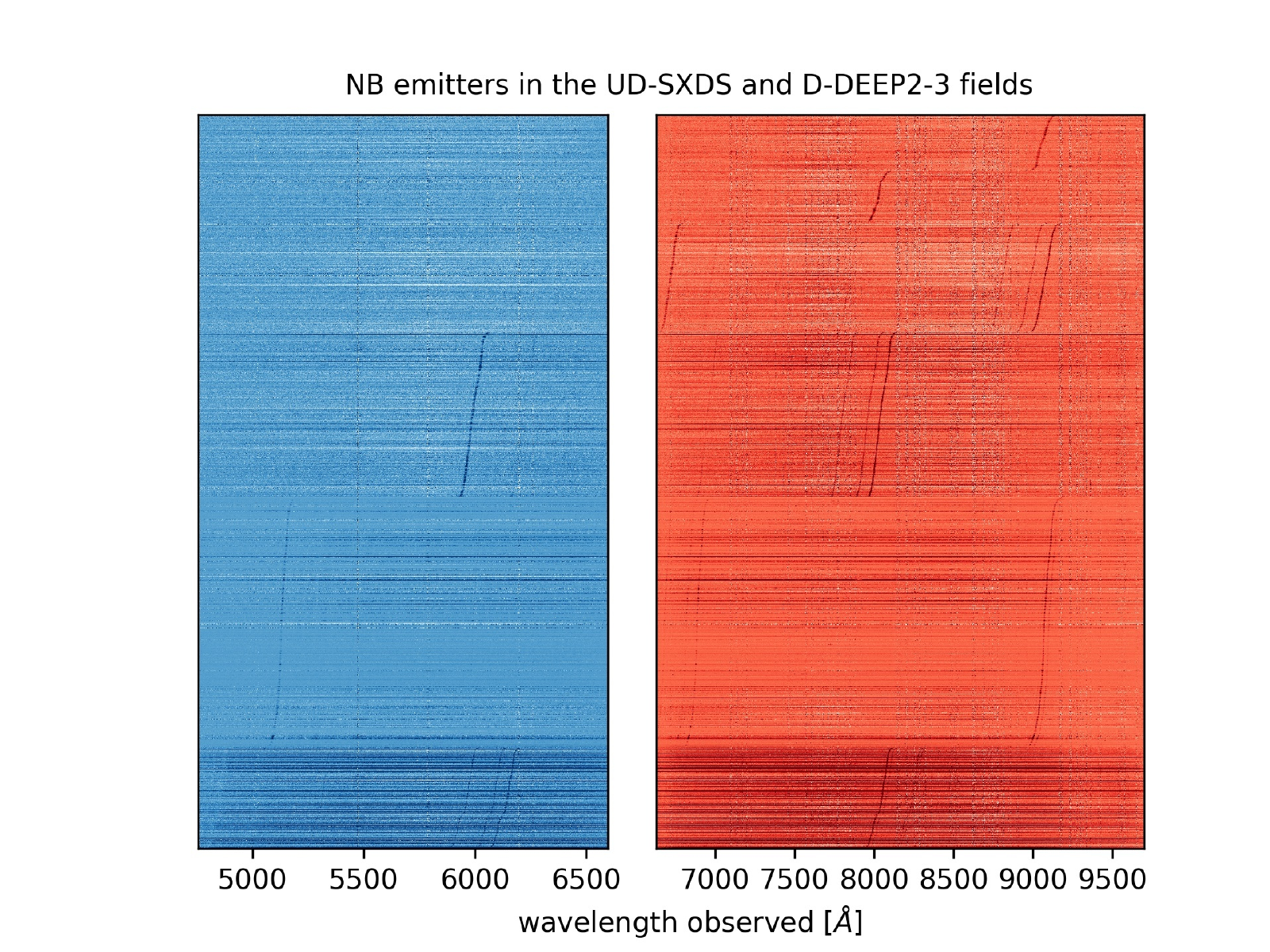} 
 \end{center}
 \caption{The AAOmega spectra of the confirmed NB816 and NB921
   emitters that are selected from the HSC-SSP PDR1 data. The left
   shows the spectra in blue arm and the right shows the spectra in
   red arm. The spectra are lined up in the order of
   redshift.}\label{fig:AAOmegaSpectra} 
\end{figure}

To confirm the NB-selected emission-line galaxies, we first search for
the emission lines in the wavelength range covered by individual NB
filters. Then, to determine the redshifts, we search for other emission
lines in the individual spectra. When the multiple emission lines are
detected at more than 3$\sigma$, the redshifts are determined and thus
secure. If a single emission line are detected, we identify the
emission line with assistance of the photometric redshifts and
colors. Note that the gratings used in the observations do not have
the spectral resolution enough to resolve the [OII] doublet and the
[OII] emitters are expected to have the only single line detected in
the wavelength covered by NB filter.  
We have confirmed 1933 emitters, indicating 64\% of success rate. 
Although some objects can be contaminants among the remaining 36\% of
the targets, many of the objects that are not confirmed by the
spectroscopy are likely to have emission lines less than the
detection limit of this observation.  
Indeed, if we focus on the emitters with emission lines brighter than
1.0$\times$10$^{-16}$ erg~s$^{-1}$~cm$^{-2}$, which is likely above the
sensitivity expected from the integration time, the success rate
increases to more than 80\%.  
Since 554 confirmed galaxies also have the spectroscopic redshifts from
the literature, we compare the redshifts from the AAOmega spectra with
those from the literature. The median of the difference is
$(\Delta z)_{\rm median}=-4.90\times10^{-4}$ and the dispersion is
$\sigma(\Delta z)=6.32\times10^{-3}$, suggesting that there is no
systematic difference in the redshifts and both redshifts are
consistent with each other.   
Figure~\ref{fig:AAOmegaSpectra} shows the
spectra taken with AAT/AAOmega+2dF. 

\section{Emission-line galaxies}
\label{sec:ELGs}

The procedures to select and identify emission-line galaxies are
basically the same as those for the PDR1 data in \citet{Hayashi2018a}. 
However, we change several points to adjust to the PDR2 data. In this
paper, we focus on the difference from the PDR1. We survey
emission-line galaxies at the 14 redshift slices as shown in
table~\ref{tbl:RedshiftSurveyed}.  
As with the PDR1 catalogs, the catalogs of the emission-line galaxies
made in this paper will be released at the HSC-SSP data release site%
\footnote{\label{foot:hscsspwebsite} \url{https://hsc-release.mtk.nao.ac.jp/}}.
The catalog from the HSC-SSP NB data will be released soon after the
paper is published. The catalog from the CHORUS data will be also
released at the same time as or after the CHORUS NB data are
published.  

\input{table3.tex}

\subsection{Selection}
\label{Selection}

\begin{figure*}
 \begin{center}
   \includegraphics[width=0.45\textwidth, bb=0 0 461 346]{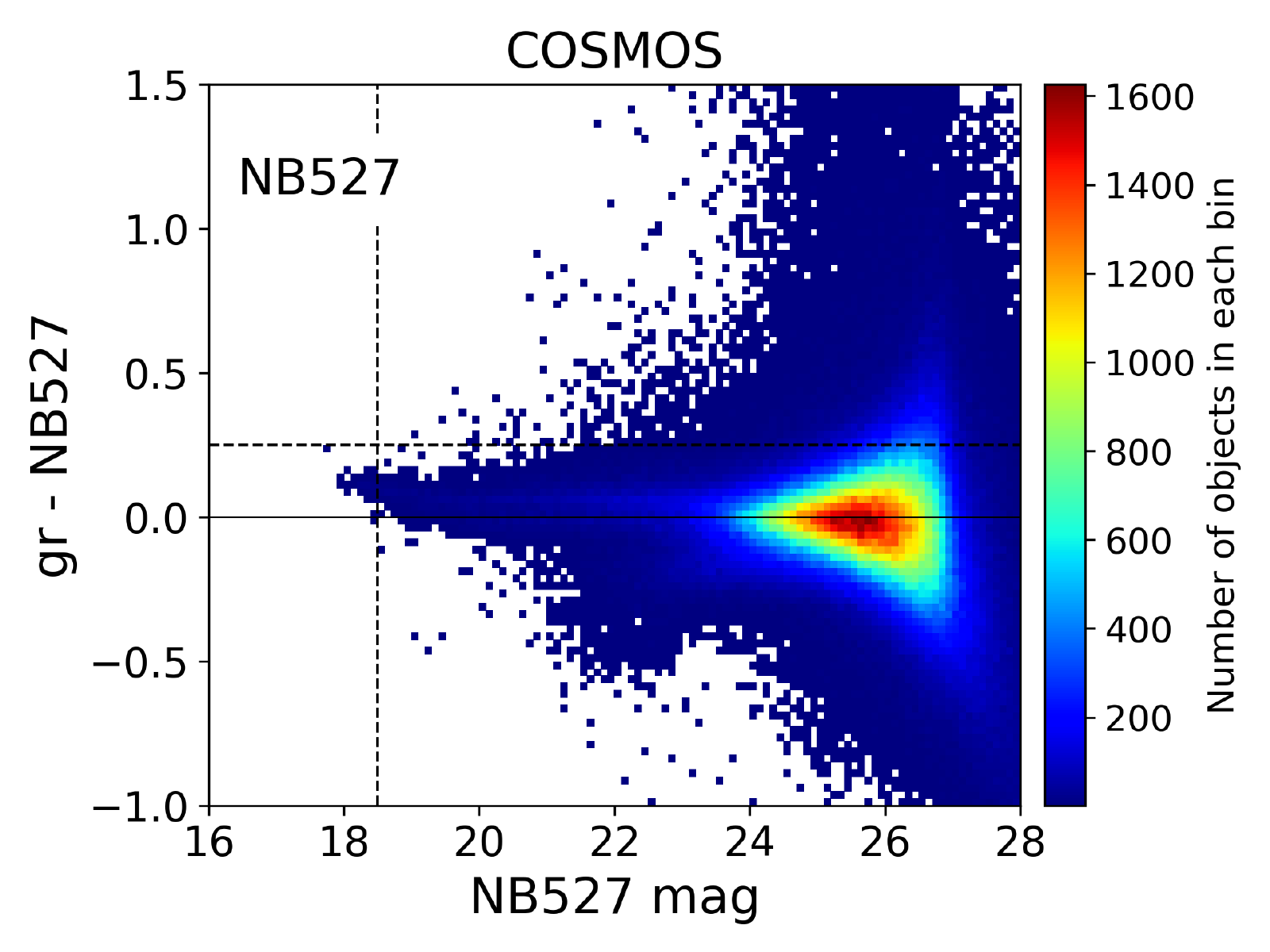} 
   \includegraphics[width=0.45\textwidth, bb=0 0 461 346]{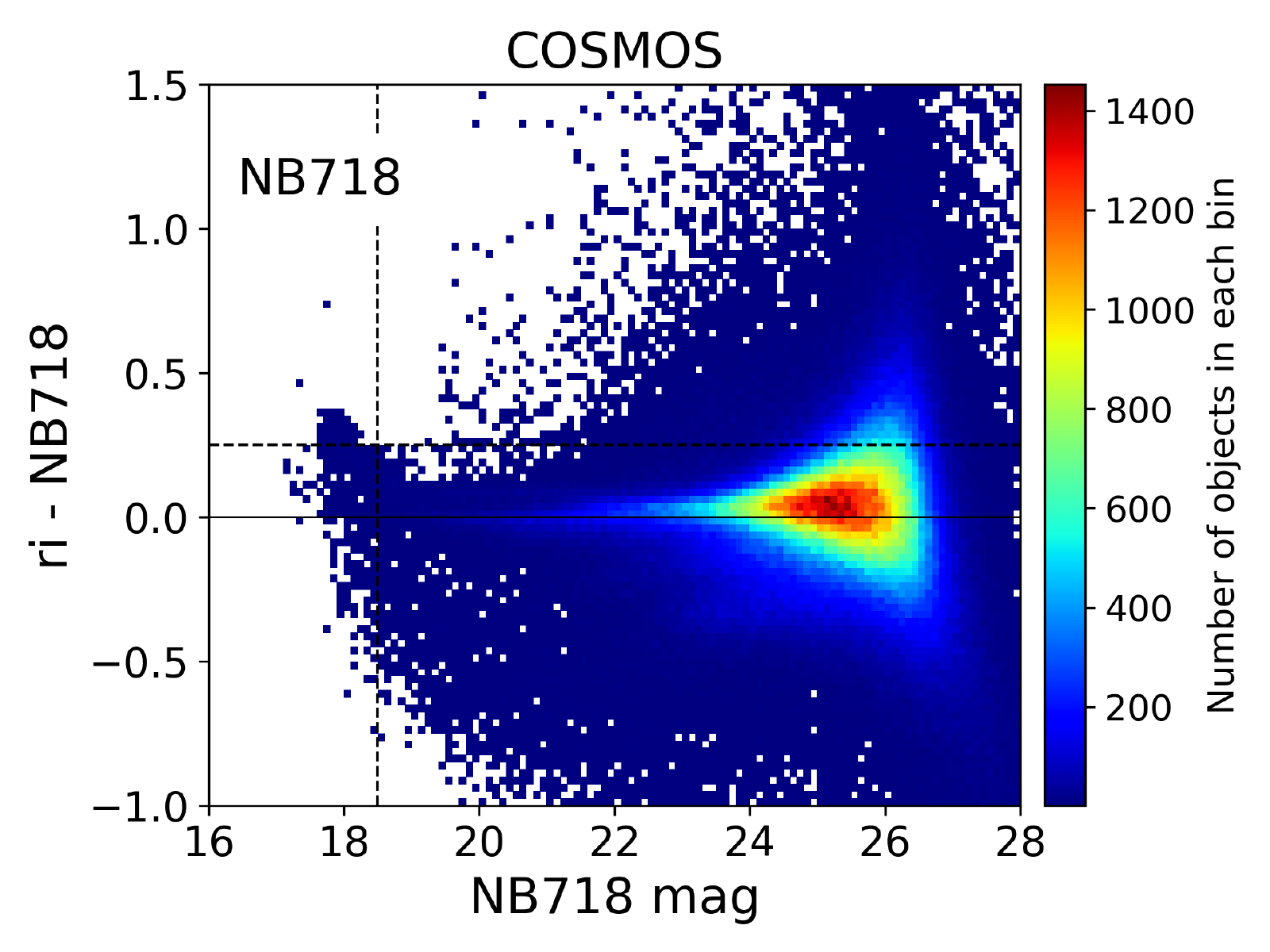} 
   \includegraphics[width=0.45\textwidth, bb=0 0 461 346]{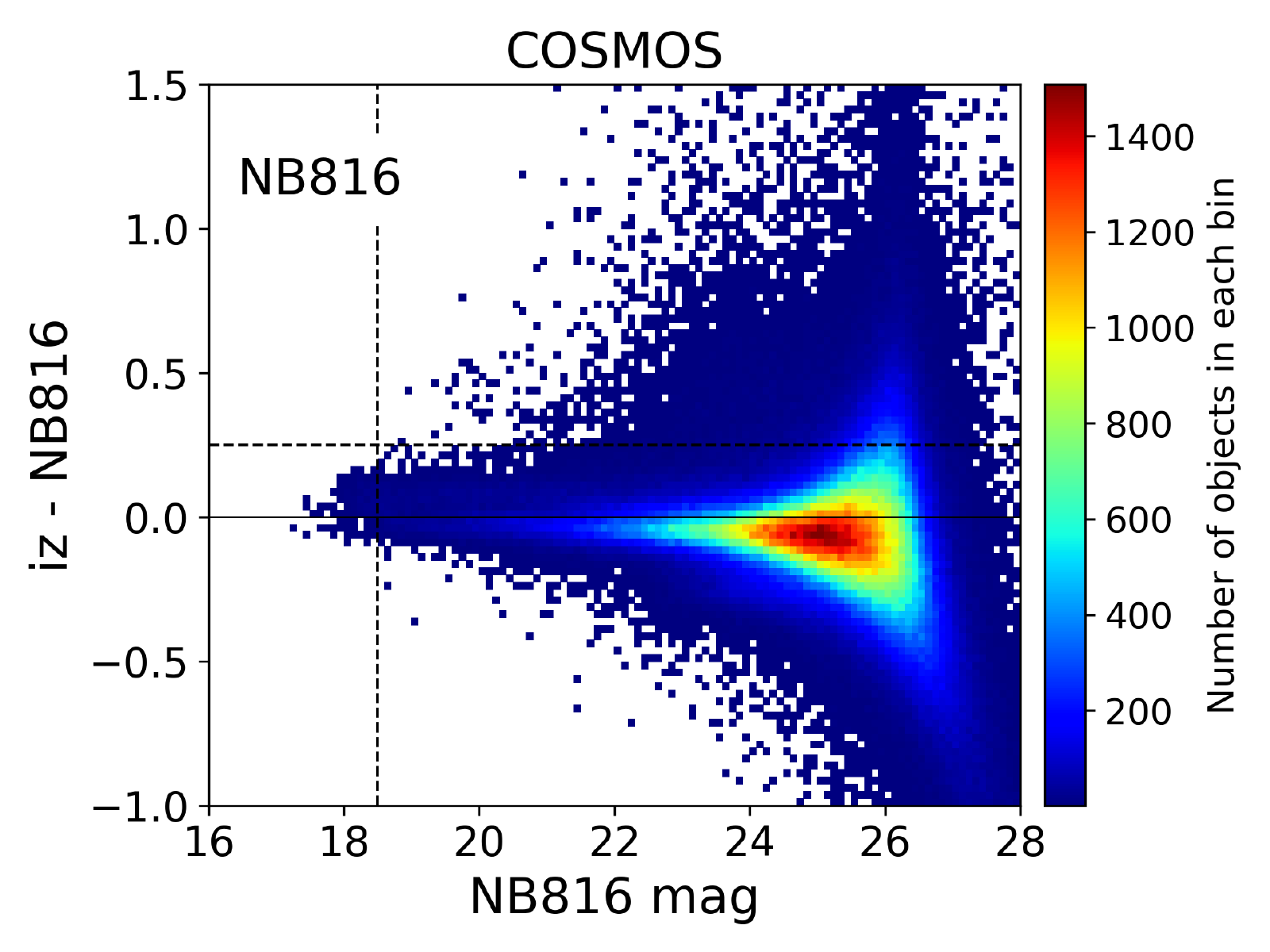} 
   \includegraphics[width=0.45\textwidth, bb=0 0 461 346]{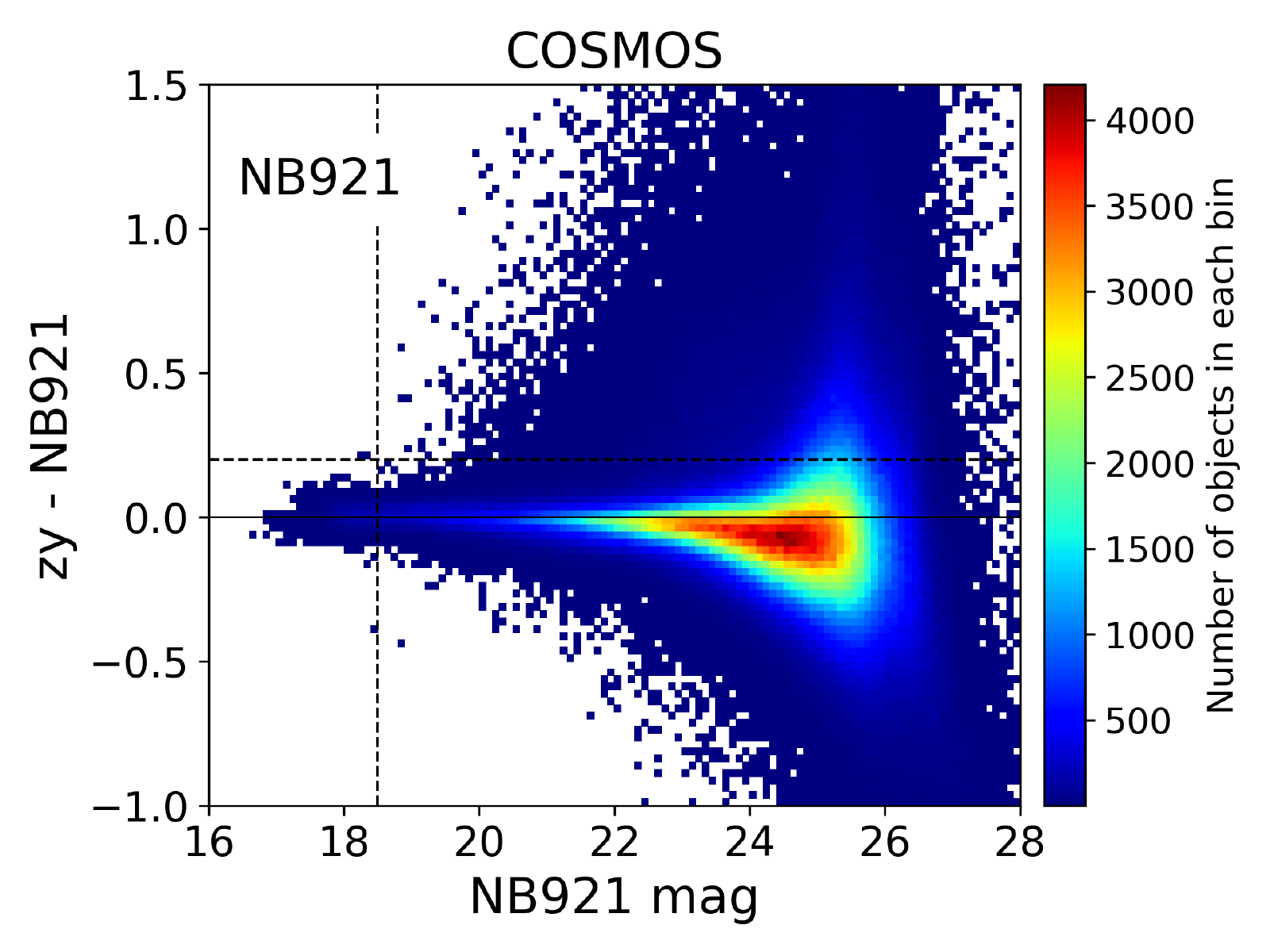} 
   \includegraphics[width=0.45\textwidth, bb=0 0 461 346]{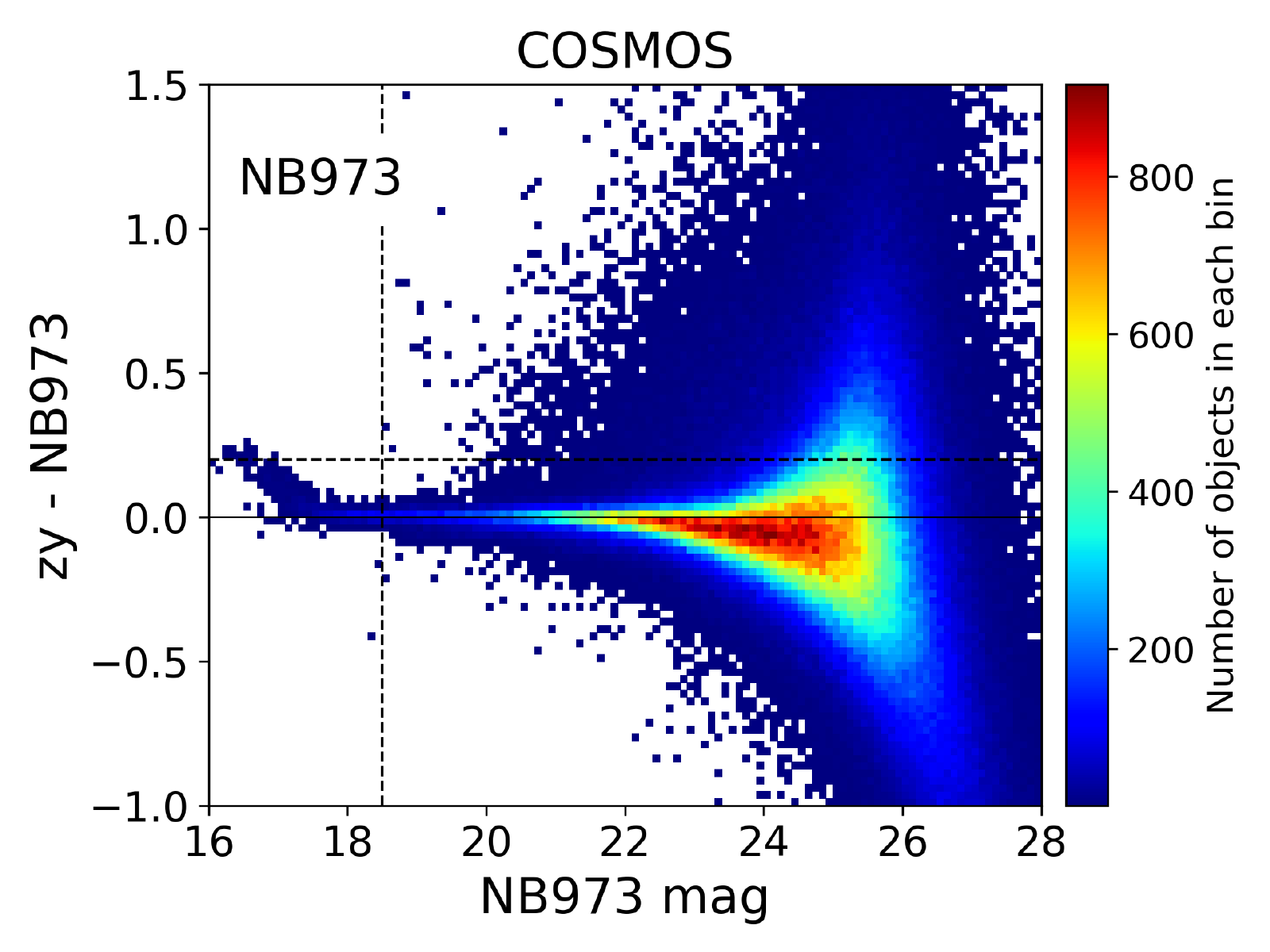} 
 \end{center}
 \caption{The distribution of objects in the COSMOS field in the
   color--magnitude diagram of BB-NB versus NB, which is color-coded
   by the number of the objects in each bin. The BB magnitude is
   corrected for the color term by interpolating with two filters and
   weights shown in table~\ref{tbl:NB_BBs_forSelection}. The dotted
   lines are the magnitude cut and color cut applied for the selection
   (table~\ref{tbl:NB_BBs_forSelection}). The solid line shows the
   color of BB-NB$=0$.}\label{fig:BBNBcolor}  
\end{figure*}

We use the NB data as well as all of the BB data to select
emission-line galaxies. 
We change the method of photometry from the model fitting with two
components ({\tt cmodel} magnitude) to the fixed aperture 
({\tt undeblended\_convolvedflux\_2\_11} magnitude) to measure the
color excess against a photometric error in BB -- NB color at a given
NB magnitude.
The aperture photometry is a measurement of flux density for each object
on the PSF matched coadd images, where the target PSF is 1.1 arcsec
and the aperture size is 1.1 arcsec. Note that the photometry is
conducted on the undeblended images, namely the detected footprints
are not deblended into each object. This is because we found that 
there are some objects having {\tt cmodel} magnitudes unnaturally
bright, which could be in part due to failure in deblending
sources. Furthermore, when we use the {\tt cmodel} magnitudes to
select emission-line galaxies, some candidates are found to
be unnaturally clustered over small area. Therefore, we decided to use
the magnitude with the small aperture on the undeblended images to 
measure the colors in the central 1.1 arcsec region of galaxies for
the selection of emission-line galaxies.    

\begin{table}
  \tbl{Filters used for the selection of emission-line galaxies. 
  }{%
    \begin{tabular}{cccccc}   
      \hline
      NB & BBs & weights & mag cut & color cut & EW$_{\rm obs}$\\ 
      \hline
      NB527 & g, r & 0.674, 0.326 & $>$18.5 & $>$0.25 & 22$\AA$\\
      NB718 & r, i & 0.079, 0.921 & $>$18.5 & $>$0.25 & 32$\AA$ \\
      NB816 & i, z & 0.631, 0.369 & $>$18.5 & $>$0.25 & 33$\AA$ \\
      NB921 & z, y & 0.643, 0.357 & $>$18.5 & $>$0.20 & 35$\AA$ \\
      NB973 & z, y & 0.052, 0.948 & $>$18.5 & $>$0.20 & 27$\AA$ \\
      \hline
  \end{tabular}}\label{tbl:NB_BBs_forSelection}
\end{table}

A slight difference in effective wavelength between NB and BB filters
requires us to correct for the color term to properly estimate the
stellar continuum underlying an emission line. In
\citet{Hayashi2018a}, we used the stellar population synthesis models
to estimate the intrinsic colors of the stellar continuum for galaxies
at redshifts that the NB filters can probe, and then correct for the
color terms using the average relation between the intrinsic BB-NB and
the observed BB colors in each redshift. This time, we linearly
interpolate to estimate the continuum flux density at the NB
wavelength by taking a weighted average of two BB magnitudes
(see also Appendix of \cite{Vilella-Rojo2015}). Note that the weights
are tuned so that the galaxies without emission lines at the
wavelength of NB filter distribute around the sequence of 
BB$_{\rm corrected}$-NB=0 with photometric errors. The weights used
are shown in table~\ref{tbl:NB_BBs_forSelection}. Figure~\ref{fig:BBNBcolor} 
shows the color--magnitude diagram of BB-NB versus NB.  

Then, we select galaxies showing more than 5$\sigma$ excess against
the photometric error in BB-NB color as candidates of emission-line
galaxies. Note that we estimate the photometric error for the
individual objects and thus a single curve of the selection criterion,  
which is often seen in studies of NB-selected emission-line galaxies,
cannot be drawn in figure~\ref{fig:BBNBcolor}. 
We also apply a BB-NB color cut to remove the bright
galaxies with intrinsic red colors in the stellar continuum. The
values of the color cut are smaller than those of PDR1, which allows
us to select emission-line galaxies with lower EW of emission
line. The limiting observed EWs are 22, 32, 33, 35, and 27 $\AA$ for
NB527, NB718, NB816, NB921, and NB973, respectively. We also apply a
magnitude cut of $>$18.5 mag to conservatively remove too bright
objects that may be saturated in HSC images. We set the magnitude
based on the color--magnitude diagram of BB-NB versus NB used for the
selection of emission-line galaxies (figure~\ref{fig:BBNBcolor}). 
Setting the magnitude cut can cause us to miss emission-line galaxies
at the very bright end \citep{Drake2013,Stroe2014,Hayashi2018a}. Since the
cut of 18.5 mag is one magnitude fainter than that in the PDR1, we must
take care of the magnitude cut to discuss the bright end of luminosity
functions in this study.  

Some emission-line galaxies with significant color excess near the
selection criterion can be missed due to the photometric errors. To
estimate the completeness of the selection, we add a photometric error
in each band randomly from a normal distribution with 1$\sigma$ sky
noise and apply the selection criteria to the magnitudes. We try 1000
iterations to see how many times the individual emission-line galaxies
meet the criteria and then estimate the selection completeness. We
multiply the selection completeness by the detection completeness in
each bands to derive the completeness of individual emission-line
galaxies. Hereafter, unless otherwise mentioned, the completeness
takes account of both detection completeness and selection one. In
this study, we limit the samples to galaxies with the completeness 
greater than 0.5.

\begin{figure}
 \begin{center}
   \includegraphics[width=0.45\textwidth, bb=0 0 461 346]{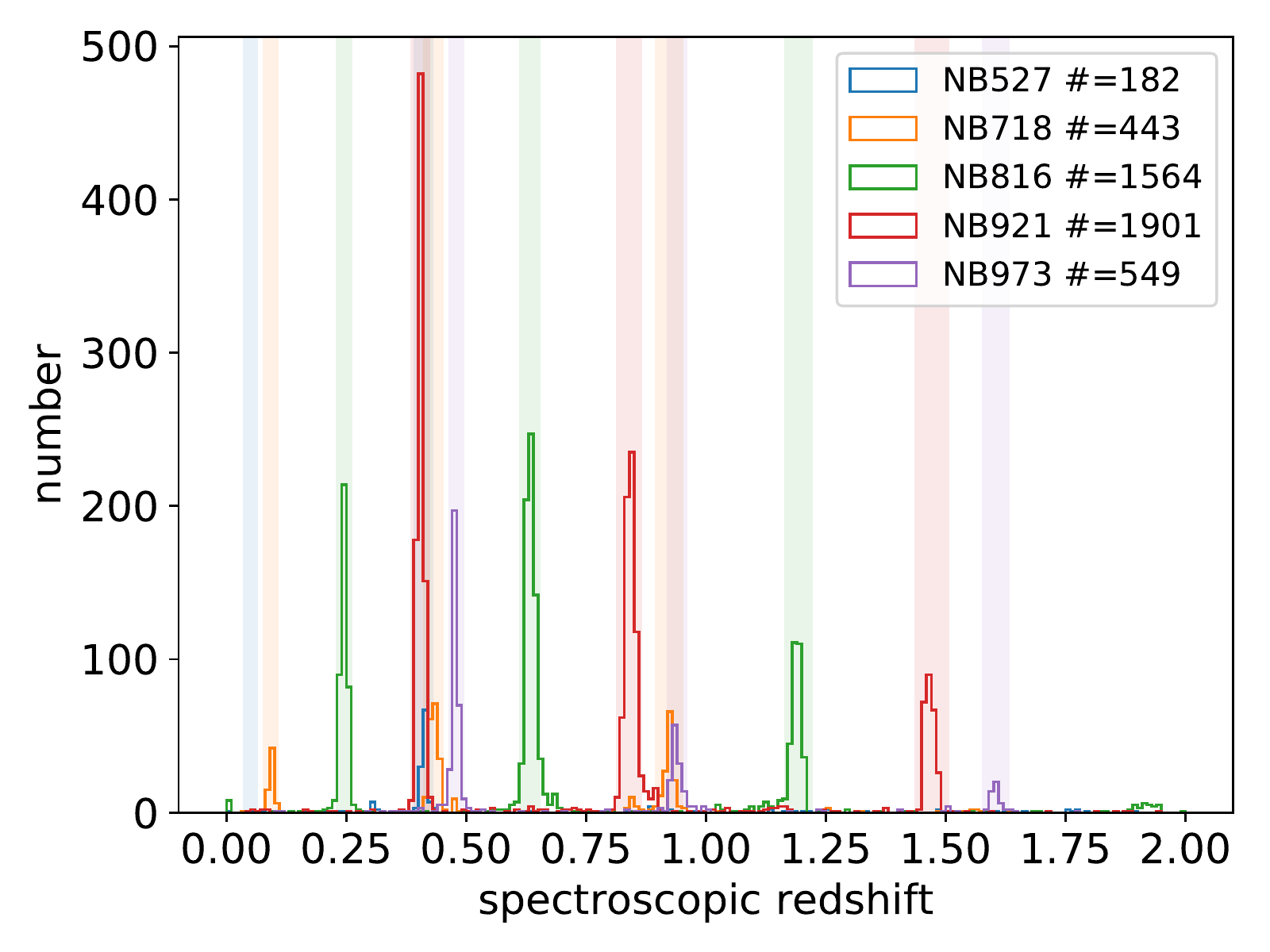} 
 \end{center}
 \caption{The distribution of spectroscopic redshift for galaxies with
   color excess in BB-NB. 
   The number of the galaxies with spectroscopic redshifts is shown in the legend.
   The regions with pale colors show the redshift ranges of
   emission-line galaxies expected by the wavelength range of each NB
   filter.}\label{fig:Nz}  
\end{figure}

Figure~\ref{fig:Nz} shows the spectroscopic redshift distribution of
emission-line galaxy candidates. The peaks of distribution are clearly
seen at the expected redshifts. The redshift distribution of the
galaxies with significant excess in the BB-NB color suggests the
validity of our selection method as well as low contaminants among the
selected galaxies. 

\subsection{Identification}
\label{sec:line_identification}

\begin{figure}
 \begin{center}
   \includegraphics[width=0.5\textwidth, bb=0 0 288 720]{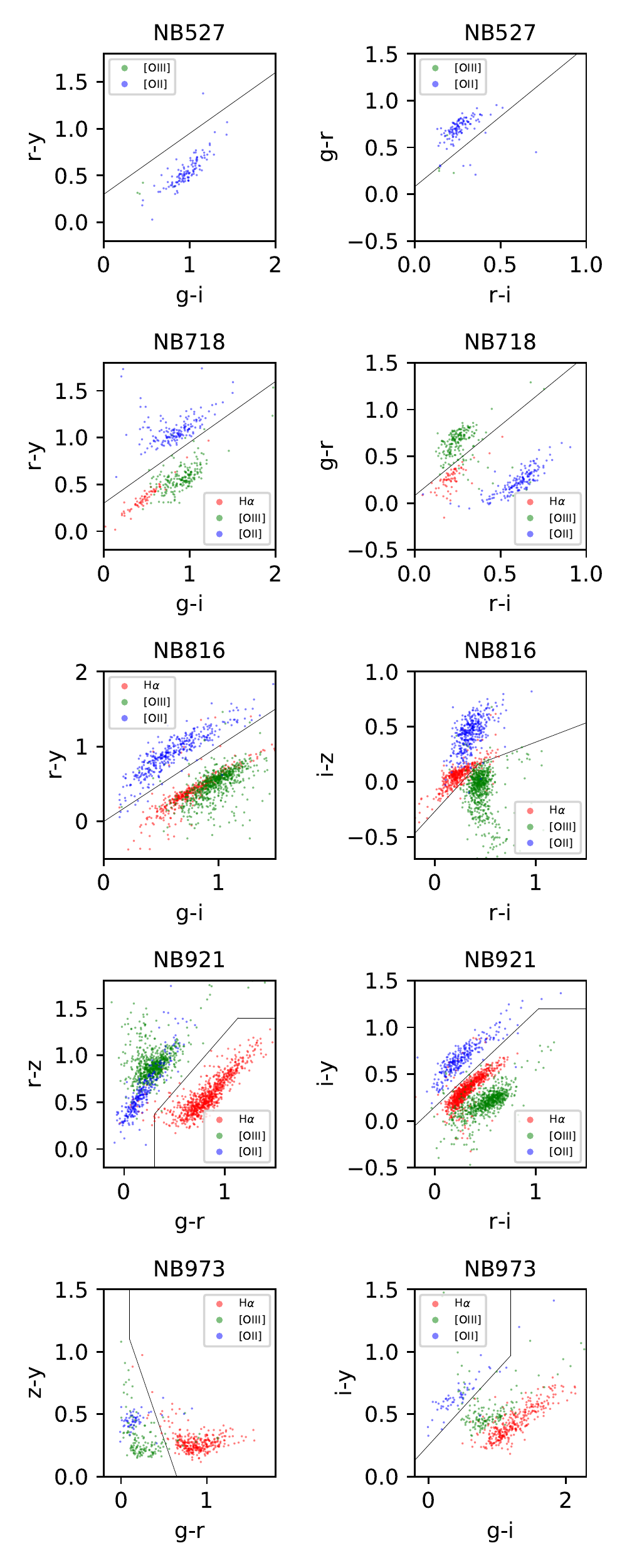} 
 \end{center}
 \caption{Two color diagrams to identify emission line galaxies based
   on the colors of galaxies. The galaxies with spectroscopic
   redshifts are plotted and color-coded based on the redshifts. 
   Note that the color selection is applied for emission-line galaxies
   for which the spectroscopic redshift is not available and
   photometric redshifts are not well determined.}\label{fig:ColorColor} 
\end{figure}

We first use spectroscopic redshifts (\S \ref{sec:specz}) to identify
the emission lines if available. If a galaxy has a spectroscopic
redshift outside of the range expected from any NB filters, we remove
it as a contaminant.  

Next, we use photometric redshifts. In the PDR2, the photometric
redshifts from two codes ({\tt DEmP} and {\tt Mizuki},
\cite{hscPhotoZ,hscPhotoZPDR2}) are available. For the NB527, NB718, and
NB973 emitters selected from CHORUS in the UD-COSMOS field, we also
use the photometric redshifts from the COSMOS2015 catalog
\citep{Laigle2016}. For the NB816 and NB921 emitters from HSC-SSP, we
use only the photometric redshifts calculated with HSC five BBs to
keep the consistency among the D/UD fields. 
We rely on the photometric redshifts with $\Delta z_{phot}/(1.0+z_{phot})\leq0.2$, 
where $\Delta z_{phot}$ is a 68\% confidence interval. When the
redshift range estimated from the width of NB filter is overlapped
with the 68\% confidence interval, the redshift is assigned to the
galaxy. In the case that all of the photometric redshifts available
are consistent in each galaxy, the emission-line galaxies are
identified based on the photometric redshift.

Finally, for galaxies not identified with spectroscopic or photometric
redshifts, we use colors to identify the emission
lines. Figure~\ref{fig:ColorColor} shows the color-color diagrams to
distinguish the emission-line from the other possibilities. The color
criteria are determined based on the colors for emission-line galaxies
with spectroscopic redshifts: In NB527, for [OII]
\begin{eqnarray*}
  g-r>1.50(r-i)+0.08
\end{eqnarray*}
otherwise, [OIII] emitters.  
In NB718, first one is for [OII] and the second is for [OIII]
\begin{eqnarray*}
  r-y>0.65(g-i)+0.30,\\
  g-r>1.50(r-i)+0.08,
\end{eqnarray*}
otherwise, H$\alpha$.
In NB816, for [OII], H$\alpha$,
\begin{eqnarray*}
  r-y & > & g-i,\\
  i-z & > & (r-i)-0.27 \lor i-z>0.35(r-i)+0.01,
\end{eqnarray*}
otherwise, [OIII].
In NB921, for H$\alpha$, [OII],
\begin{eqnarray*}
  r-z\leq1.4 & \land & g-r\geq0.3 \land r-z\leq1.24(g-r),\\
  i-y>1.2    & \lor  & i-y>1.02(r-i)+0.15,
\end{eqnarray*}
otherwise, [OIII].
In NB973, for H$\alpha$, [OII],
\begin{eqnarray*}
  g-r\geq0.1 & \land & z-y>2.0(g-r)+1.30,\\
  g-i\leq1.2 & \land & i-y>0.6(g-i)+0.25,
\end{eqnarray*}
otherwise, [OIII].

\input{table5.tex}

There is H$\beta$ emission line near [OIII]. We remove H$\beta$
emission-line galaxies based on the spectroscopic and photometric
redshifts (see also \cite{Hayashi2018a}). It is impossible to
distinguish H$\beta$ from [OIII] based on the color-color
diagrams. Therefore, some of [OIII] emission-line galaxies identified
by the colors can be a contaminant of H$\beta$ emission-line
galaxy. However, figures~\ref{fig:AAOmegaSpectra} and \ref{fig:Nz}
show that contamination of H$\beta$ emission-line galaxies in the
sample of [OIII] emission-line galaxies is small (see also
\cite{Sobral2015,Khostovan2016}). There are [NII] doublet lines close
to H$\alpha$ line, which implies that all of the emission lines can
enter the NB filter simultaneously. Although [NII] lines are weaker
than or comparable to H$\alpha$ at most (e.g., \cite{BPT1981}), we
discuss the contribution of [NII] lines to the measurement of
H$\alpha$ emission-line flux in section~\ref{sec:LFs}.   

To summarize, we identify 75,377 emission-line galaxies with
completeness more than 0.5 using NB816 and NB921 from the HSC-SSP and
NB527, NB718, and NB973 from the CHORUS. The numbers of each
population are listed in table~\ref{tbl:SummaryNBEsample}. 

\subsection{Contamination}

We succeed in reducing the contamination rate
compared to that in the PDR1 catalog. This is within expectations that
figure~\ref{fig:Nz} demonstrates. As with the PDR1 emitter catalogs,
we investigate the contamination rate by applying the photo-$z$ and
color selections to galaxies confirmed with spectroscopic
redshifts. Note that here we do not use the spectroscopic redshifts
obtained by the follow-up spectroscopy
(\S~\ref{sec:specz_Subaru}--\ref{sec:specz_AAT}) for NB emitter
candidates selected from the HSC-SSP PDR1 data. That is, we use the
only spectroscopic redshifts (\S~\ref{sec:specz_literature}) for the
targets unbiased towards the NB emitters to estimate the contamination
rate.  

Among the galaxies with spec-$z$ that meet the photo-$z$ or color
selection, more than 90\% of the galaxies have the spectroscopic
redshifts expected from the wavelength of the NB filter, except for
NB921 [OII] emitters and NB718 H$\alpha$ emitters. The NB921 [OII]
emitters that meets the criteria is 81.5\%, while that for the NB718
H$\alpha$ emitters is 84.2\%. As a result, we estimate the
contamination rate to be less than 10\% for most cases and $\sim20$\%
at most, which suggests that the contamination rate of the emitter
catalogs from the PDR2 is much improved compared with the PDR1
catalogs (13--43\%, \cite{Hayashi2018a}).    

\subsection{Emission-line galaxies with an AGN}
\label{sec:Xray}

In the PDR1 catalog, a small fraction of emission-line galaxies
($\sim$0.1\%) have counterparts in the X-ray. However, we
were not able to fully reject the possibility that
excluding point sources from the PDR1 catalog results in the small
fraction of the X-ray sources in the emitter sample. Since we do not
exclude point sources explicitly to avoid the possible selection bias
in the PDR2 catalogs, we revisit the X-ray counterparts in the
emission-line galaxies selected by the PDR2 data by using the catalog
of 4016 X-ray sources selected from the 4.6Ms X-ray data by the
Chandra COSMOS Legacy Survey \citep{Marchesi2016,Civano2016}.  
Additionally, we find counterparts in a catalog of 2937 radio-selected
AGNs from the VLA-COSMOS 3 GHz Large Project \citep{Delvecchio2017}. 

We use 3534 X-ray sources that have an optical/IR counterpart located
in the UD-COSMOS field. Then, we match 14,754 emission-line galaxies
in the UD-COSMOS field with a completeness larger than 0.5 to the X-ray
sources. As a results, 66 emission-line galaxies have an X-ray
counterpart ($\sim$0.45\%). Although the fraction is slightly larger
than that from the PDR1 catalog, the fraction of emission-line
galaxies detected in the X-rays is undoubtfully small. Among 66 X-ray
sources, 13 sources are H$\alpha$ emitters, 20 sources are [OIII]
emitters, and 33 sources are [OII] emitters. 
Next, 2639 radio-selected AGNs in the UD-COSMOS field are matched with
the emission-line galaxies. After removing the 21 radio sources that
are also matched X-ray sources, there remains 30 emission-line
galaxies found as a radio-selected AGN. Among them, 3 sources are
H$\alpha$ emitters, 11 sources are [OIII] emitters, and 16 sources are
[OII] emitters. As a result, about 0.65\%\ of the emission-line
galaxies have an AGN selected in X-ray and/or radio.
As shown in table~\ref{tbl:RedshiftSurveyed}, the [OII] emitters tend
to be at higher redshifts, and thus the luminosities of emission-line are
likely larger than those of H$\alpha$ emitters at lower
redshifts. Although some emitters have the luminosity of emission line
larger than $L^*$ derived in \S~\ref{sec:LFs}, many emitters have the
luminosity lower than $L^*$. Therefore, the emission-line galaxies
with the X-ray/radio counterpart do not necessarily dominate the bright end
of luminosity function. 

As shown in figure~\ref{fig:AAOmegaSpectra}, our follow-up
spectroscopy succeeds in detecting multiple emission lines from
the emitters. We note that the AAT/AAOmega spectroscopic data can be 
used to investigate further the fraction of AGN in the H$\alpha$
emitters at $z=0.246$ and 0.404 based on the BPT diagram with
H$\beta$, [OIII], H$\alpha$, and [NII] emission lines
\citep{BPT1981}. We will address this in the forthcoming paper. 

\subsection{Stellar mass}
As with the PDR1 catalog \citep{Hayashi2018a}, stellar mass for the
emission-line galaxies is calculated by spectral energy distribution
(SED) fit with five HSC BB data at the fixed redshift using the code
with Bayesian priors ({\tt Mizuki}: \cite{Tanaka2015}). If available,
the redshift is fixed to a spectroscopic redshift, otherwise, the
redshift is assumed to that estimated from the central wavelength of
NB. The model SED templates of galaxies are generated by the code of
\citet{BruzualCharlot2003}, and nebular emission lines are taken into
account. Solar metallicity and the extinction curve of
\citet{Calzetti2000} are adopted. Readers should refer to Section 3.6
of \citet{Hayashi2018a} for more details.

\subsection{Spatial distribution}

\begin{figure*}
 \begin{center}
   \includegraphics[width=0.32\textwidth, bb=0 0 461 461]{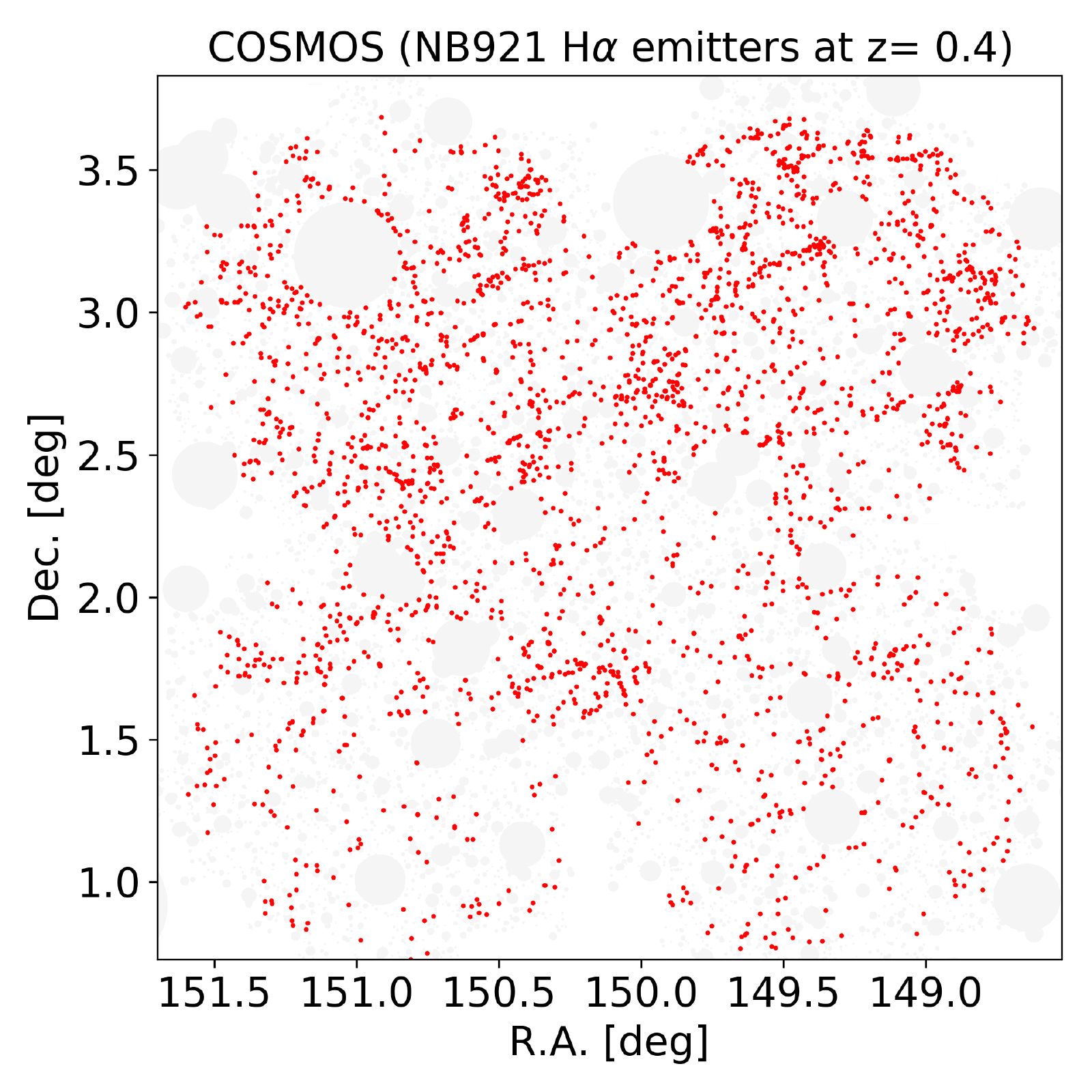} 
   \includegraphics[width=0.32\textwidth, bb=0 0 461 461]{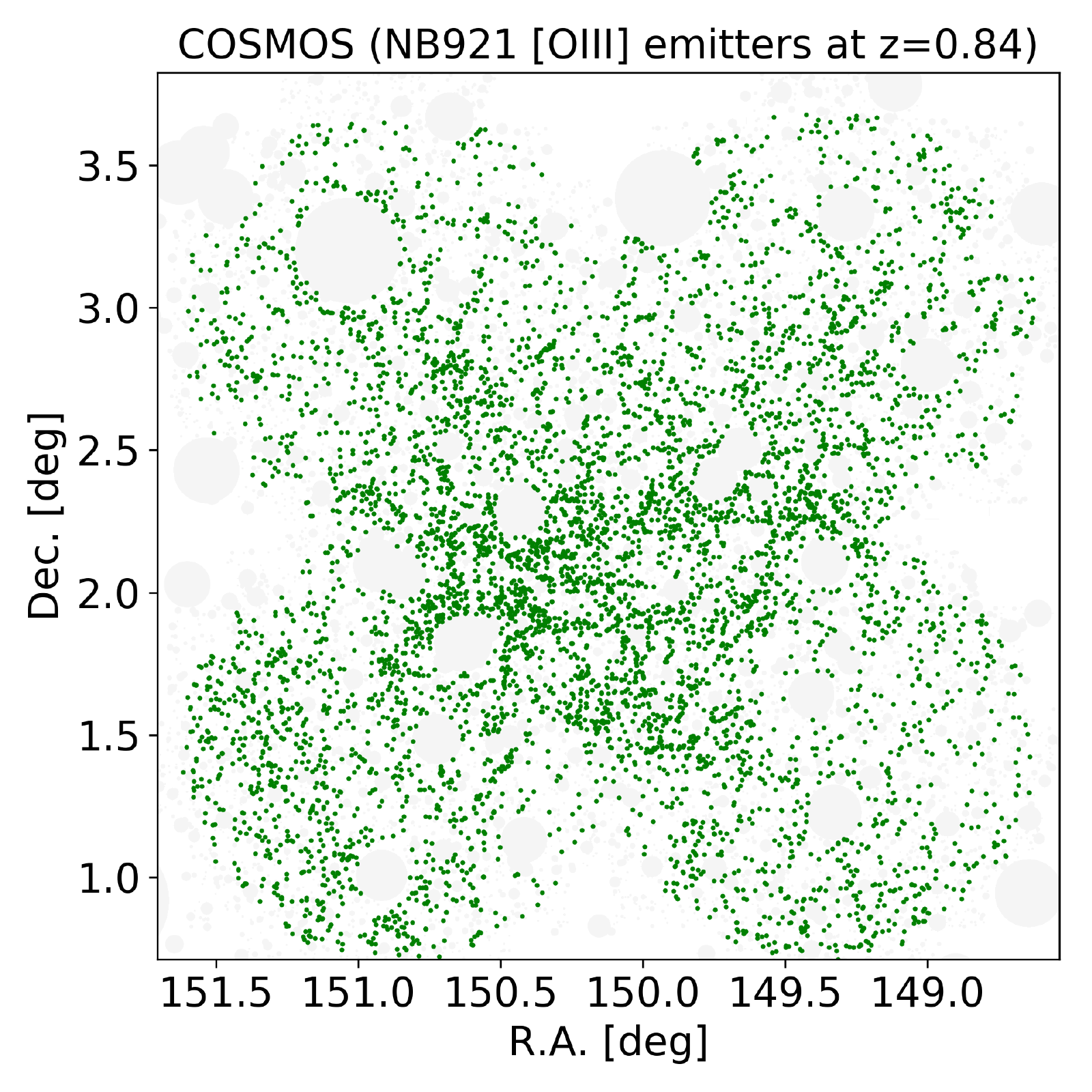} 
   \includegraphics[width=0.32\textwidth, bb=0 0 461 461]{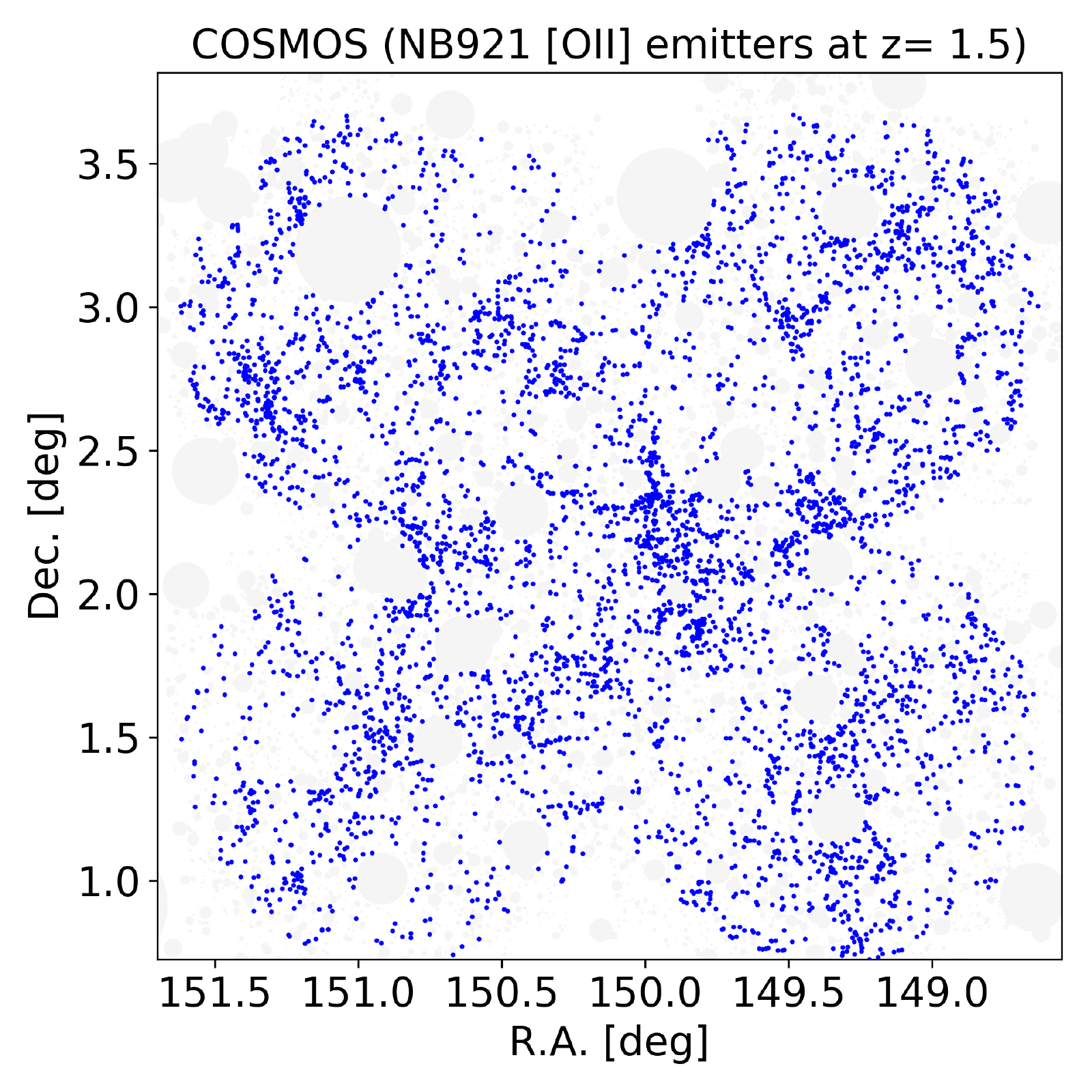} 
 \end{center}
 \caption{The spatial distribution of NB921-selected H$\alpha$,
   [OIII], and [OII] emission-line galaxies in the COSMOS field. The
   masked regions are shown in gray. The distribution of the other
   emission-line galaxies selected in this study is shown in
   Appendix~\ref{appendix:Map}.}\label{fig:MapNB921COSMOS}  
\end{figure*}

Figure~\ref{fig:MapNB921COSMOS} shows the spatial distributions of
H$\alpha$, [OIII], and [OII] emission-line galaxies in the COSMOS
field. The spatial distributions of emission-line galaxies in the
other fields are shown in the figures in
Appendix~\ref{appendix:Map}. The distributions are consistent with
those from the PDR1 catalogs \citep{Hayashi2018a}.  

Recently, several studies report the large-scale structures revealed
by the large spectroscopic surveys at redshifts matched with those of
our NB emitters (table~\ref{tbl:RedshiftSurveyed}). 
\citet{Paulino-Afonso2018} show a super cluster at $z=0.84$ confirmed
by VLT/VIMOS spectroscopy, whose redshift is covered by NB921 [OIII]
emitters.  
\citet{Hasinger2018} show a redshift spike at $z=1.458$ confirmed by
Keck/DEIMOS spectroscopy, whose redshift is covered by NB921 [OII]
emitters.  
\citet{Boehm2020} spectroscopically confirm two overdensities of NB921
[OII] emitters at $z=1.47$ selected from the HSC-SSP PDR1 data using
VLT/KMOS. 
Our samples of emission-line galaxies reproduce all of the structures
spectroscopically confirmed, suggesting the effectiveness of NB
imaging survey to reveal the large-scale structures at specific
redshifts.  

\section{Emission-line luminosity functions}
\label{sec:LFs}

\begin{figure*}
 \begin{center}
   \includegraphics[width=0.45\textwidth, bb=0 0 461 346]{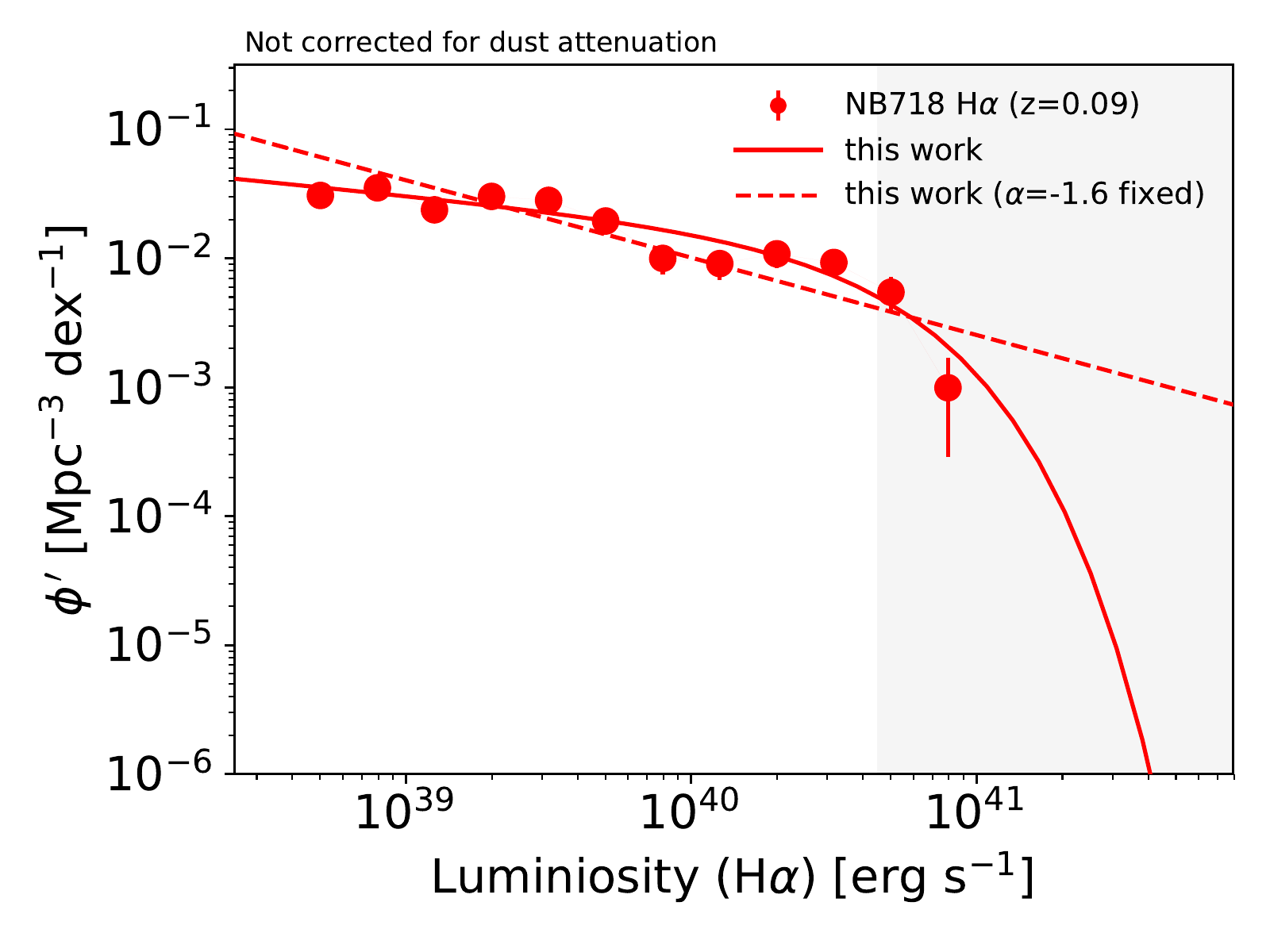} 
   \includegraphics[width=0.45\textwidth, bb=0 0 461 346]{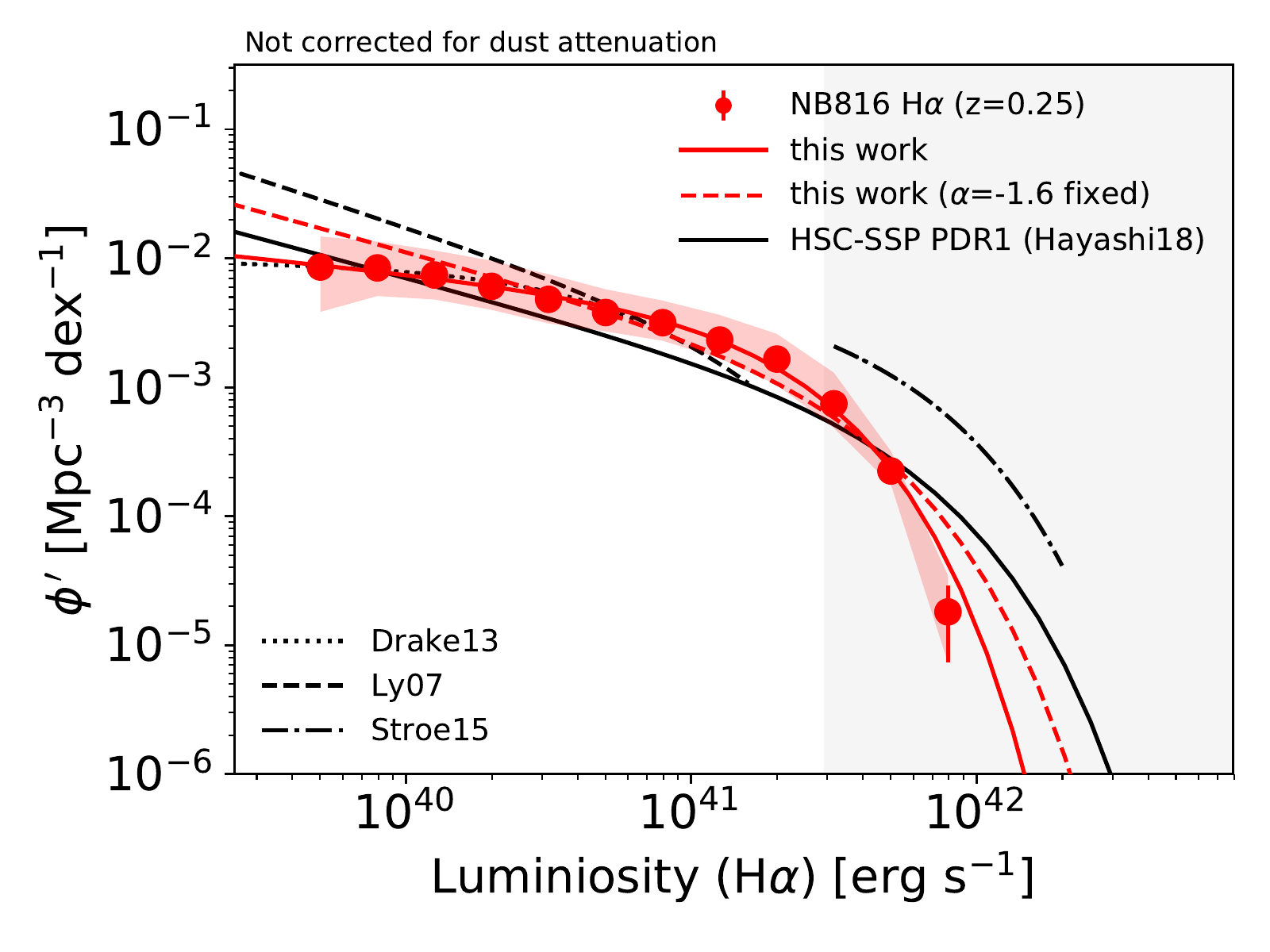} 
   \includegraphics[width=0.45\textwidth, bb=0 0 461 346]{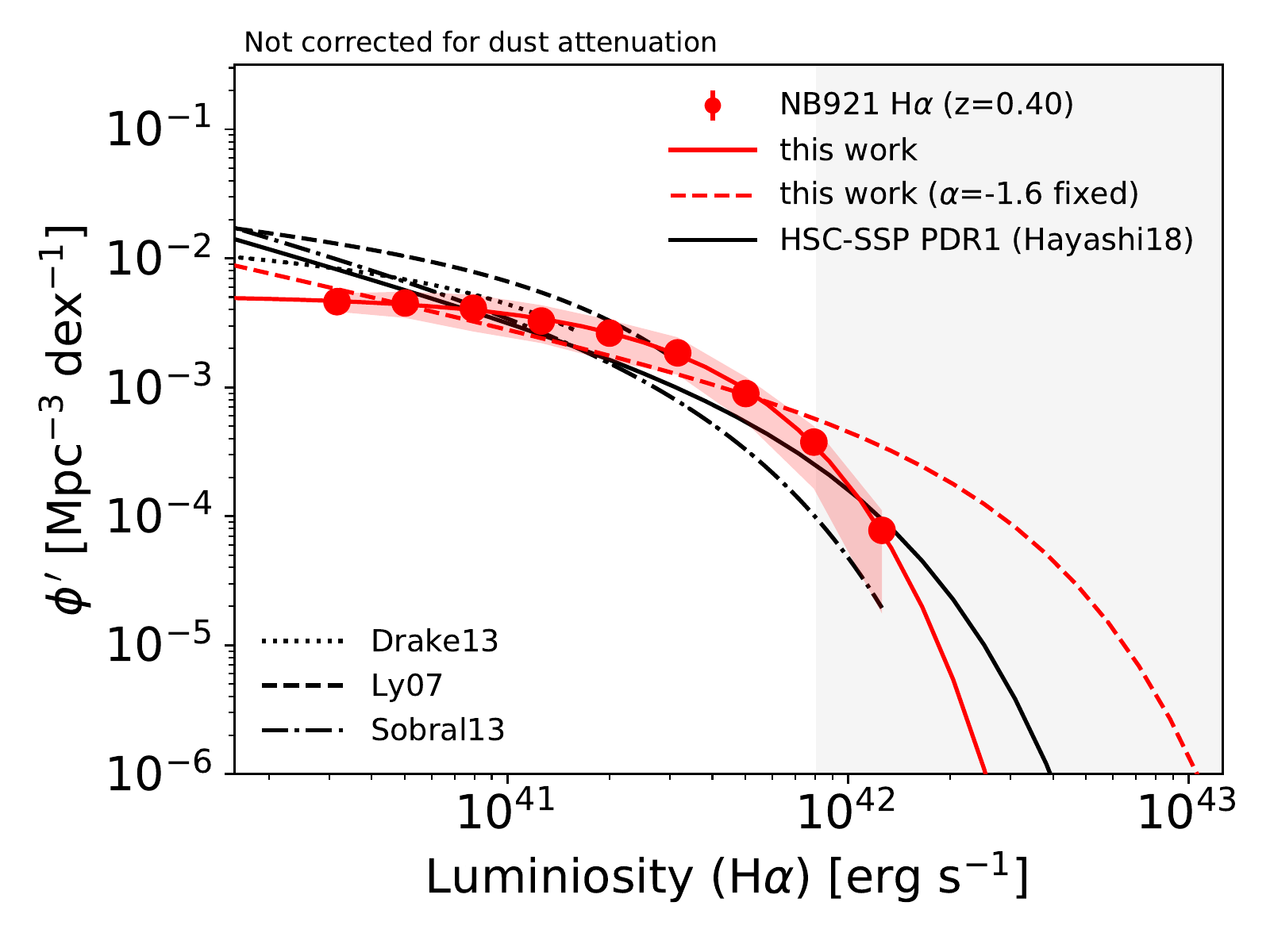} 
   \includegraphics[width=0.45\textwidth, bb=0 0 461 346]{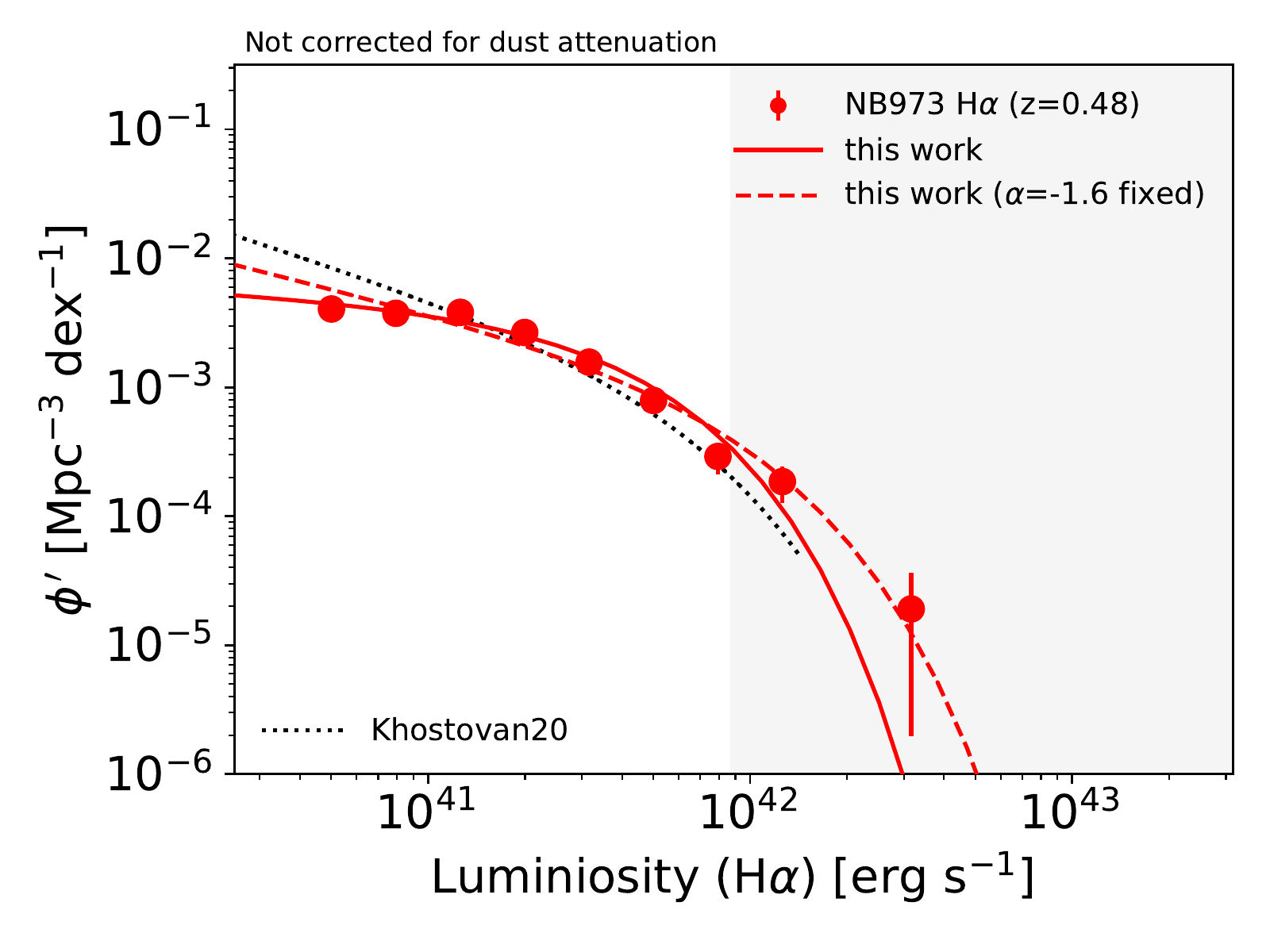} 
 \end{center}
 \caption{The luminosity functions of H$\alpha$ emission-line
   galaxies. The filled circles show the measurement with equation
   (\ref{eq:LFcorrected}) and the error bars show the Poisson
   error. The pale red region shows the variation in the four discrete
   areas of HSC-SSP D/UD layer. Note that since the NB527, NB718, and
   NB973 data from the CHORUS survey are available in the UD-COSMOS
   field only, the pale region cannot be drawn for the luminosity
   functions of the emitters selected from the three NB data. The red
   line is a Schechter function fitted to the measurement with free
   parameters of $L^*$, $\phi^{\prime *}$ and $\alpha$. The dashed line is also
   a Schechter function fitted with $\alpha$ fixed to be $-1.6$. The
   black line is the result from HSC-SSP PDR1 by \citet{Hayashi2018a}.
   The black dotted line, dashed line, and dash-dotted line are the
   luminosity functions from the literature
   \citep{Ly2007,Drake2013,Sobral2013,Stroe2015,Khostovan2020}, where
   the luminosity range covering the data in each study is shown.
   The gray area shows the luminosity range where the completeness
   could be lower than what we estimate because of the magnitude cut
   of 18.5 mag. 
}\label{fig:LF_ha}     
\end{figure*}

\begin{figure*}
 \begin{center}
   \includegraphics[width=0.45\textwidth, bb=0 0 461 346]{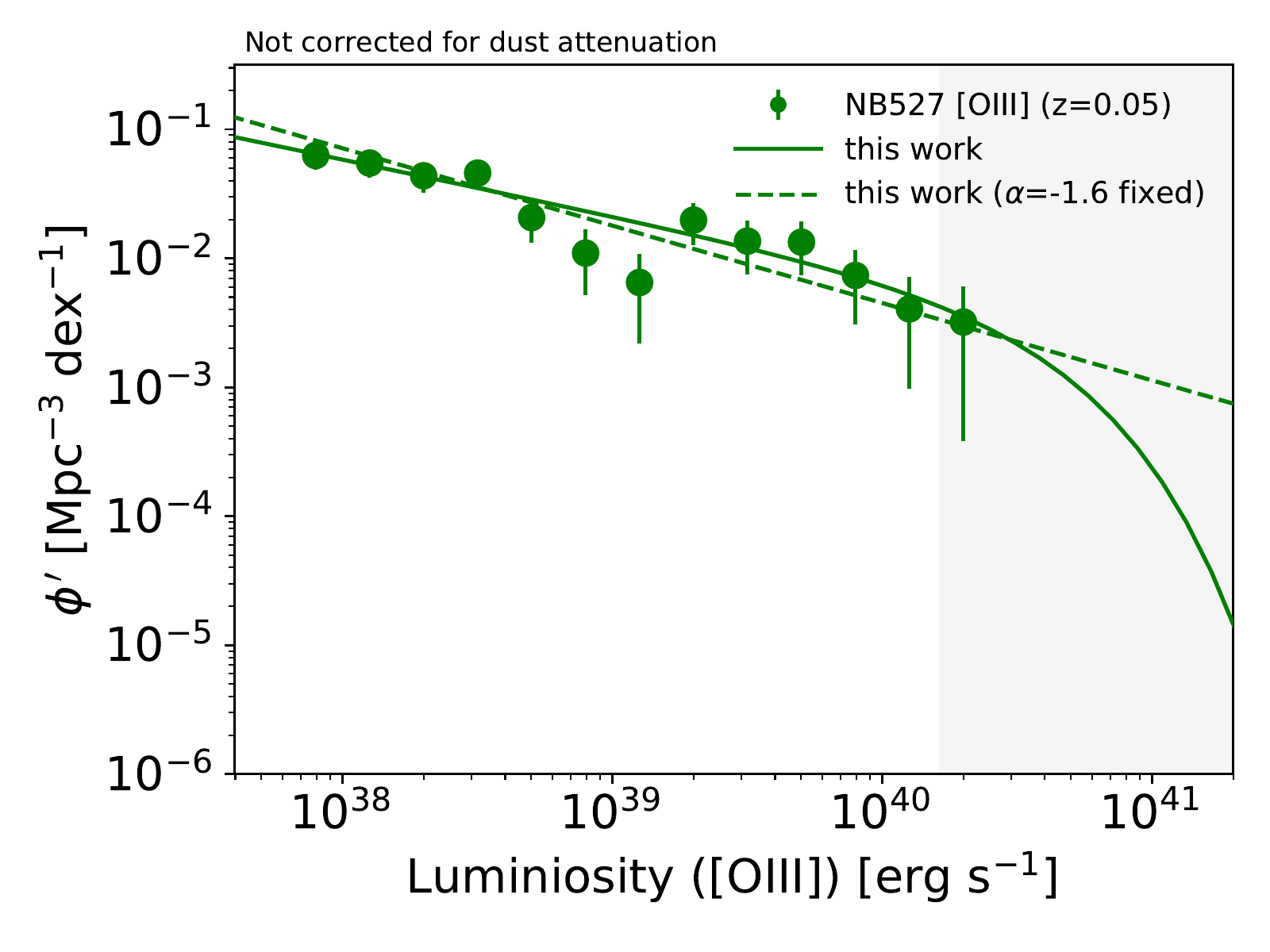} 
   \includegraphics[width=0.45\textwidth, bb=0 0 461 346]{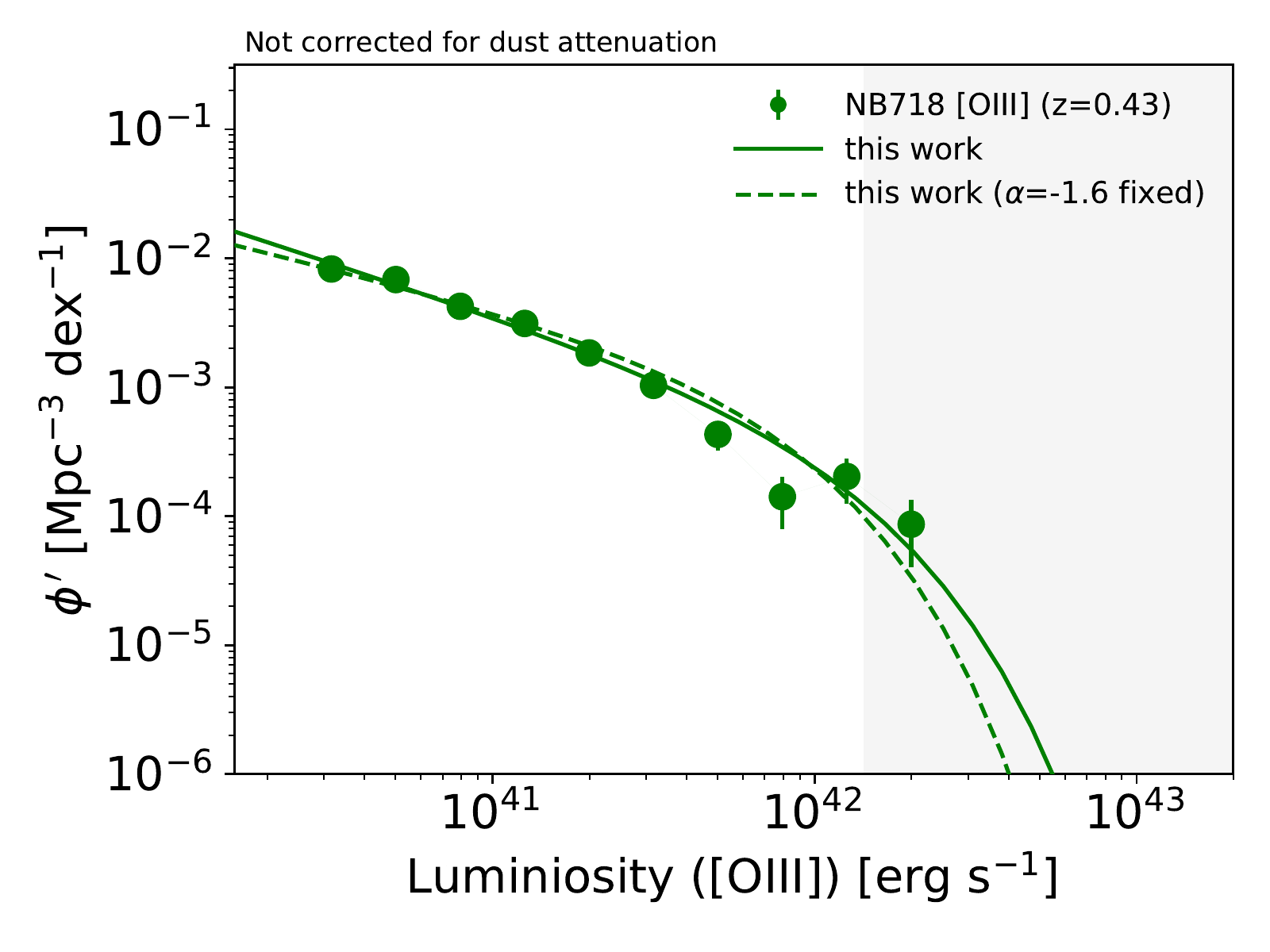} 
   \includegraphics[width=0.45\textwidth, bb=0 0 461 346]{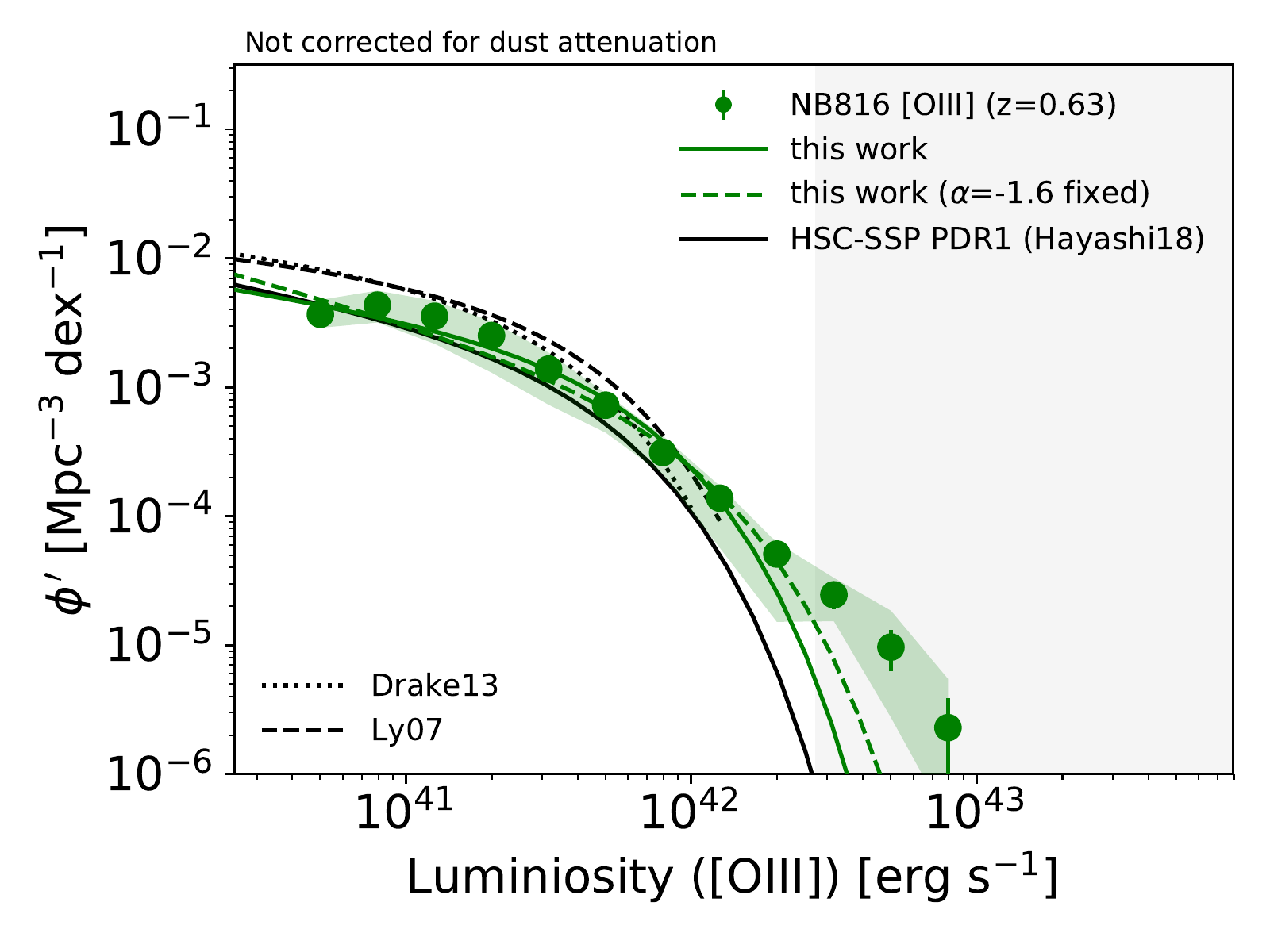} 
   \includegraphics[width=0.45\textwidth, bb=0 0 461 346]{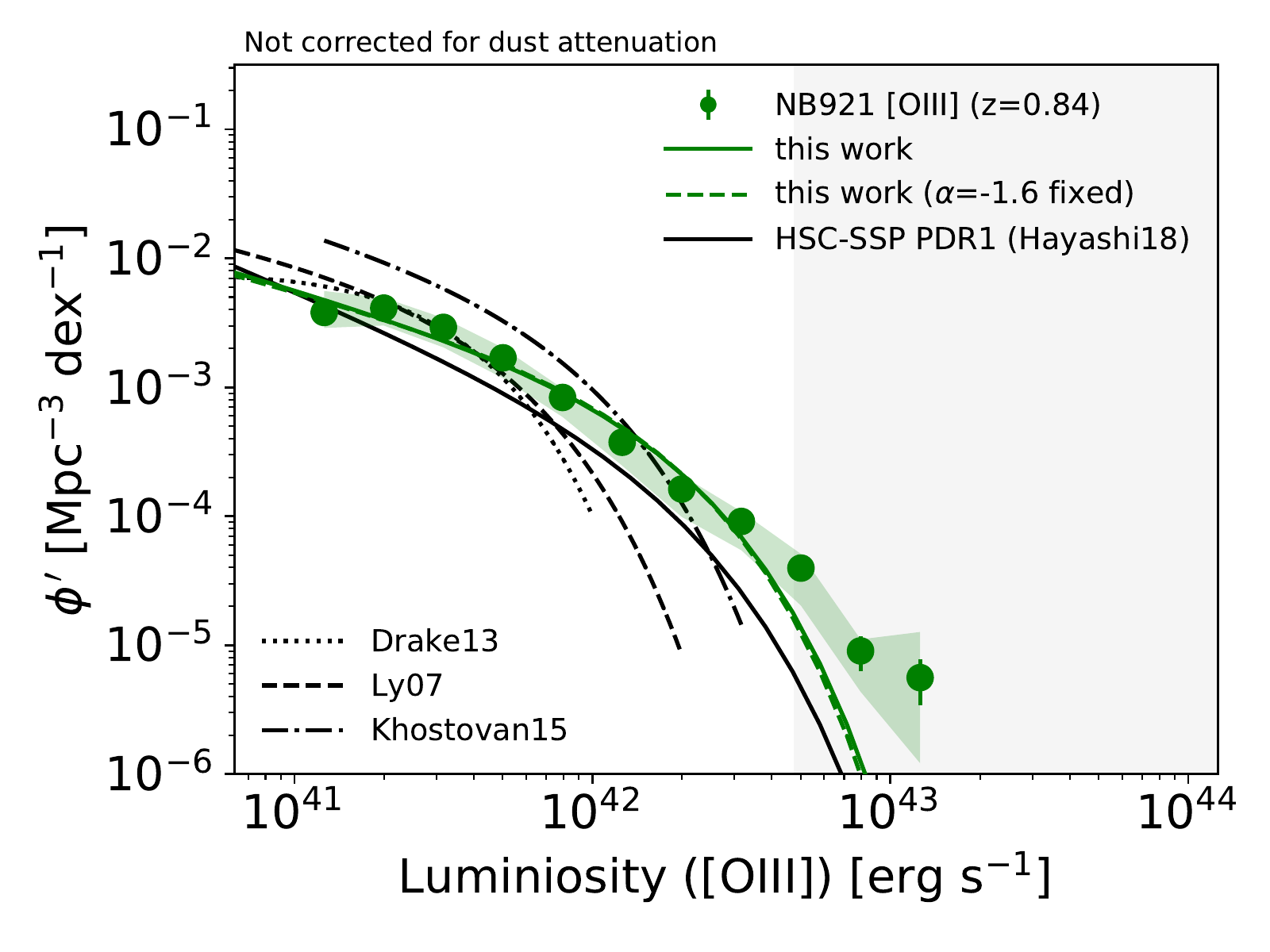} 
   \includegraphics[width=0.45\textwidth, bb=0 0 461 346]{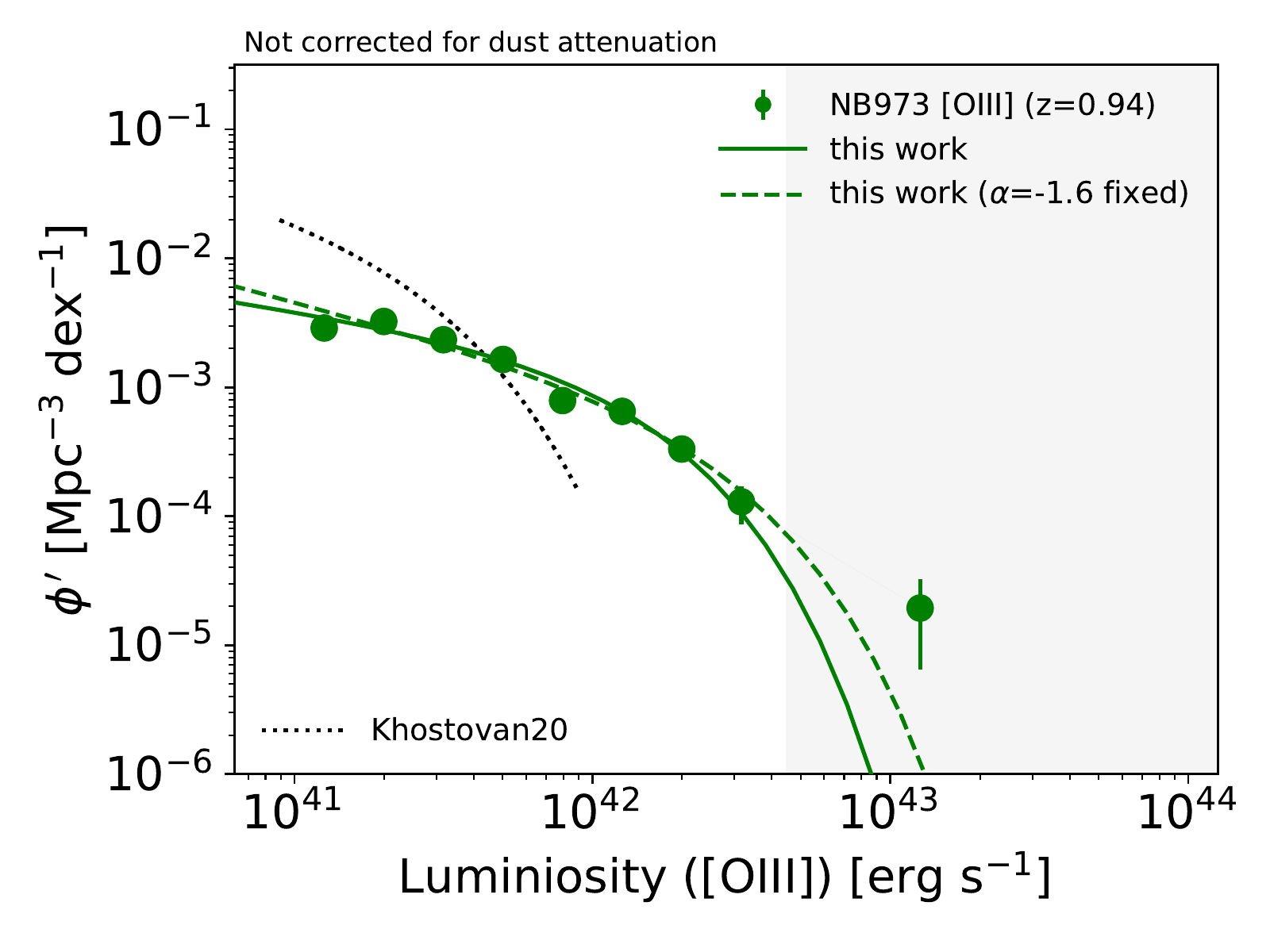} 
 \end{center}
 \caption{The same as fig~\ref{fig:LF_ha}, but for [OIII]
   emission-line galaxies. The luminosity functions from the
   literature are also shown for comparison \citep{Ly2007,Drake2013,Khostovan2015,Khostovan2020}.
 }\label{fig:LF_oiii}  
\end{figure*}

\begin{figure*}
 \begin{center}
   \includegraphics[width=0.45\textwidth, bb=0 0 461 346]{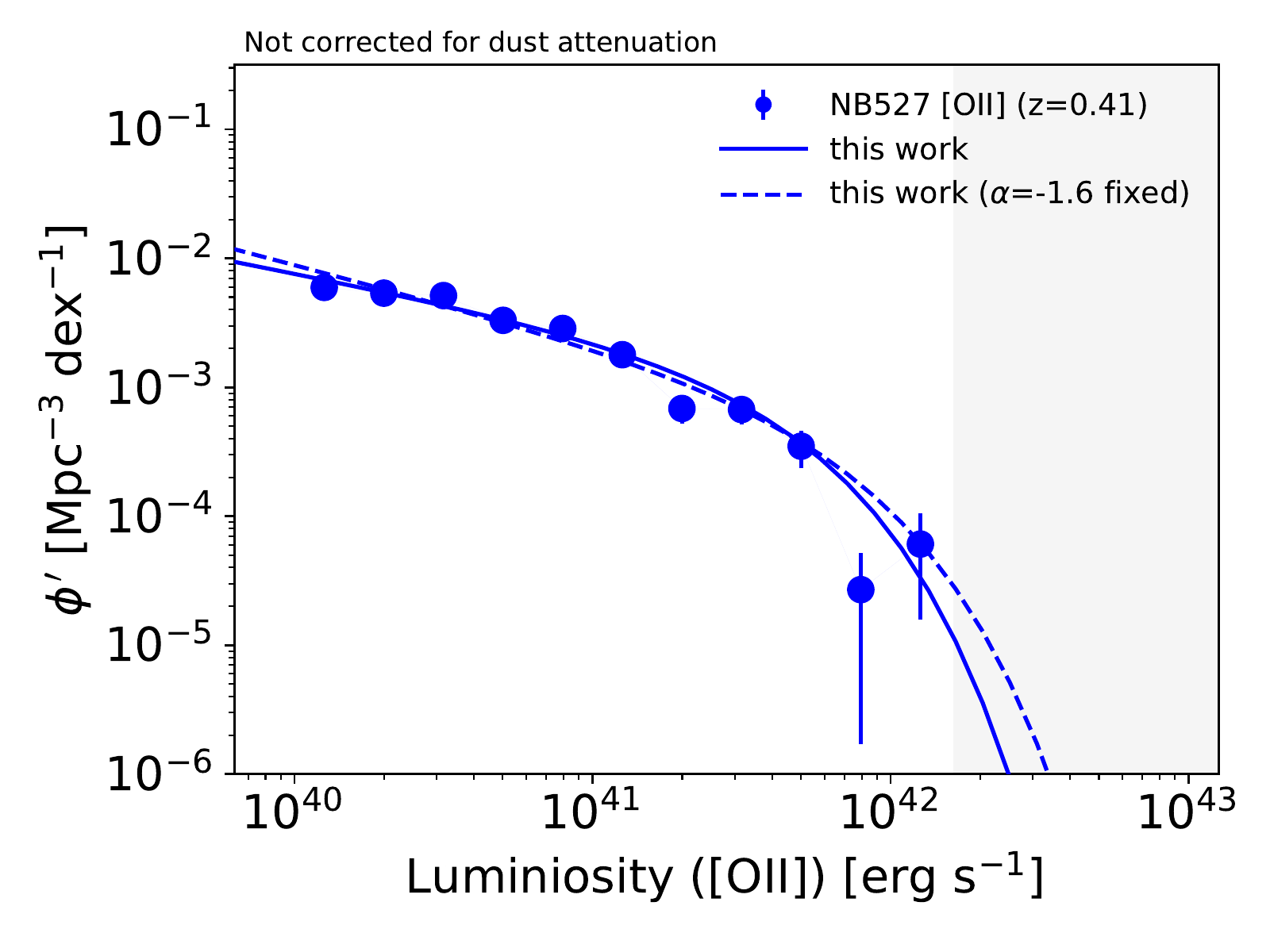} 
   \includegraphics[width=0.45\textwidth, bb=0 0 461 346]{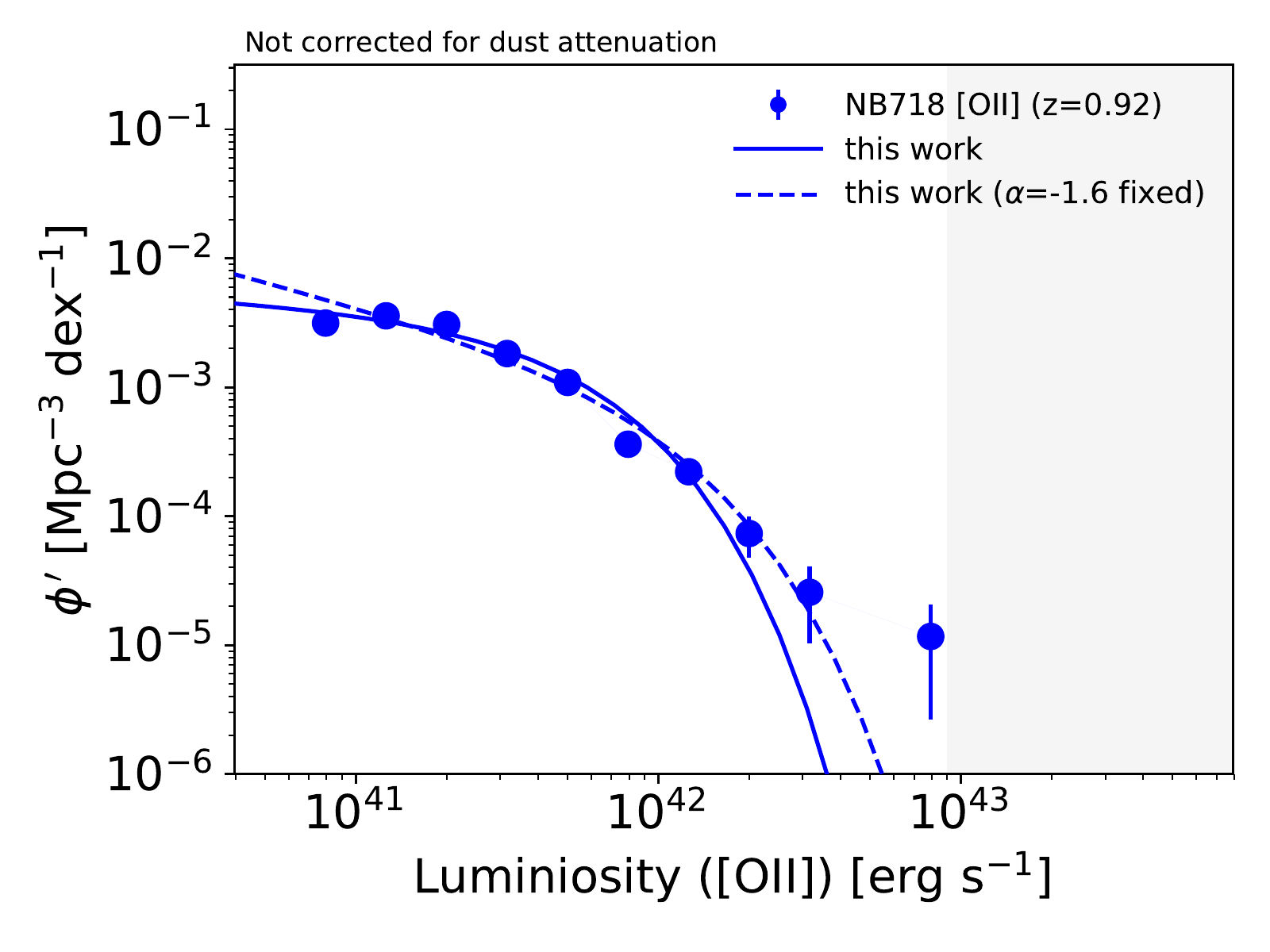} 
   \includegraphics[width=0.45\textwidth, bb=0 0 461 346]{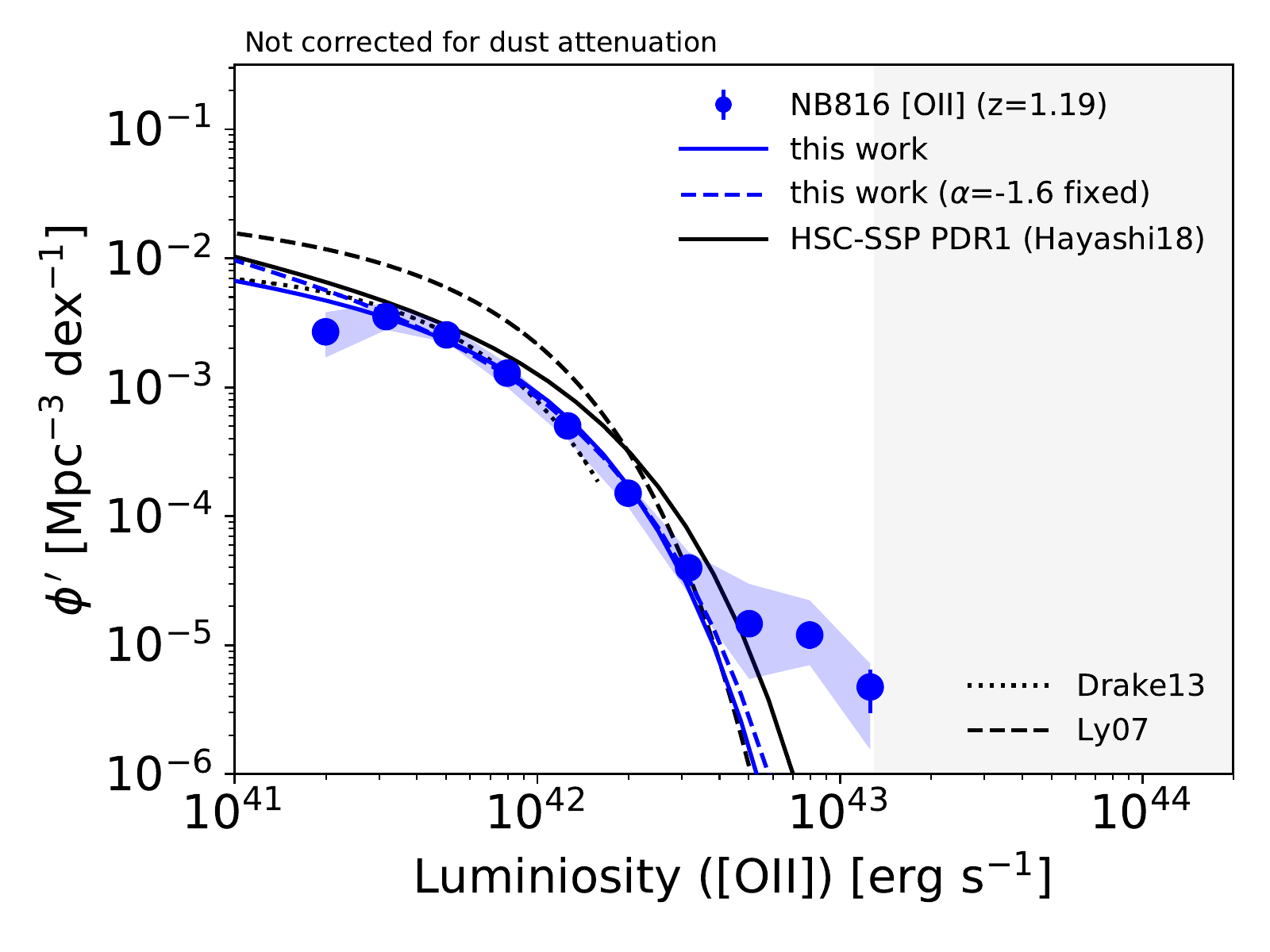} 
   \includegraphics[width=0.45\textwidth, bb=0 0 461 346]{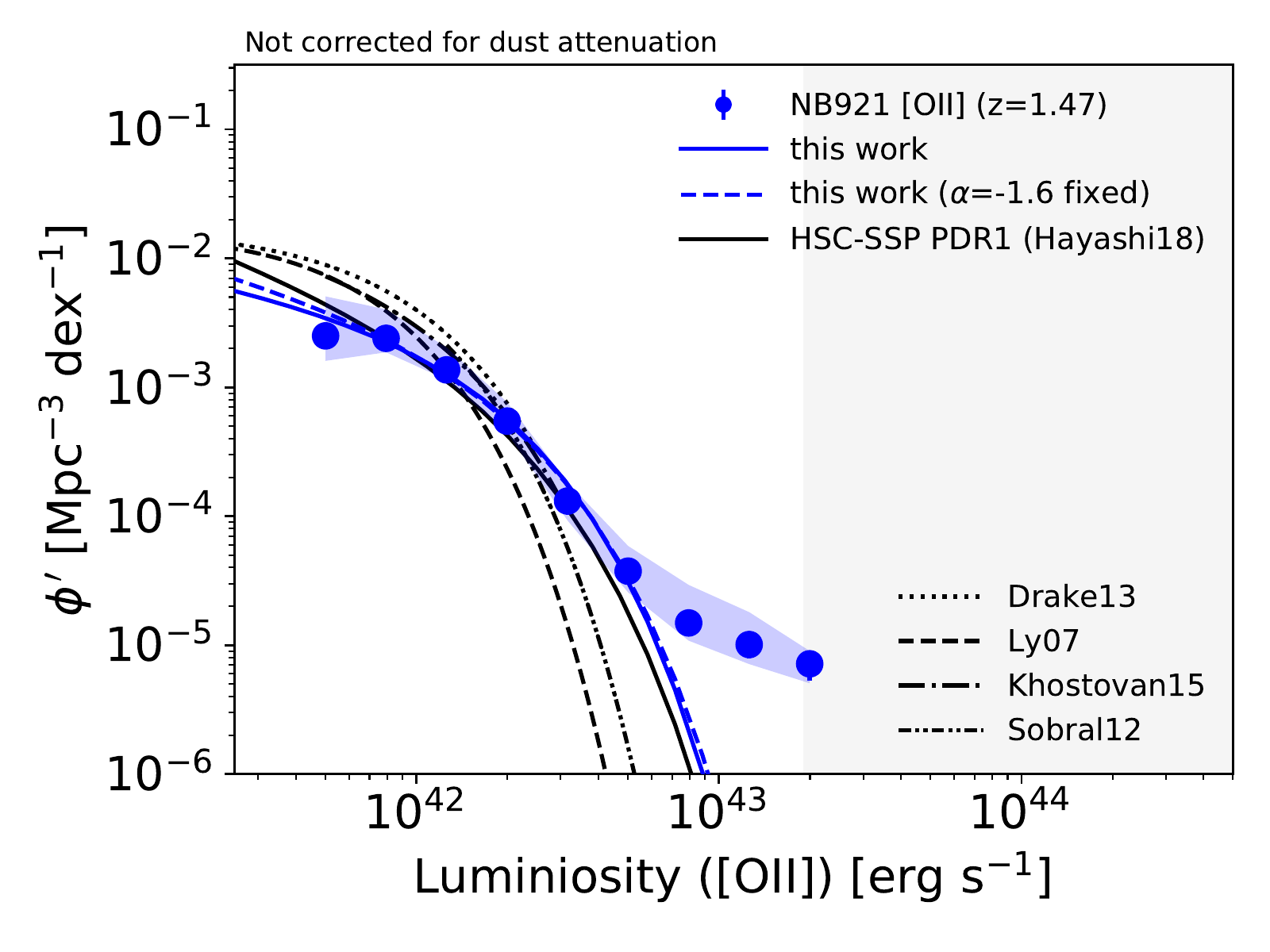} 
   \includegraphics[width=0.45\textwidth, bb=0 0 461 346]{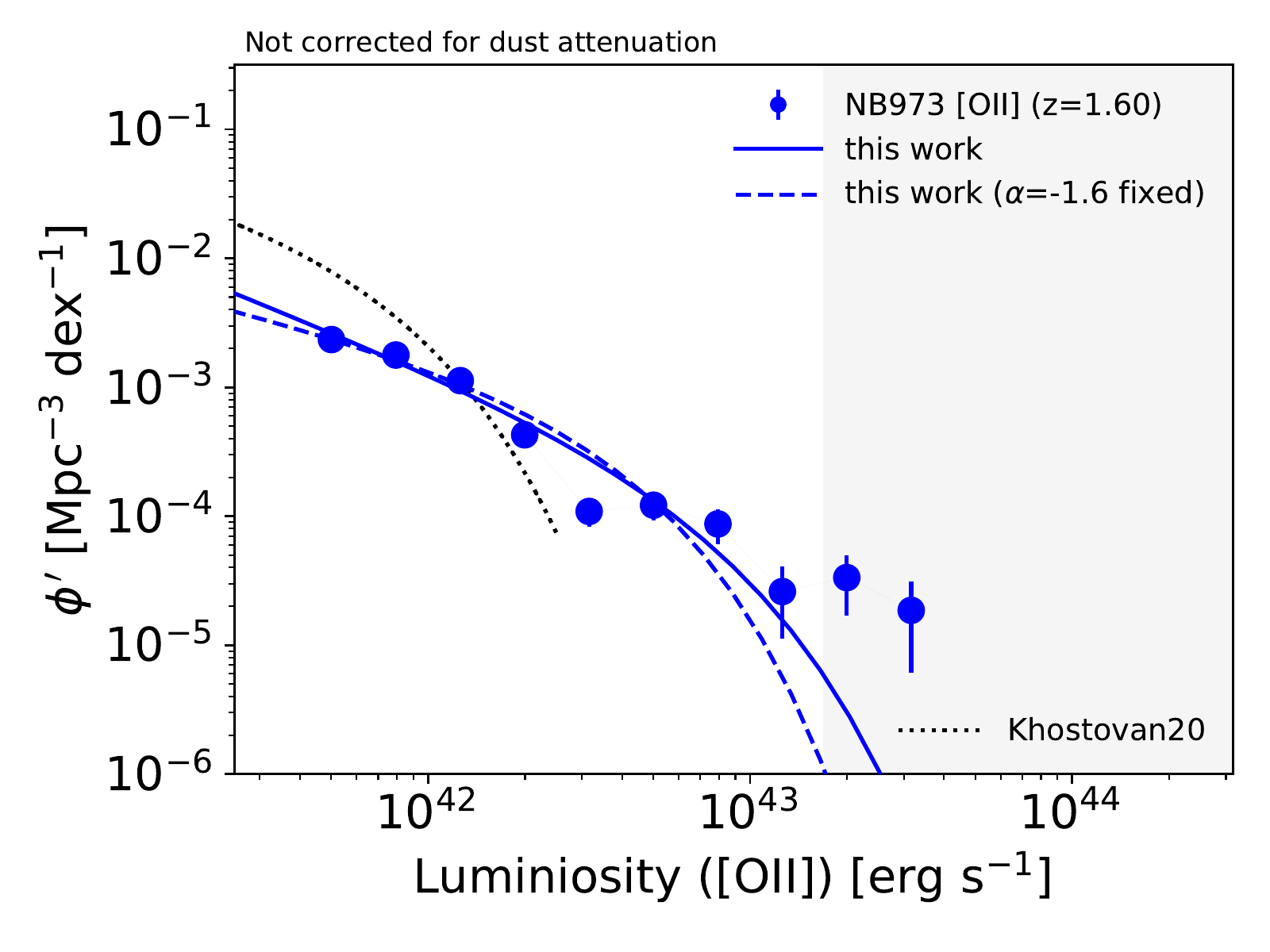} 
 \end{center}
 \caption{The same as fig~\ref{fig:LF_ha}, but for [OII] emission-line
   galaxies. The luminosity functions from the literature are also
   shown for comparison \citep{Ly2007,Drake2013,Sobral2012,Khostovan2015,Khostovan2020}. 
}\label{fig:LF_oii}  
\end{figure*}

\begin{figure*}
 \begin{center}
   \includegraphics[width=0.45\textwidth, bb=0 0 461 346]{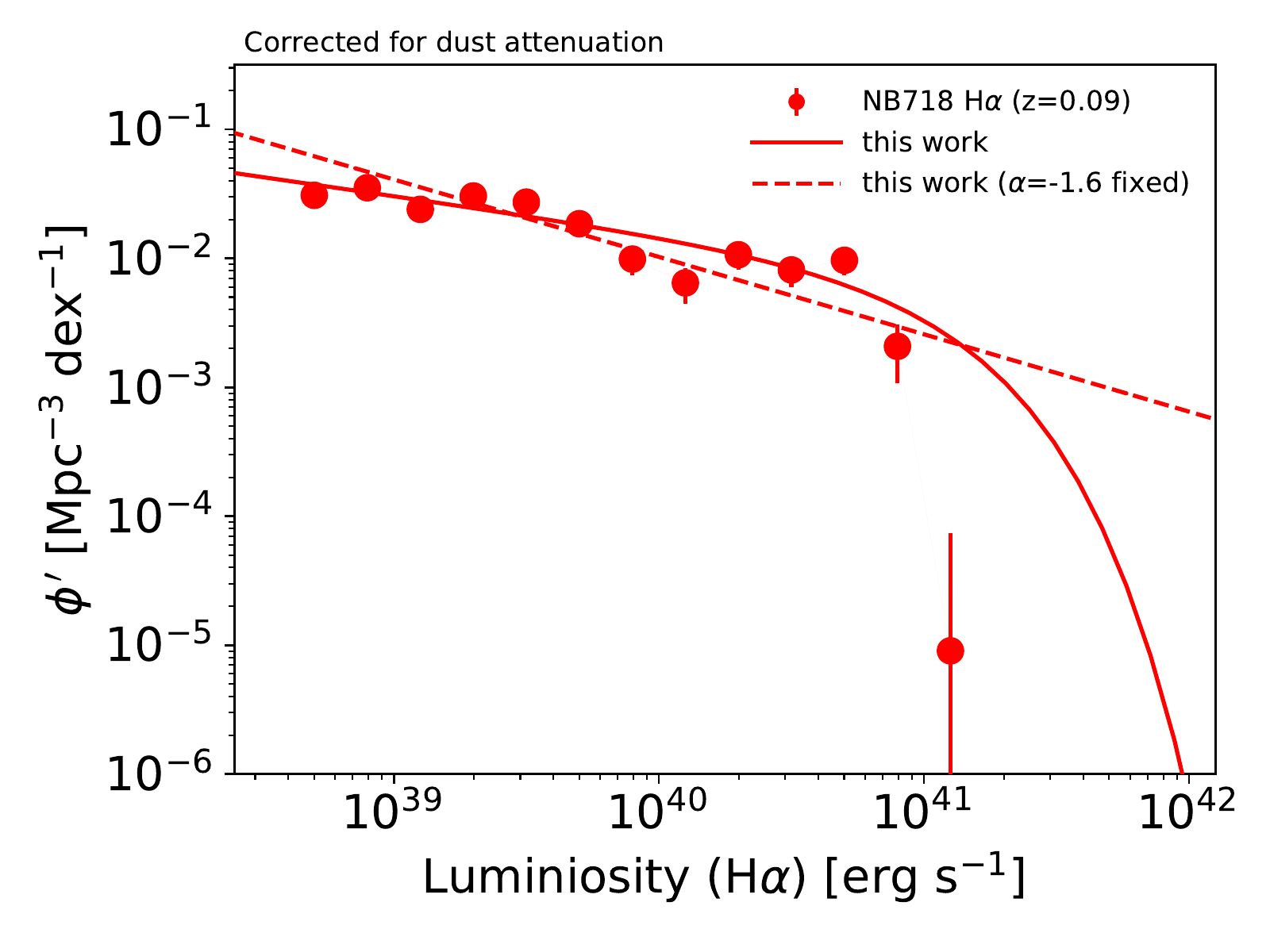} 
   \includegraphics[width=0.45\textwidth, bb=0 0 461 346]{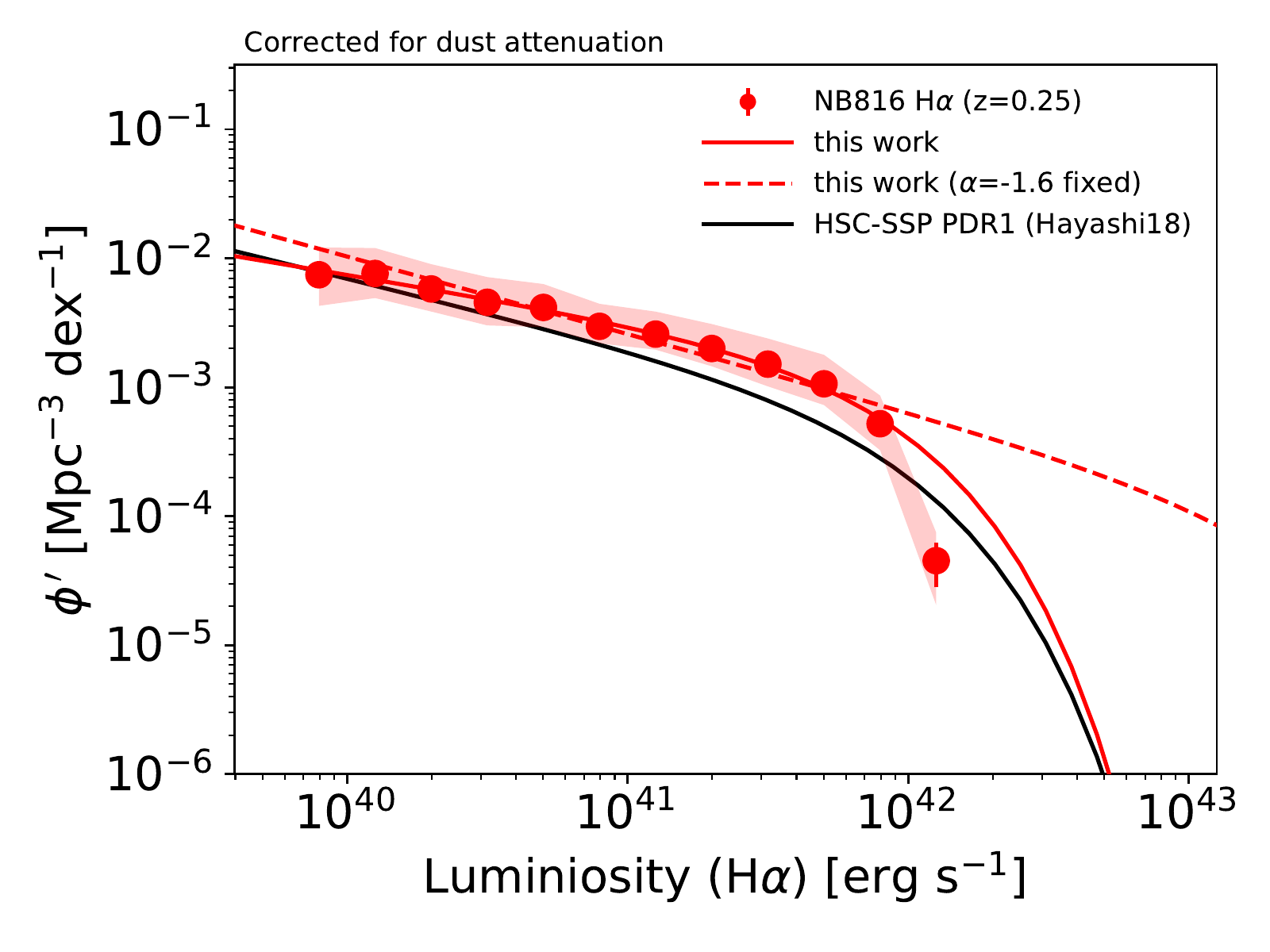} 
   \includegraphics[width=0.45\textwidth, bb=0 0 461 346]{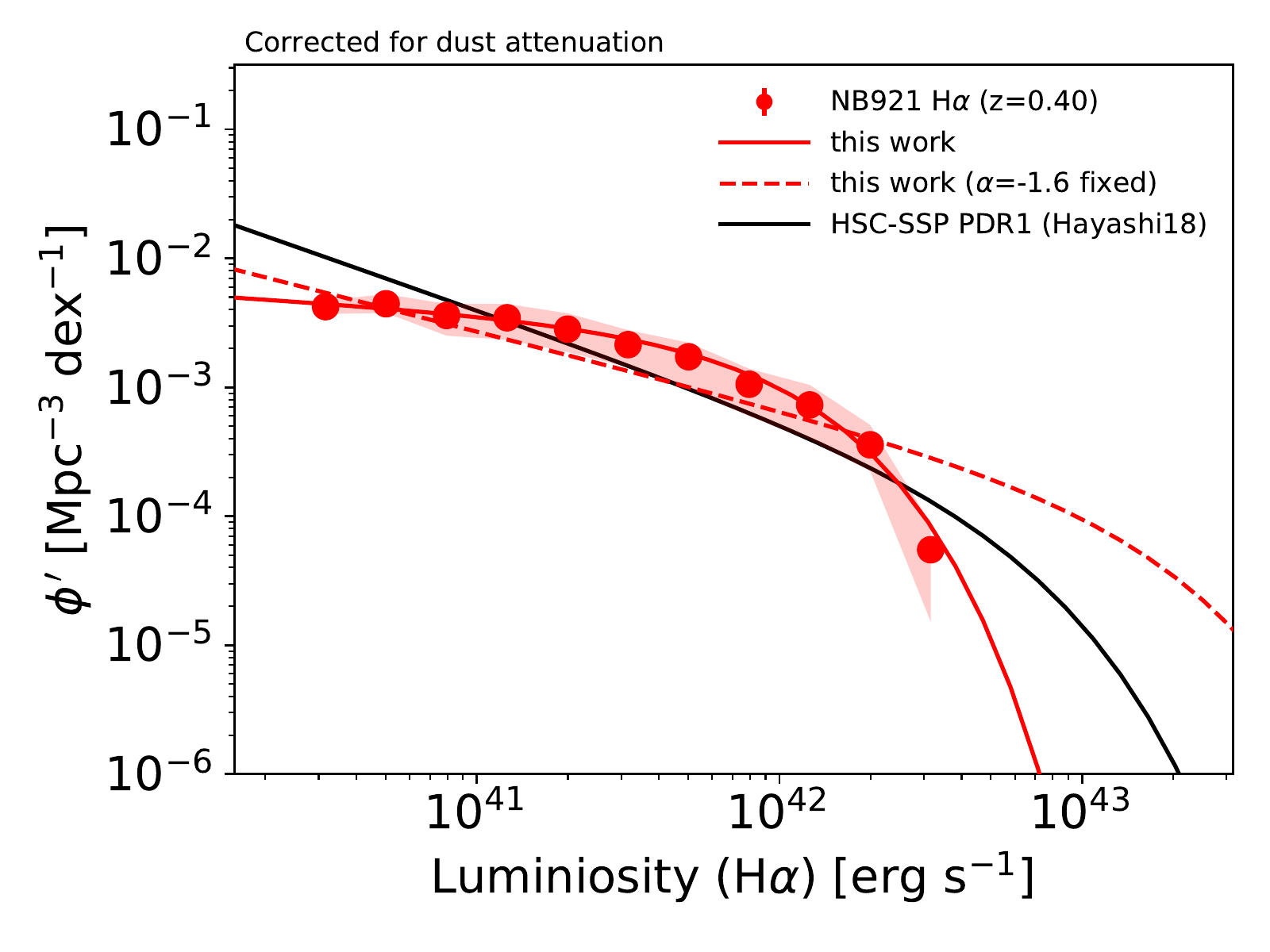} 
   \includegraphics[width=0.45\textwidth, bb=0 0 461 346]{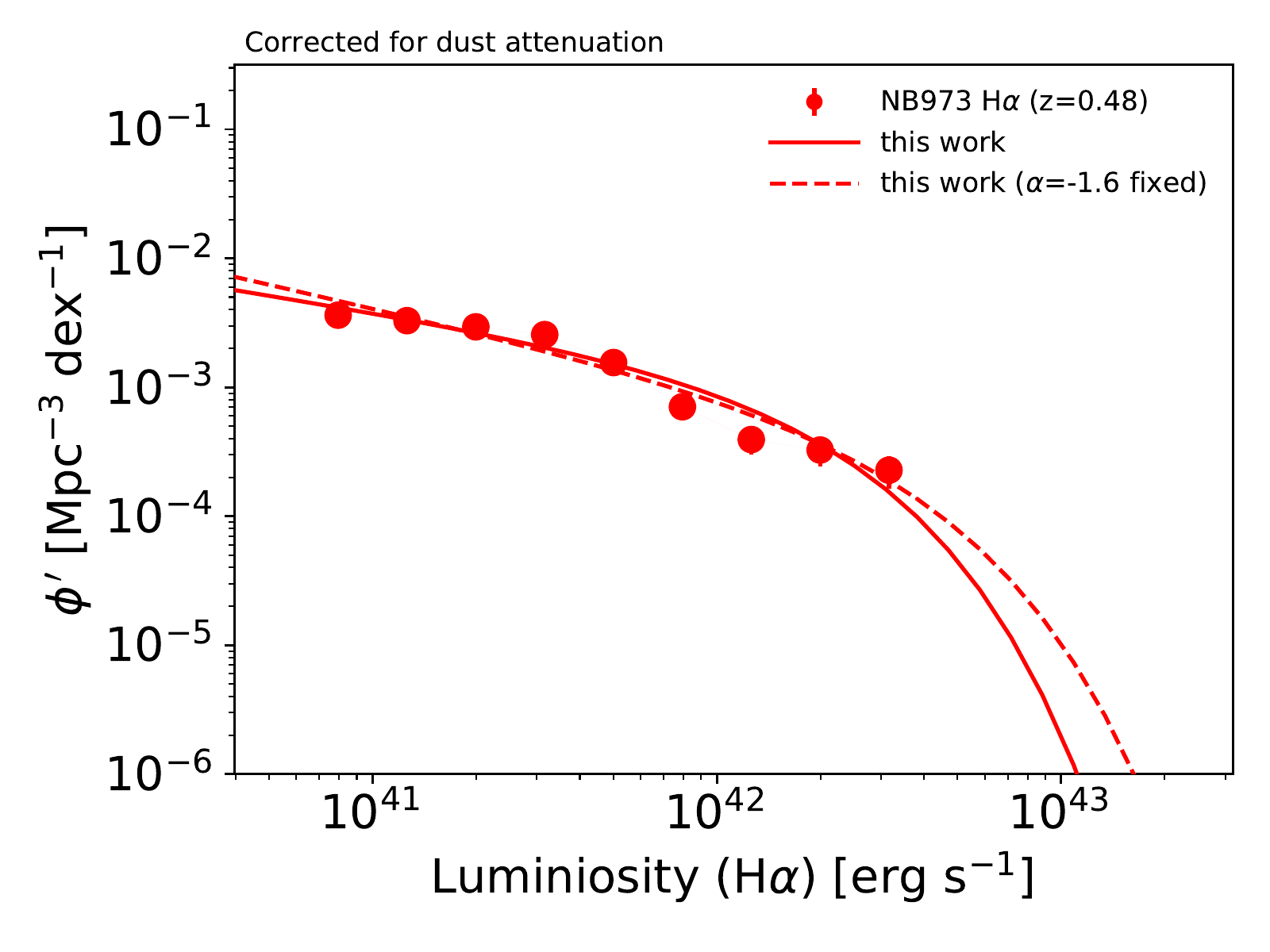} 
 \end{center}
 \caption{The same as fig~\ref{fig:LF_ha}, but the luminosity
   functions are corrected for dust attenuation. See the text for the
   details of how to correct for the dust attenuation.}\label{fig:LFcorrected_ha}  
\end{figure*}

\begin{figure*}
 \begin{center}
   \includegraphics[width=0.45\textwidth, bb=0 0 461 346]{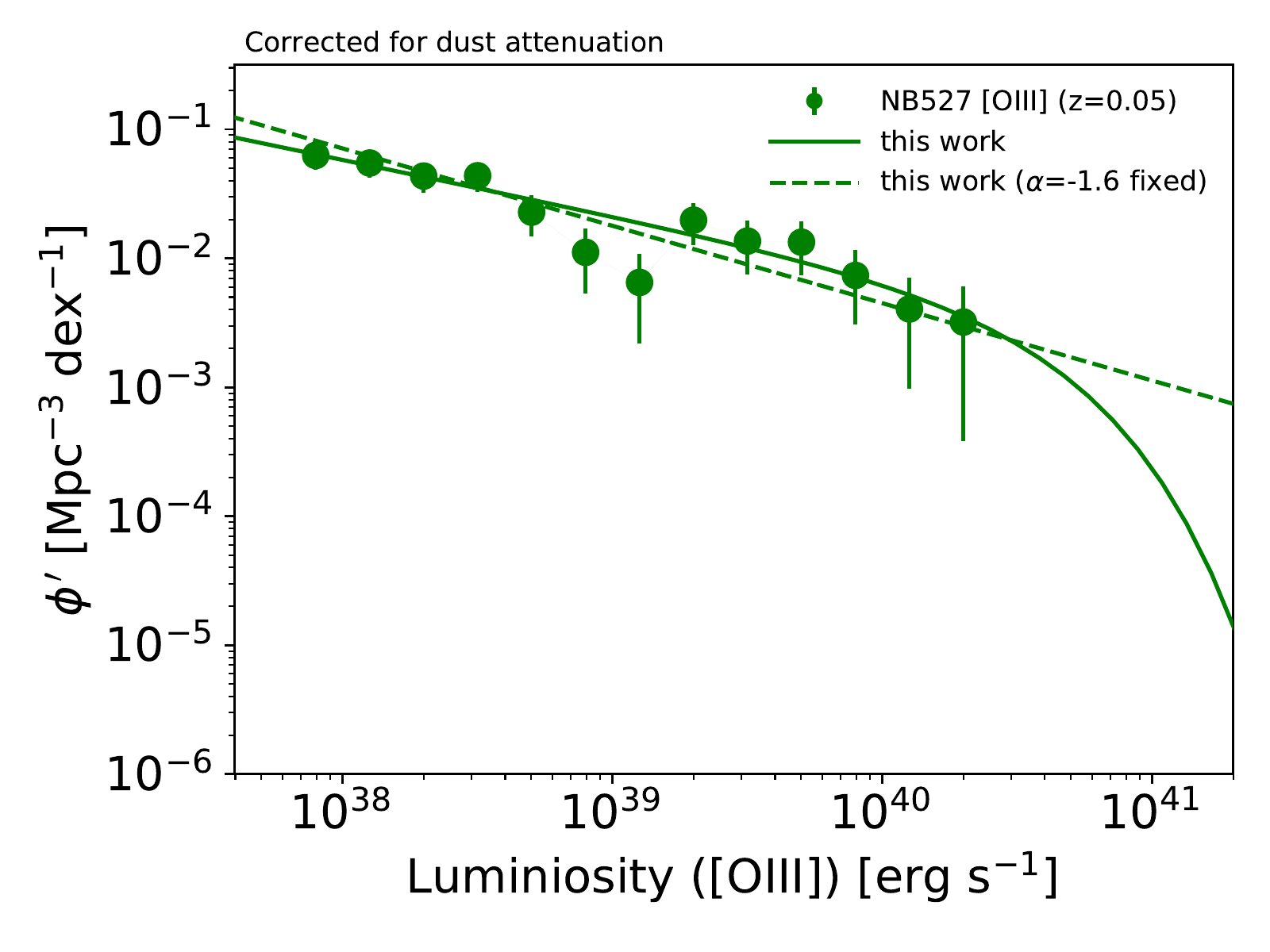} 
   \includegraphics[width=0.45\textwidth, bb=0 0 461 346]{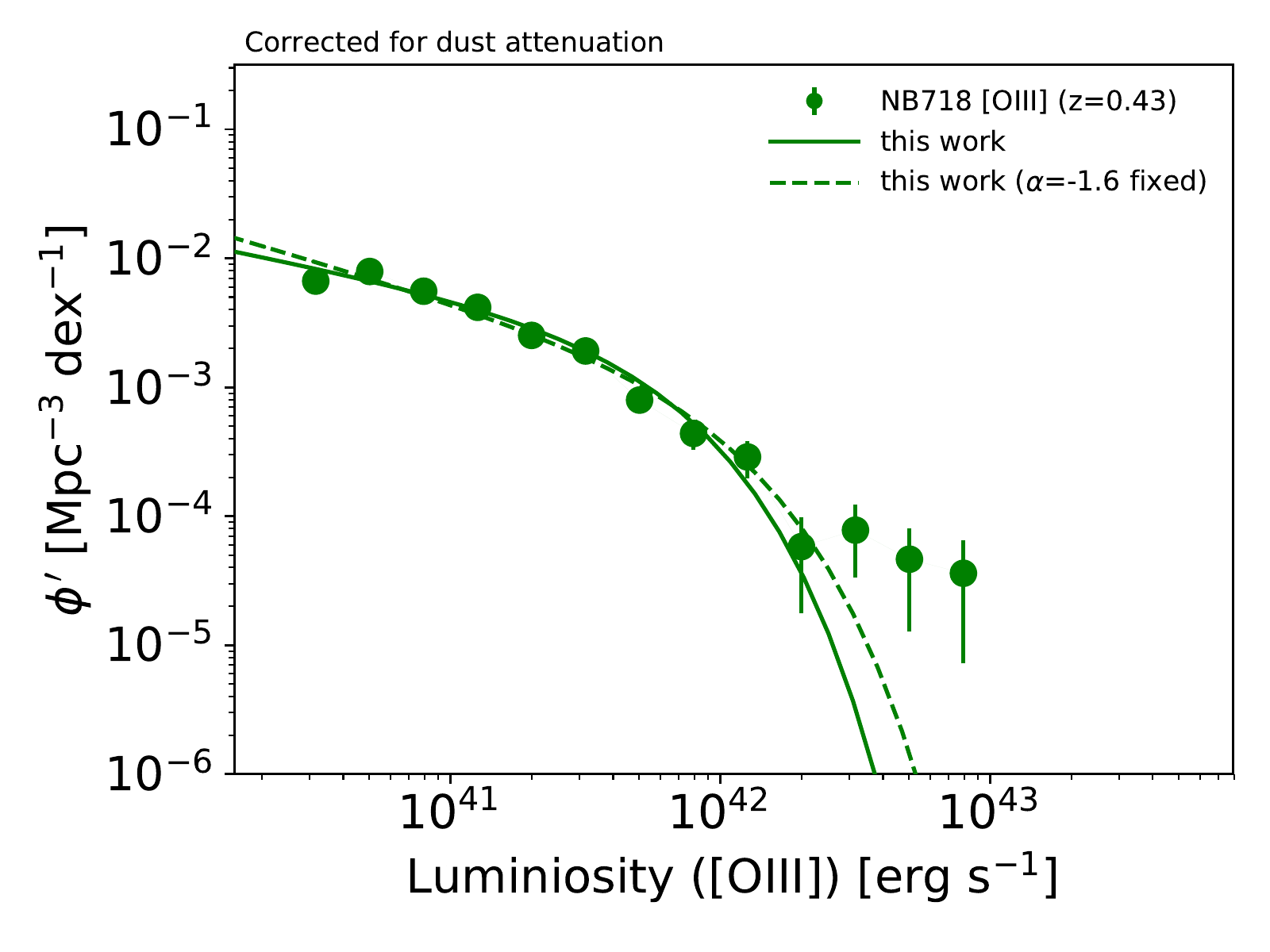} 
   \includegraphics[width=0.45\textwidth, bb=0 0 461 346]{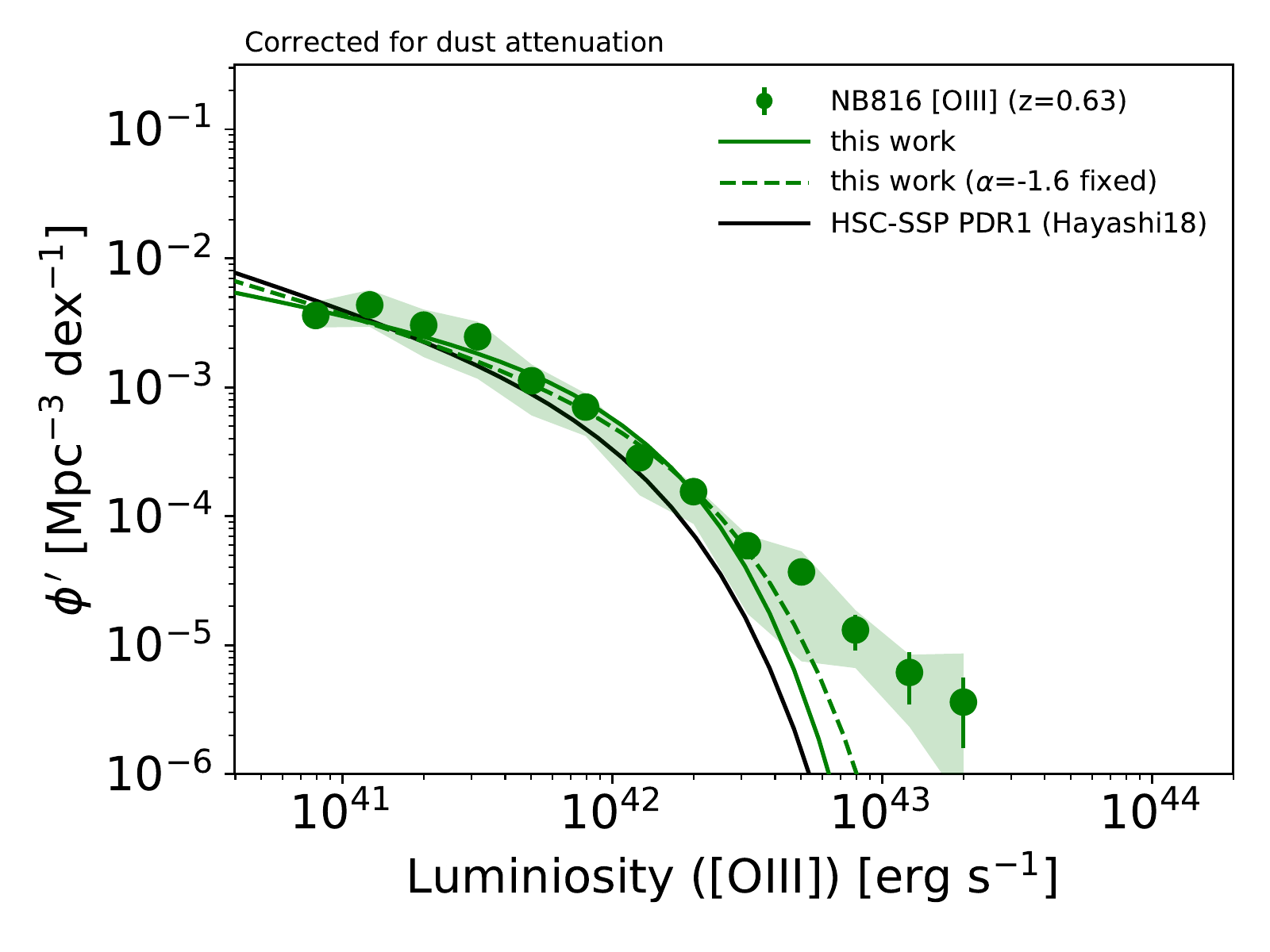} 
   \includegraphics[width=0.45\textwidth, bb=0 0 461 346]{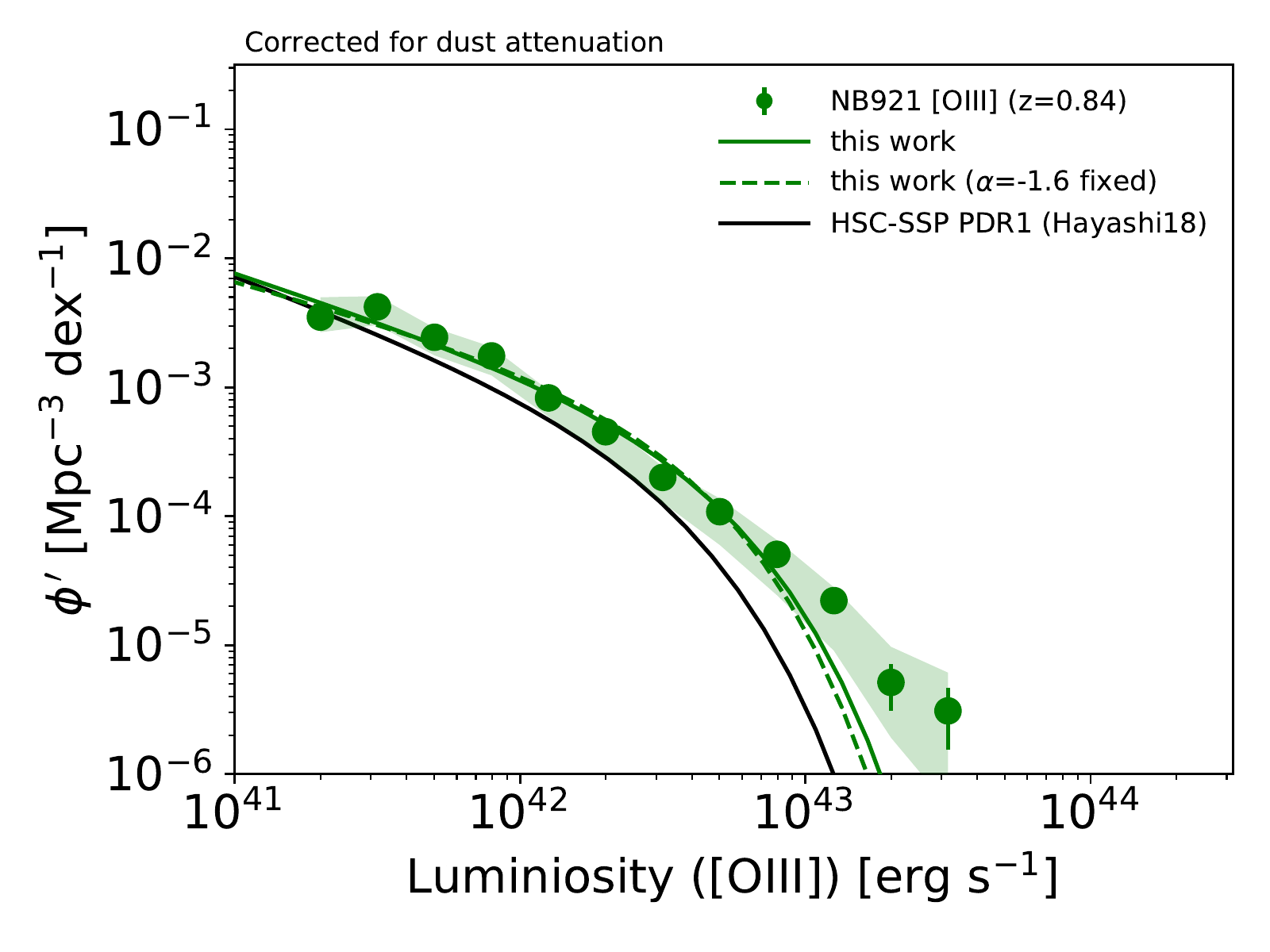} 
   \includegraphics[width=0.45\textwidth, bb=0 0 461 346]{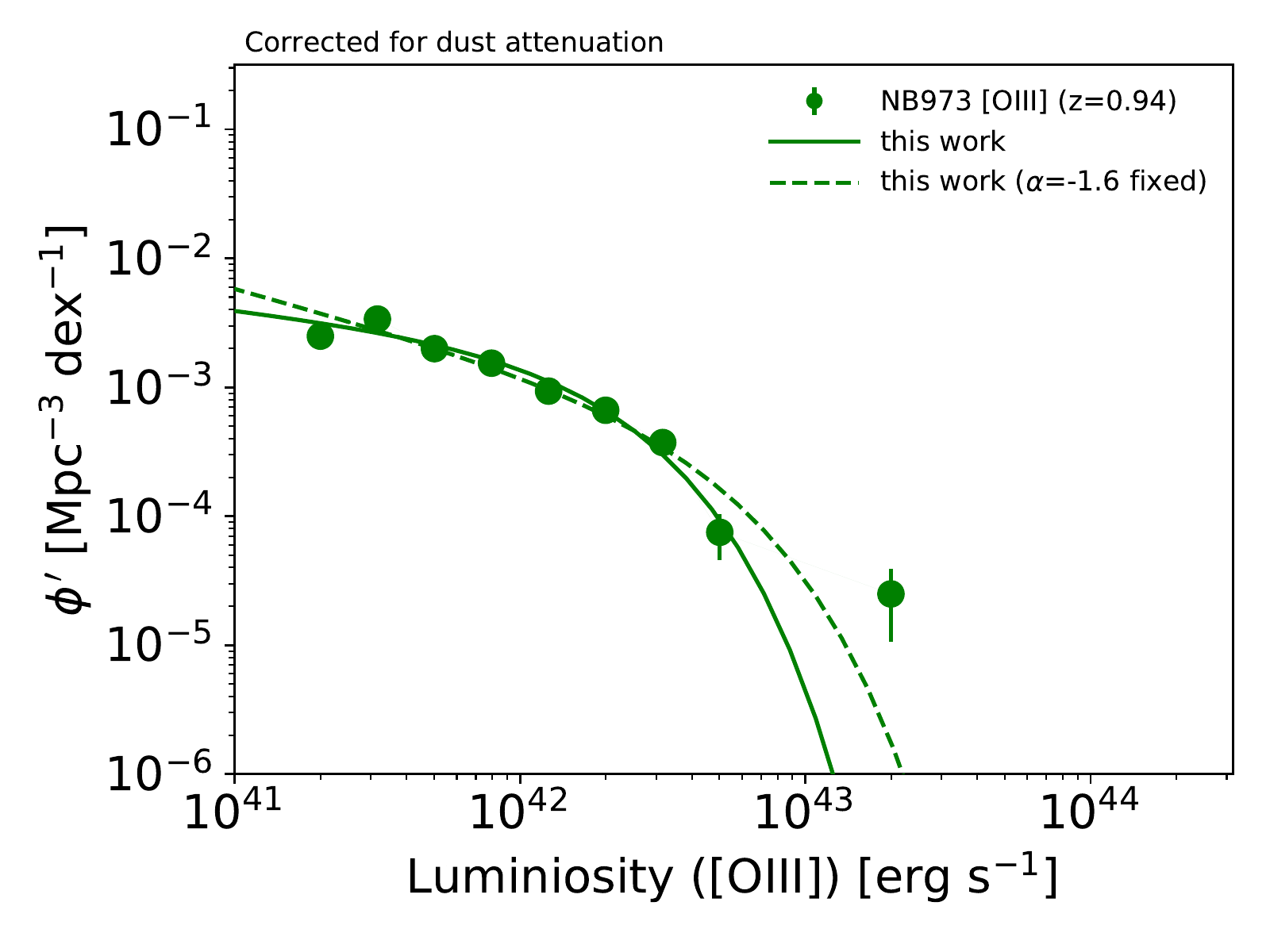} 
 \end{center}
 \caption{The same as fig~\ref{fig:LFcorrected_ha}, but for [OIII] emission-line galaxies.}\label{fig:LFcorrected_oiii}  
\end{figure*}

\begin{figure*}
 \begin{center}
   \includegraphics[width=0.45\textwidth, bb=0 0 461 346]{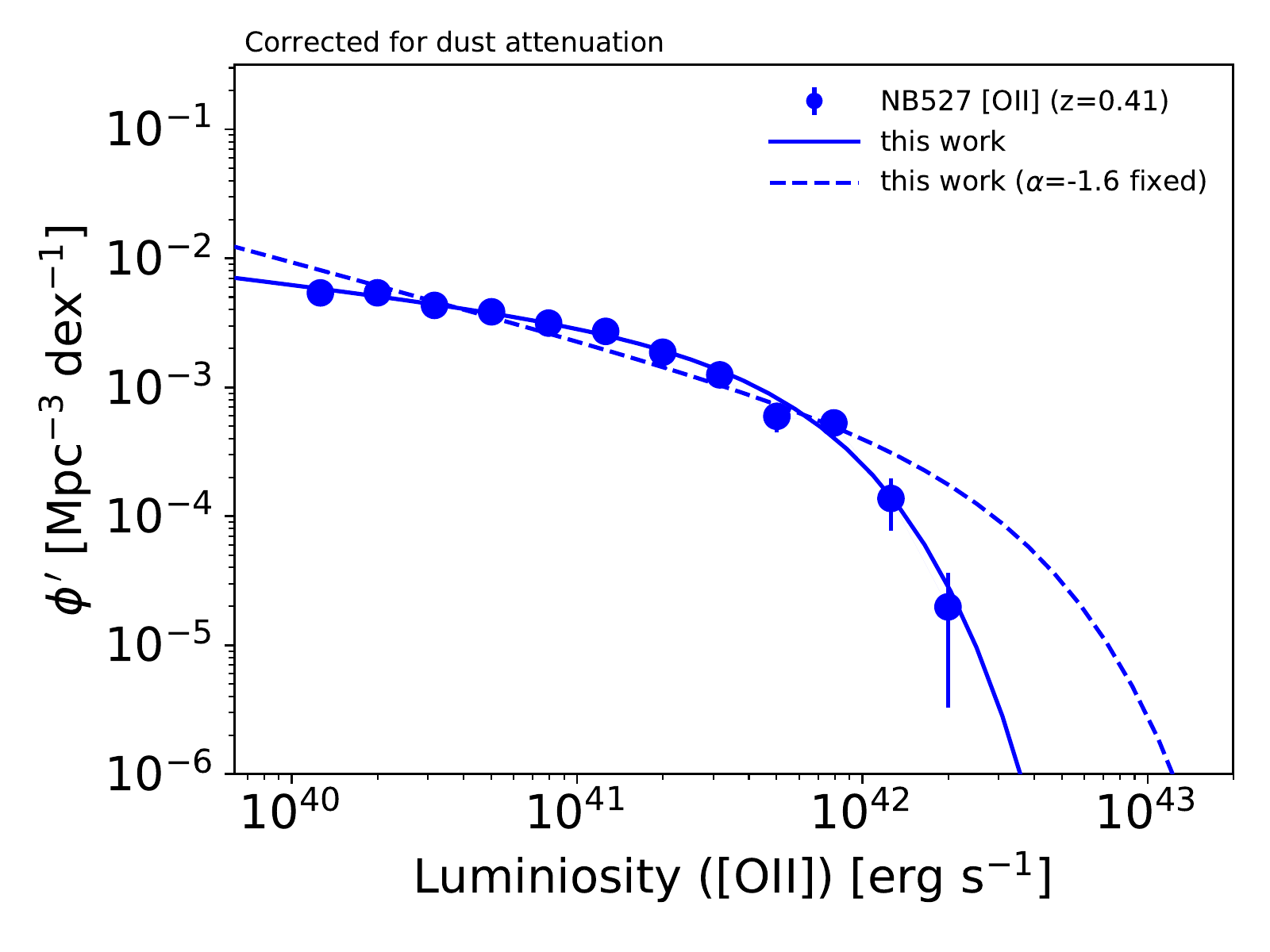} 
   \includegraphics[width=0.45\textwidth, bb=0 0 461 346]{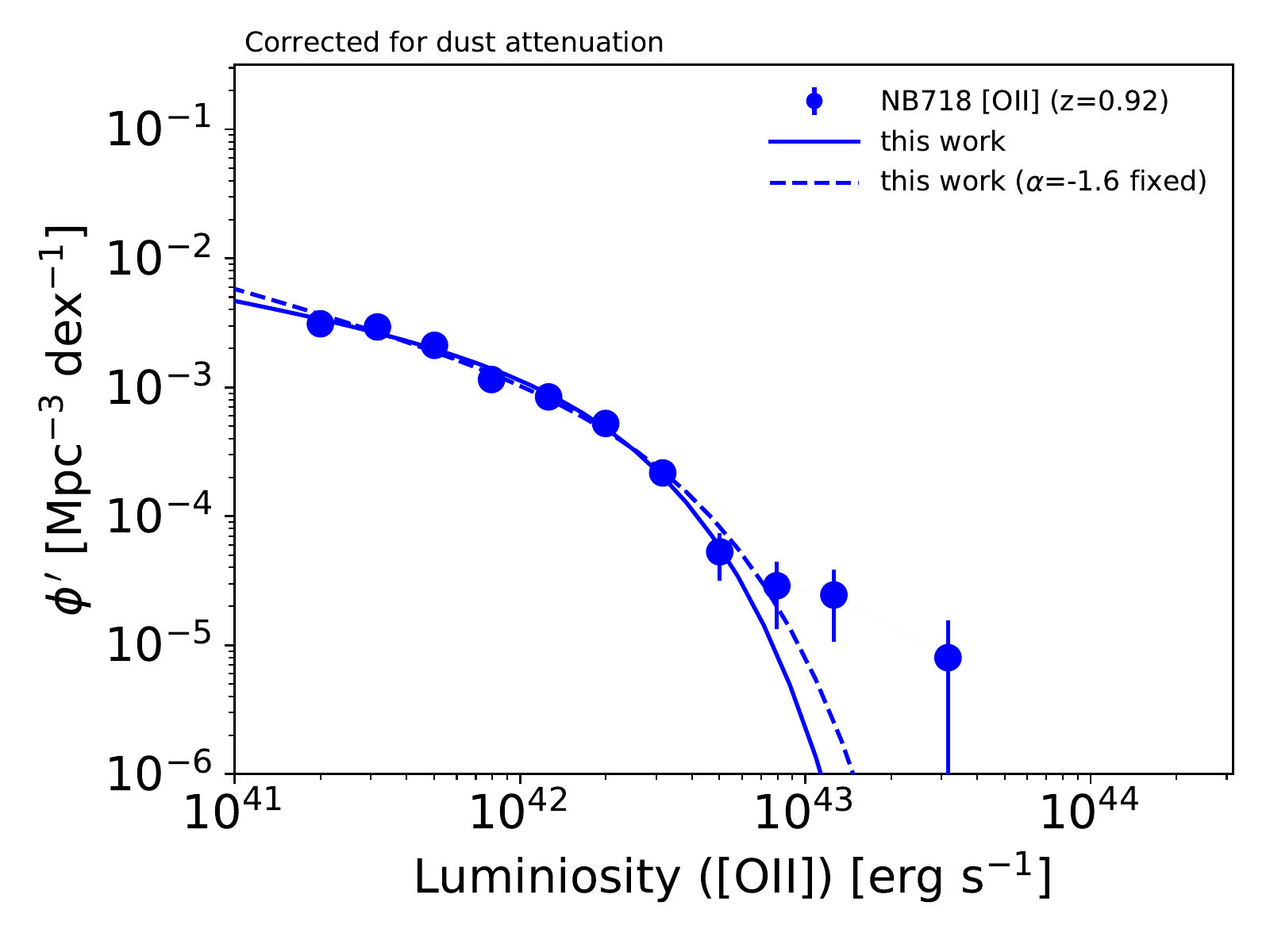} 
   \includegraphics[width=0.45\textwidth, bb=0 0 461 346]{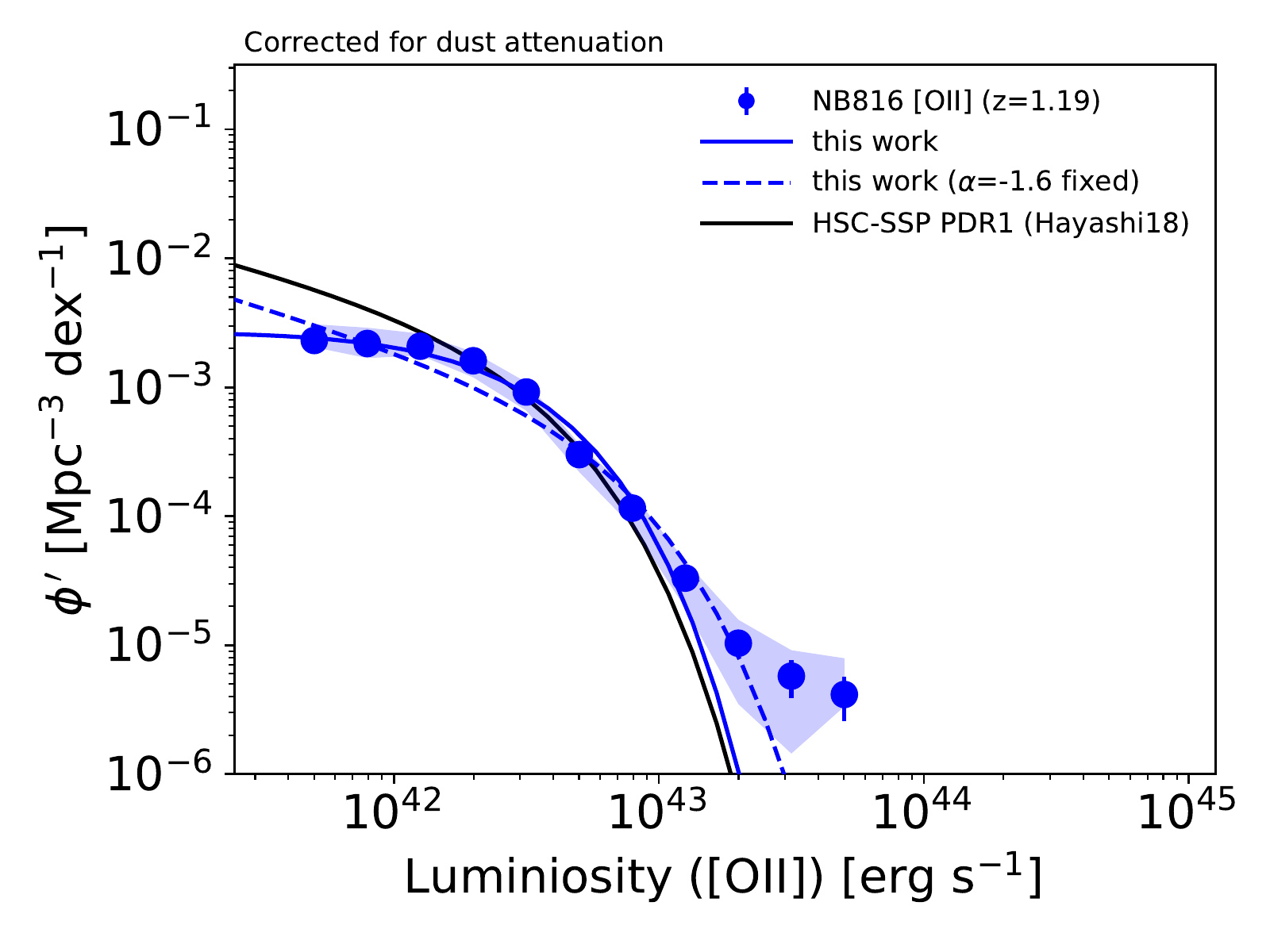} 
   \includegraphics[width=0.45\textwidth, bb=0 0 461 346]{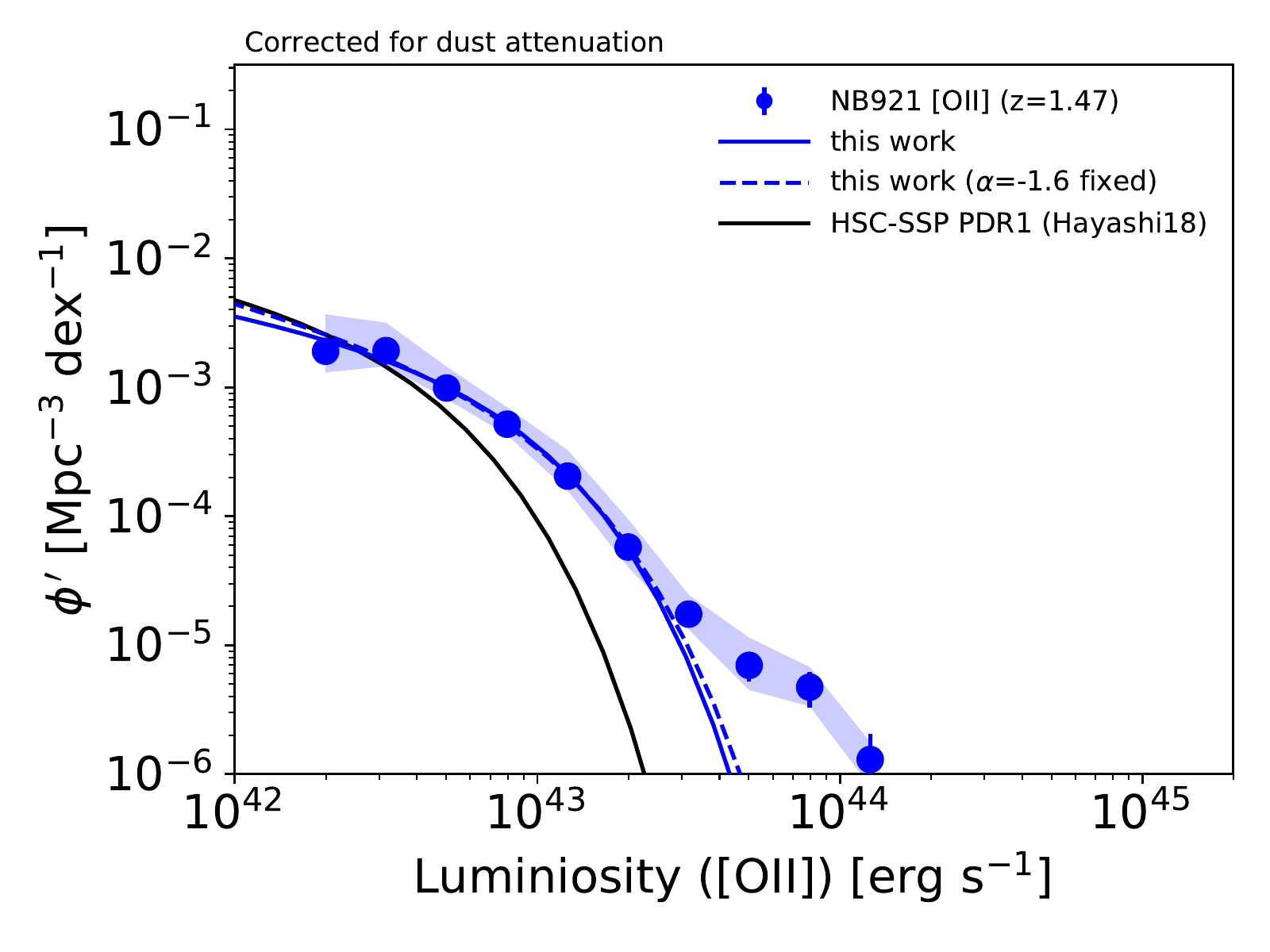} 
   \includegraphics[width=0.45\textwidth, bb=0 0 461 346]{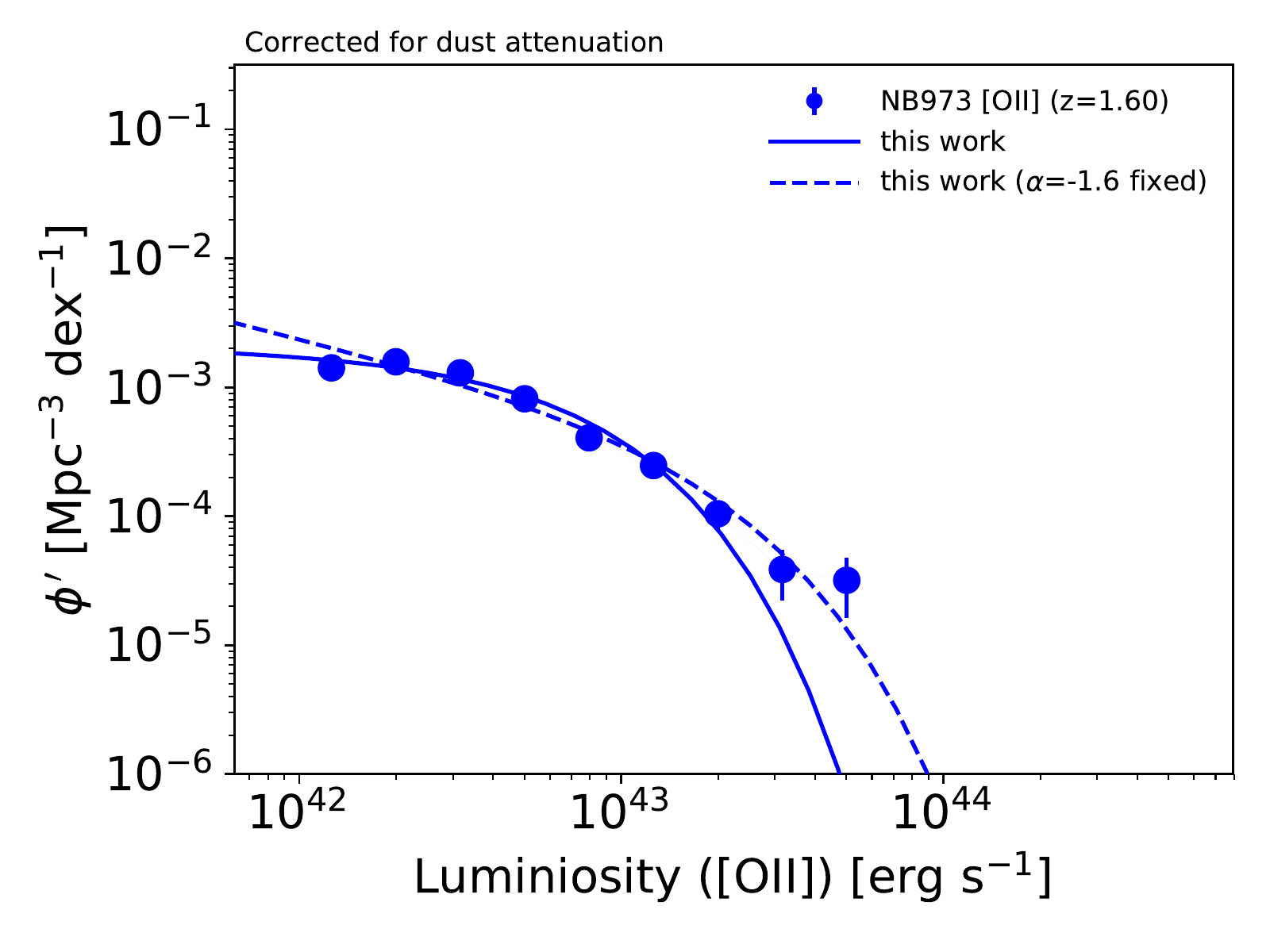} 
 \end{center}
 \caption{The same as fig~\ref{fig:LFcorrected_ha}, but for [OII] emission-line galaxies.}\label{fig:LFcorrected_oii}  
\end{figure*}

\input{table6.tex}

Luminosity function is often expressed with the form of a Schechter function
\citep{Schechter1976}:
\begin{eqnarray*}
\phi(L)dL = \phi^* \left(\frac{L}{L^*}\right)^{\alpha} \exp(-\frac{L}{L^*})~d\left(\frac{L}{L^*}\right),
\end{eqnarray*}
or 
\begin{eqnarray*}
\phi^\prime(L)d(\log L) = \phi^{\prime *} \left(\frac{L}{L^*}\right)^{\alpha+1} \exp(-\frac{L}{L^*})~\ln10~ d(\log L),
\end{eqnarray*}
where $L^*$, $\phi^*$ (or $\phi^{\prime *}$) and $\alpha$ are Schechter parameters.
$L^*$ is a characteristic luminosity, $\phi^*$ (or $\phi^{\prime *}$)
is a normalization density at $L^*$, and $\alpha$ is an exponential
power of the function dominant at faint end. The normalization
densities in the two forms are related by $\phi^* = \ln10~\phi^{\prime *}$.

We derive the luminosity functions in the same way as
\citet{Hayashi2018a}. First, we derive fluxes of the emission lines
from the NB and BB photometry. The BB--NB color and the total flux
density in NB are required to calculate emission-line flux using the
equation (7) of \citet{Hayashi2018a}. We use the 
{\tt undeblended\_convolvedflux\_2\_11} magnitudes for the colors and
the {\tt cmodel} magnitudes for the total flux density. Then, the 
fluxes are converted to luminosities using the spectroscopic redshifts
if available, otherwise, redshifts estimated from the central
wavelength of NB. In the following section, we compare the luminosity
functions of emission-line galaxies at different redshifts and discuss
the redshift evolution. As shown in table~\ref{tbl:NB_BBs_forSelection}, 
the limiting observed EWs (thus rest-frame EWs as well) are different
between the NB data. For a fair comparison, hereafter, we only use the H$\alpha$
emission-line galaxies with a rest-frame EW larger than 29.3$\AA$ at
each redshift, the [OIII] emitters with a rest-frame EW larger than
22.3$\AA$, and the [OII] emitters with a rest-frame EW larger than 16.6$\AA$.  

For H$\alpha$ emitters, not only H$\alpha$ emission line but also
[NII] doublet enter the NB filter simultaneously due to small
separation in wavelength between the lines. We estimate the
contribution of [NII] line to the fluxes measured with NB and BB
photometry, based on the line flux ratio of the local galaxies with a
given stellar mass and line luminosity (see Appendix 2 of
\cite{Hayashi2018a}). The sample of the local galaxies are selected
from the SDSS so that the galaxies have equivalent widths of emission
lines similar to the NB emission-line galaxies selected in this
study. The other line ratios (i.e., H$\alpha$/H$\beta$) for the local
galaxies are used later as well for correction for dust attenuation.    
We notice that we should check whether the line ratios of the local
galaxies are valid for the emission-line galaxies at higher redshifts
up to $z\sim1.6$. We will investigate the line ratios of the
emission-line galaxies using the spectra taken by our follow-up
observations (\S~\ref{sec:specz_Subaru}--\ref{sec:specz_AAT}) in the
forthcoming paper. In this paper, we assume that the emission-line
galaxies selected in this study follow the same relations in the line
ratios as the local SDSS galaxies. 

Since the transmission curve of HSC NB filters is not a perfect
top-hat (Figure~\ref{fig:filters}), the survey volume for individual
emission line galaxies is dependent on the luminosities. Because only
emission lines with fluxes observed above a limiting flux can be
detected, galaxies with an intrinsically larger flux can be selected
even at wavelengths with lower transmission, which means that
intrinsically brighter emission-line galaxies can be selected over a
wider range of redshifts than fainter emission-line galaxies.  
For a given luminosity, the minimum and maximum redshifts where the
emission line can be observed at more than 5 $\sigma$ are calculated
based on the filter transmission curve and a limiting flux. Then, the
redshift range is converted to a comoving volume. 

The luminosity function of emission-line galaxies is derived according to
the $V_{max}$ method:
\begin{eqnarray}
  \label{eq:LF}
  \phi_{obs}^\prime(\log L) = \sum_{i}\frac{1}{V_{max} \cdot f_c \cdot \Delta(\log L)},
\end{eqnarray}
where $i$ is for individual galaxies with $\log L\pm0.5\Delta(\log L)$, 
$V_{max}$ is the survey volume and $f_c$ is completeness taking
account of both selection and detection completeness. In the PDR1
catalog, we assumed that the emission line is observed at the central
wavelength of the NB filter. This assumption can result in
underestimating the line luminosities of galaxies at redshifts away
from the filter center. 
In this study, we correct the luminosity function,
$\phi_{obs}^\prime(\log L)$, derived from equation (\ref{eq:LF}) 
for an impact of filter response function to obtain intrinsic
luminosity function, $\phi^\prime(\log L)$
(see also \cite{Sobral2009,Sobral2013,Sobral2018} for correction of
filter response function). Here, we assume a uniform redshift
distribution within a redshift range corresponding to the wavelength
range of the NB filter, because the survey area is large enough to
overcome the field-to-field variance. For individual emission-line
galaxy with a given observed luminosity, the distribution of intrinsic
luminosities expected within the possible redshift range can be
estimated from the filter response function under the assumption of
the uniform redshift distribution. Then, by combining with the
observed luminosity function derived from equation (\ref{eq:LF}), the
probability distribution, $P_i(\log L)$, of the individual expected
intrinsic luminosities is estimated for each object;  
\begin{eqnarray*}
  P_i(\log L) & = & \phi_{obs}^\prime(\log L)/ \int \phi_{obs}^\prime(\log L^\prime) d(\log L^\prime) .
\end{eqnarray*}
By taking account of the probability distribution of the intrinsic
luminosities, the luminosity function can be derived as follows;
\begin{eqnarray*}
  \label{eq:LFcorrected}
  \phi^\prime(\log L) & = & \sum_i \frac{P_i(\log L)}{V_{max} \cdot f_c \cdot \Delta(\log L)} .
\end{eqnarray*}
The correction allows us to derive the bright end of luminosity
functions more properly, because there is the possibility that the
observed luminosity can be underestimated due to the nonuniformity on
the filter response function. 

Finally, we correct the luminosities for dust attenuation to derive
the intrinsic luminosity functions. To do that, we use the relation
between the line luminosity, stellar mass and Balmer decrement (the
ratio of H$\alpha$ to $\beta$) in the SDSS sample of the emitter
analogs. Note that we assume the \citet{Cardelli1989} extinction curve
and the intrinsic H$\alpha$/H$\beta$ ratio of 2.86 for Case B
recombination under an electron temperature of $T_e=10^4$ K and
electron density of $n_e=10^2$ cm$^{-3}$ \citep{Osterbrock1989},   

Figures~\ref{fig:LF_ha}--\ref{fig:LF_oii} show the observed luminosity
functions for H$\alpha$, [OIII], and [OII] emitters at each redshift. 
Figures~\ref{fig:LFcorrected_ha}--\ref{fig:LFcorrected_oii} show the
dust-corrected intrinsic luminosity functions for H$\alpha$, [OIII],
and [OII] emitters at each redshift. For the emission-line galaxies
selected from the HSC-SSP data (i.e., NB816 and NB921), the variation
of the measurements of the luminosity function in the four discrete
areas of HSC-SSP D/UD layer is shown by the pale color region. While a
Schechter function fitted to the measurement with free parameters of
$L^*$, $\phi^{\prime *}$ and $\alpha$ is shown by solid line, the dashed line
is a result fitted with $\alpha$ fixed to be $-1.6$. In the fitting,
we take account of the field variation as an uncertainty for the
emitters selected from NB816 or NB921 data, while the Poisson error is
taken into account for the other emitters.  
The best-fit Schechter parameters are summarized in table~\ref{tbl:SchechterParameters}. 

The luminosity functions in this study are almost consistent with our
previous study with the PDR1 data \citep{Hayashi2018a} that is shown by
the black solid line. Note that since we select emission-line galaxies
with an EW lower than those by our previous study (\S~\ref{sec:ELGs}),
the selection of emission-line galaxies from the PDR2 data is not
perfectly the same as our study with the PDR1 data as well as the
other previous studies \citep{Ly2007,Sobral2012,Drake2013,Sobral2013,Stroe2015,Khostovan2015,Khostovan2020}. 
However, it would be worth comparing the results between this study
and the others. As mentioned in \S~\ref{Selection}, the bright end of
the LF in this study can be affected by the magnitude cut applied for
the selection. When a galaxy has BB-NB color applied for the color cut
(table~\ref{tbl:NB_BBs_forSelection}), the NB magnitude of 18.5
corresponds to an observed line flux of 2.7, 2.1, 1.6, 1.4, and 1.0
$\times10^{-15}$ erg s$^{-1}$ cm$^{-2}$ for NB527, NB718, NB816,
NB921, and NB973 emitters, respectively. This can miss a fraction of
emitters with luminosities corresponding to a few of luminosity bins
in the bright end of luminosity function. 
The gray area in figures~\ref{fig:LF_ha}--\ref{fig:LF_oii} shows the
luminosity range affected by the magnitude cut, indicating that the
bright end of the LFs in the gray area can be incomplete. 
It is also true that number density of objects with NB magnitude
brighter than 18.5 sharply decreases (figures~\ref{fig:NumCountSSP}
and \ref{fig:NumCountCHORUS}). At higher redshifts, galaxies with NB
magnitude of $<18.5$ would be rarer. On the other hand, the 
correction for the filter response function allows us to estimate more
realistic bright end of the LF. Each of the effects can impact on the
bright end of LF from the opposite point of view.

For the observed H$\alpha$ LFs (figure~\ref{fig:LF_ha}), the faint end
slope is flatter than the PDR1 result as well as the results from the
other previous studies, which makes the observed LF more consistent
with the LF expected from a semi-analytic model \citep{Ogura2020}.  
For the observed [OIII] LFs (figure~\ref{fig:LF_oiii}), the number
density is larger than that of the PDR1 LF. 
While the point sources were excluded from the selection in the PDR1
study, we do not exclude the point sources explicitly in this
study. AGNs can be included in the sample of [OIII] emitters, compared
with the PDR1 sample that shows a low AGN contamination
fraction. Indeed, the fraction of the emitters that have an AGN
selected in X-ray and/or radio from the PDR2 data increases by a
factor of $\sim6$ to be $\sim0.65$\% (\S~\ref{sec:Xray}).   
Although the [OIII] luminosity function at $z=0.63$ is consistent with
the previous studies within the uncertainty, there is a discrepancy
between our results and the previous studies for the [OIII] luminosity
functions at $z=0.84$ and 0.94. The difference in the bright end can
affect the parameters of the fitted Schechter function. Also, note
that \citet{Khostovan2015} investigate luminosity functions of
H$\beta$+[OIII]. However, the fraction of H$\beta$ is likely to be
small in the samples \citep{Sobral2015,Khostovan2016,Khostovan2020}.   
Since the contribution of H$\beta$ becomes relatively larger in
fainter luminosity range \citep{Sobral2015,Khostovan2016}, the
difference of the treatment of H$\beta$ can affect the slope of the
faint end of luminosity function. Follow-up spectroscopy for all
[OIII] emitters is required to completely distinguish H$\beta$ from
[OIII]. 
For the observed [OII] LFs (figure~\ref{fig:LF_oii}), the LF of the 
NB921-selected [OII] emitters is in good agreement with the PDR1
result, while the number density of the NB816-selected [OII] emitters
is lower than the PDR1 sample. As discussed in section 3.4 of
\citet{Hayashi2018a}, the non-negligible contaminants of red galaxies
are likely to be included in the PDR1 sample of NB816 [OII]
emitters. The NB816 [OII] sample is more sensitive to the effect than
the NB921 sample due to the relative position in wavelength between NB
and nearest BB filter \citep{Hayashi2018a}. 
The [OII] luminosity function at $z=1.19$ is in agreement with the
result of \citet{Drake2013} and the bright end of \citet{Ly2007}.
The number density at $L^{*}$ in our [OII] luminosity function at $z=1.47$
is consistent with the previous studies, however, there is a large
variation in the number density at the luminosity range of $>L^{*}$.
The [OII] luminosity function at $z=1.60$ shows a large discrepancy
from the previous study of \citet{Khostovan2020}. Our luminosity
function has a flatter faint end as well as more prominent excess of
the bright end. 
Compared with the fitted Schechter function, the excess of the bright
end is also seen in the LFs of [OII] emitters.
The wider HSC survey tends to find the rare bright galaxies, which
results in the excess of the bright end in the luminosity function of
[OII] as well as [OIII]. 
Such excess is not seen in the LFs of H$\alpha$ emitters. As discussed
in section~\ref{sec:Xray}, some (but not all) emitters with an AGN can
contribute to the excess of the bright end of the LF. 

For the luminosity functions corrected for dust extinction, it is not
easy to fairly compare the results in this study with those from the
previous studies, because there is a difference in the correction for
filter response function and dust attenuation, which can be the other
factors to cause the difference between the studies. 

\section{Discussions}
\label{sec:discussions}

As shown in table~\ref{tbl:RedshiftSurveyed}, in this study, the wide
range of the 14 specific redshifts are covered by H$\alpha$, [OIII],
and [OII] emission lines between $z=1.6$ and $z=0.05$. Since the
luminosities of the emission lines are sensitive to SFR of galaxies,
this allows us to discuss the evolution of star-formation activity of galaxies
from $z=1.6$ down to $z=0.05$. 
However, because the correlation between the intensity of [OIII] and
SFR is not as strong as H$\alpha$ and [OII] in particular at lower
redshifts as mentioned below (e.g., \cite{Moustakas2006}), we do not
use [OIII] emission-line galaxies in this section.  

\subsection{Redshift evolution of luminosity functions}

\begin{figure}
 \begin{center}
   \includegraphics[width=0.5\textwidth, bb=0 0 461 346]{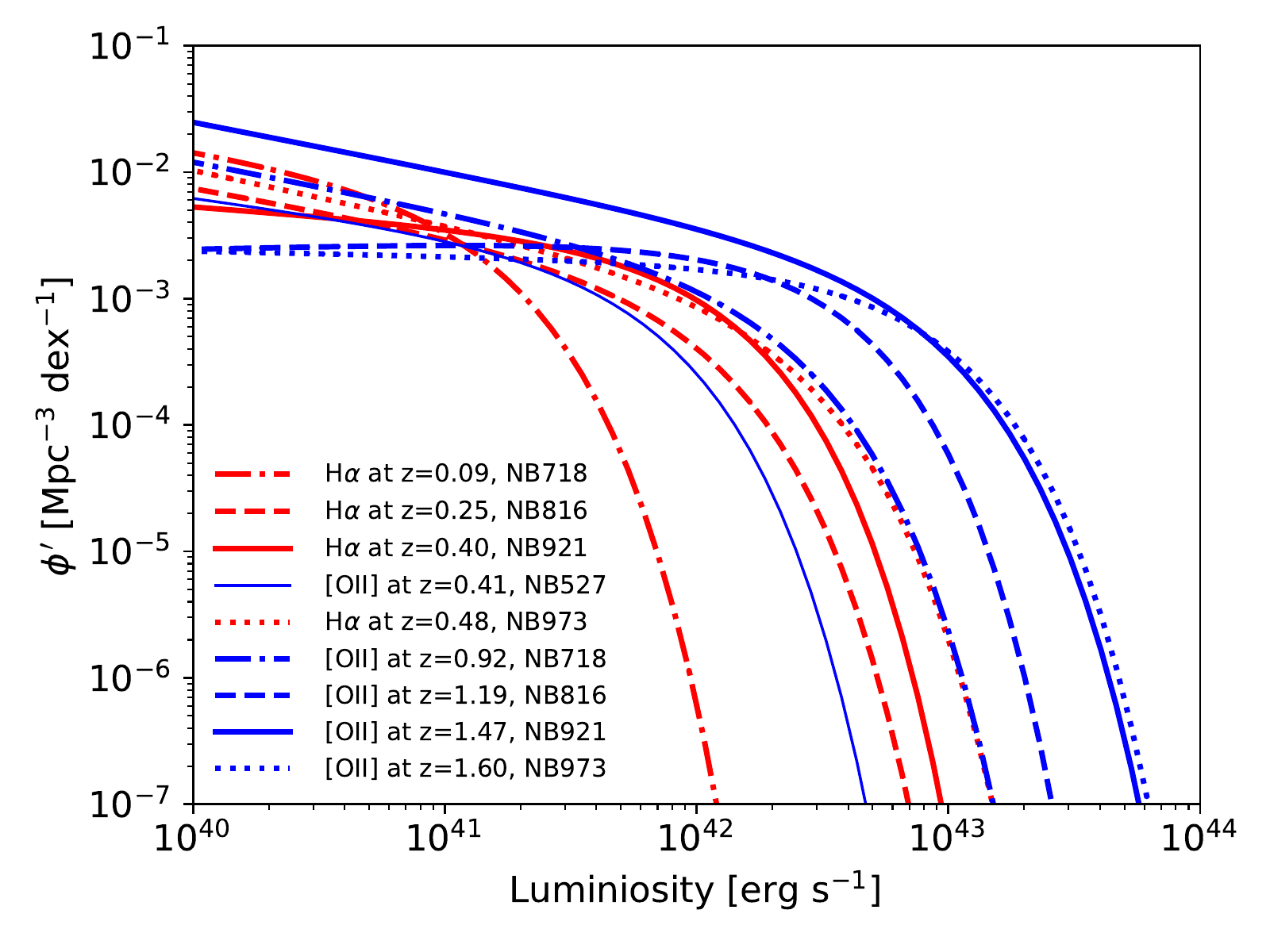} 
 \end{center}
 \caption{The evolution of the dust-corrected luminosity functions of
   H$\alpha$ and [OII] emission-line galaxies. The Schechter functions
   fitted to the measurement are shown for H$\alpha$ in red and [OII]
   in blue. The luminosity functions from NB527, NB718, NB816,
   NB921, and NB973 are shown by thin solid line, dash-dotted line, dashed
   line, thick solid line, and dotted line, respectively.}\label{fig:EvolutionLFs}
\end{figure}

The intrinsic luminosity of H$\alpha$ emitted from HII regions is
sensitive to SFR of galaxies (e.g., \cite{Kennicutt1998,Moustakas2006}).  
It is known that the luminosity of [OII] is also a usable indicator of
SFR in galaxies not only in the local Universe but also at high
redshifts, although the relation between the luminosity and SFR in
[OII] emission can depend on the physical condition of the nebular gas
such as metallicity and ionization state and thus is more complicated
rather than H$\alpha$ (e.g., \cite{Kennicutt1998,Kewley2004,Moustakas2006,Hayashi2013,Hayashi2015}).
Therefore, the evolution of the luminosity functions of H$\alpha$ and
[OII] emission-line galaxies reflects how the star-formation
activities of galaxies have changed with time.

Figure~\ref{fig:EvolutionLFs} shows the comparison between the
individual luminosity functions, where H$\alpha$ and [OII] lines are
shown in red and blue, respectively, and the luminosity functions
based on each NB data are shown with the same style. The luminosity
function of emission-line galaxies at higher redshifts is likely to
have larger $L^*$. Most of the luminosity function seem to have
similar slope, $\alpha$, of faint end. This is consistent with the
well-known picture that galaxies have star formation activities
peaking at $z\sim2$, often called the cosmic noon, and then the
activities gradually decrease towards the present day (e.g., \cite{MadauDickinson2014}). 

In the next section, to further discuss the evolution of
star-formation activities at $z\lesssim1.6$ more quantitatively, we
investigate the cosmic star formation rate densities.     

\subsection{Cosmic star formation rate densities at $z<1.6$}

\begin{figure}
 \begin{center}
   \includegraphics[width=0.5\textwidth, bb=0 0 463 346]{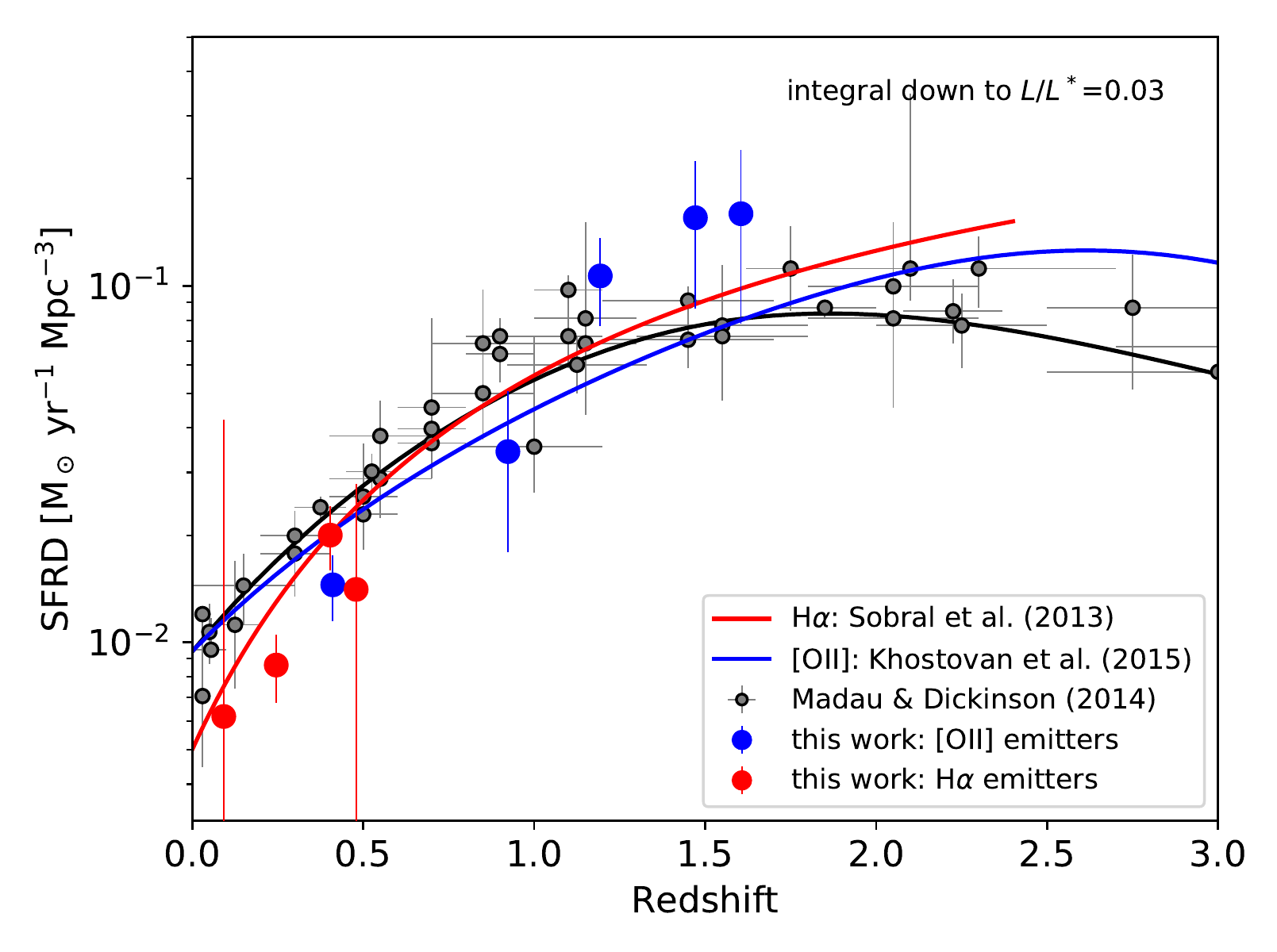} 
 \end{center}
 \caption{The cosmic star formation rate density. The SFR densities
   estimated from H$\alpha$ are shown in red, and those from [OII] are
   shown in blue. To estimate the SFR densities, the dust-corrected
   luminosity functions are integrated down to $L/L^*=0.03$, and then
   the luminosity densities are converted to the SFR densities using
   the relations given by \citet{Kennicutt1998}. The gray points are
   the results of the previous studies 
   \citep{Wyder2005,Schiminovich2005,Robotham2011,Cucciati2012,Dahlen2007,Reddy2009,Sanders2003,Takeuchi2003,Magnelli2011,Magnelli2013,Gruppioni2013}, 
   which are compiled by \citet{MadauDickinson2014}.
   The red and blue curves are the cosmic star-formation histories
   derived from H$\alpha$ and [OII] SFR, respectively
   \citep{Sobral2013,Khostovan2015}. 
 }\label{fig:CSFRD}    
\end{figure}

To derive the star formation rate density ($\rho_{SFR}$), we estimate
the luminosity density ($\rho_L$) from the dust-corrected luminosity
function by integrating $L \cdot \phi(L)dL$ down to a luminosity of
$L$. The integral is expressed with the upper incomplete gamma
function:  
\begin{eqnarray*}
\rho_L & = & \int_{L}^{\infty} L^\prime~\phi(L^\prime)~dL^\prime \\  
     & = & \int_{L/L^*}^{\infty} L^\prime~\phi^*\left(\frac{L^\prime}{L^*}\right)^{\alpha} \exp(-\frac{L^\prime}{L^*})~d(\frac{L^\prime}{L^*}) \\  
     & = & L^*~\phi^*\int_{L/L^*}^{\infty}\left(\frac{L^\prime}{L^*}\right)^{\alpha+1} \exp(-\frac{L^\prime}{L^*})~d(\frac{L^\prime}{L^*})\\
     & = & L^*~\phi^*~\Gamma(\alpha+2,L/L^*) \\
 ({\rm or} & = & L^*~\ln10~\phi^{\prime *}~\Gamma(\alpha+2,L/L^*) ).
\end{eqnarray*}
Here, we integrate the function down to $L/L^*=0.03$, which is the
same lower limit of the luminosity as the previous studies compiled by
\citet{MadauDickinson2014}. Then, the luminosity densities derived from
H$\alpha$ and [OII] lines are converted to the SFR densities using the
relations given by \citet{Kennicutt1998}. 
Note that we rescale the SFR densities to those with the \citet{Chabrier2003} IMF.

Figure \ref{fig:CSFRD} shows the SFR density as a function of
redshift. Our results are consistent with the previous studies. Each
data point is from the compilation by \cite{MadauDickinson2014} and
the black curve is the cosmic star-formation history fitted to the
SFRDs derived from UV+IR SFR. In the comparison, the same IMF
(\cite{Chabrier2003}) is assumed and the luminosity functions are
integrated down to $L/L^*=0.03$. At $z\sim0.4$, the SFR densities can
be derived from both H$\alpha$ and [OII] lines, and the measurements from 
two lines are consistent with each other within the uncertainty. The
SFR densities estimated from H$\alpha$ emitters at $z=0.092$ and 0.479
and [OII] emitters at $z=0.411$ and 0.923 are lower than the previous
studies. The emitters are selected  with NB527, NB718 and NB973 data
from the CHORUS survey that are available only in the UD-COSMOS field
of 1.37 deg$^2$. This may suggest that the survey volume for the
galaxies is not sufficient to overcome the cosmic variance. 
Also, as shown in figure~\ref{fig:LF_ha}, our luminosity functions of
H$\alpha$ emitters tend to have a flatter faint end than the previous
studies, which can result in the lower luminosity (i.e., SFR)
density. 
We note that we use the criterion of $>5\sigma$ color excess for the
selection of emission-line galaxies, while most previous studies apply
the 3$\sigma$ color excess. Although we correct the luminosity
function for the detection and selection completeness, our
conservative criterion can make a faint end slope flatter.  
Furthermore, since it become more difficult to select emission-line
galaxies with a lower EW at lower redshifts, the selection
completeness of the emission-line galaxies with an EW around the limit
can be lower and thus we may miss a fraction of the emission line
galaxies with a smaller EW at lower redshifts. However, note that we
apply the same EW cut to the emission-line galaxies used in the
discussion for a fair comparison.  

Several other studies on the NB-selected emission-line galaxies
also study the cosmic SFR density (e.g.,
\cite{Ly2011,Sobral2013,Khostovan2015,Ramon-Perez2019,Khostovan2020,Harish2020}).
The red and blue curves are the cosmic star-formation histories fitted
to the SFRDs derived from H$\alpha$ SFR and [OII] SFR
\citep{Sobral2013,Khostovan2015}, where the luminosity functions are
integrated down to $L/L^*=0$. 
All of the previous studies reach a consistent conclusion that the SFR
densities gradually increases from the local Universe toward the
redshift $\sim2$. On the other hand, the SFR densities estimated from
H$\alpha$ line at $z\lesssim0.6$ seems to be slightly lower than those
for the other SFR indicators such as UV or IR \citep{Sobral2013,Khostovan2020,Harish2020}. 
The systematic difference is seen in our measurements as well, especially in
lower redshifts. Since the NB imaging can survey only the small
redshift range, the larger coverage is essential to overcome the
field-to-field variance with enough survey volume. Even though the
recent imaging surveys have covered larger and larger area, it is
still only a few square degrees at the most in many cases. The cosmic
variance can be a formidable factor even for survey covering a few
square degrees. Since our H$\alpha$ emitters at $z\sim0.246$ and 0.404
are surveyed over $\sim$ 16 deg$^2$, the impact of the cosmic variance
would be relatively small. Nevertheless, the comparison with the
cosmic SFR density \citep{MadauDickinson2014} suggests that other
factors such as SFR indicators (i.e., time-scale of SF) and dust
correction can also influence the measurement of the cosmic SFR
density. To avoid the factors causing the possible systematic
differences in the measurements, a comprehensive survey with large
enough volume ($\gtrsim5\times10^{5}$ Mpc$^3$) at a wide range of redshifts would be required. 

\section{Conclusions}
\label{sec:conclusions}

We update the catalogs of emission-line galaxies at $z\lesssim1.6$
selected with narrowband imaging data taken with Hyper Suprime-Cam
(HSC) on the Subaru Telescope. The NB816 and NB921 data are from the
second Public Data Release (PDR2) of Subaru Strategic Program (SSP) of
HSC, and the NB527, NB718, and NB973 data are from the Subaru open-use
intensive program, Cosmic HydrOgen Reionization Unveiled with Subaru (CHORUS)
survey. The data from the HSC-SSP are available over the effective
area of $\sim16$ deg$^2$ in the Deep and UltraDeep layer, while the
data from the CHORUS survey are available over 1.37 deg$^2$ of the
COSMOS field in the UltraDeep layer.  

We use the data from the five NB filters to select H$\alpha$, [OIII], and [OII]
emission-line galaxies at 14 specific redshifts ranging from
$z\sim1.6$ down to $z\sim0.05$. Furthermore, 2,019 emission-line
galaxies selected with the HSC-SSP PDR1 data are confirmed by our
follow-up spectroscopic observations. These newly available data
allow us to improve the catalogs of the emission-line galaxies. We
find that compared with the PDR1 catalogs, the emission-line galaxies
with lower equivalent width are selected and there is a smaller
fraction of contaminants. As with our PDR1 catalogs, the PDR2 catalogs
of the emitters will be published in the HSC-SSP Public Data Release
website\footnotemark[\ref{foot:hscsspwebsite}]   

Using the catalogs of emission-line galaxies selected at small
redshift intervals (i.e., the redshifts surveyed by the five NB
filters are separated by $\Delta z\sim0.2$), we show the redshift
evolution of luminosity function for the emission-line galaxies. The
characteristic luminosity increases for emission-line galaxies at 
higher redshifts, suggesting that the luminosities of emission lines
at each redshift follow the cosmic star formation rate density (SFRD).  
By integrating the luminosity functions and then converting the
luminosities of H$\alpha$ and [OII] emission lines to the star
formation rates, we discuss the SFRD. The cosmic SFRD increases
monotonically from the local Universe towards $z\sim1.6$, which are
consistent with the previous studies. All of the results shown in this
paper demonstrate that the catalogs of emission-line galaxies at
various redshifts are useful to investigate the evolution of the
star-forming galaxies at $z\lesssim1.6$.  
Follow-up observations, e.g., optical and/or near infrared
spectroscopy and radio observation, will be the next step to study
evolution of physical condition and gas content in individual
star-forming galaxies since the cosmic noon.


\begin{ack}

We thank the anonymous referee for providing constructive comments and
suggestions.
We are grateful to Prof.~Chris Lidman and Dr.~Anais M$\rm\ddot{o}$ller
for the instruction and support in the observations at AAT. 
RS acknowledges the financial support of KAKENHI (19K14766)
Grant-in-Aid for Early-Career Scientists through the Japan Society for
the Promotion of Science (JSPS).   

This work is based on data collected at the Subaru Telescope and
retrieved from the HSC data archive system, which is operated by
Subaru Telescope and Astronomy Data Center (ADC) at NAOJ. Data
analysis was in part carried out with the cooperation of Center for
Computational Astrophysics (CfCA), NAOJ. The NB718 and NB816 filters
were supported by Ehime University and the NB527, NB921 and NB973
filters were supported by KAKENHI (24244018 and 23244025) Grant-in-Aid
for Scientific Research (A) through the JSPS. This work is based in
part on data acquired at the Anglo-Australian Telescope, under program
A/2018B/01. We acknowledge the traditional owners of the land on which
the AAT stands, the Gamilaraay people, and pay our respects to elders
past and present.  

The Hyper Suprime-Cam (HSC) collaboration includes the astronomical
communities of Japan and Taiwan, and Princeton University.  The HSC
instrumentation and software were developed by the National
Astronomical Observatory of Japan (NAOJ), the Kavli Institute for the
Physics and Mathematics of the Universe (Kavli IPMU), the University
of Tokyo, the High Energy Accelerator Research Organization (KEK), the
Academia Sinica Institute for Astronomy and Astrophysics in Taiwan
(ASIAA), and Princeton University.  Funding was contributed by the
FIRST program from the Japanese Cabinet Office, the Ministry of
Education, Culture, Sports, Science and Technology (MEXT), the Japan
Society for the Promotion of Science (JSPS), Japan Science and
Technology Agency  (JST), the Toray Science  Foundation, NAOJ, Kavli
IPMU, KEK, ASIAA, and Princeton University.  

This paper makes use of software developed for the Large Synoptic
Survey Telescope. We thank the LSST Project for making their code
available as free software at \url{http://dm.lsst.org}.

The Pan-STARRS1 Surveys (PS1) and the PS1 public science archive have
been made possible through contributions by the Institute for
Astronomy, the University of Hawaii, the Pan-STARRS Project Office,
the Max Planck Society and its participating institutes, the Max
Planck Institute for Astronomy, Heidelberg, and the Max Planck
Institute for Extraterrestrial Physics, Garching, The Johns Hopkins
University, Durham University, the University of Edinburgh, the
Queen’s University Belfast, the Harvard-Smithsonian Center for
Astrophysics, the Las Cumbres Observatory Global Telescope Network
Incorporated, the National Central University of Taiwan, the Space
Telescope Science Institute, the National Aeronautics and Space
Administration under grant No. NNX08AR22G issued through the Planetary
Science Division of the NASA Science Mission Directorate, the National
Science Foundation grant No. AST-1238877, the University of Maryland,
Eotvos Lorand University (ELTE), the Los Alamos National Laboratory,
and the Gordon and Betty Moore Foundation. 
\end{ack}

\appendix 

\section{Validation of CHORUS NB data}
\begin{figure*}
 \begin{center}
   \includegraphics[width=0.45\textwidth, bb=0 0 461 346]{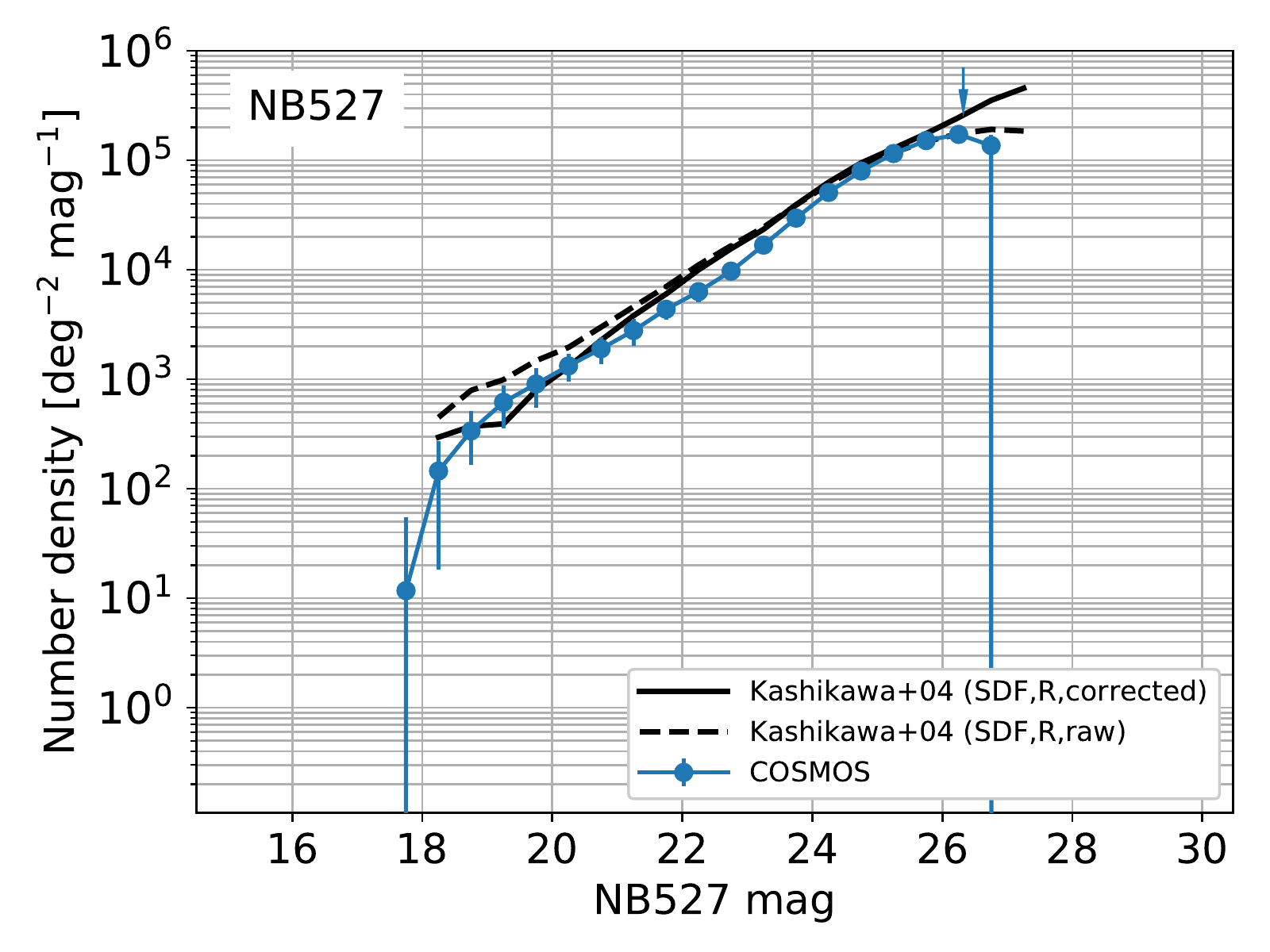} 
   \includegraphics[width=0.45\textwidth, bb=0 0 461 346]{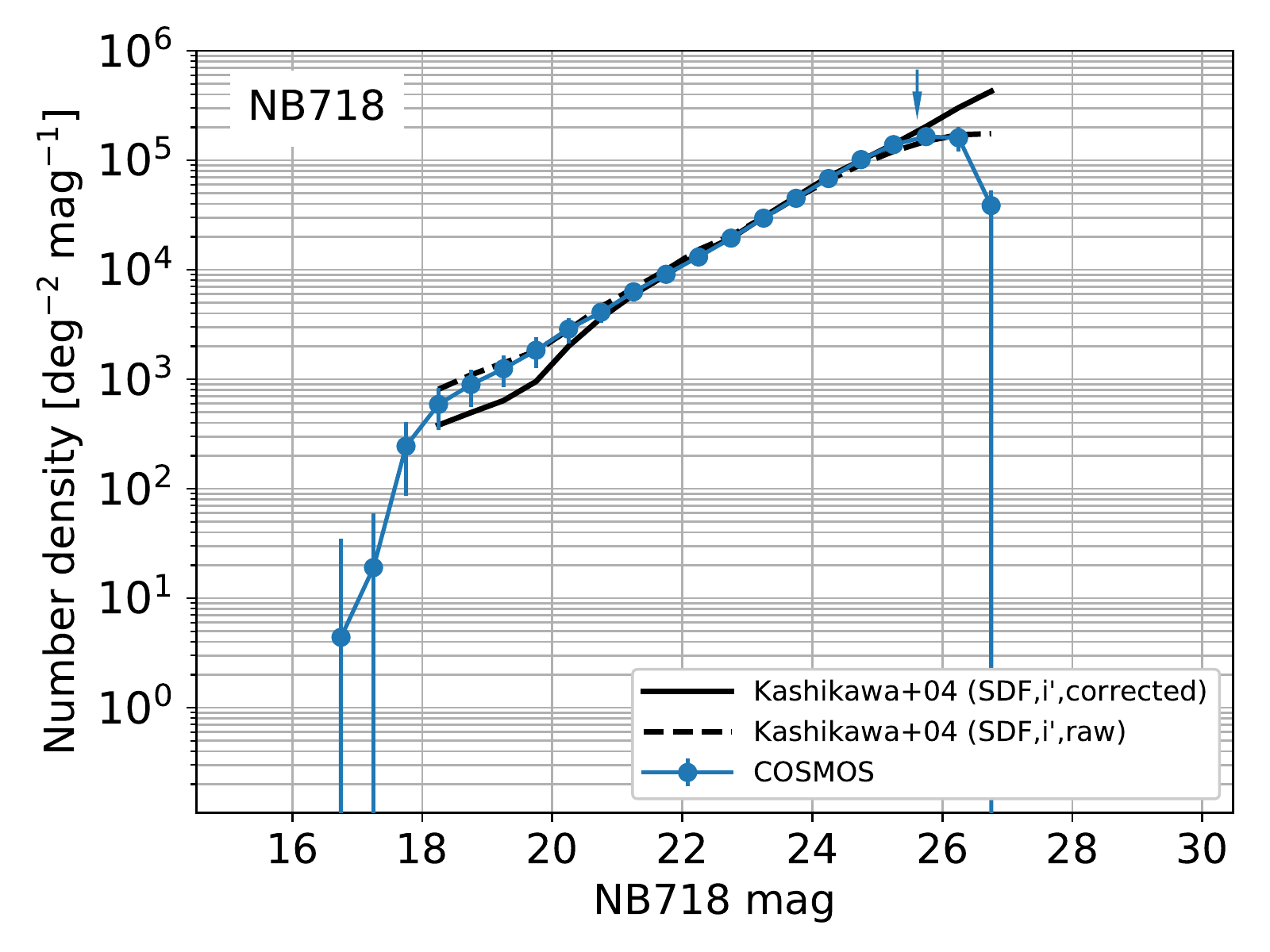} 
   \includegraphics[width=0.45\textwidth, bb=0 0 461 346]{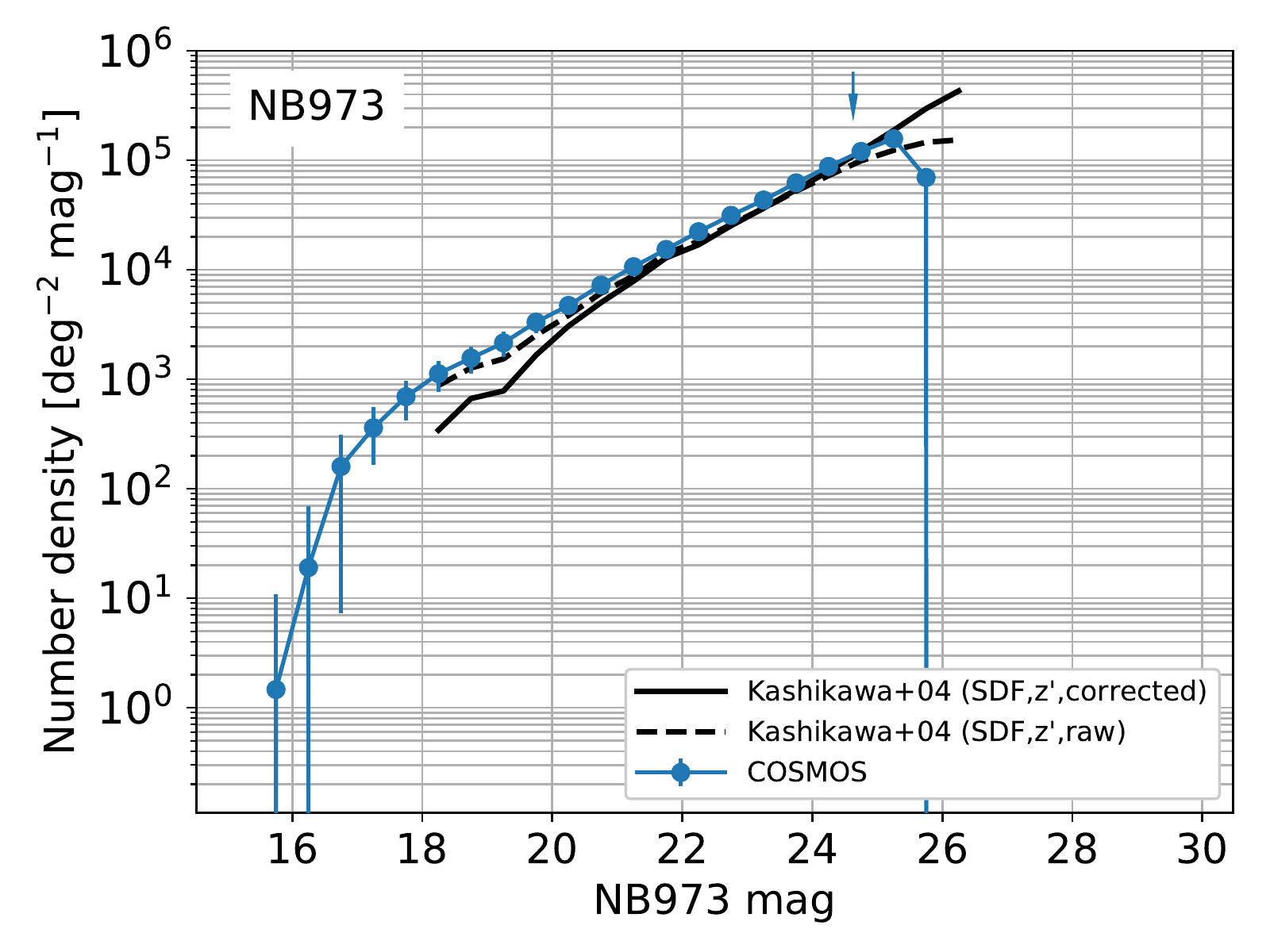} 
 \end{center}
 \caption{The same as figure~\ref{fig:NumCountSSP}, but for the number counts of sources detected in NB527, NB718 and NB973.}\label{fig:NumCountCHORUS}
\end{figure*}

\begin{figure*}
 \begin{center}
   \includegraphics[width=0.45\textwidth, bb=0 0 461 346]{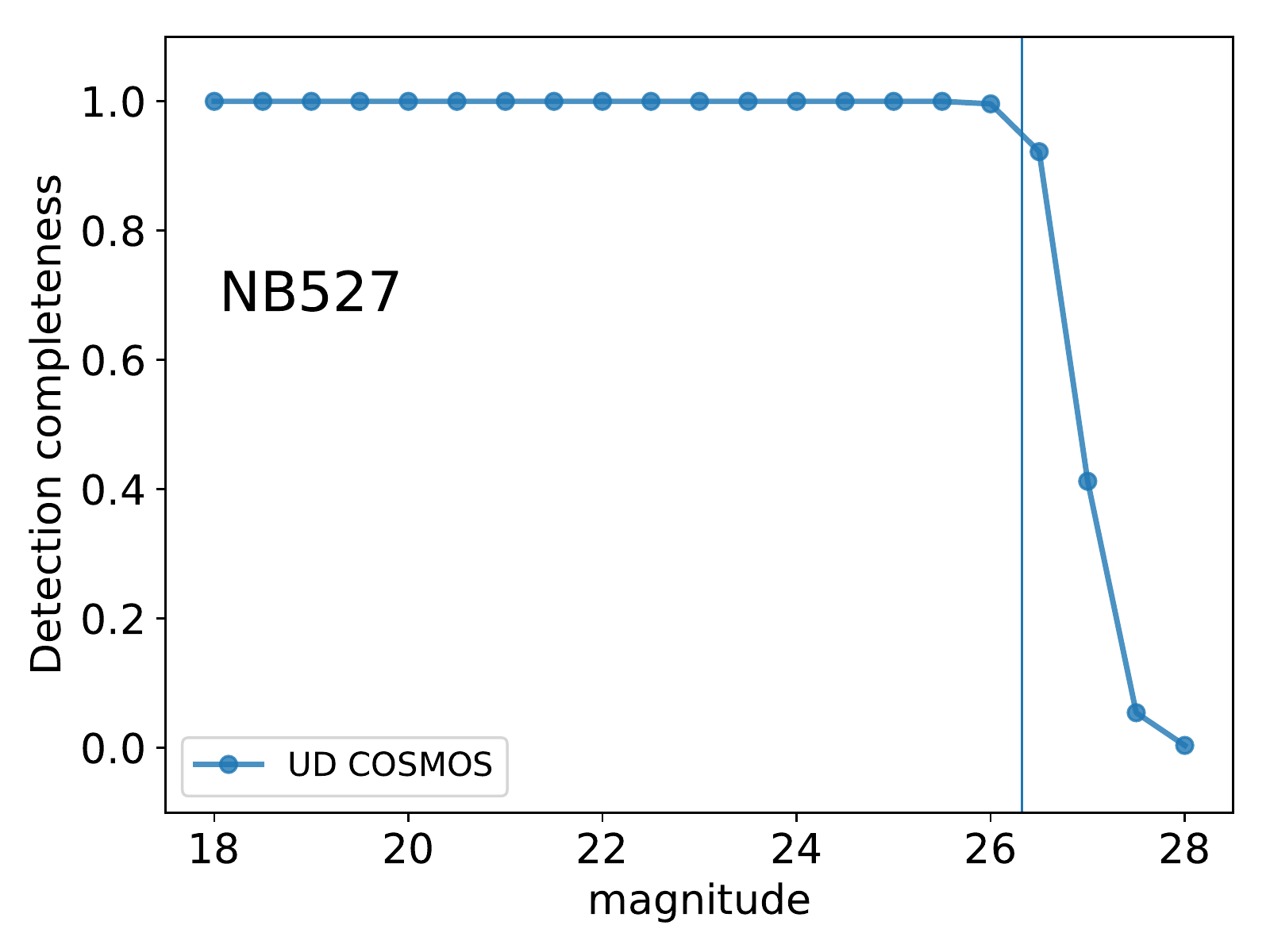} 
   \includegraphics[width=0.45\textwidth, bb=0 0 461 346]{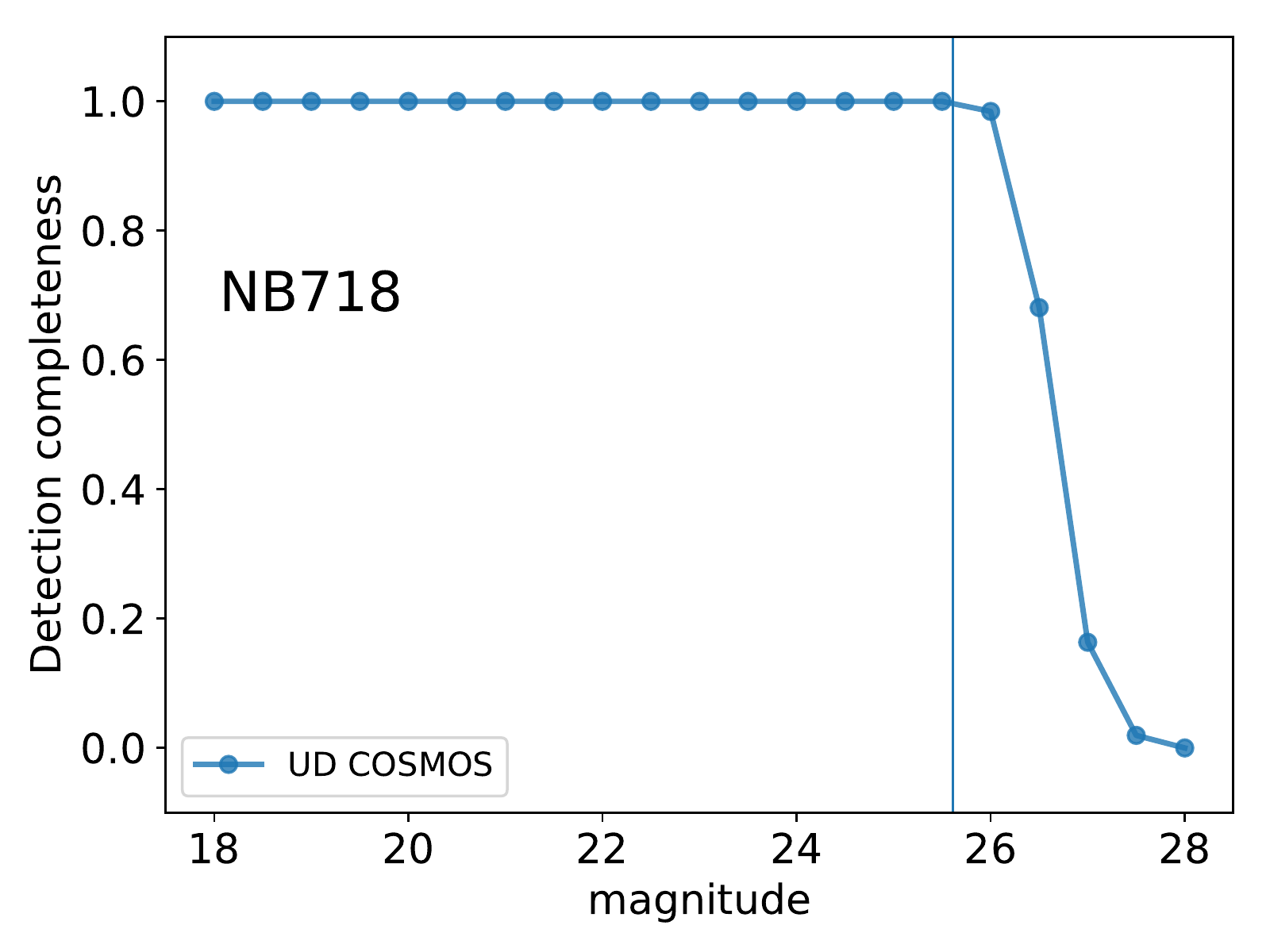} 
   \includegraphics[width=0.45\textwidth, bb=0 0 461 346]{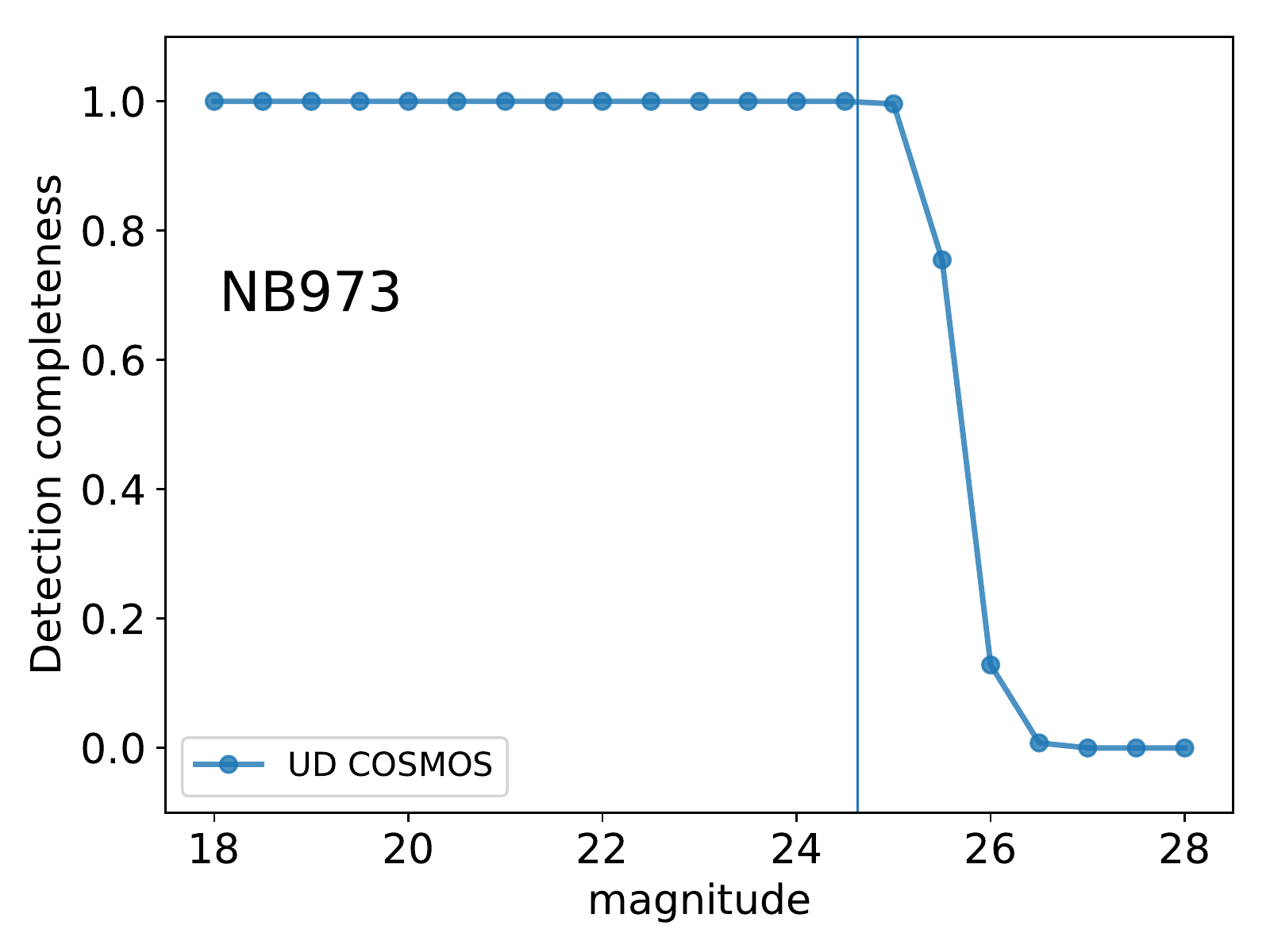} 
 \end{center}
 \caption{The same as figure~\ref{fig:DetectionCompletenessSSP}, but for the detection completeness for NB527, NB718 and NB973 filters.}\label{fig:DetectionCompletenessCHORUS}
\end{figure*}

\begin{figure*}
 \begin{center}
   \includegraphics[width=0.45\textwidth, bb=0 0 461 346]{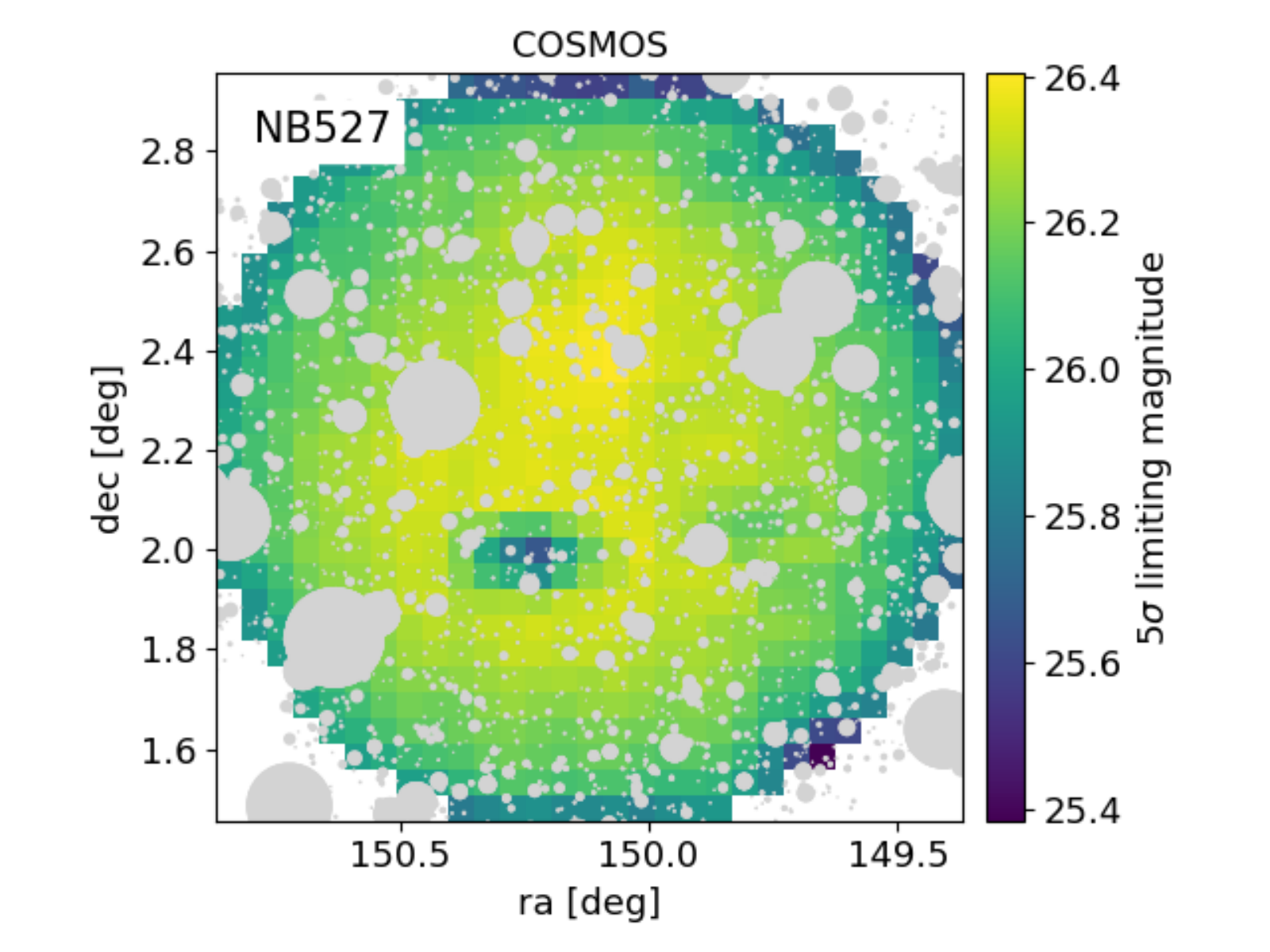} 
   \includegraphics[width=0.45\textwidth, bb=0 0 461 346]{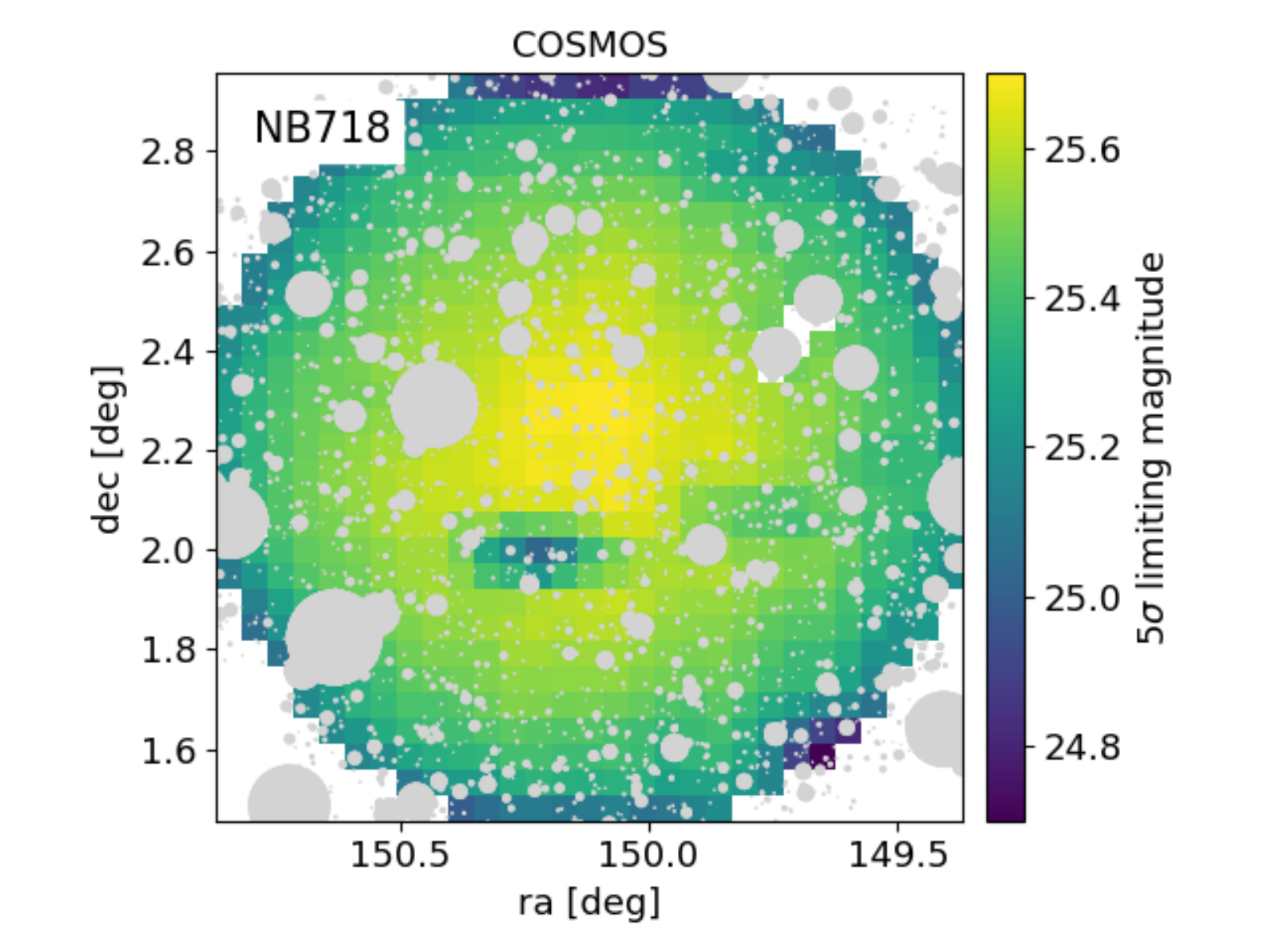} 
   \includegraphics[width=0.45\textwidth, bb=0 0 461 346]{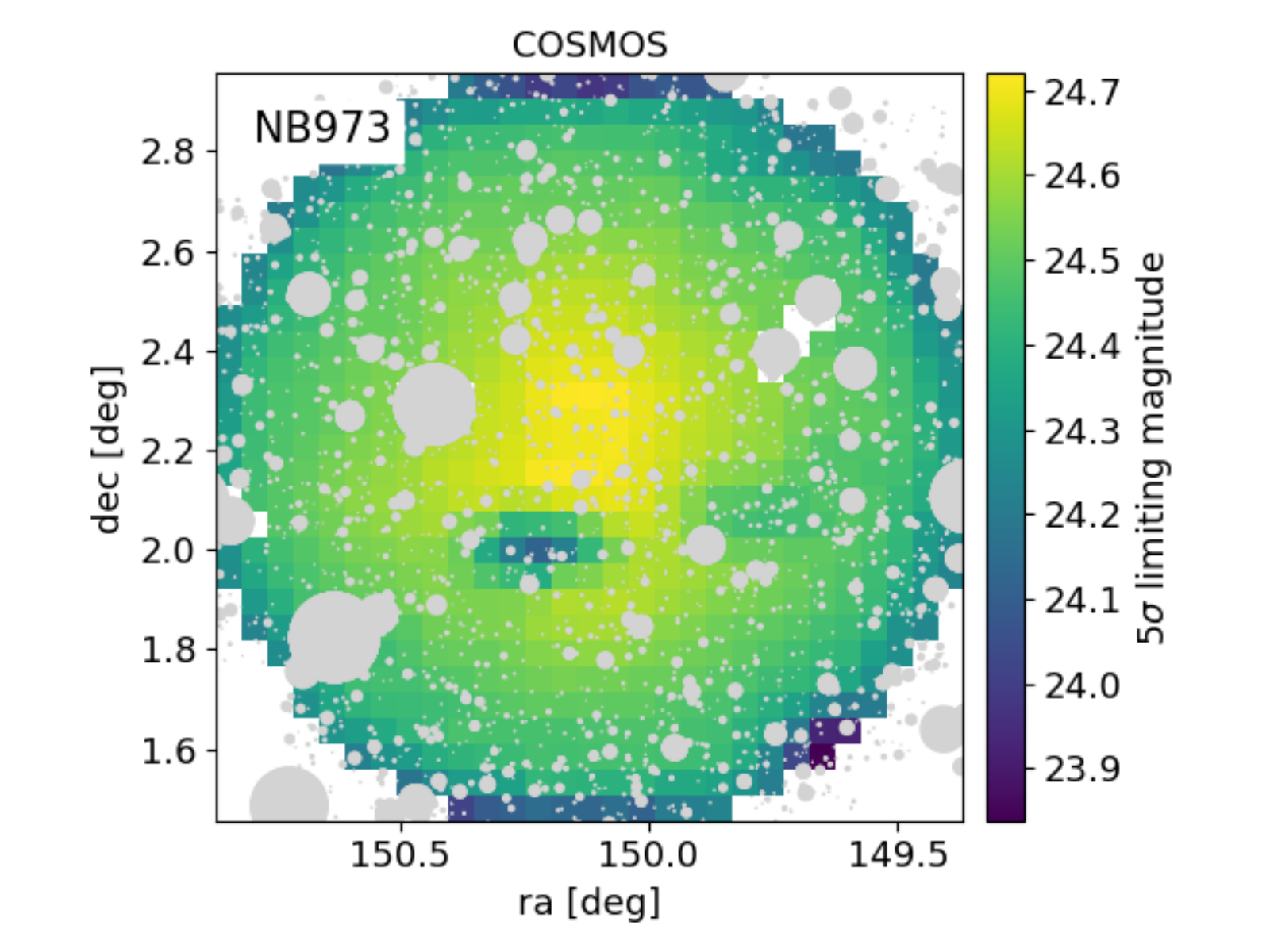} 
 \end{center}
 \caption{The same as figure~\ref{fig:LimMagSSP}, but for the map of 5$\sigma$ limiting magnitudes in NB527, NB718 and NB973.}\label{fig:LimMagCHORUS}
\end{figure*}

We investigate the number counts, the detection completeness, and the
$5\sigma$ limiting magnitude in the three NB filters from the CHORUS
survey in the same manner as for the SSP data. The results are shown
in figures~\ref{fig:NumCountCHORUS}--\ref{fig:LimMagCHORUS}.

\section{Spatial distribution of emitters}
\label{appendix:Map}

\begin{figure*}
 \begin{center}
   \includegraphics[width=0.32\textwidth, bb=0 0 461 461]{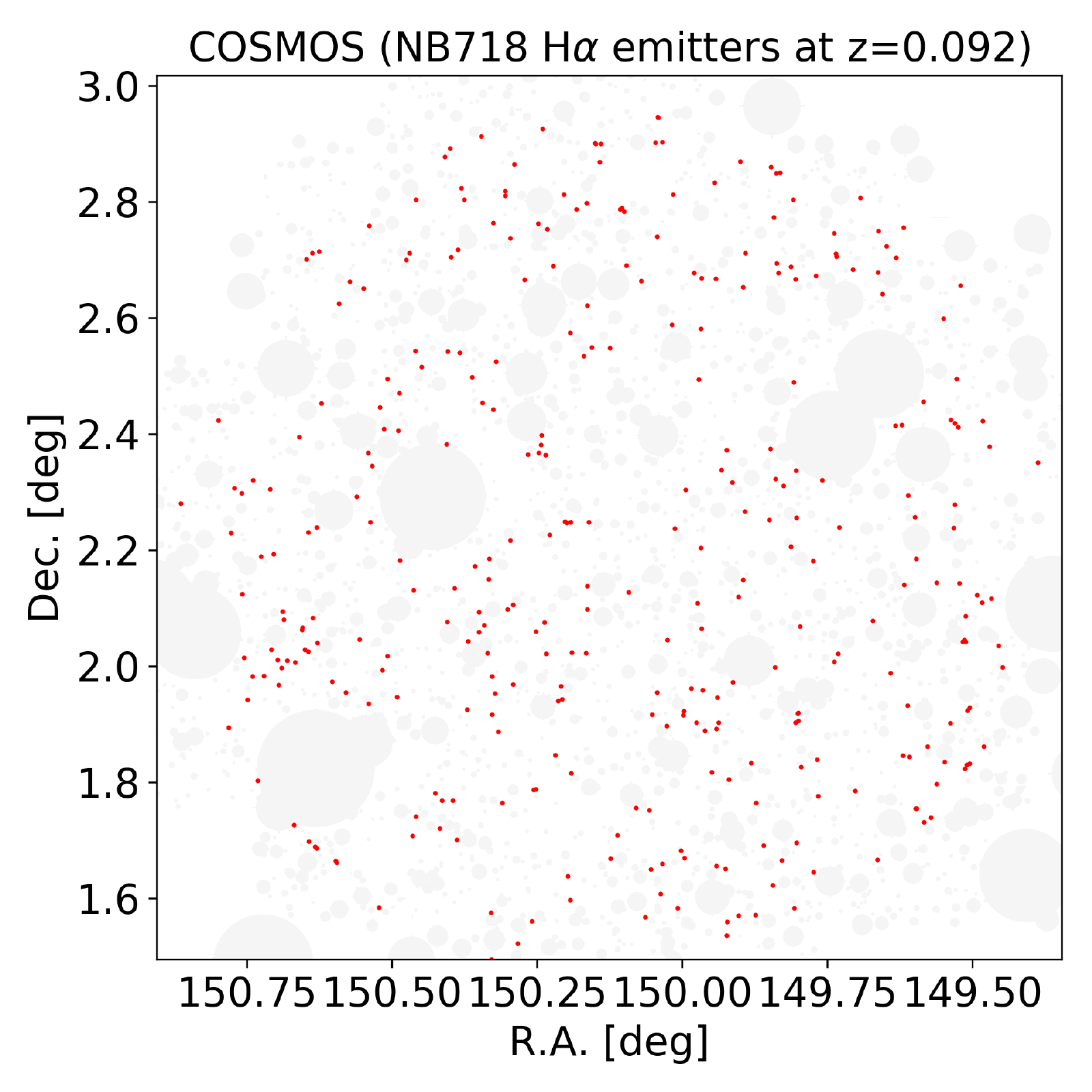} 
   \includegraphics[width=0.32\textwidth, bb=0 0 461 461]{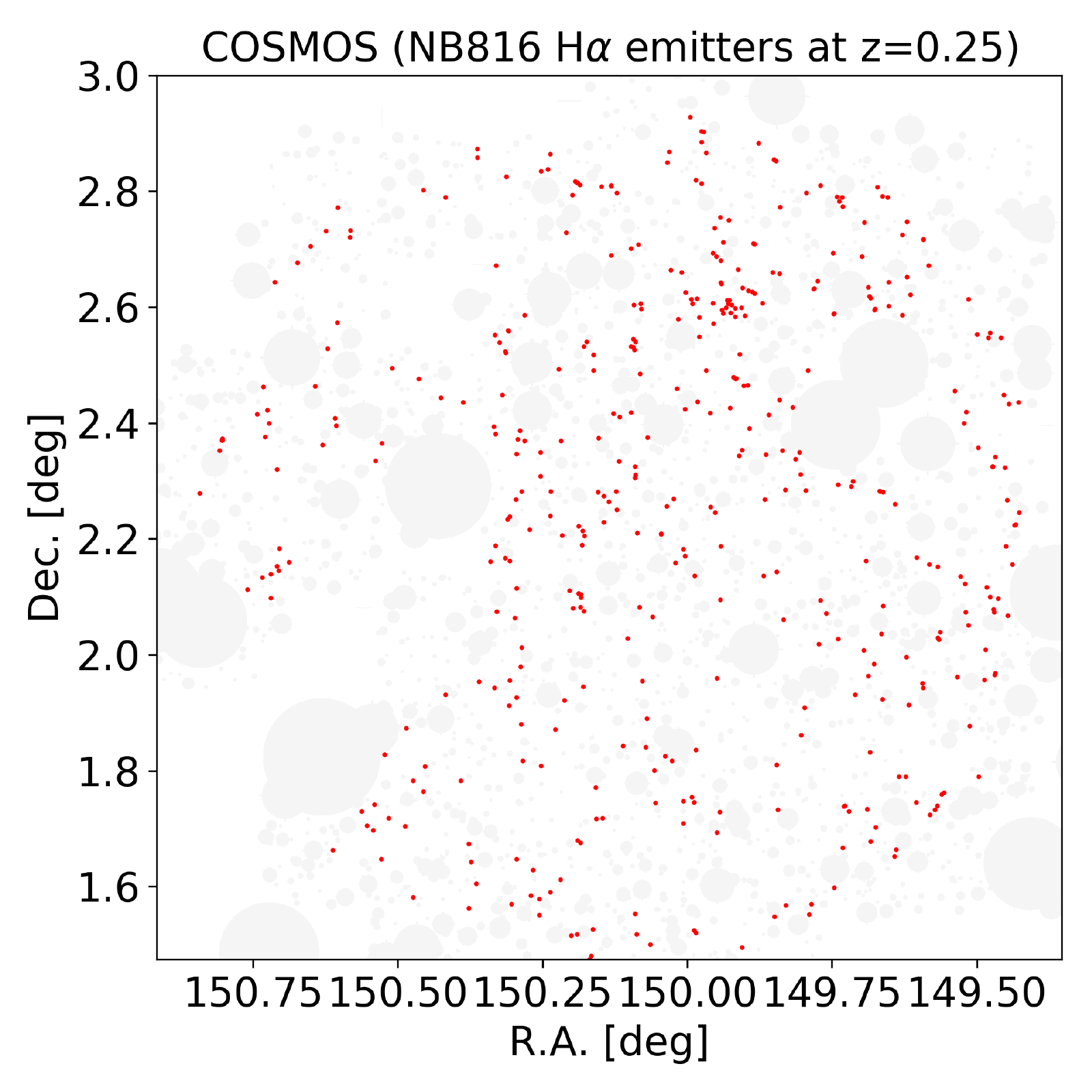} 
   \includegraphics[width=0.32\textwidth, bb=0 0 461 461]{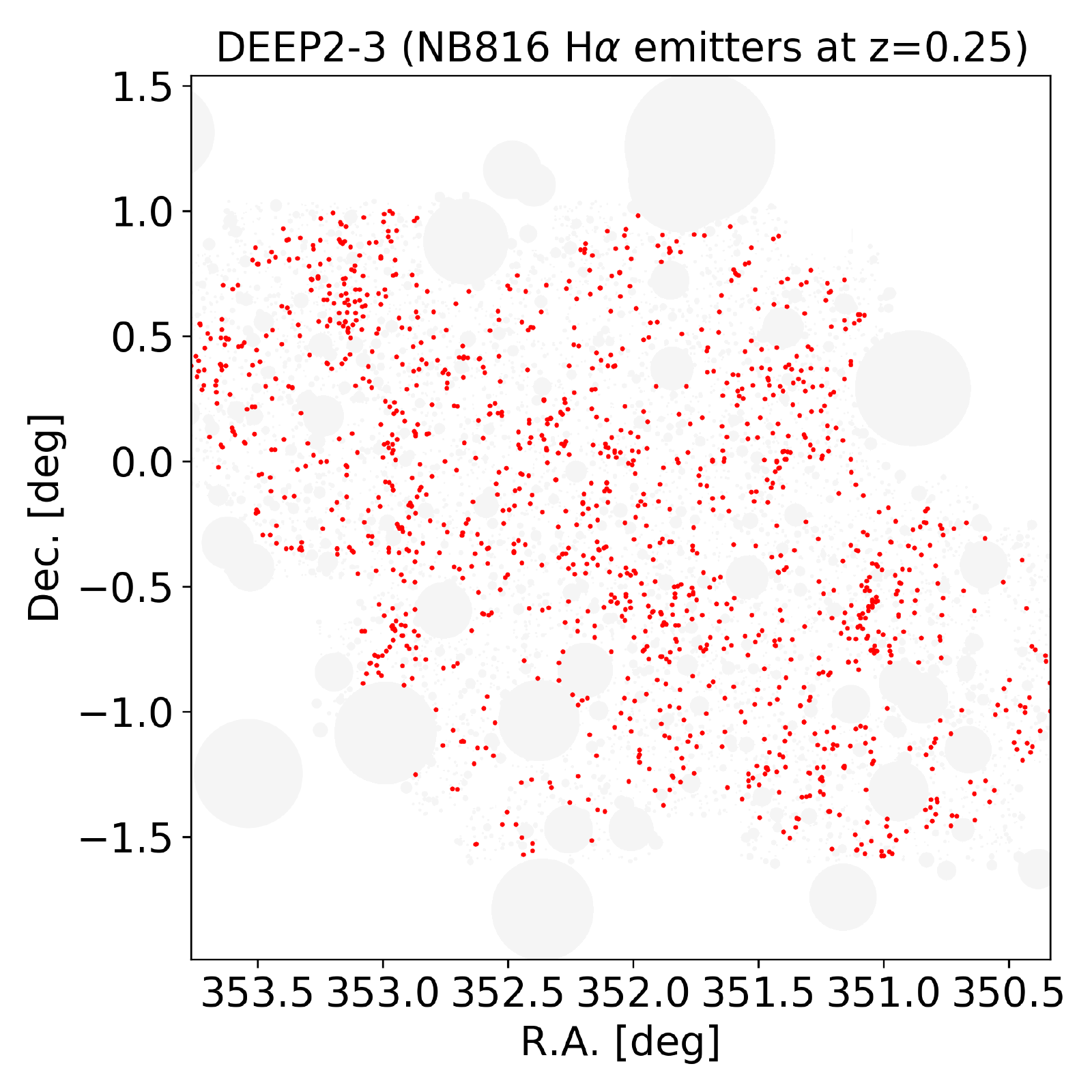} 
   \includegraphics[width=0.32\textwidth, bb=0 0 461 461]{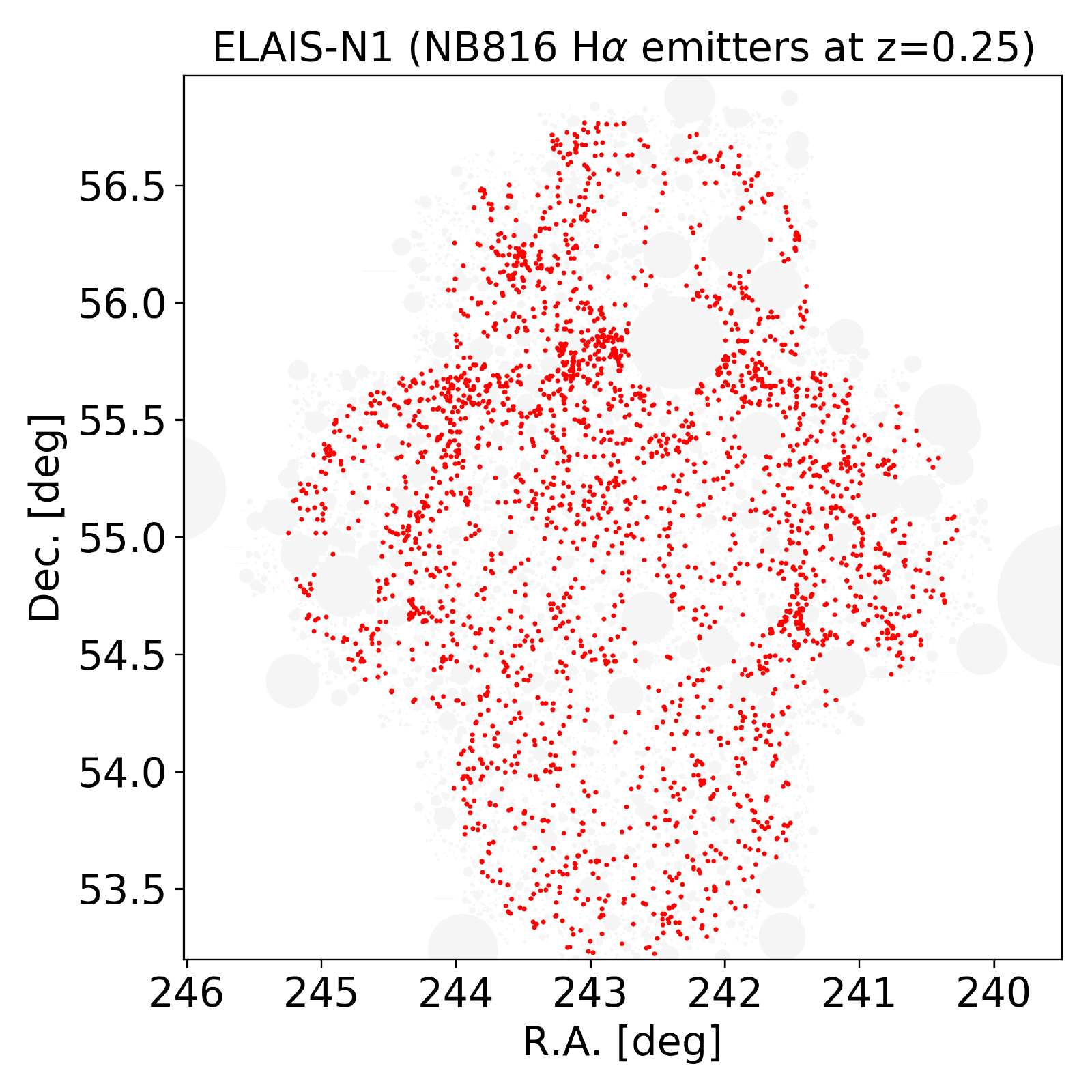} 
   \includegraphics[width=0.32\textwidth, bb=0 0 461 461]{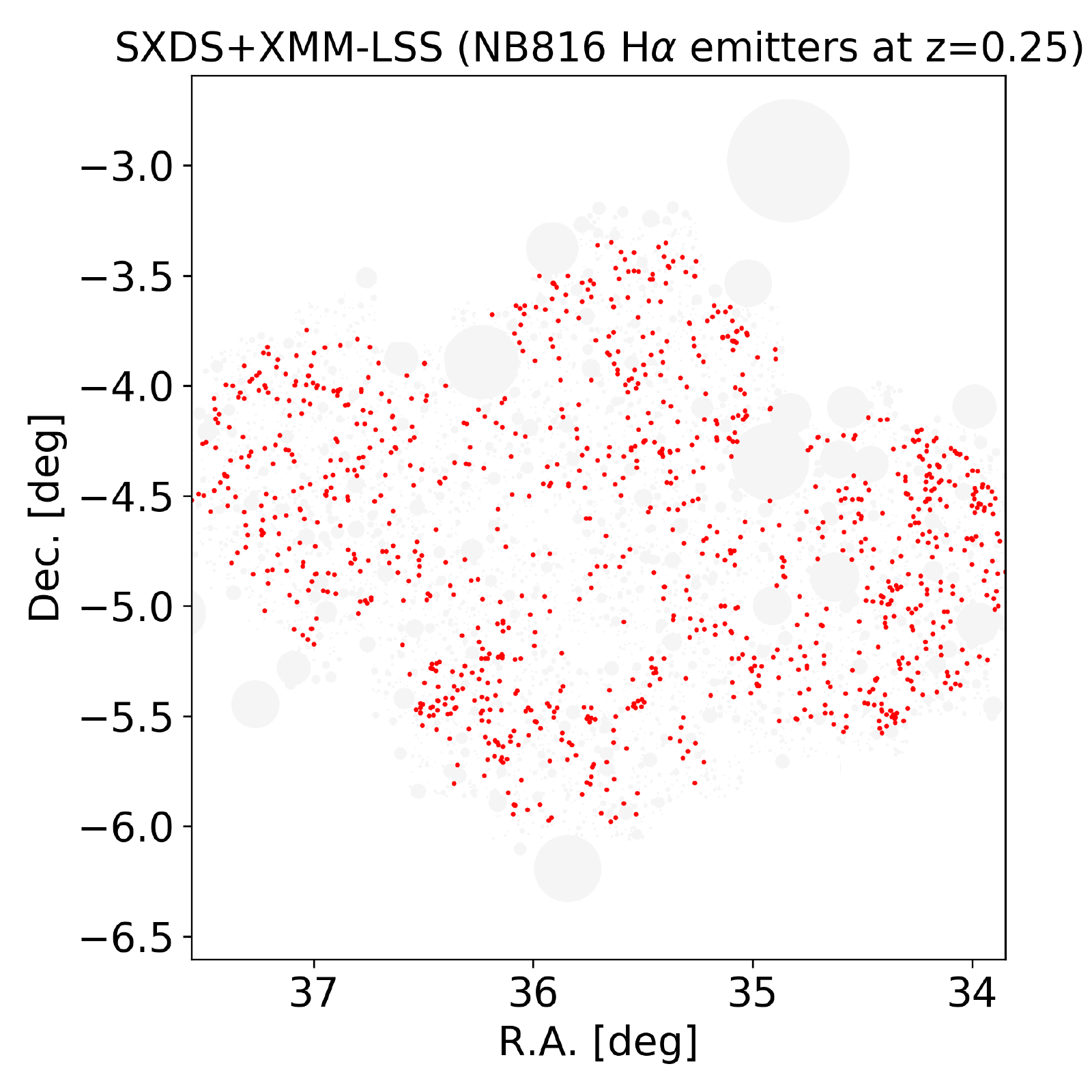} 
   \includegraphics[width=0.32\textwidth, bb=0 0 461 461]{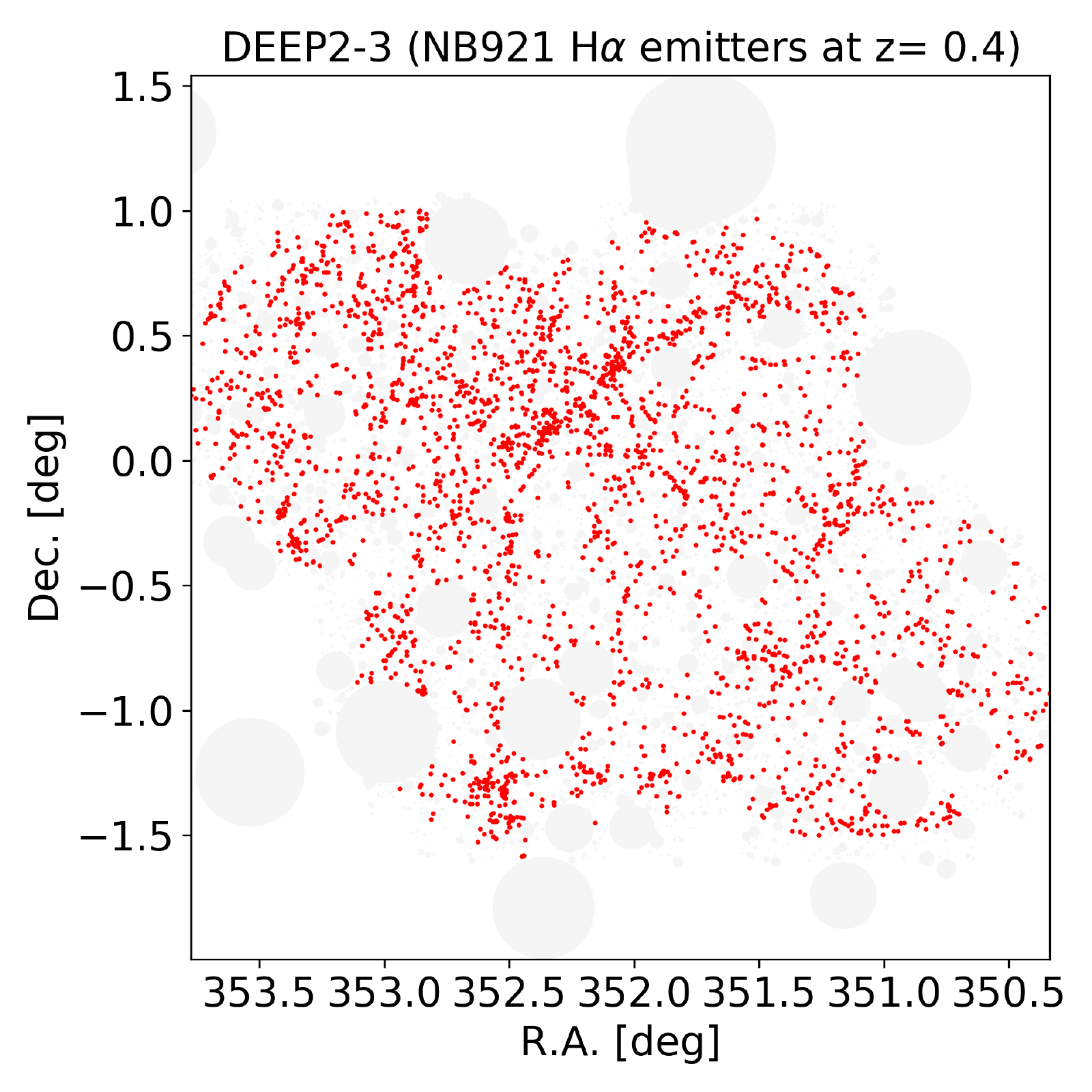} 
   \includegraphics[width=0.32\textwidth, bb=0 0 461 461]{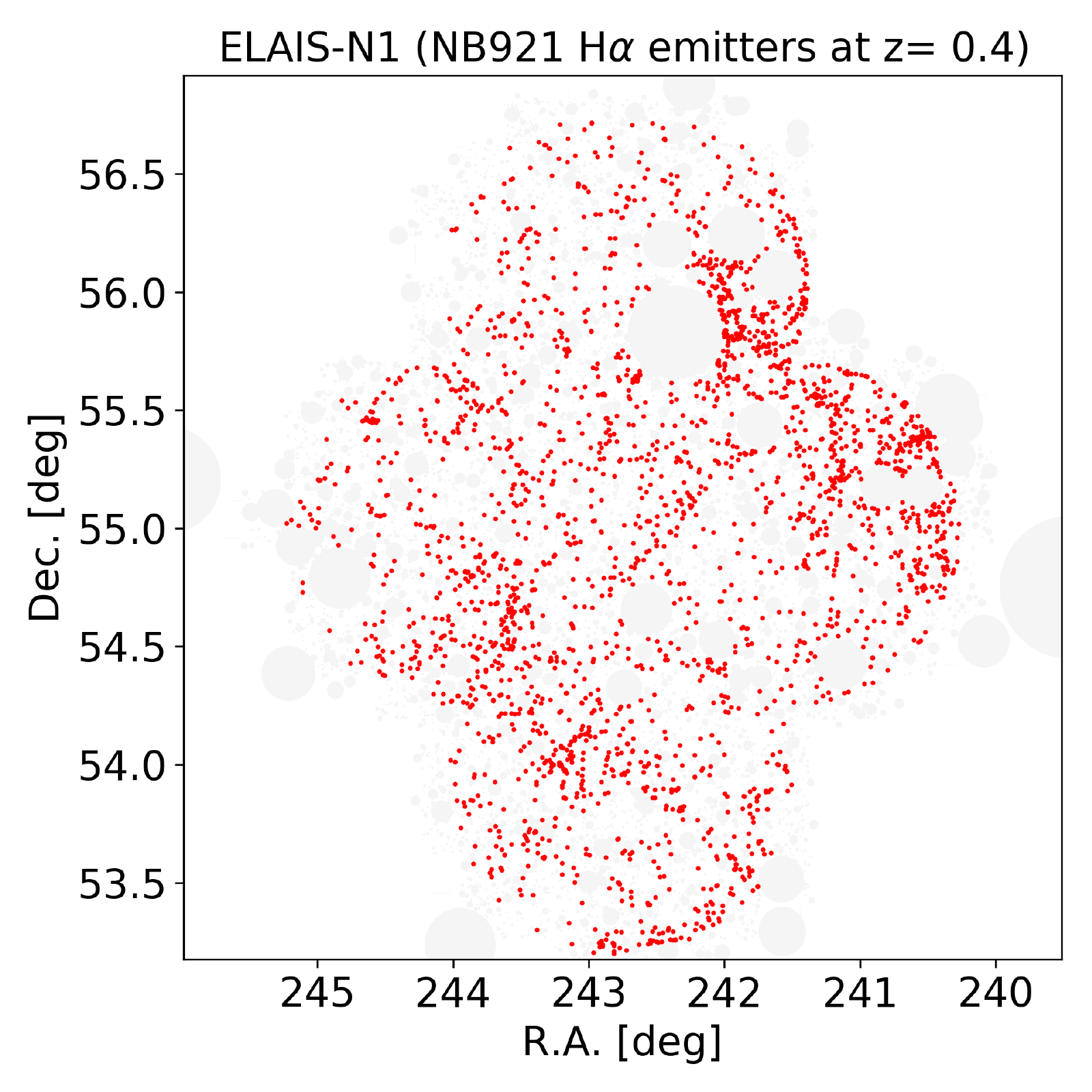} 
   \includegraphics[width=0.32\textwidth, bb=0 0 461 461]{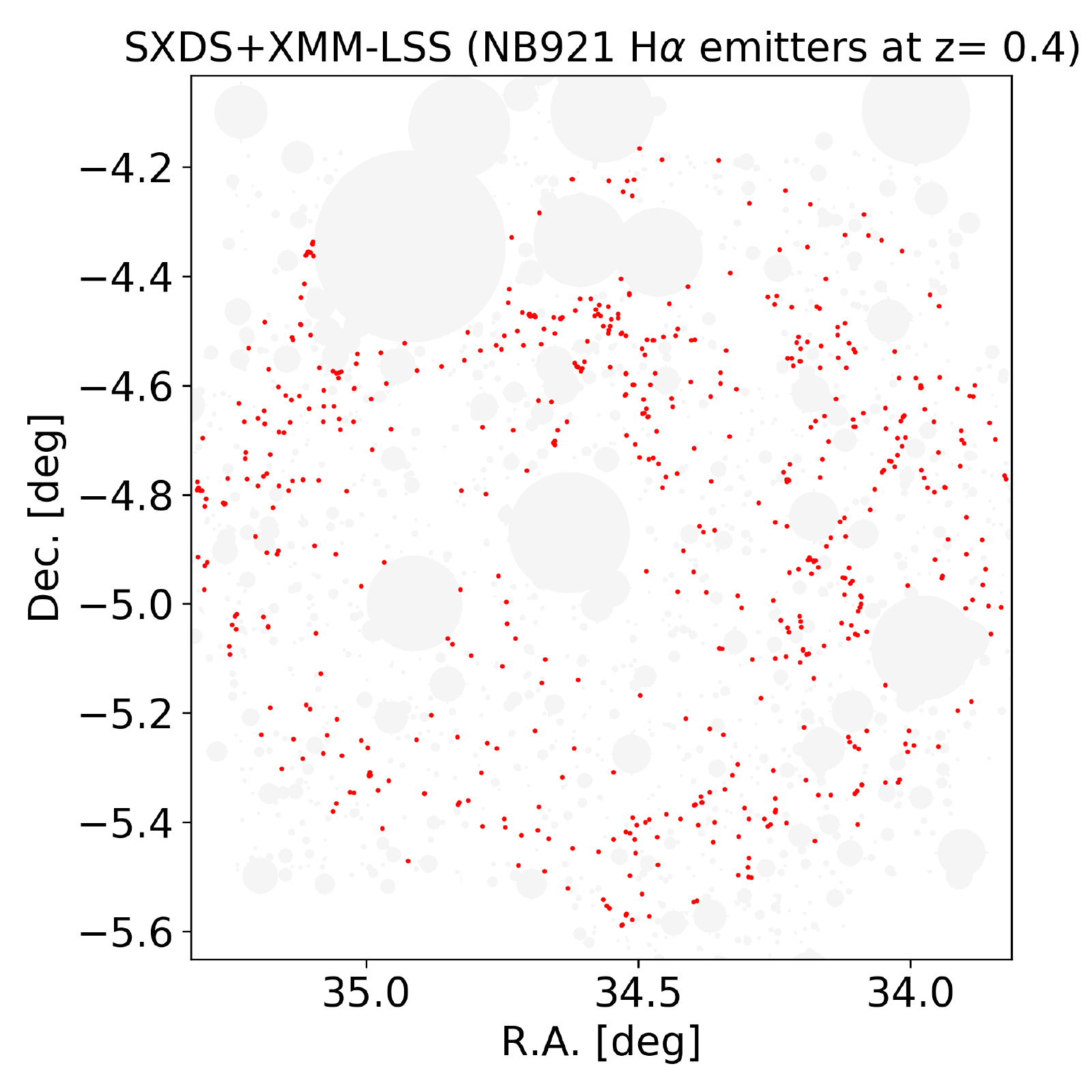} 
   \includegraphics[width=0.32\textwidth, bb=0 0 461 461]{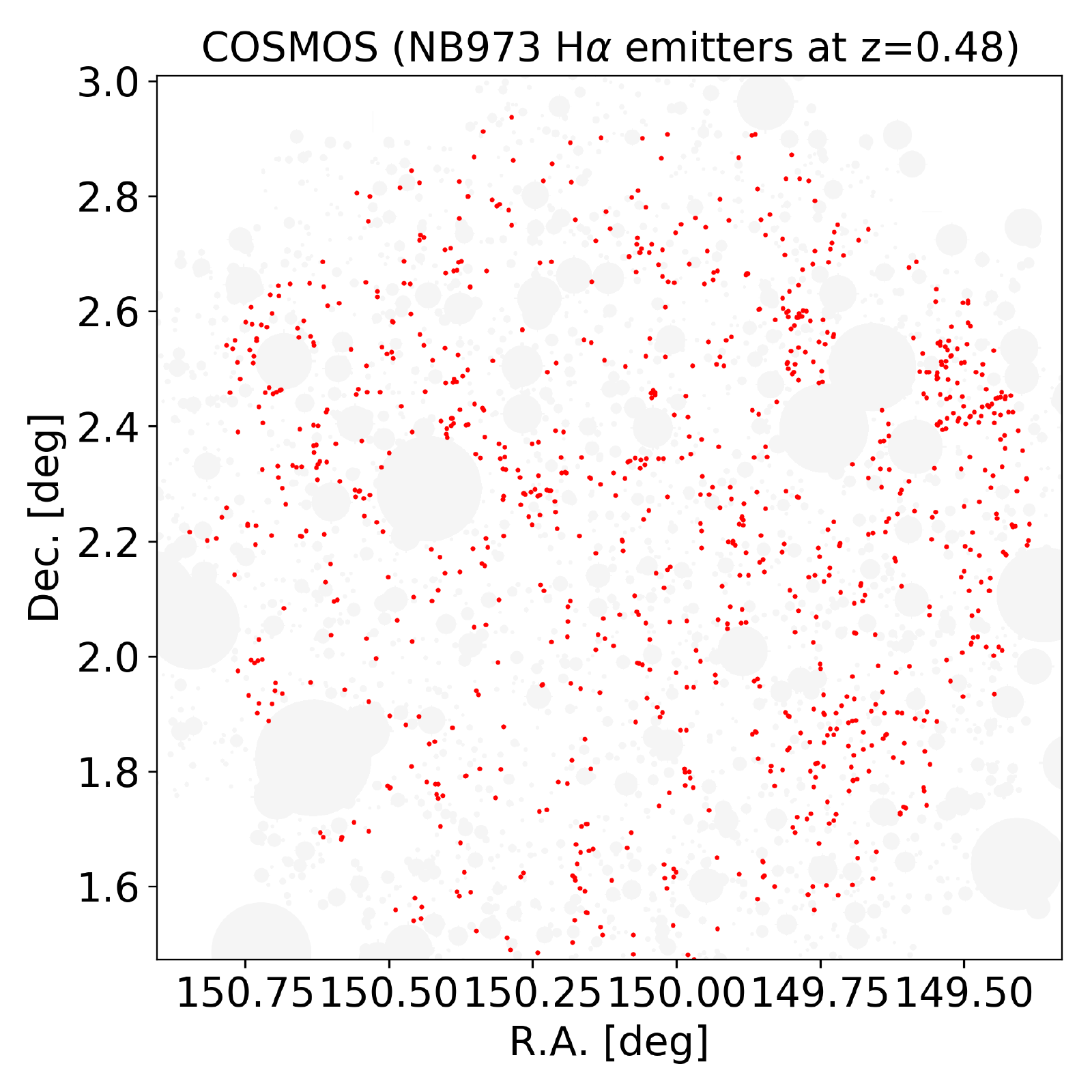} 
 \end{center}
 \caption{The spatial distribution of H$\alpha$ emission-line galaxies.}\label{fig:MapHAE}  
\end{figure*}

\begin{figure*}
 \begin{center}
   \includegraphics[width=0.32\textwidth, bb=0 0 461 461]{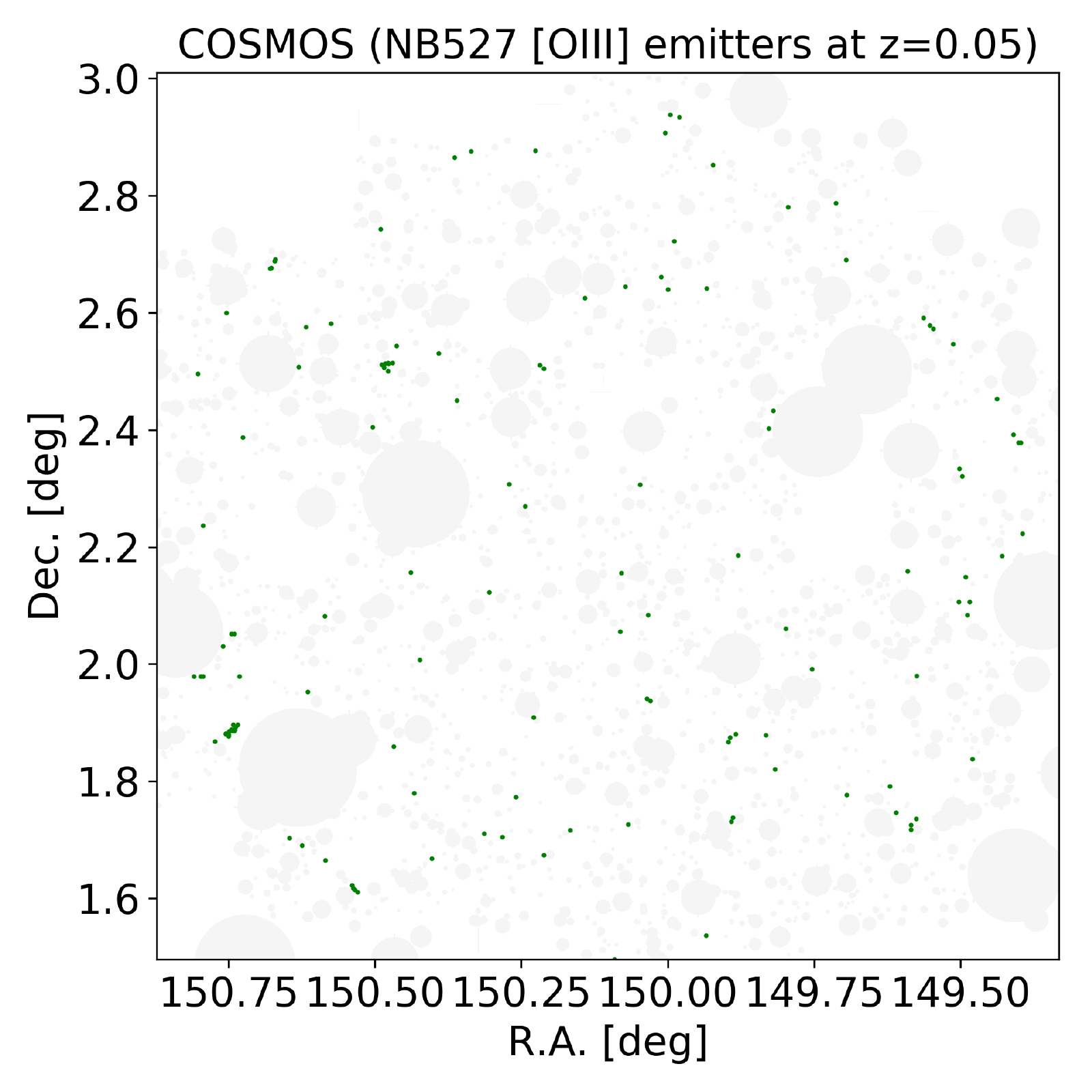} 
   \includegraphics[width=0.32\textwidth, bb=0 0 461 461]{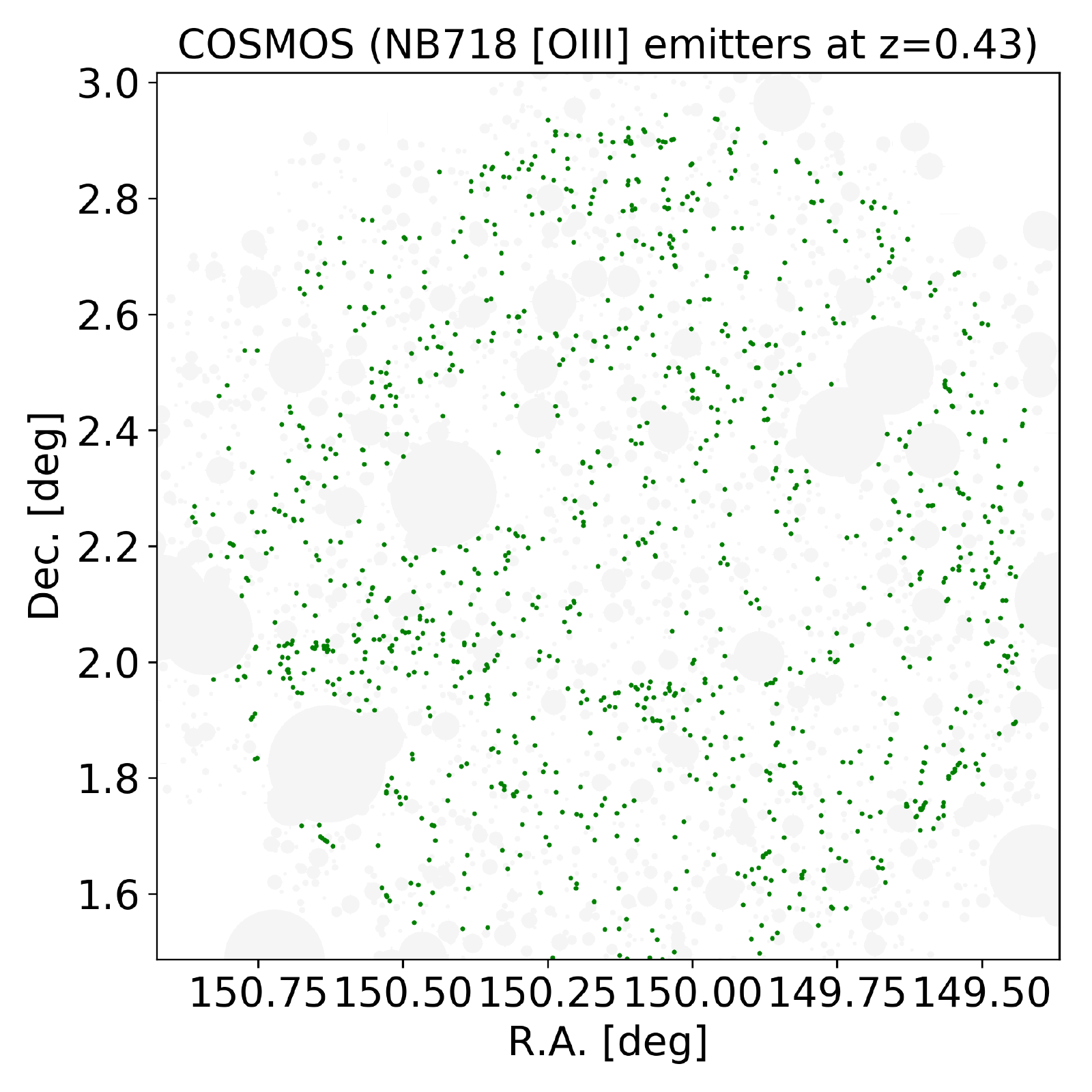} 
   \includegraphics[width=0.32\textwidth, bb=0 0 461 461]{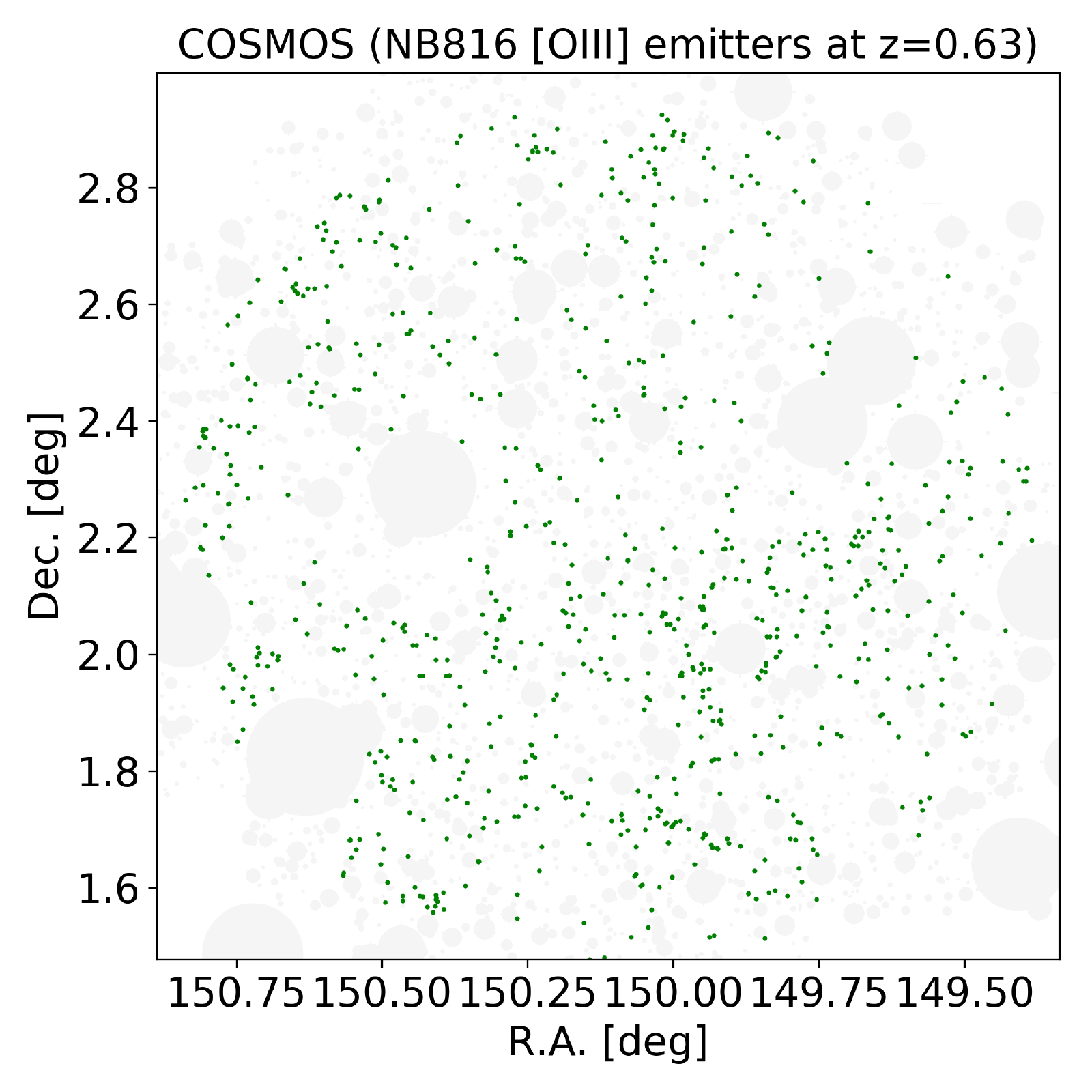} 
   \includegraphics[width=0.32\textwidth, bb=0 0 461 461]{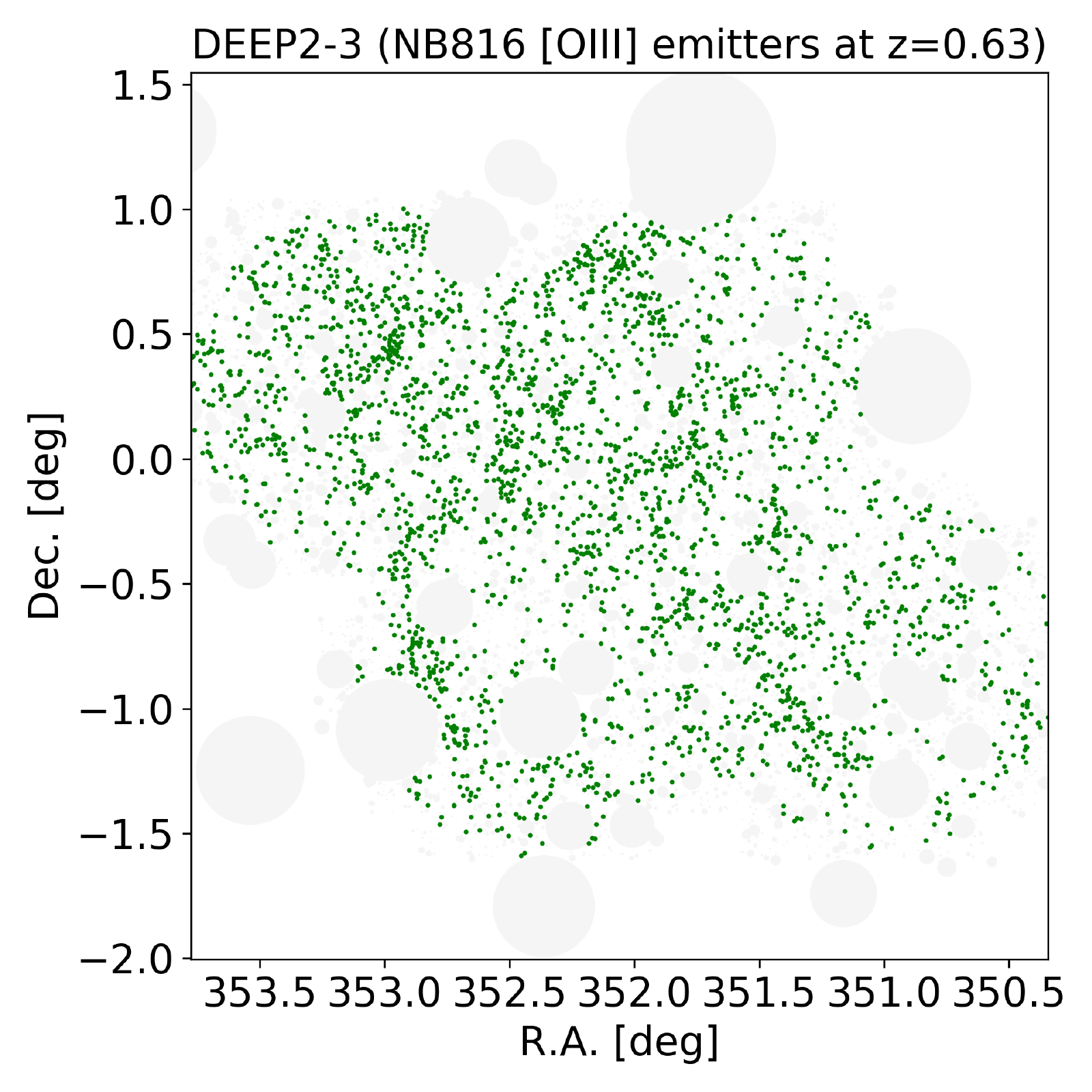} 
   \includegraphics[width=0.32\textwidth, bb=0 0 461 461]{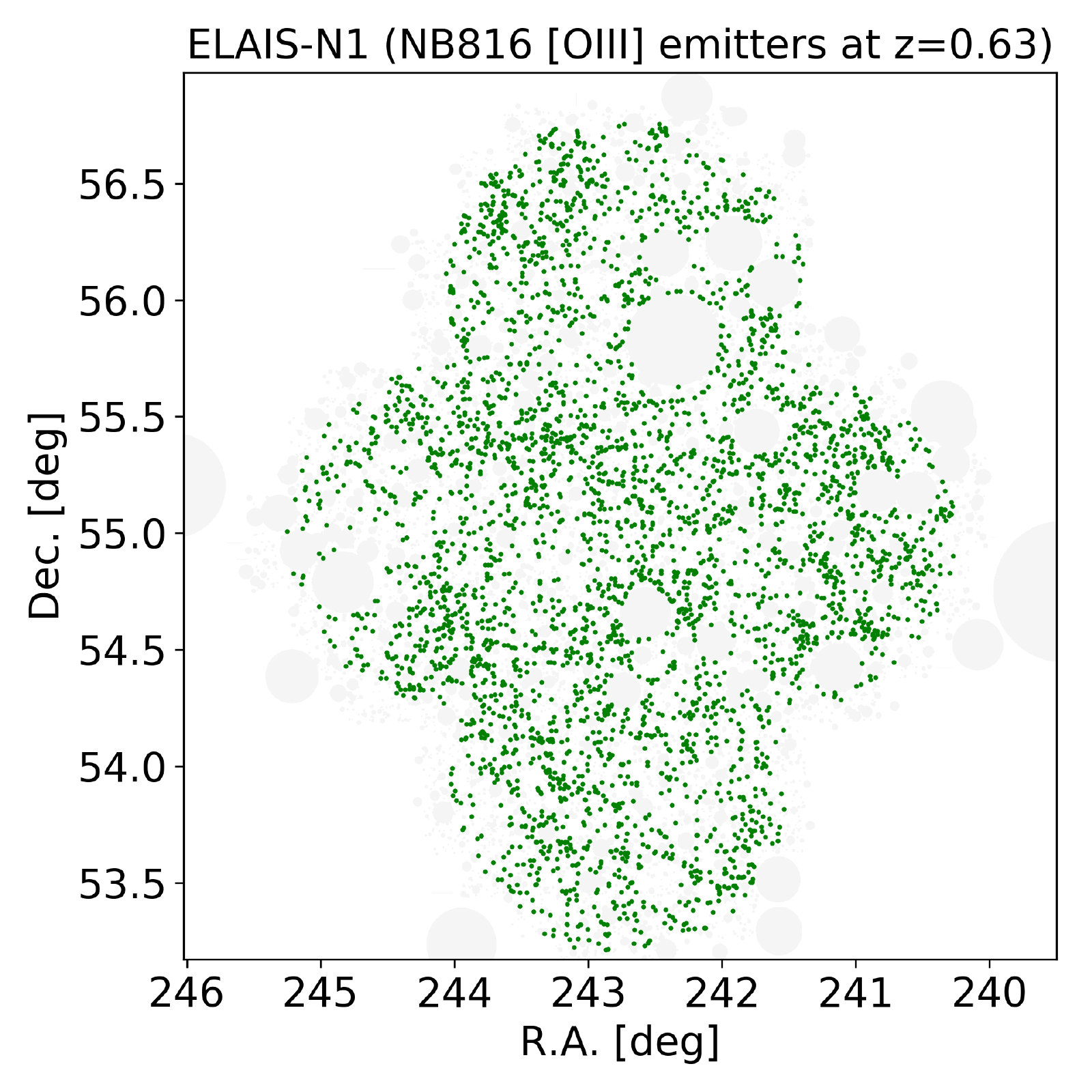} 
   \includegraphics[width=0.32\textwidth, bb=0 0 461 461]{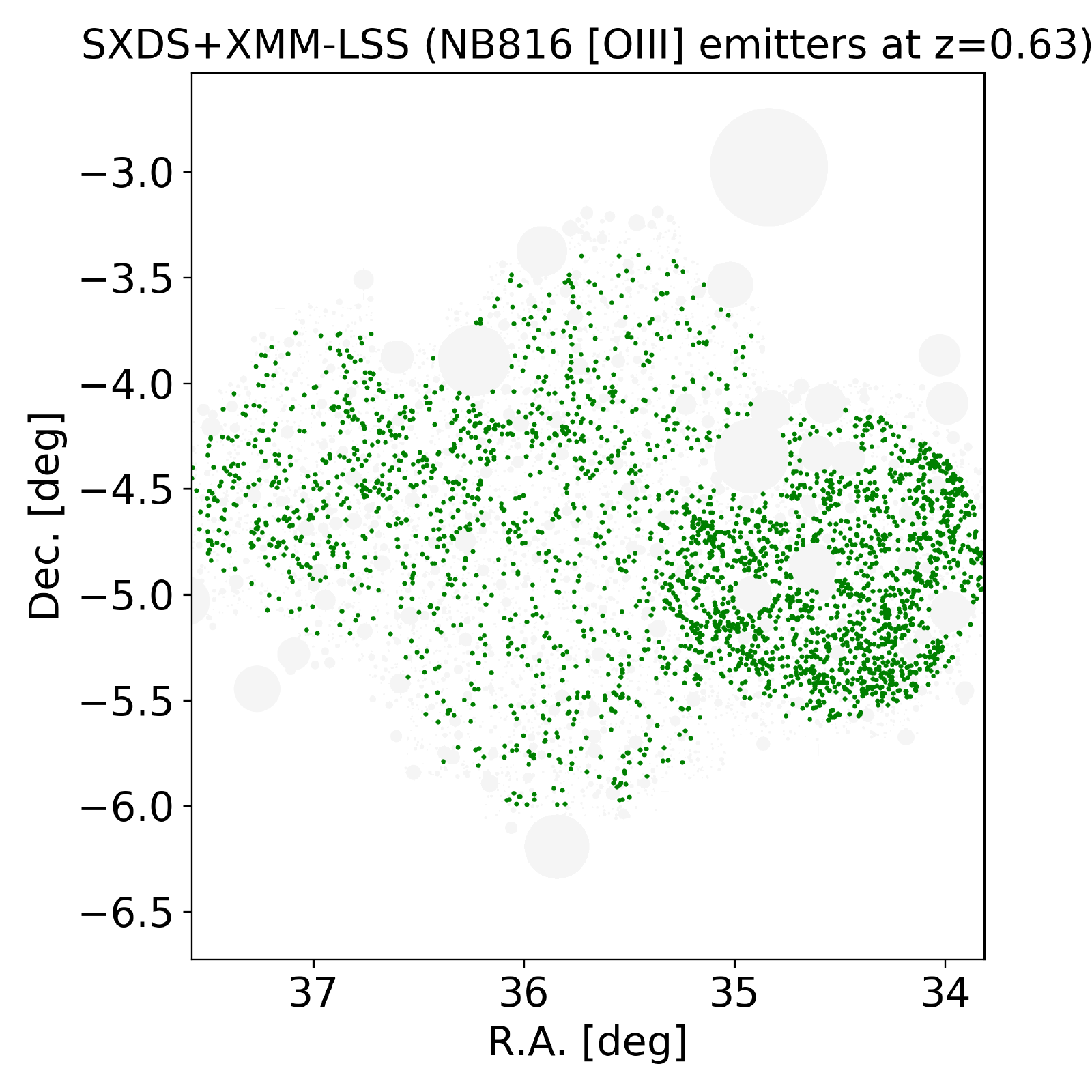} 
   \includegraphics[width=0.32\textwidth, bb=0 0 461 461]{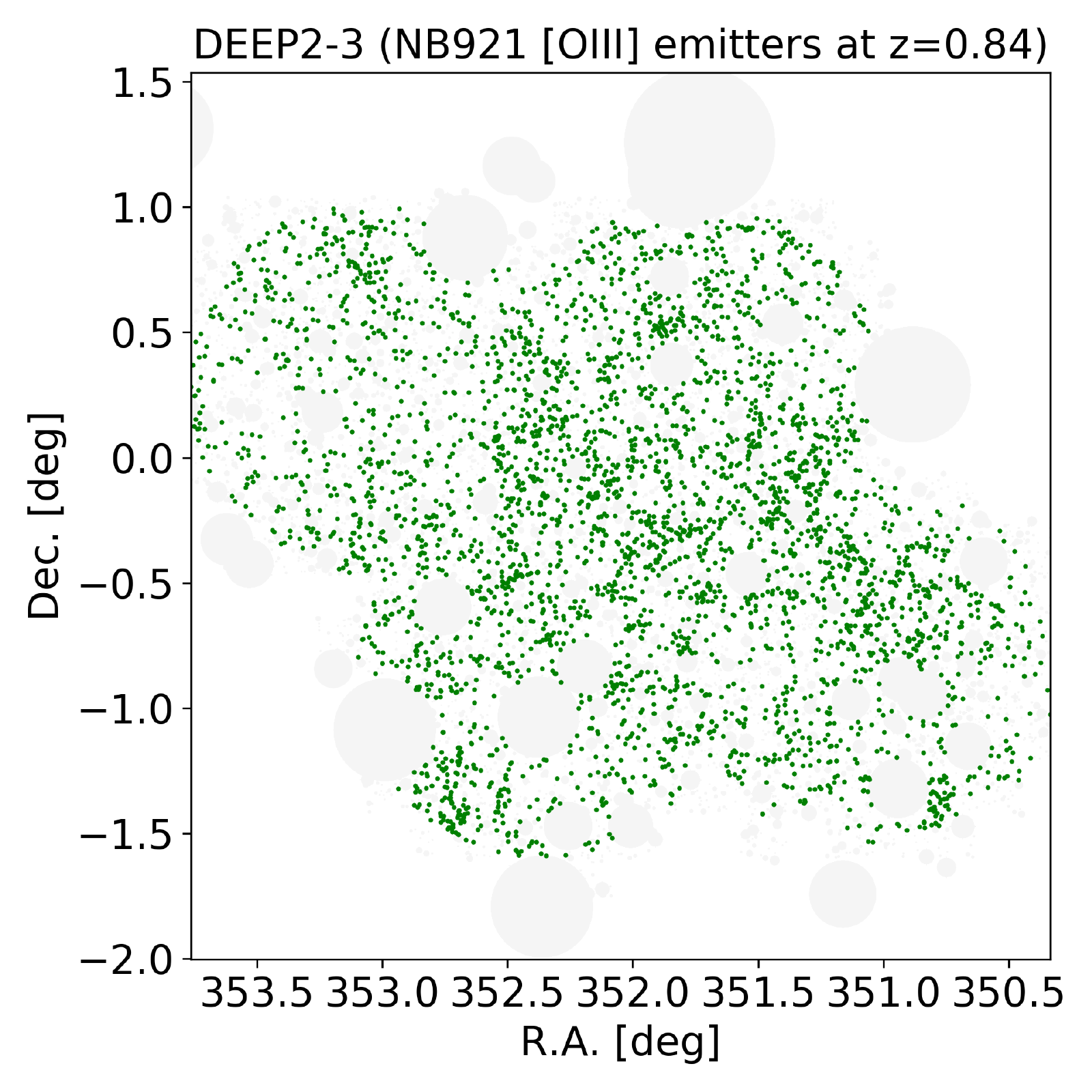} 
   \includegraphics[width=0.32\textwidth, bb=0 0 461 461]{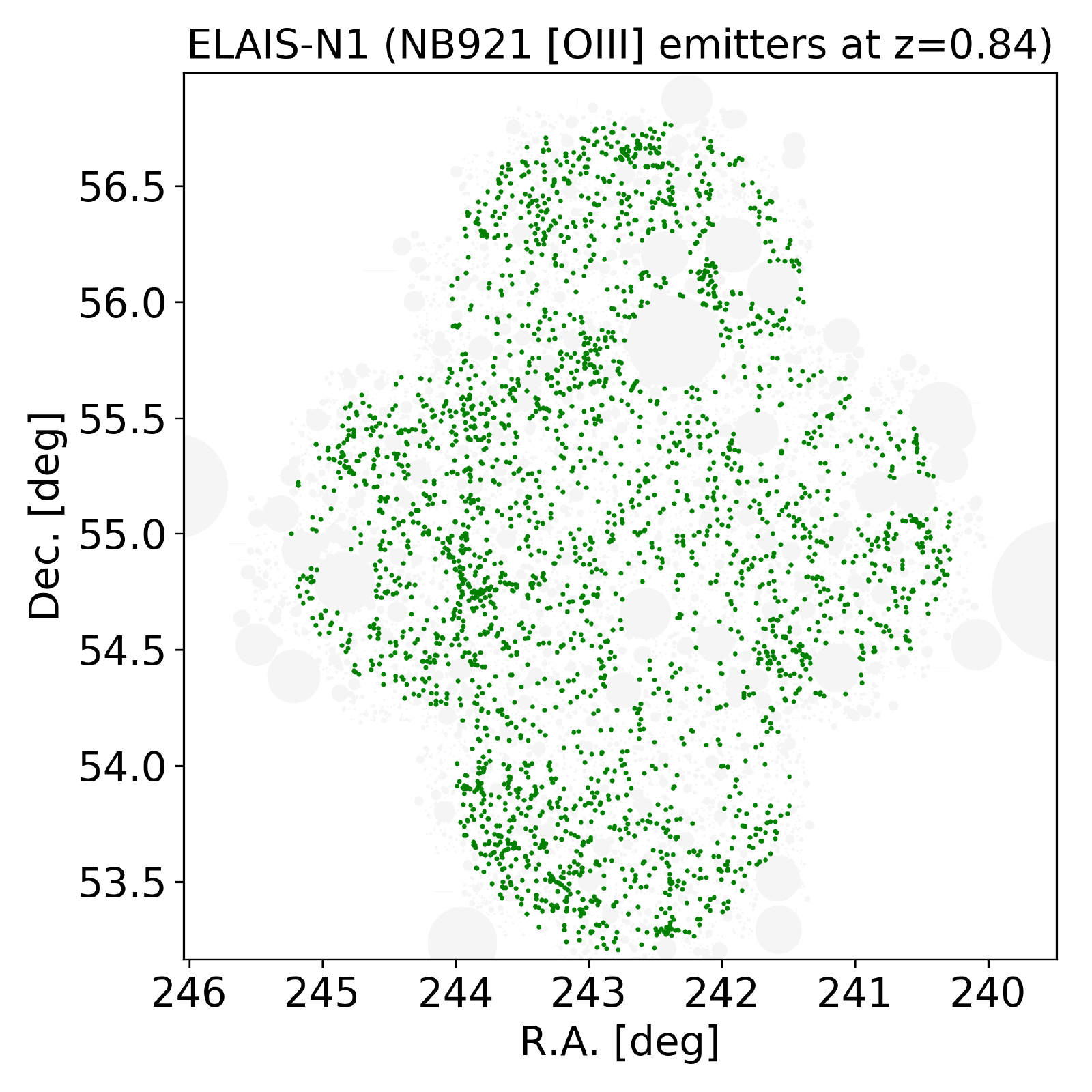} 
   \includegraphics[width=0.32\textwidth, bb=0 0 461 461]{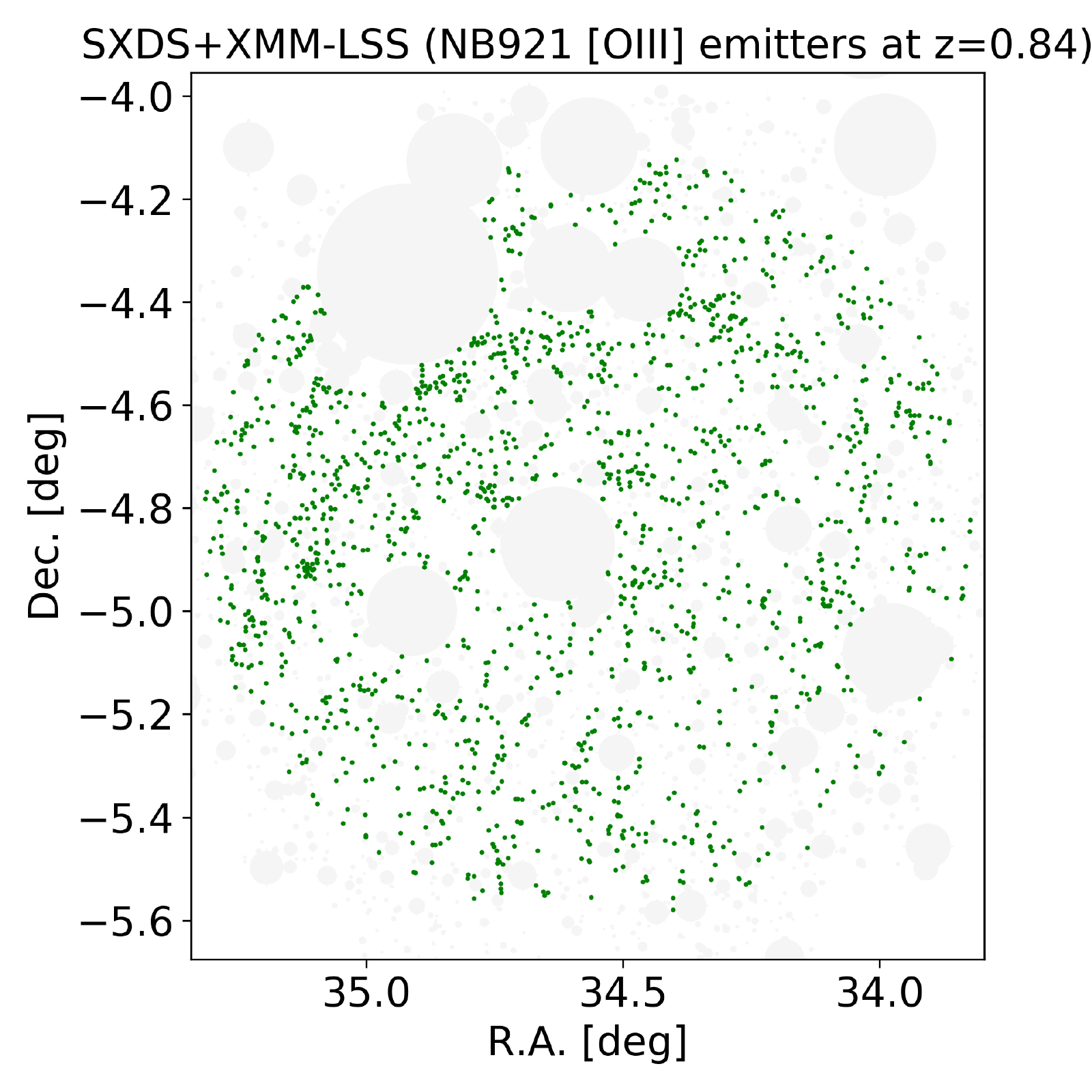} 
   \includegraphics[width=0.32\textwidth, bb=0 0 461 461]{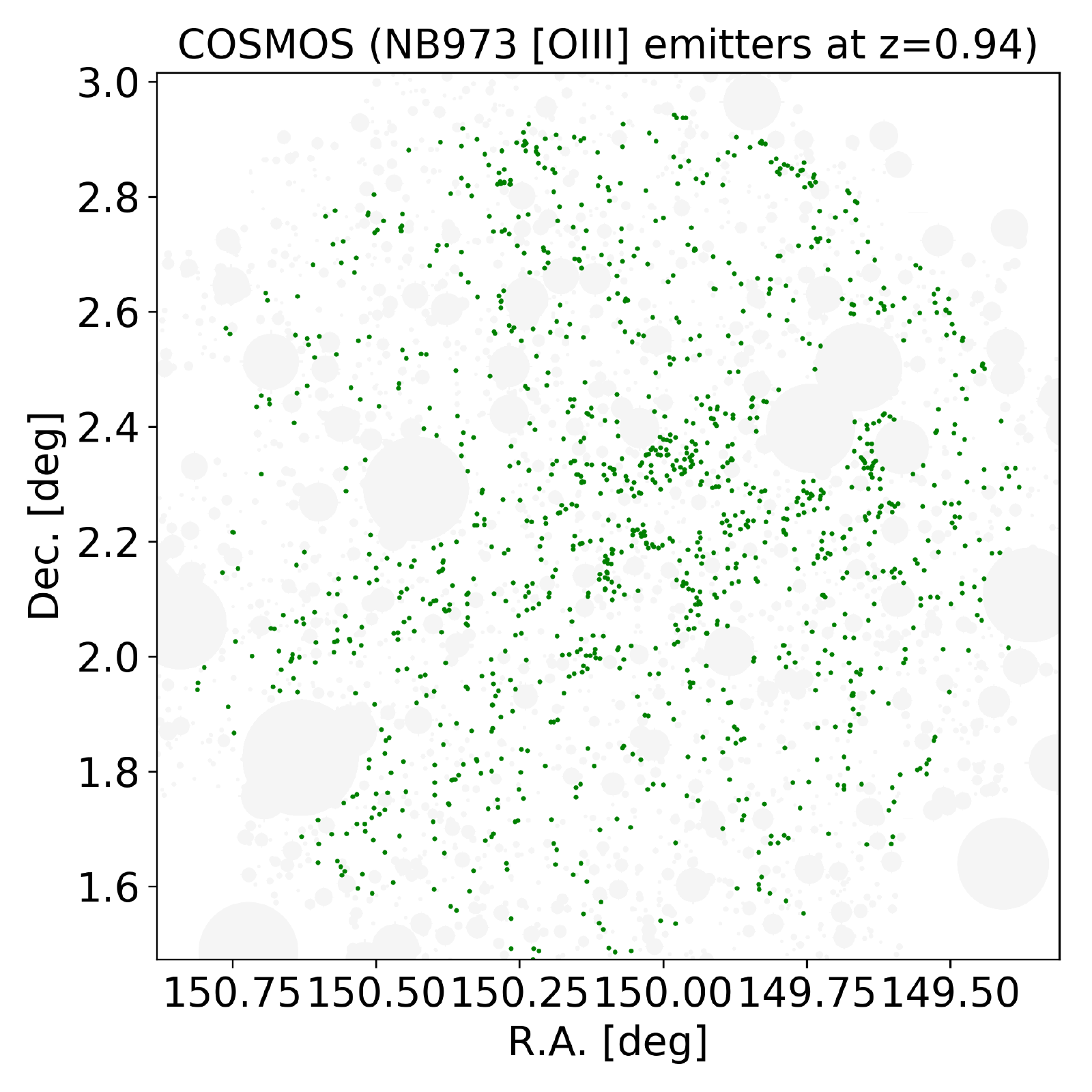} 
 \end{center}
 \caption{The spatial distribution of [OIII] emission-line galaxies.}\label{fig:MapO3E}  
\end{figure*}

\begin{figure*}
 \begin{center}
   \includegraphics[width=0.32\textwidth, bb=0 0 461 461]{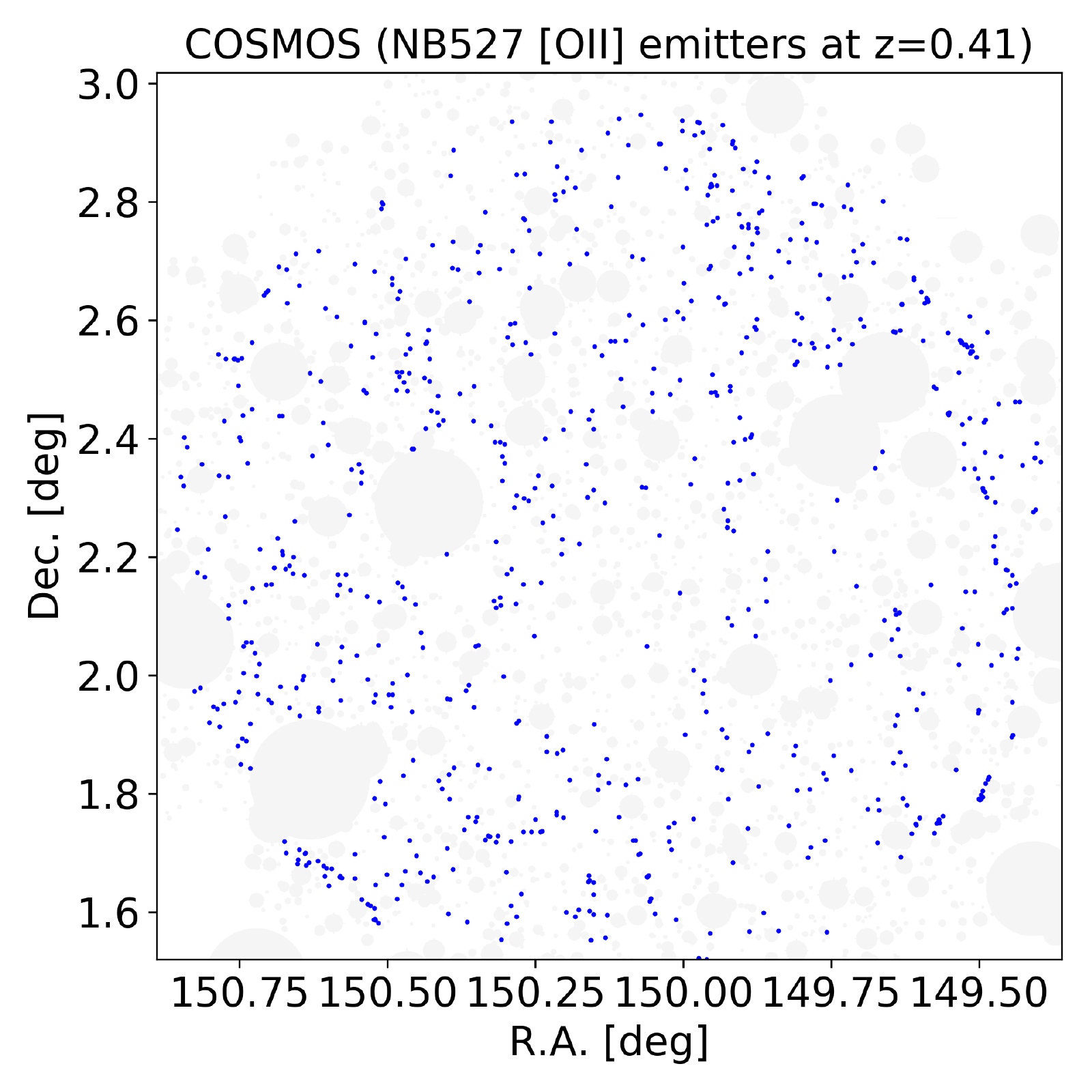} 
   \includegraphics[width=0.32\textwidth, bb=0 0 461 461]{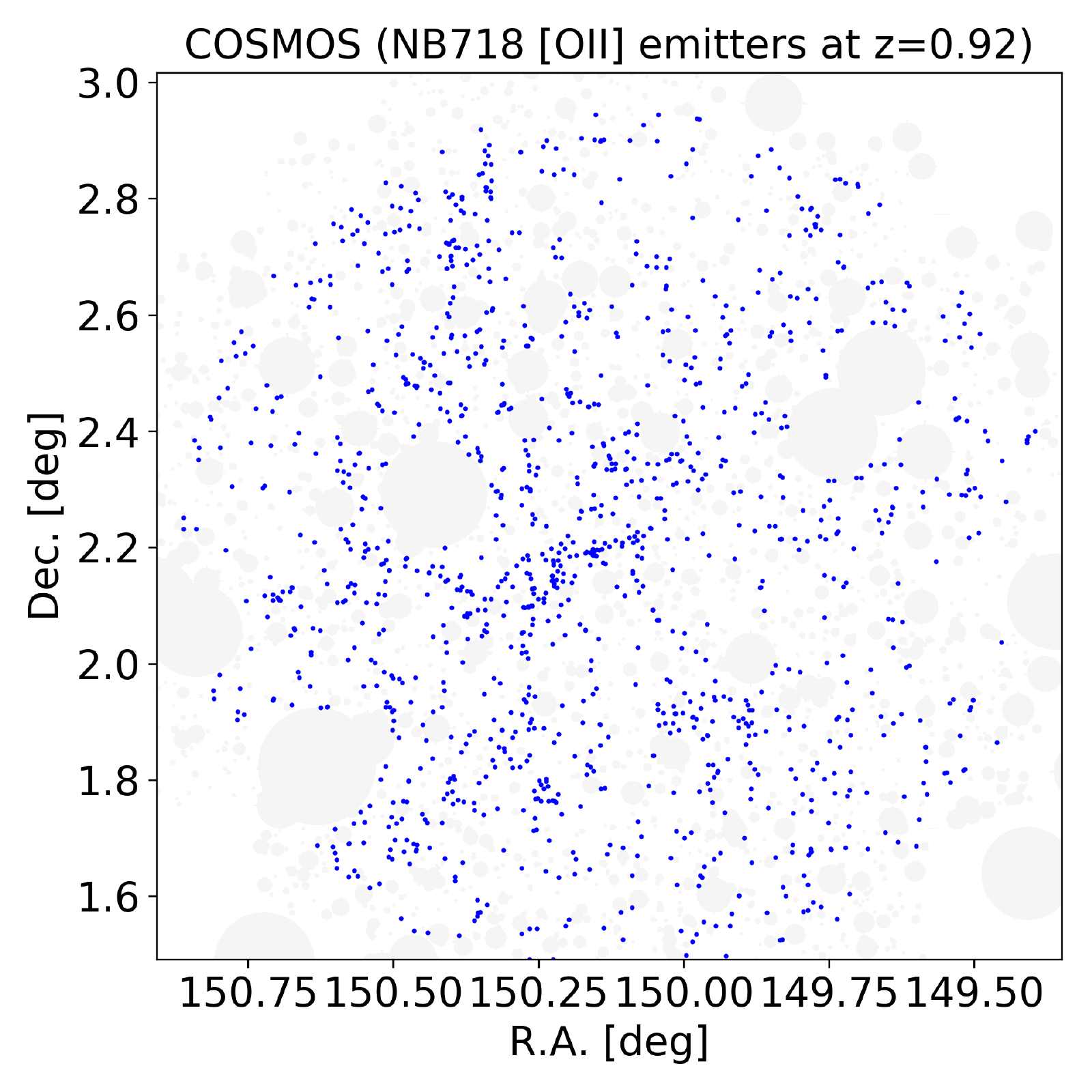} 
   \includegraphics[width=0.32\textwidth, bb=0 0 461 461]{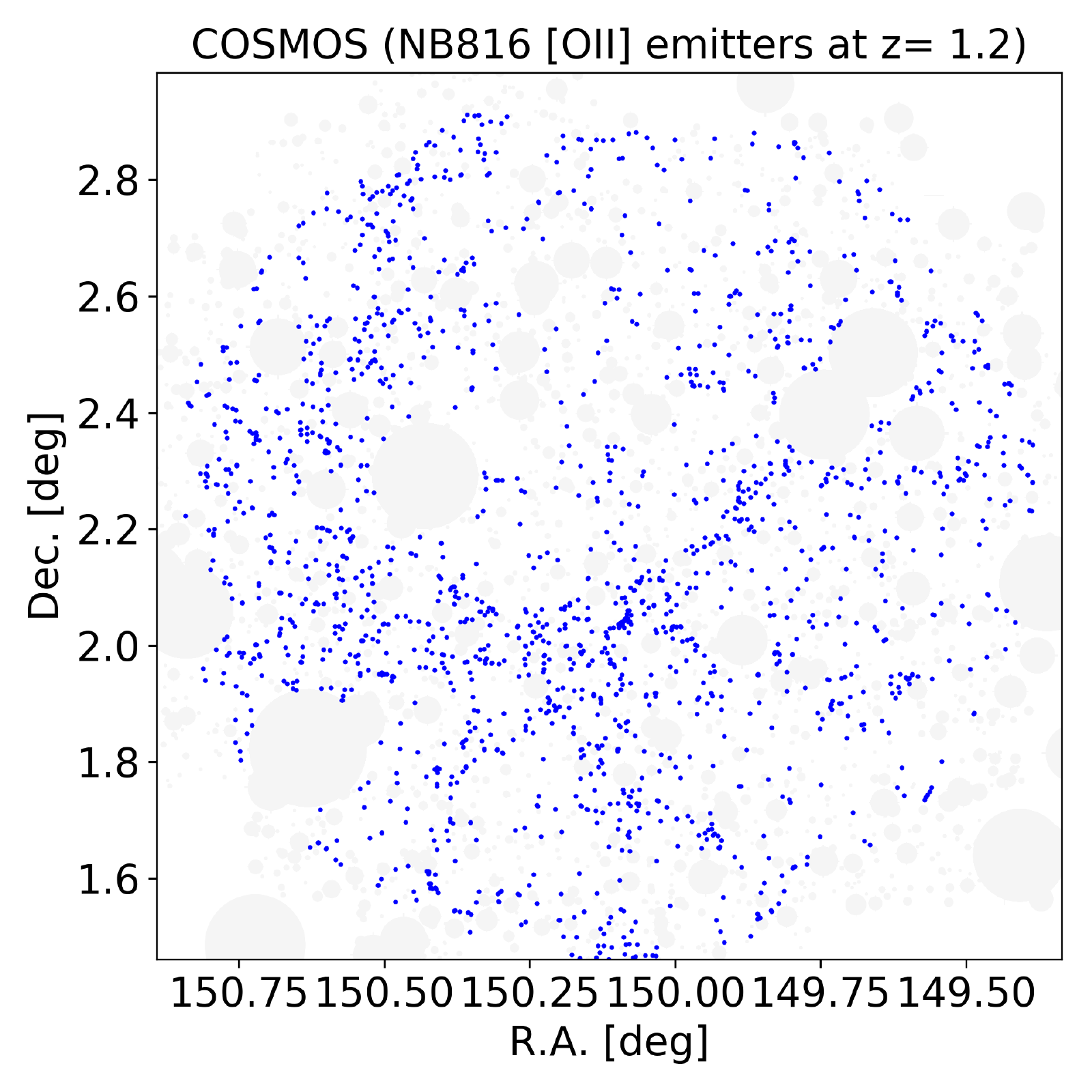} 
   \includegraphics[width=0.32\textwidth, bb=0 0 461 461]{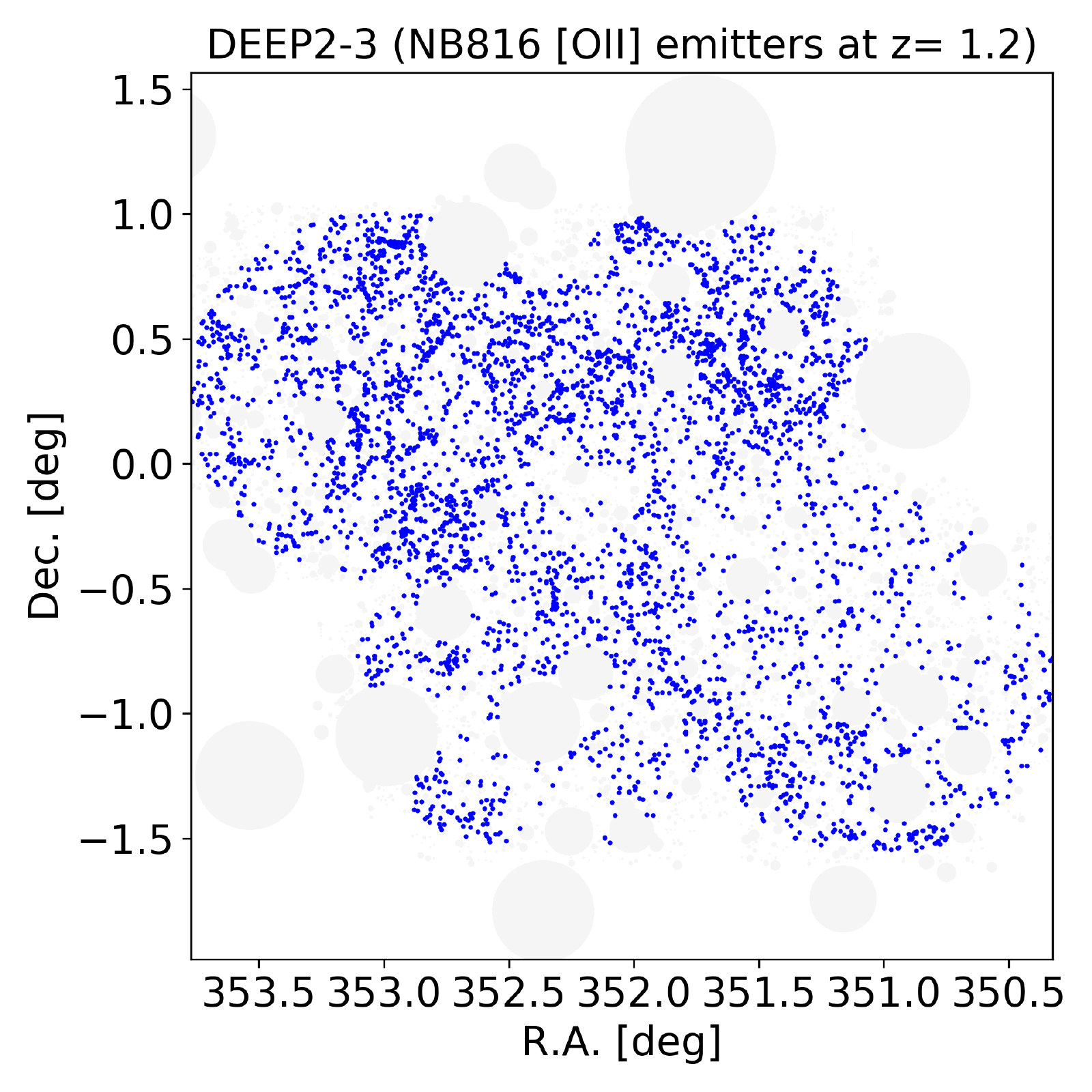} 
   \includegraphics[width=0.32\textwidth, bb=0 0 461 461]{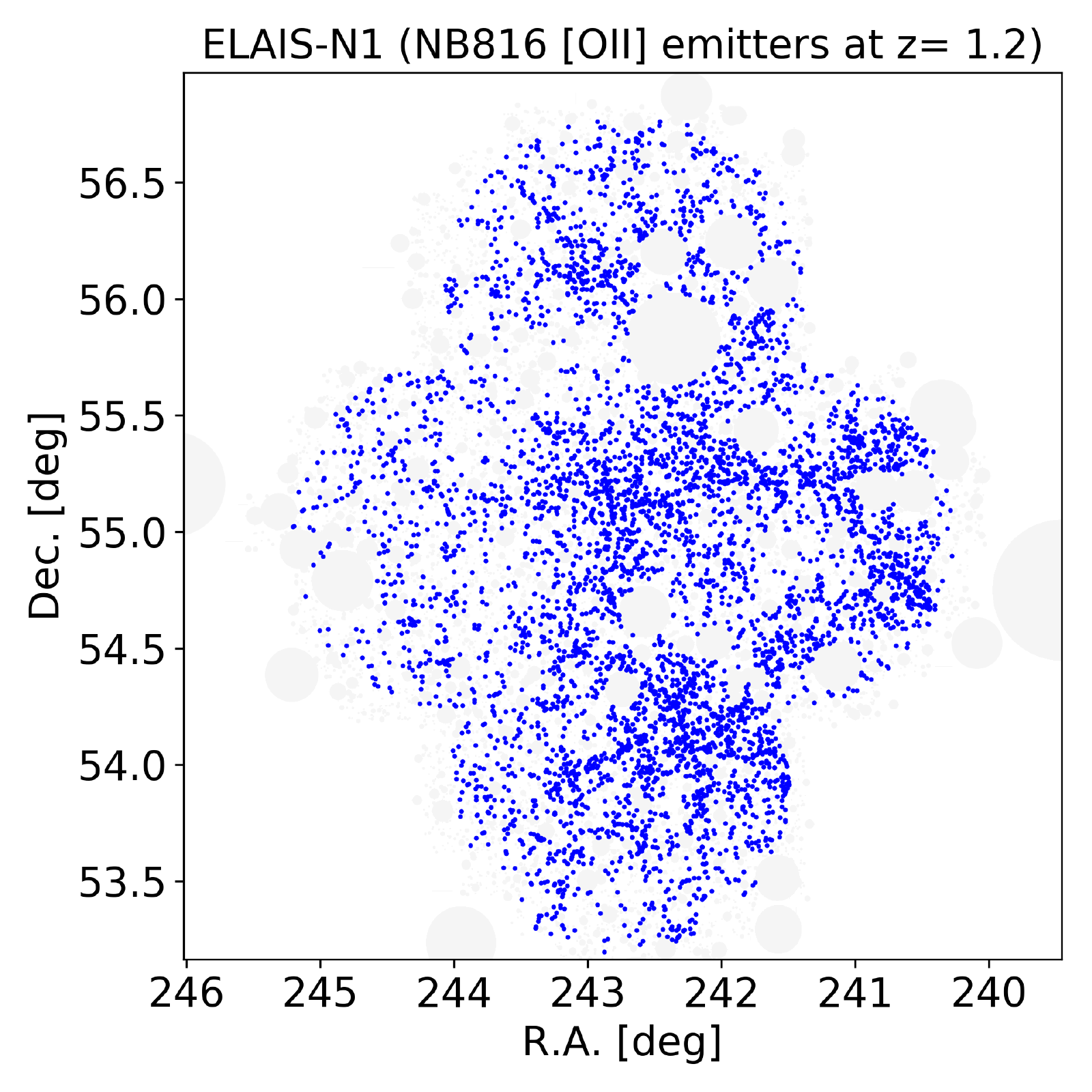} 
   \includegraphics[width=0.32\textwidth, bb=0 0 461 461]{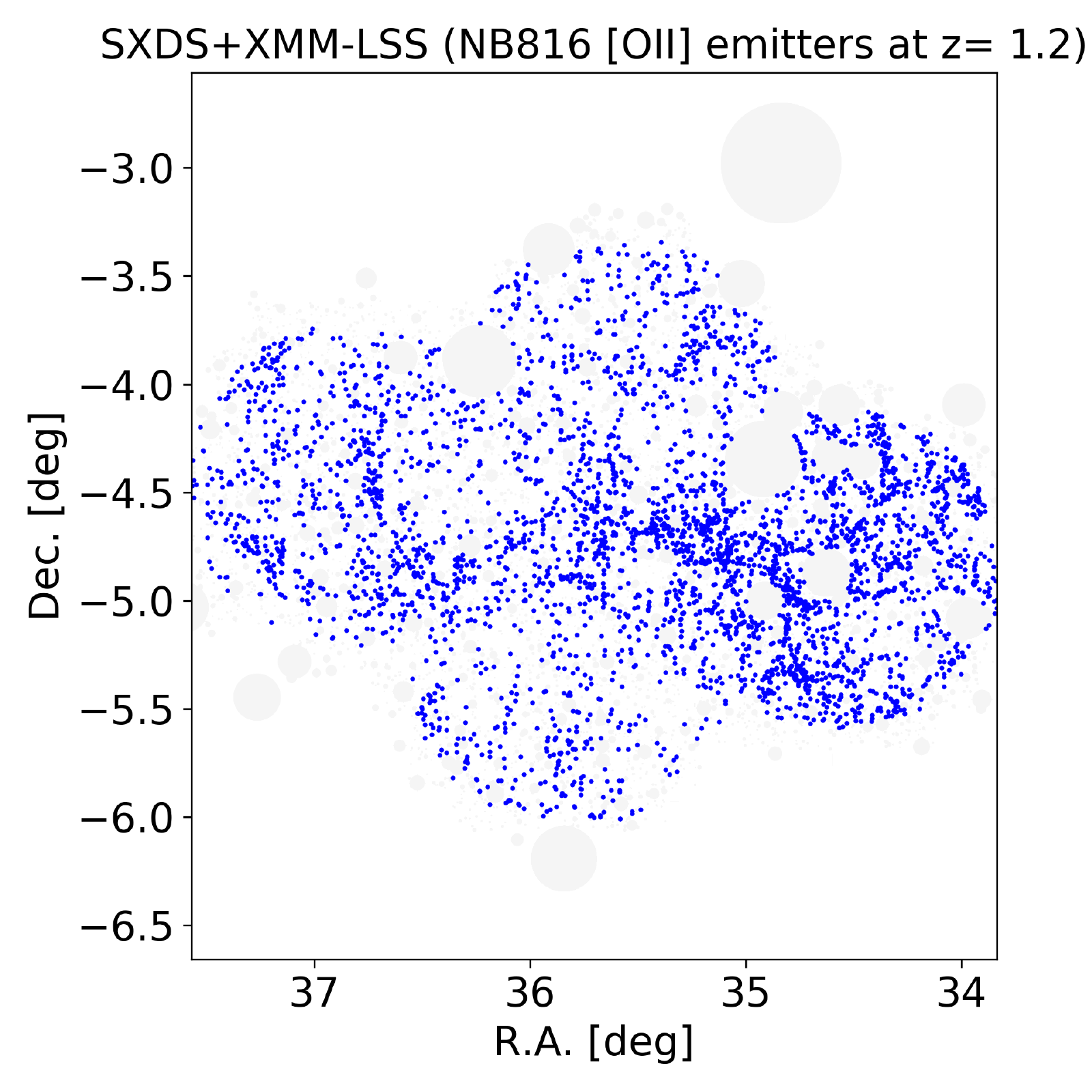} 
   \includegraphics[width=0.32\textwidth, bb=0 0 461 461]{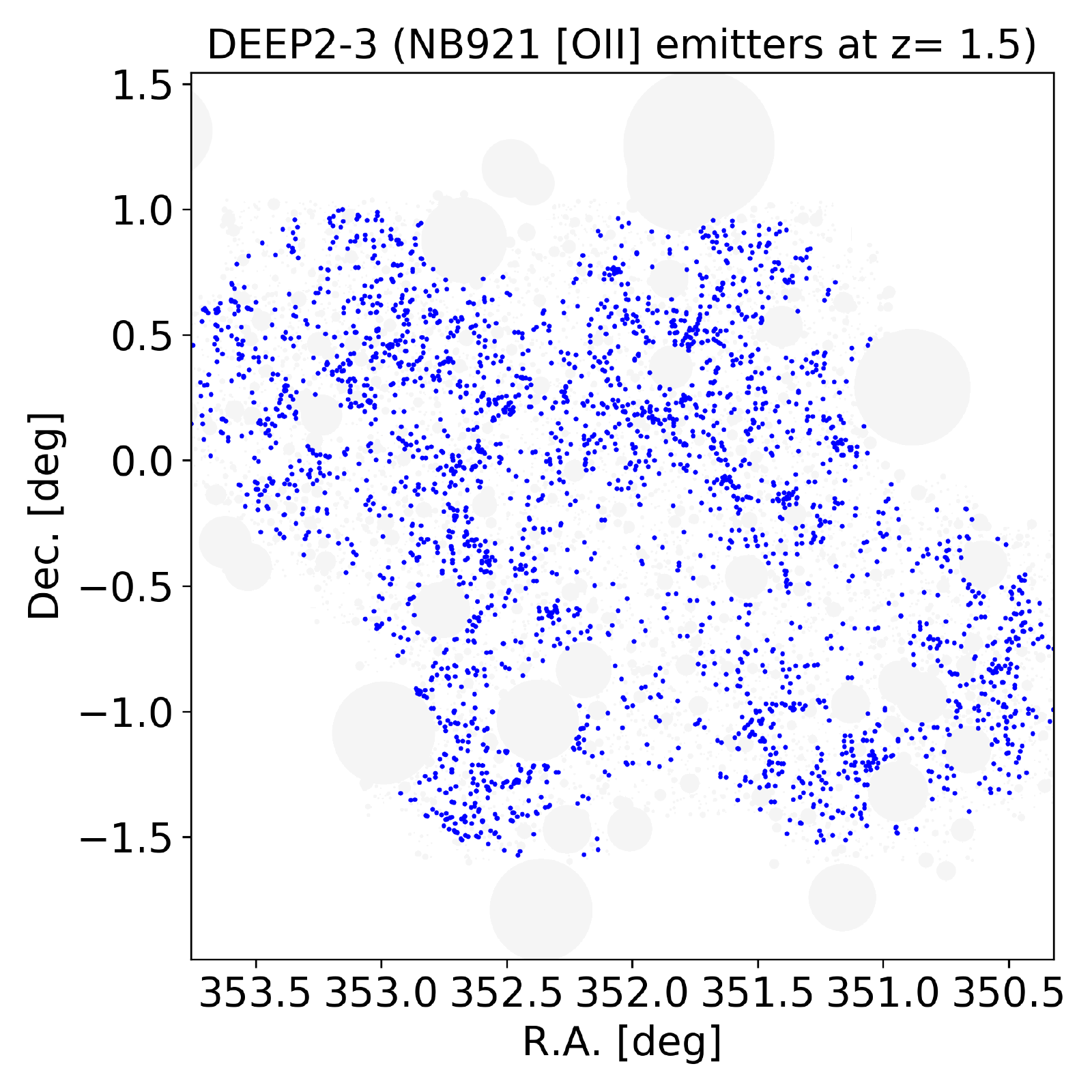} 
   \includegraphics[width=0.32\textwidth, bb=0 0 461 461]{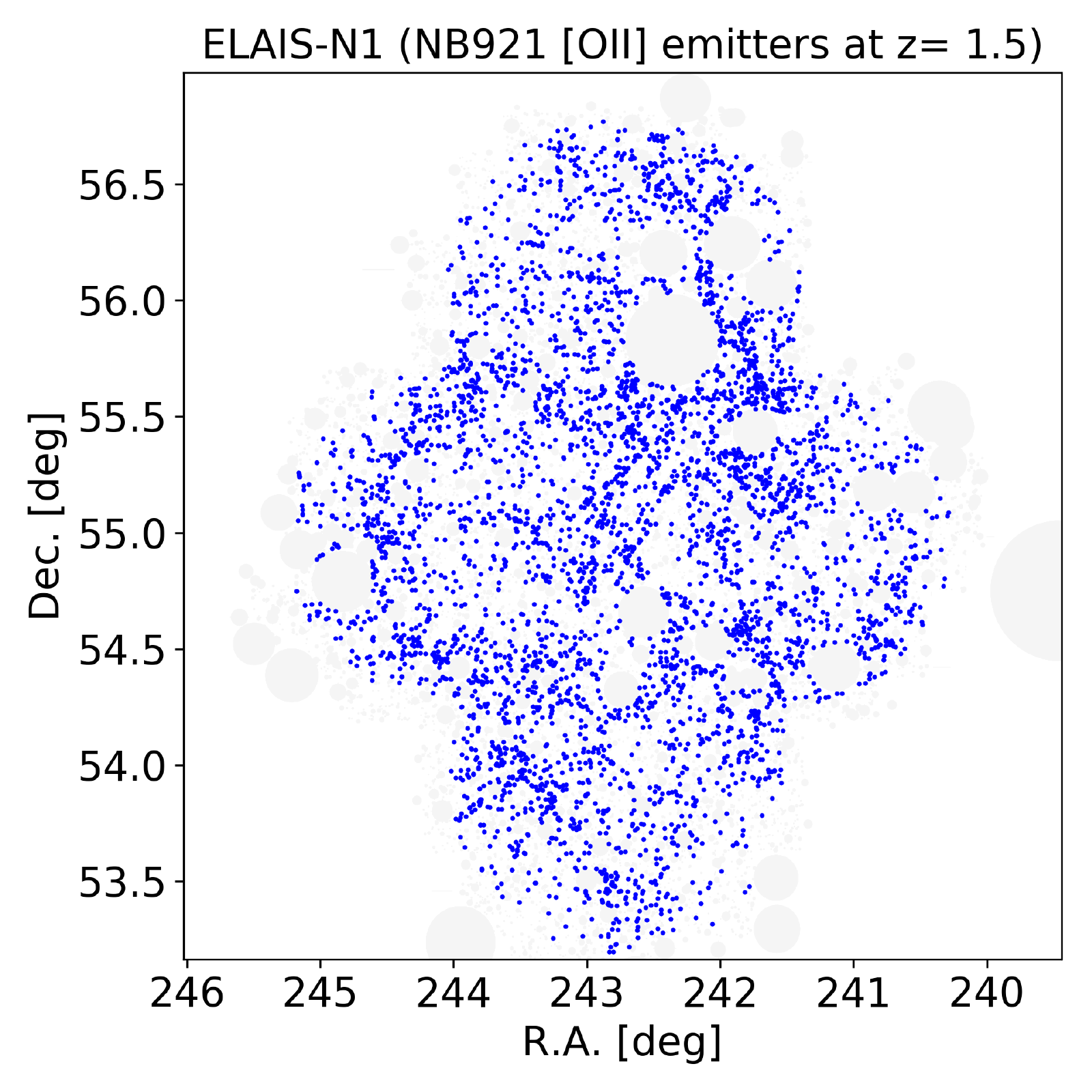} 
   \includegraphics[width=0.32\textwidth, bb=0 0 461 461]{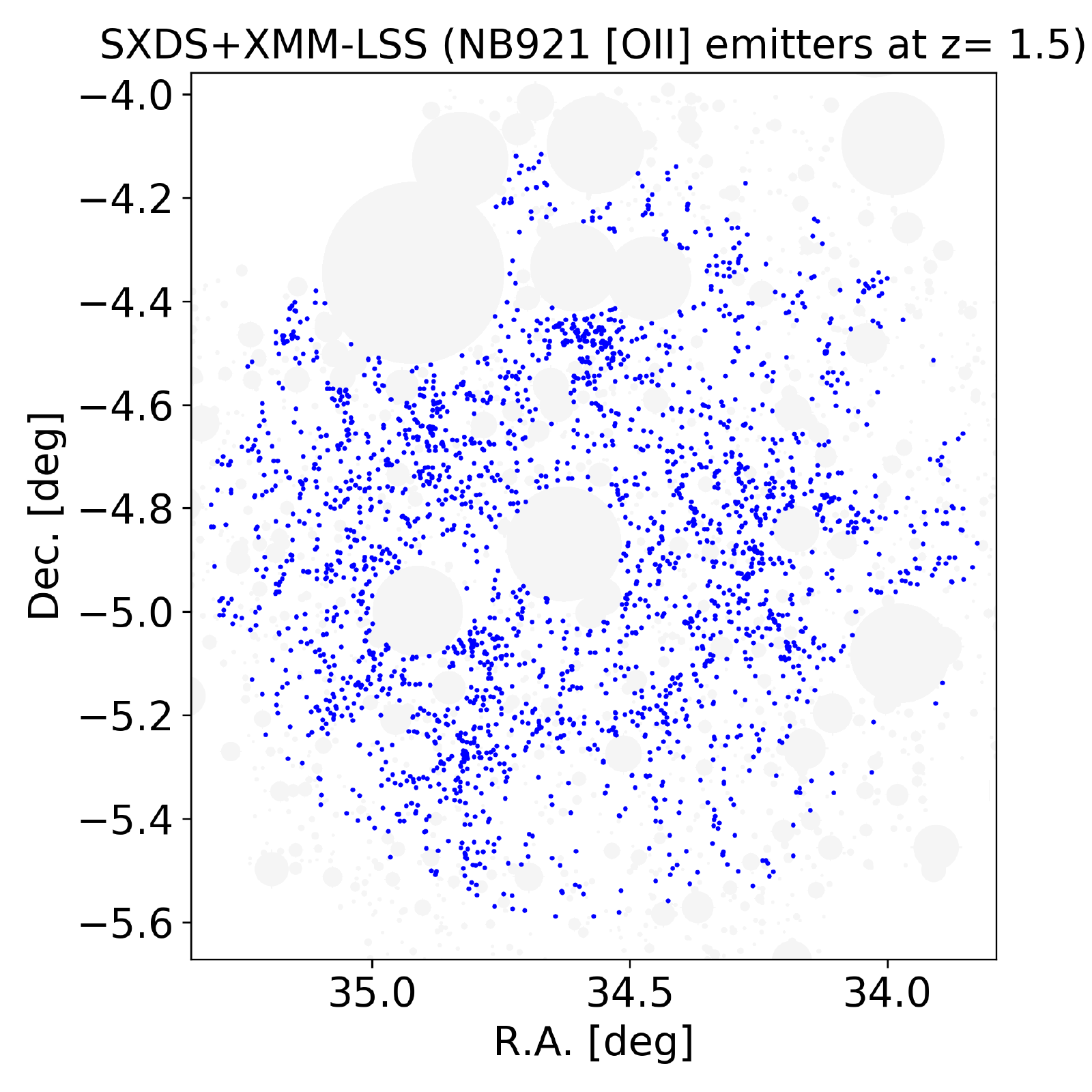} 
   \includegraphics[width=0.32\textwidth, bb=0 0 461 461]{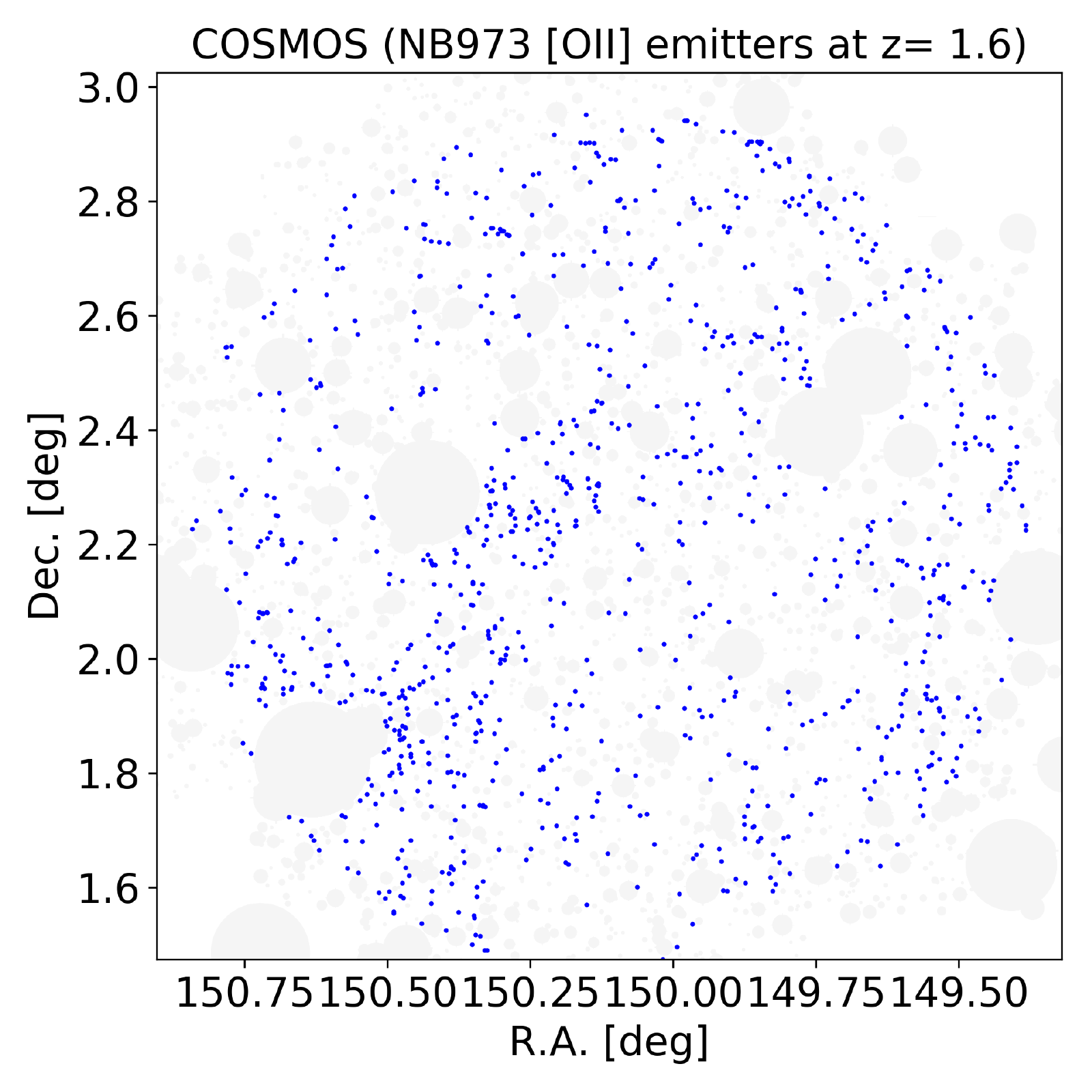} 
 \end{center}
 \caption{The spatial distribution of [OII] emission-line galaxies.}\label{fig:MapO2E}  
\end{figure*}

Figures~\ref{fig:MapHAE} -- \ref{fig:MapO2E} show the spatial
distribution of H$\alpha$, [OIII], and [OII] emission-line galaxies.   

\end{document}

%% file: table1.tex
\begin{table*}
\tbl{The NB filters used in this study. The area shows the effective area that is not masked with the flags we apply and bright object masks (\S~\ref{sec:data.ssp}). The depth is 5$\sigma$ limiting magnitude in the representative 9 patch regions of each field (see table~\ref{tbl:9patches}), which is measured from a standard deviation of 2 arcsec aperture photometry in random sky positions in the image with PSF matched to 1.1 arcsec.}{%
\footnotesize
\begin{tabular}{lcccccccccccccc}
\hline
  &  &  &  & \multicolumn{11}{c}{Deep / UltraDeep layer}  \\\cline{5-15}
\multirow{2}{*}{NB filter} & \multirow{2}{*}{$\lambda_c$} & \multirow{2}{*}{$\Delta\lambda$} & \multirow{2}{*}{AREA\footnotemark[$\dag$]} & \multicolumn{2}{c}{SXDS+XMM-LSS} && \multicolumn{2}{c}{COSMOS} && \multicolumn{2}{c}{ELAIS-N1} && \multicolumn{2}{c}{DEEP2-3} \\\cline{5-6}\cline{8-9}\cline{11-12}\cline{14-15}
  &  &  &  & area & depth\footnotemark[$\ddag$] && area & depth\footnotemark[$\ddag$] && area & depth &&  area & depth \\
  & ($\AA$) & ($\AA$) & (deg$^2$) & (deg$^2$) & (mag) && (deg$^2$) & (mag) && (deg$^2$) & (mag) && (deg$^2$) & (mag)\\
\hline
NB527\footnotemark[$*$] & 5261 & 79 & 1.37 (1.76) & -- & -- && 1.37 (1.76) & 26.32 && -- & -- && -- & -- \\
NB718\footnotemark[$*$] & 7170 & 111 & 1.37 (1.76) & -- & -- && 1.37 (1.76) & 25.61 && -- & -- && -- & -- \\
NB816& 8177 & 113 & 16.28 (21.02) & 5.16 (6.34) & 25.43 && 1.37 (1.76) & 25.58 && 4.79 (6.42) & 24.90 && 4.97 (6.49) & 24.82\\
NB921& 9214 & 135 & 16.79 (22.09) & 1.32 (1.76) & 25.25 && 5.78 (7.50) & 25.39 && 4.79 (6.42) & 24.76 && 4.91 (6.41) & 24.59\\
NB973\footnotemark[$*$] & 9711 & 108 & 1.37 (1.76) & -- & -- && 1.37 (1.76) & 24.63 && -- & -- && -- & -- \\
\hline
\end{tabular}}\label{tbl:NBdataArea}
\begin{tabnote}
\footnotemark[$*$] The data are from CHORUS survey, otherwise the data are from HSC-SSP.\\
\footnotemark[$\dag$] The area in parenthesis shows the area before masking with bright object masks.\\
\footnotemark[$\ddag$] In the SXDS+XMM-LSS and COSMOS fields, the depth is measured in the UD layer.\\
\end{tabnote}
\end{table*}

%% file: table3.tex
\begin{table}
\tbl{Redshifts of emission-line galaxies surveyed with the NB filters.}{%
\footnotesize
\begin{tabular}{cccc}
\hline
 redshift & redshift range & line & NB \\
\hline
0.050 & 0.042 -- 0.058 & [OIII] & NB527 \\
0.092 & 0.084 -- 0.101 & H$\alpha$ & NB718 \\
0.246 & 0.237 -- 0.254 & H$\alpha$ & NB816 \\
0.404 & 0.393 -- 0.414 & H$\alpha$ & NB921 \\
0.411 & 0.400 -- 0.422 & [OII] & NB527 \\
0.432 & 0.421 -- 0.443 & [OIII] & NB718 \\
0.479 & 0.471 -- 0.488 & H$\alpha$ & NB973 \\
0.633 & 0.621 -- 0.644 & [OIII] & NB816 \\
0.840 & 0.826 -- 0.853 & [OIII] & NB921 \\
0.923 & 0.908 -- 0.938 & [OII] & NB718 \\
0.939 & 0.928 -- 0.950 & [OIII] & NB973 \\
1.193 & 1.178 -- 1.208 & [OII] & NB816 \\
1.471 & 1.453 -- 1.489 & [OII] & NB921 \\
1.605 & 1.590 -- 1.619 & [OII] & NB973 \\
\hline
\end{tabular}}\label{tbl:RedshiftSurveyed}
\end{table}

%% file: table5.tex
\begin{table*}
\tbl{Summary of the emission-line galaxies with completeness more than 0.5.}{%
\begin{tabular}{cccccccc}
\hline
 \multirow{2}{*}{line} & \multirow{2}{*}{filter} & \multirow{2}{*}{redshift} &  \multirow{2}{*}{number} & \multicolumn{4}{c}{field} \\
\cline{5-8}
  &  &  &  & SXDS+XMM-LSS & COSMOS & ELAIS-N1 & DEEP2-3 \\
\hline
$\rm [OIII]$ & NB527 & 0.050 & 130 & -- & 130 & -- & -- \\
H$\alpha$ & NB718 & 0.092 & 331 & -- & 331 & -- & -- \\
H$\alpha$ & NB816 & 0.246 & 5400 & 1053 & 436 & 2603 & 1308 \\
H$\alpha$ & NB921 & 0.404 & 8532 & 571 & 2554 & 2479 & 2928 \\
$\rm [OII]$ & NB527 & 0.411 & 729 & -- & 729 & -- & -- \\
$\rm [OIII]$ & NB718 & 0.432 & 1075 & -- & 1075 & -- & -- \\
H$\alpha$ & NB973 & 0.479 & 919 & -- & 919 & -- & -- \\
$\rm [OIII]$ & NB816 & 0.633 & 10323 & 2910 & 748 & 3823 & 2842 \\
$\rm [OIII]$ & NB921 & 0.840 & 14647 & 1678 & 6376 & 2872 & 3721 \\
$\rm [OII]$ & NB718 & 0.923 & 1238 & -- & 1238 & -- & -- \\
$\rm [OIII]$ & NB973 & 0.939 & 1211 & -- & 1211 & -- & -- \\
$\rm [OII]$ & NB816 & 1.193 & 15301 & 3885 & 1666 & 5579 & 4171 \\
$\rm [OII]$ & NB921 & 1.471 & 14586 & 2460 & 4878 & 4512 & 2736 \\
$\rm [OII]$ & NB973 & 1.605 & 955 & -- & 955 & -- & -- \\
\hline
\end{tabular}}\label{tbl:SummaryNBEsample}
\end{table*}

%% file: table6.tex
\begin{table*}
\tbl{Parameters of the Schechter function fitted to the luminosity function.}{%
\footnotesize
\begin{tabular}{ccccccccccc}
\hline
 \multirow{2}{*}{line} & \multirow{2}{*}{redshift}  & \multirow{2}{*}{NB} && \multicolumn{3}{c}{No correction for dust attenuation} &&  \multicolumn{3}{c}{Corrected for dust attenuation} \\
\cline{5-7}\cline{9-11}
  & & && $\log\left(\frac{\phi^{\prime *}}{\rm Mpc^{-3}~dex^{-1}}\right)$ & $\log\left(\frac{L^*}{\rm erg~s^{-1}}\right)$ & $\alpha$ && $\log\left(\frac{\phi^{\prime *}}{\rm Mpc^{-3}~dex^{-1}}\right)$ & $\log\left(\frac{L^*}{\rm erg~s^{-1}}\right)$ & $\alpha$ \\
\hline
O3E & 0.050 & NB527 && -2.73$\pm$0.53 & 40.60$\pm$0.75 & -1.44$\pm$0.10 && -2.73$\pm$0.50 & 40.60$\pm$0.71 & -1.43$\pm$0.09 \\
HAE & 0.092 & NB718 && -2.23$\pm$0.19 & 40.66$\pm$0.19 & -1.22$\pm$0.10 && -2.48$\pm$0.36 & 41.06$\pm$0.51 & -1.29$\pm$0.12 \\
HAE & 0.246 & NB816 && -2.75$\pm$0.04 & 41.27$\pm$0.02 & -1.22$\pm$0.04 && -3.16$\pm$0.07 & 41.89$\pm$0.07 & -1.36$\pm$0.03 \\
HAE & 0.404 & NB921 && -2.65$\pm$0.02 & 41.47$\pm$0.02 & -1.00$\pm$0.03 && -2.91$\pm$0.07 & 41.98$\pm$0.05 & -1.14$\pm$0.05 \\
O2E & 0.411 & NB527 && -3.17$\pm$0.17 & 41.58$\pm$0.13 & -1.44$\pm$0.10 && -3.01$\pm$0.07 & 41.70$\pm$0.06 & -1.26$\pm$0.04 \\
O3E & 0.432 & NB718 && -3.62$\pm$0.28 & 42.04$\pm$0.19 & -1.80$\pm$0.10 && -2.91$\pm$0.20 & 41.72$\pm$0.15 & -1.40$\pm$0.14 \\
HAE & 0.479 & NB973 && -2.77$\pm$0.14 & 41.57$\pm$0.10 & -1.12$\pm$0.17 && -3.29$\pm$0.26 & 42.25$\pm$0.20 & -1.42$\pm$0.17 \\
O3E & 0.633 & NB816 && -3.02$\pm$0.20 & 41.70$\pm$0.12 & -1.34$\pm$0.17 && -3.14$\pm$0.21 & 41.98$\pm$0.12 & -1.39$\pm$0.16 \\
O3E & 0.840 & NB921 && -3.33$\pm$0.29 & 42.15$\pm$0.16 & -1.65$\pm$0.21 && -3.58$\pm$0.29 & 42.55$\pm$0.17 & -1.72$\pm$0.17 \\
O2E & 0.923 & NB718 && -2.79$\pm$0.16 & 41.66$\pm$0.11 & -1.11$\pm$0.22 && -3.14$\pm$0.15 & 42.23$\pm$0.10 & -1.39$\pm$0.13 \\
O3E & 0.939 & NB973 && -3.12$\pm$0.22 & 42.10$\pm$0.16 & -1.33$\pm$0.19 && -3.04$\pm$0.24 & 42.24$\pm$0.18 & -1.23$\pm$0.24 \\
O2E & 1.193 & NB816 && -2.72$\pm$0.17 & 41.84$\pm$0.10 & -1.30$\pm$0.36 && -2.86$\pm$0.10 & 42.39$\pm$0.06 & -0.95$\pm$0.16 \\
O2E & 1.471 & NB921 && -2.79$\pm$0.17 & 42.08$\pm$0.11 & -1.40$\pm$0.42 && -3.06$\pm$0.14 & 42.80$\pm$0.09 & -1.39$\pm$0.25 \\
O2E & 1.605 & NB973 && -3.98$\pm$0.92 & 42.79$\pm$0.53 & -1.98$\pm$0.38 && -3.10$\pm$0.15 & 42.81$\pm$0.12 & -1.04$\pm$0.23 \\
\hline
\end{tabular}}\label{tbl:SchechterParameters}
\end{table*}